\documentclass[aps,pra,twocolumn,notitlepage,showpacs,superscriptaddress,nofootinbib,amsmath,amsthm,amssymb,UTF8,floatfix]{revtex4-2}

\usepackage{bm}
\usepackage{nameref}
\usepackage[utf8]{inputenc}
\usepackage{amsmath,amsfonts,amssymb}
\usepackage{amsthm}
\usepackage{graphicx}
\usepackage{hyperref}
\usepackage{physics}
\usepackage[ruled,vlined]{algorithm2e}
\usepackage{setspace}
\usepackage[normalem]{ulem}
\usepackage{xcolor}

\newcommand{\infiniteSizeLimitEta}{0.88}          %
\newcommand{\infiniteSizeLimitOneOverEta}{1.13}   %
\newcommand{\infRtwo}{6.83 \times 10^{-5}} %

\newcommand{\infiniteSizepFirst}{15}

\newcommand{\infiniteSizeLastpError}{2.2}
\newcommand{\infiniteSizeLastp}{160}

\newcommand{\lastOptp}{80}
\newcommand{\LastN}{30}
\newcommand{\FirstN}{2}

\newcommand{\finiteDepthScaling}{1.3}

\newcommand{\finiteCScaling}{0.72}
\newcommand{\oneOverfiniteCScaling}{1.39}

\hypersetup{colorlinks=true, linkcolor=blue!80!black, urlcolor=blue!80!black, citecolor=blue!80!black}

\makeatletter
\newtheorem*{rep@theorem}{\rep@title}
\newcommand{\newreptheorem}[2]{%
\newenvironment{rep#1}[1]{%
 \def\rep@title{#2 \ref{##1}}%
 \begin{rep@theorem}}%
 {\end{rep@theorem}}}
\makeatother

\newtheorem{definition}{Definition}
\newtheorem{lemma}{Lemma}
\newtheorem{theorem}{Theorem}
\newreptheorem{theorem}{Theorem}
\newtheorem{proposition}{Proposition}
\newreptheorem{proposition}{Proposition}
\newtheorem{corollary}{Corollary}
\newreptheorem{corollary}{Corollary}
\allowdisplaybreaks

\definecolor{darkgreen}{rgb}{0.0, 0.6, 0.0}

\usepackage{xcolor}

\newcommand{\rev}[1]{#1}

\newcommand{\maxcut}{MAXCUT\xspace}

\usepackage{subfiles}
\usepackage{bibunits}
\let\oldaddcontentsline\addcontentsline
\newcommand{\stoptocentries}{\renewcommand{\addcontentsline}[3]{}}
\newcommand{\starttocentries}{\let\addcontentsline\oldaddcontentsline}

\begin{document}

\begin{bibunit}[apsrev4-2]
\stoptocentries

\title{Spin--Boson Mapping of the Quantum Approximate Optimization Algorithm}

\author{Sami Boulebnane}
\thanks{Equal contribution.}
\author{Abid Khan}
\thanks{Equal contribution.}
\author{Minzhao Liu}
\thanks{Email: \href{minzhao.liu@jpmchase.com}{minzhao.liu@jpmchase.com}}
\affiliation{Global Technology Applied Research, JPMorganChase, New York, NY 10001, USA}
\author{Jeffrey~Larson}
\affiliation{Mathematics and Computer Science Division, Argonne National Laboratory, Lemont, IL 60439, USA}
\author{Dylan Herman}
\affiliation{Global Technology Applied Research, JPMorganChase, New York, NY 10001, USA}
\author{Ruslan Shaydulin}
\thanks{Email: \href{ruslan.shaydulin@jpmchase.com}{ruslan.shaydulin@jpmchase.com}}
\author{Marco Pistoia}
\affiliation{Global Technology Applied Research, JPMorganChase, New York, NY 10001, USA}

\date{April 22, 2026} %
\begin{abstract}
The Quantum Approximate Optimization Algorithm (QAOA) achieves monotonically improving performance with circuit depth $p$, yet the study of the high-depth regime has been obstructed by the exponential in $p$ cost of existing exact evaluation techniques.
In this Letter, we prove that, in the infinite-size limit, the depth-$p$ QAOA state for the Sherrington--Kirkpatrick (SK) model converges to the state of a spin coupled to $p$ bosonic modes.
We simulate the spin--boson system using matrix product states and provide numerical evidence that QAOA obtains a $(1-\epsilon)$ approximation to the optimal energy of the SK model with circuit depth $\mathcal{O}(n/\epsilon^{\infiniteSizeLimitOneOverEta})$ in the average case. 
The modest computational cost of our approach allows us to optimize QAOA parameters and observe that QAOA achieves $\varepsilon\lesssim\infiniteSizeLastpError\%$ at $p=\infiniteSizeLastp$ in the infinite-size limit, extending far beyond $p\leq 20$ accessible to prior exact methods.
Our mapping provides a many-body route to study and optimize high-depth QAOA in regimes previously inaccessible to exact evaluation.
\end{abstract}

\maketitle

Optimization problems are ubiquitous in science and industry and are often computationally hard to solve. As a result, there is a growing interest in utilizing quantum computers to potentially speed up the solution of optimization problems. One promising general-purpose optimization algorithm is the Quantum Approximate Optimization Algorithm (QAOA), for which there exists evidence of polynomial or exponential speed up over \emph{specialized} classical algorithms for certain problems~\cite{qaoa_ksat,qaoa_labs,2503.12789,qaoa_near_symmetric_optimization_problems}.

One way to show quantum advantage is by proving that the quantum algorithm scales better than any classical algorithm, as is the case for Grover's algorithm for unstructured search~\cite{Grover1996}. More commonly, classical lower bounds are out of reach and the argument for quantum advantage rests on the lack of efficient classical algorithms, as is the case for quantum advantage in factoring \cite{shor1994algorithms} and solving optimal polynomial intersection with decoded quantum inteferometry (DQI) \cite{jordan2024optimization}. Importantly, even in the absence of analytical bounds on the quantum algorithm's performance, specialized numerical schemes have been developed to obtain guarantees on the performance of the quantum algorithm at large scales beyond brute-force classical simulation, allowing us to benchmark against existing classical algorithms and search for instances of quantum speedup. This type of argument is used for both DQI and QAOA and is a crucial tool in our exploration of quantum advantage \cite{huang2025vast}.

Formalisms for evaluating the performance of QAOA in the infinite-size limit allow comparison against classical algorithms \cite{basso_et_al:LIPIcs.TQC.2022.7}. Since the performance of QAOA improves monotonically with the depth $p$ of the algorithm, a common approach is to evaluate the performance at $p$ as large as possible (e.g. $p=20$ \cite{basso_et_al:LIPIcs.TQC.2022.7}) and compare against classical algorithms. The central challenge with such schemes is the exponential growth of the cost of evaluating the QAOA performance with depth $p$. We overcome this challenge by identifying an equivalence between the infinite-size limit of QAOA state and the state of a spin--boson system, and simulating the latter using matrix product states at modest cost. Our technique evaluates QAOA performance; simulating the QAOA circuit and sampling still require a quantum computer.

The goal of QAOA is to optimize a classical objective function $C(\bm{z})$ over $n$-bit strings $\bm{z} = (z_1,\ldots,z_n)\in \{+1,-1\}^n$. Promoting this objective function to an operator $C$, such that
\begin{equation}
    C\ket{\bm{z}} = C(\bm{z})\ket{\bm{z}},
\end{equation}
where $\ket{\bm z}$ is a computational basis state, QAOA then seeks to find a low-cost-bitstring using a quantum circuit characterized by an integer parameter $p \geq 1$ and generating quantum states of the form:
\begin{equation}
\ket{\bm\gamma, \bm\beta,C} = U(B, \beta_p) U(C, \gamma_p) \cdots U(B, \beta_1) U(C, \gamma_1) \ket{s}.
\end{equation}
Here, $\ket{s}$ represents the uniform superposition over computational basis states, the operator $U(C, \gamma) = e^{-i\gamma C}$ is diagonal in the computational basis, and $U(B, \beta) = e^{-i\beta B}$, where $B = \sum_j X_j$ is the sum of single-qubit Pauli $X$ operators. The angles $\bm\gamma = (\gamma_1, \ldots, \gamma_p)$ and $\bm\beta = (\beta_1, \ldots, \beta_p)$ are classical parameters that define the QAOA state. When the state $\ket{\bm\gamma, \bm\beta, C}$ is measured in the computational basis, it yields a bit string with a cost function value whose expected cost is $\bra{\bm\gamma, \bm\beta, C} C \ket{ \bm\gamma, \bm\beta, C}$. 

In this Letter, we study QAOA applied to the SK model, which is a classical spin system with all-to-all couplings between $n$ spins with the energy given by
\begin{equation}
C^{\text{SK}}(\bm{z}) = \frac{1}{\sqrt{n}}\sum_{j<k} J_{jk}z_{j}z_{k},
\end{equation}
where $J_{jk}\sim\mathcal{N}(0,1)$. The SK model is encoded on qubits by the cost operator $C^{\text{SK}} = \frac{1}{\sqrt{n}}\sum_{j<k} J_{jk}Z_{j}Z_{k},$ where $Z_j$ is a Pauli $Z$ operator on qubit $j$. The SK model was originally proposed to study disordered magnets~\cite{Sherrington1975} but has since become a paradigmatic model for understanding the behavior of complex energy landscapes. The SK model and its generalizations have been used to understand random combinatorial optimization problems~\cite{Gamarnik2021,Mezard2002} and neural networks~\cite{Amit1985,engel2001statistical,nishimori2001statistical}, among other systems~\cite{Villani2009}.

In the infinite-size limit, the lowest energy for typical instances reaches the so-called Parisi value~\cite{PhysRevLett.43.1754,PhysRevE.65.046137,Binder1986}
\begin{equation}
    P_{*} = -\lim_{n\to\infty}\min_{\bm{z}}\frac{C^{\text{SK}}(\bm{z})}{n} =0.763166\ldots\,\,.
\end{equation}
Finding the exact ground state of the SK model is an NP-hard optimization problem in the worst case~\cite{Barahona1982} and in practice, general-purpose heuristics take super-polynomial time~\cite{Aramon2019,Leleu2021,Sankar2024}. There are algorithms that provably find $(1-\epsilon)$-approximations based on Hessian ascent~\cite{jekel2025potential} and message passing~\cite{montanari2019} using time exponential in $1/\epsilon$. The lack of provable exact solvers, combined with its empirical average case hardness, motivates the study of quantum algorithms for the SK model.

\begin{figure}[!ht]
\centering
\includegraphics[width=\linewidth]{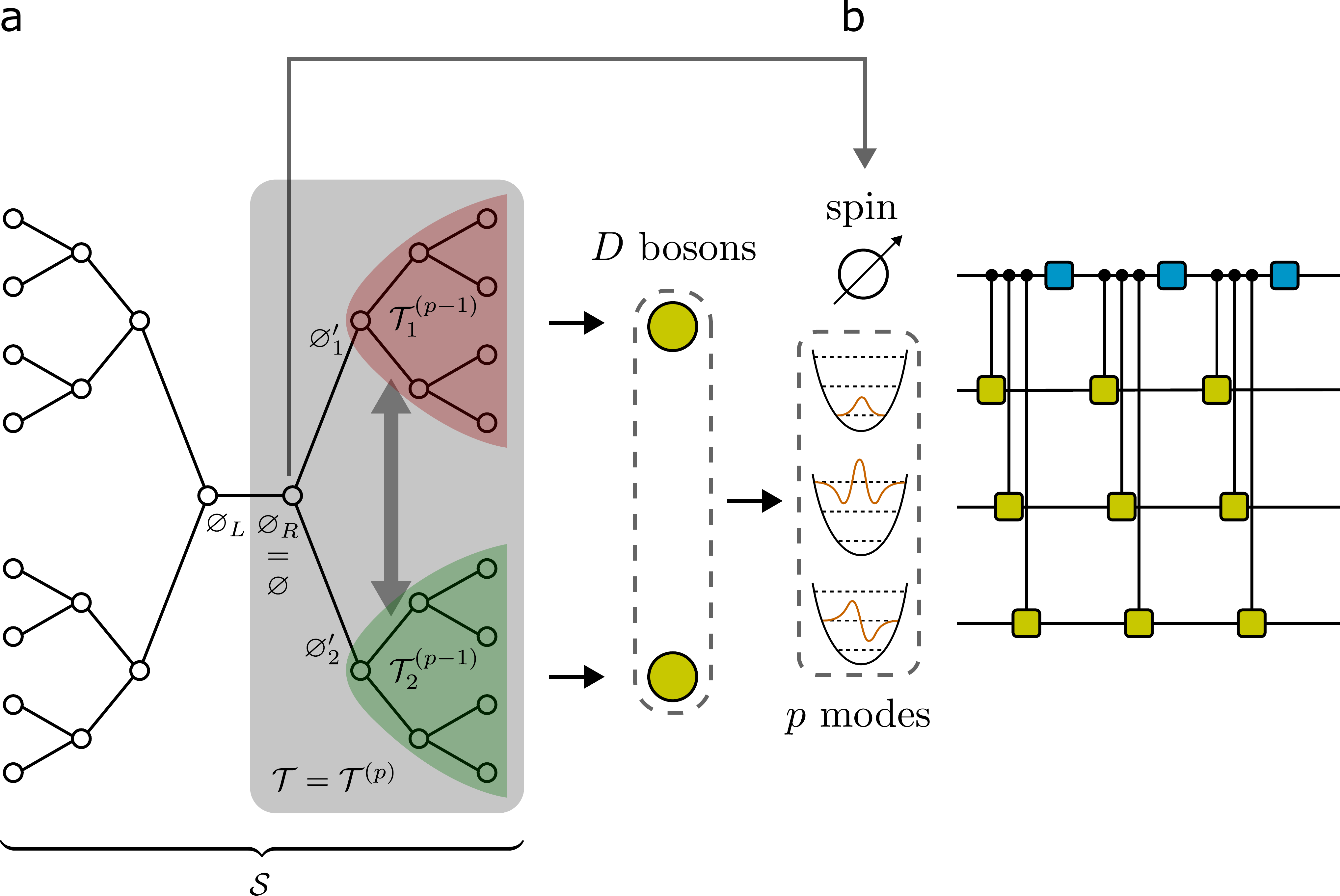}
\caption{
 \textbf{Equivalence between QAOA states.} \textbf{a}, The two-sided tree $\mathcal{S}$ used to compute the SK QAOA energy density. We specifically consider $p=3$ and $D=2$. Each node is a qubit and each edge indicates $ZZ$ interactions. Half of it is the tree $\mathcal{T}$. The QAOA state on $\mathcal{T}$ is symmetric under exchange of subtrees with roots $\varnothing_j'$. \textbf{b}, The evolution required to construct the QAOA state on $\mathcal{T}$ in the $D\rightarrow\infty$ limit. The root qubit $\varnothing$ maps to a spin, and each subtree maps to a boson or vacuum. The bosonic modes are initialized to the vacuum state. The blue gates are $X$ rotations and the yellow gates are spin-controlled bosonic-mode displacements.}
\label{fig:qaoa_state}
\end{figure}

QAOA has been conjectured to efficiently find a $(1-\epsilon)$-approximation to the SK model with depth $p$ that is approximately linear in $1/\epsilon$~\cite{Farhi_2022,basso_et_al:LIPIcs.TQC.2022.7}. Specifically, Refs.~\cite{Farhi_2022,Basso_2022} exactly computed the instance-averaged infinite-size QAOA energy density\begin{equation}\label{eq:nu_general_definition}
    \nu_p(\bm\gamma,\bm\beta) = \lim_{n\rightarrow\infty}\mathbb{E}_J\bra{\bm\gamma, \bm\beta, C^{\text{SK}}} C^{\text{SK}} / n \ket{ \bm\gamma, \bm\beta, C^{\text{SK}}},
\end{equation}
where the expectation is taken over random choices of SK couplings $J$, up to $p=11$ and $p=20$ respectively. The exponential complexity in $p$ prevented Refs.~\cite{Farhi_2022,Basso_2022} from studying higher depths. With the spin--boson mapping introduced in this work (Fig.~\ref{fig:qaoa_state}), we compute $\nu_p$ for much larger $p$, showing convergence of QAOA energy to $P_{*}$.

To establish the spin--boson equivalence, we first consider QAOA state on a tree graph. Below, for \rev{$D$-regular} graph $G=(V,E)$ with edges $E$, we denote the cost Hamiltonian by $C^{G}=\frac{1}{\sqrt{D}}\sum_{(u,v)\in E}Z_u Z_v$ and the mixing Hamiltonian by $B^{G} = \sum_{v\in V} X_v$. 
We choose this problem due to the following result from Ref.~\cite{basso_et_al:LIPIcs.TQC.2022.7}:
\begin{theorem}[Theorem 1 of Ref.~\cite{basso_et_al:LIPIcs.TQC.2022.7}]\label{th:qaoa_maxcut_converges_to_sk}
For all $p$ and all parameters $\bm\gamma, \bm\beta$, the infinite-size disorder-averaged QAOA energy density on the SK model is equal to the $D\rightarrow\infty$ limit of $ZZ$ expectation value for QAOA state on a $D$-ary two-sided tree $\mathcal{S}$ of Fig. \ref{fig:qaoa_state}\textbf{a}:
\begin{equation}
\nu_p(\bm\gamma,\bm\beta)=\lim_{D\rightarrow\infty}\bra{\bm\gamma, \bm\beta, C^{\mathcal{S}}} Z_{L} Z_{R} \ket{ \bm\gamma, \bm\beta, C^{\mathcal{S}}}.
\end{equation}
\end{theorem}

In Ref.~\cite{basso_et_al:LIPIcs.TQC.2022.7}, the tree graph arises from the observation that random regular graphs are locally tree-like with high-probability. Consequently, QAOA energy on sufficiently large random regular graphs is equivalent to the QAOA energy on the tree graph. This locality argument has been extended to show equivalence between QAOA energy on a broad range of sparse constraint satisfaction problems and spin glasses~\cite{Basso_2022}, suggesting a path to generalizing the spin-boson mapping to other problems.

Our main result is that, in the $D\rightarrow\infty$ limit, the QAOA state on the one-sided tree $\mathcal{T}$ (one half of the two-sided tree $\mathcal{S}$) converges (in norm) to a spin--boson state, and the spin’s two-time autocorrelation function coincides with the disorder-averaged QAOA energy density.
We now provide an intuitive picture, with detailed proofs deferred to
the Supplemental Material (Sec.~VI).

The depth-$p$ tree $\mathcal{T}=\mathcal{T}^{(p)}$ consists of the root vertex $\varnothing$ and $D$ copies of depth-$(p-1)$ subtree $\mathcal{T}^{(p-1)}$ with roots $\varnothing'_i$ for $i\in[D]$. 
For the QAOA state $\ket{ \bm\gamma, \bm\beta, C^{\mathcal{T}}}$, permutations of these $D$ subtrees leave the state invariant. This permutation symmetry provides a natural identification of $D$ subtrees with $D$ bosons.

Further, for the $t$-th QAOA layer,
\begin{align}
&U(B^{\mathcal{T}},\beta_t)U(C^{\mathcal{T}},\gamma_t)=e^{-i\beta_t B^{\mathcal{T}}}e^{-i\gamma_t C^{\mathcal{T}}}\\
=&e^{-i\beta_t X_{\varnothing}}\left(e^{-i\beta_t B^{\mathcal{T}^{(p-1)}}}e^{-i\gamma_t Z_{\varnothing}Z_{\varnothing'}/\sqrt{D}}e^{-i\beta_t C^{\mathcal{T}^{(p-1)}}}\right)^{\otimes D}.
\end{align}
Therefore, the root $\varnothing'_i$ of each subtree $\mathcal{T}^{(p-1)}$ interacts with the full tree root $\varnothing$ via $e^{-i\gamma_t Z_{\varnothing}Z_{\varnothing'}/\sqrt{D}}$.

At each QAOA layer $t$ we insert, on the root qubit only, the computational-basis resolution of the identity, $I_{\varnothing}=\sum_{z_{\varnothing}^{[t]}=\pm 1}\lvert z_{\varnothing}^{[t]}\rangle\langle z_{\varnothing}^{[t]}\rvert,$
thereby rewriting the dynamics as a sum over root-spin histories $\boldsymbol{z}_{\varnothing}=(z_{\varnothing}^{[1]},\ldots,z_{\varnothing}^{[p]})$. For a fixed history $\boldsymbol{z}_{\varnothing}$, the coupling between the root and the root $\varnothing'_i$ of subtree $i$ at layer $t$ reduces to
$$
e^{-i\gamma_t z_{\varnothing}^{[t]} Z_{\varnothing'_i}/\sqrt{D}}
= 1 - \frac{i\gamma_t z_{\varnothing}^{[t]}}{\sqrt{D}}\, Z_{\varnothing'_i} + \mathcal{O}(D^{-1}).
$$
Thus each layer contributes either the identity or a single $Z$ insertion on $\varnothing'_i$ with amplitude $\mathcal{O}(D^{-1/2})$. Across $p$ layers, this yields one term with no insertions and amplitude $\mathcal{O}(1)$, and $p$ terms with a single $Z$ insertion at some $t\in[p]$ and amplitude $\mathcal{O}(D^{-1/2})$; terms with two or more insertions vanish as $D\to\infty$ (for fixed $p$).

Denote by $\lvert\Psi^{(p-1)}\rangle$ the subtree state with no $Z$ insertion and by $\lvert\Psi^{(p-1)}_t\rangle$ the subtree state with a single $Z$ insertion at layer $t$. In the large-$D$ expansion above, the subtree state is of the form
\begin{align}
\mathcal{O}(1)\vert\Psi^{(p-1)}\rangle+\sum_{t\in[p]}\mathcal{O}(D^{-1/2})\vert\Psi^{(p-1)}_t\rangle+\mathcal{O}(D^{-1}).\label{eqn:subtree_state}
\end{align}
Identifying $\lvert\Psi^{(p-1)}\rangle$ with the vacuum and $\lvert\Psi^{(p-1)}_t\rangle$ with a single excitation in bosonic mode $t$ yields, for each subtree, a $p$-mode single-particle manifold. Permutation symmetry over subtrees then promotes this to a description in terms of a spin coupled to $p$ bosonic modes. In particular, the state converges to a well-defined limit with a finite number of particles as $D\rightarrow\infty$.
\begin{reptheorem}{th:convergence_of_tree_state_to_spin_boson}[Informal]
    The state $\ket{ \bm\gamma, \bm\beta, C^{\mathcal{T}}}$ strongly converges to a limit in $\mathbb{C}^2 \otimes \mathbf{L}_2\left(\mathbf{R}\right)^{\otimes p}$ as $D \to \infty$:
\begin{align}
   \ket{ \bm\gamma, \bm\beta, C^{\mathcal{T}}} & \xrightarrow[D \to \infty]{} \ket{\Phi^{\left(p\right)}}\\
    \ket{\Phi^{\left(p\right)}} & := \sum_{\substack{z_{\varnothing} \in \{1, -1\}\\\bm{n} \in \mathbf{N}^p}}c\left(z_{\varnothing}, \bm{n}\right)\ket{z_{\varnothing}} \otimes \ket{\bm{n}},
\end{align}
with an appropriate choice of $c$. The summation is over all natural integer tuples $\bm{n} = \left(n_1, \ldots, n_p\right)$, defining the corresponding number states $\ket{\bm{n}}=\ket{n_1, \ldots, n_p}$.
\end{reptheorem}

To see the particle number is finite, we note that the probability that each subtree state is not vacuum (i.e. in state $\vert\Psi^{(p-1)}_t\rangle$) is $\mathcal{O}(D^{-1})$ as is evident from Eq. \ref{eqn:subtree_state}. Having $D$ subtrees means the total particle number is $\mathcal{O}(1)$ for fixed $p$. Therefore, we can truncate the Fock-space dimension to a finite number $d$.

This result establishes the existence of a corresponding bosonic state but does not describe how to obtain it. We now present the procedure to obtain the state, with proof deferred to
the Supplemental Material (Sec.~I). We first initialize the spin to the $\vert+\rangle$ state and the bosonic modes to the vacuum state. Then, for each layer of QAOA, single-mode displacements controlled by the spin are applied to all bosonic modes, followed by a rotation $\exp(-i\beta_t X)$ of the spin state. This is repeated for all $p$ layers. The evolution of the spin--boson state is illustrated in Fig.~\ref{fig:qaoa_state}\textbf{b}.

Evolving this spin--boson system would incur $\mathcal{O}(d^p)$ cost via statevector simulation. 
Fortunately, we find that $d\lesssim10$ is enough to accurately evolve the spin--boson system, as the error vanishes exponentially in $d$ and the $p$ dependence is
moderate (see the Supplemental Material,
Sec.~V). Additionally, we find that it has relatively low entanglement across the system, which can be exploited using the matrix product state (MPS). The success of these truncations may be of independent interest as it sheds light on the fundamental aspects of QAOA behavior in the infinite-size limit.

\begin{figure}
\centering
\includegraphics[width=0.8\linewidth]{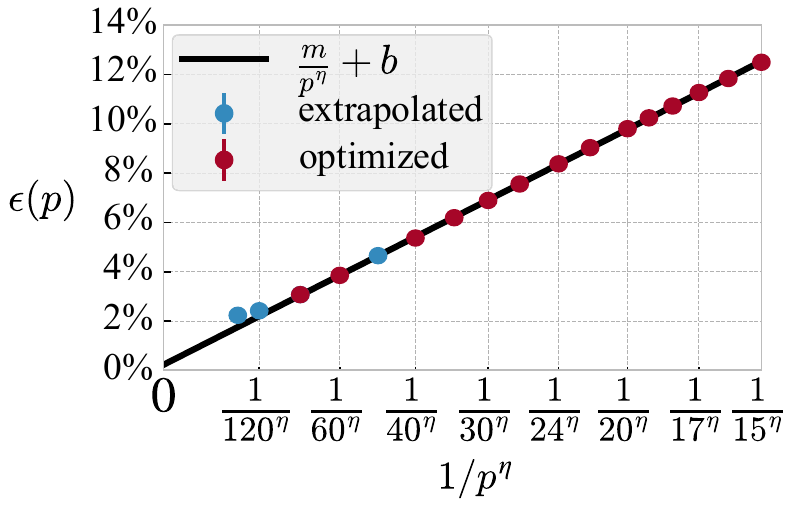}
\caption{Normalized QAOA energy deviation $\epsilon(p)$ from the Parisi value. The QAOA parameters are optimized up to $p=80$. The parameters for $p\in \{120,160\}$ are extrapolated from $p=\lastOptp$ but not optimized. The x-axis is rescaled to $1/p^\eta$, where $\eta\approx\infiniteSizeLimitEta$ is chosen to maximize the $R^2$ value of the linear regression described in Eq.~\ref{eq:regress}.}
\label{fig:parisi}
\end{figure}

\begin{figure*}
    \centering
    \includegraphics[width=\linewidth]{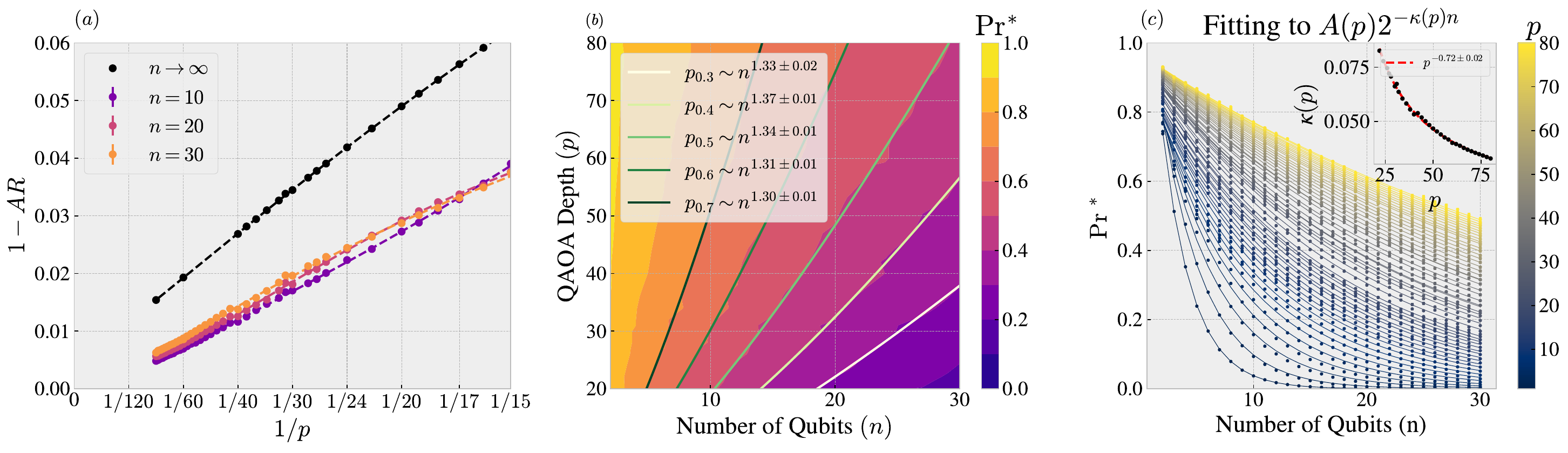}
    \caption{QAOA on finite-sized instances using parameters from the infinite-size limit. (a) Average approximation ratio approaches 1 as QAOA depth $p$ grows regardless of $n$. (b) QAOA depth $p$ required to achieve constant success probability ${\Pr}^*$. (c) Success probability decays exponentially with system size $n$ at fixed $p$. We fit the exponent $\kappa(p)$ vs $p$ in the inset.
    }
    \label{fig:finite}
\end{figure*}

Using our spin--boson mapping, we numerically investigate how the relative error $\epsilon(p)$ approaches zero by fitting the points $(p, \nu_p)$ to the equation
\begin{equation}\label{eq:regress}
    \epsilon(p) = 1 - \frac{\nu_{p}}{P_{*}} = \frac{m}{p^\eta}+b.
\end{equation}
We fit the data in Fig.~\ref{fig:parisi} from $p=\infiniteSizepFirst$ to $p=\lastOptp$ and find
$\eta = 0.88 \pm 0.03,m = 1.34 \pm 0.10, b = 0.002  \pm 0.003$ with a goodness of fit of $1-R^2 = \infRtwo $. The error bars are a 99.7\% confidence interval (up to 3 standard deviations from the mean), obtained by bootstrapping.

Although we cannot rule out an obstruction from approaching the Parisi value arbitrarily, we observe QAOA converging to the Parisi value empirically. Therefore, an interesting direction for future work is to prove that QAOA can achieve arbitrarily small $\epsilon$ with depth $poly(1/\epsilon)$. In the Supplemental Material (Sec.~III), we also fit to the function
$\epsilon(p) = m/(p^\eta + c)$ as in Ref.~\cite{leoTQC2022talk}.

The infinite-size-limit energy is a good predictor of QAOA performance for typical large-size instances due to the concentration over disorder established by Ref.~\cite{Farhi_2022}. Consequently, QAOA with infinite-size-optimized angles should perform well on random finite-size instances. We validate this by generating random SK model instances of sizes ranging from $N = \FirstN$ to $N = \LastN$ ($2{,}000$ instances per size), with results presented in Fig.~\ref{fig:finite}.

The performance on a given instance for a given depth $p$ is measured by the approximation ratio, defined as:
\begin{align}
    \mathrm{AR} & := \frac{E\left(\bm\gamma, \bm\beta\right) - E_{\mathrm{min}}}{E_{\mathrm{max}} - E_{\mathrm{min}}},
\end{align}
with $E\left(\bm\gamma, \bm\beta\right)$ the bitstring-averaged energy produced by QAOA for the instance, and $E_{\mathrm{min}}$ $(E_{\mathrm{max}})$ the minimum (maximum) energy values for the instance. We use the approximation rather than the raw energy because the typical optimal energy of a finite-size SK instance differs from the infinite-size optimal energy (Parisi value). As a result, the raw QAOA energies may be challenging to compare between different sizes. 

The approximation ratio achieved by QAOA as a function of $p$ (for selected instance sizes) is shown in Fig.~\ref{fig:finite}a. The  average  approximation across instances improves with increasing $p$ when using infinite-size optimal parameters with no instance-specific optimization, despite instances being of modest sizes. We plot the predicted approximation ratio in the $n \to \infty$ limit, equal to $1 - \nu_p\left(\bm{\gamma}, \bm{\beta}\right)/P_{*}$, for reference. This shows that using angles optimized in the infinite-size limit is a practical way to run QAOA as an approximate optimizer.

Next, we evaluate the performance of QAOA as an exact solver. In Fig.~\ref{fig:finite}b, we show how QAOA depth $p$ has to grow to achieve a constant probability of obtaining an optimal bitstring ${\Pr}^*$. Our results suggest that with depth $p\propto n^{\varsigma}$ for $\varsigma\approx \finiteDepthScaling$, QAOA achieves a constant success probability ${\Pr}^*$. We remark that there is a small variation in the values of $\varsigma$ obtained for different ${\Pr}^*$; we attribute it to the modest size of our dataset and finite-size effects. For reference, using ${\Pr}^*=0.5$ gives a prediction of the circuit depth to solve the SK model with QAOA exactly of $\approx\infiniteSizeLimitEta\cdot n^{\finiteDepthScaling}$. 

We can alternatively estimate the $p$ scaling by examining the decay of the success probability with $n$ for a fixed $p$. This is shown in Fig.~\ref{fig:finite}c. Results for small $p$ provide evidence for an exponential decay of the success probability with $n$. For each $p$, we fit the success probability to $A\left(p\right)2^{-\kappa\left(p\right)n}$. We observe that the exponent satisfies $\kappa(p)\propto p^{-\finiteCScaling}$. This suggests that the depth $p\propto n^{1/\finiteCScaling}\approx n^{\oneOverfiniteCScaling}$ is sufficient to achieve a constant ${\Pr}^*$, in agreement with Fig.~\ref{fig:finite}b.

\section*{Discussion}

Our results introduce an equivalence between the limiting behavior of QAOA and dynamics of a spin--boson system. This connection opens the door to the application of techniques from nonequilibrium quantum many-body physics to the study of quantum optimization algorithms. The developed equivalence introduces a surprising duality between space ($p$ bosonic modes) and time ($p$ QAOA layers), which merits further investigation. We believe several optimization problems beyond the Sherrington--Kirkpatrick model could benefit from a similar mapping, allowing a study of QAOA at larger depth than allowed by exact evaluation. 
Our mapping is likely to generalize straightforwardly to higher-order spin glasses for which Ref.~\cite{Basso_2022} gives exact analytic formulae similar to those for the Sherrington--Kirkpatrick model considered in this work.
Another direction could be to extend our mapping and state construction procedure to graphs of large but finite degrees. While a bosonic interpretation of the QAOA state exists in this case due to the exchange symmetry, efficient sequential construction of this state might prove more challenging.

The observed empirical performance of QAOA is better than prior results for classical algorithms for SK. First, we are not aware of any classical algorithms that achieve a $(1-\epsilon)$ approximation in $poly(n,1/\epsilon)$ time. It may be possible to extend Refs.~\cite{montanari2019,jekel2025potential} to show a polynomial dependence on $1/\epsilon$. In the Supplemental Material
(Sec.~VIII), we try to improve the complexity. Ref.~\cite{montanari2019} reports the running time of $C(\epsilon)n^2$. We show parallelization improves it by a factor of $n$ and that $C(\epsilon)=\mathcal{O}\left(\frac{e^{\mathcal{O}(1/\epsilon^4)}}{\epsilon^{16}}\right)$, falling short of a polynomial dependence. We note the folklore belief that this complexity could be improved to $\mathcal{O}\left(n/\epsilon^2\right)$ and the lack of detailed empirical investigation which could show better complexity than the provable bounds. Second, we are not aware of any classical algorithms that solve the SK model \emph{exactly} in time $poly(n)$. This appears even more challenging through extensions of Refs.~\cite{montanari2019,jekel2025potential}. We remark that our results on QAOA as an exact solver are obtained by simulating the full QAOA state for small instances; further work is needed to validate that the behavior holds at higher $n$.

While the SK model is known to be NP-hard in the worst case~\cite{Barahona1982}, the average-case hardness of SK is less well-understood. Certain classes of algorithms have been ruled out as approaches for finding the ground state of the SK model in the average case. For example, no stable algorithm can approximately sample from the SK Gibbs measure at small temperatures~\cite{alaoui2024samplingsherringtonkirkpatrickgibbsmeasure}. Computing the partition function of the SK model is \#P-hard~\cite{Gamarnik2021}. Furthermore, the SK model is chaotic under small perturbations in couplings~\cite{chatterjee2009disorderchaosmultiplevalleys}, suggesting that low-degree polynomial (LDP) algorithms should not succeed in solving it. The exact and efficient solution of the SK model by QAOA, however, is not ruled out by these results. For example, QAOA with $p=poly(n)$ is not LDP. We hope that our results suggesting that the SK model is easy to solve in the average case provide valuable insights into the behavior of spin glasses and therefore are interesting outside the field of quantum computation.

\nocite{Lykov2023,zenodo_data}
\putbib[bibliography]

\section*{Competing Interests}
The authors declare no competing interests.

\section*{Data Availability}
All optimized angles for a depth-$p$ QAOA for the SK model in the
infinite-sized limit can be found in
QOKit~\cite{Lykov2023}: \url{https://github.com/jpmorganchase/QOKit}.
The full data presented in this work is available in
Ref.~\cite{zenodo_data}.
Detailed derivations and additional numerical results are provided
in the Supplemental Material.

\section*{Acknowledgments}
The authors thank Leo Zhou, Kunal Marwaha, and James Sud for helpful comments on a draft of the current manuscript and Shouvanik Chakrabarti, Helmut Katzgraber, Sergey Knysh, and Hidetoshi Nishimori for helpful discussion of the SK model.
The authors thank Katherine Klymko for helpful feedback on the numerical experiments and for the support of executions on Perlmutter supercomputer.
The authors thank Shree Hari Sureshbabu, Zichang He, Atithi Acharya, and
Brajesh Gupt for useful discussions of the QAOA parameter optimization and classical simulation of QAOA and the spin--boson system. The authors thank their colleagues at the
Global Technology Applied Research center of JPMorganChase for support and
helpful discussions. 

\section*{Funding}
This material is based upon work supported by the U.S.~Department of Energy, Office of Science, National Quantum Information Science Research Centers.
This research used resources of the Argonne Leadership Computing Facility, a U.S. Department of Energy (DOE) Office of Science user facility at Argonne National Laboratory and is based on research supported by the U.S. DOE Office of Science-Advanced Scientific Computing Research Program, under Contract No. DE-AC02-06CH11357.
This research used resources of the National Energy Research Scientific Computing Center (NERSC), a Department of Energy Office of Science User Facility using NERSC award DDR-ERCAP0034098.

\section*{Author Contributions}
S.B. and R.S.~conceived the project. 
S.B.~conceived and developed the theoretical analysis proving the equivalence between simulating QAOA energy and a spin--boson system.
S.B. implemented a prototype of tensor network simulation.
A.K.~improved the tensor network simulation, scaled it up to supercomputers and conducted the numerical experiments.
M.L.~helped improved the tensor network simulation. 
J.L.~developed and interfaced the numerical optimizer. 
D.H. analyzed the classical message-passing algorithm of Ref.~\cite{montanari2019}.
S.B., A.K, M.L, J.L., D.H., R.S, and M.P. contributed to the technical discussions and writing of the manuscript.

\section*{Disclaimer}
This paper was prepared for informational purposes with contributions from the Global Technology Applied Research center of JPMorganChase. This paper is not a product of the Research Department of JPMorganChase or its affiliates. Neither JPMorganChase nor any of its affiliates makes any explicit or implied representation or warranty and none of them accept any liability in connection with this paper, including, without limitation, with respect to the completeness, accuracy, or reliability of the information contained herein and the potential legal, compliance, tax, or accounting effects thereof. This document is not intended as investment research or investment advice, or as a recommendation, offer, or solicitation for the purchase or sale of any security, financial instrument, financial product or service, or to be used in any way for evaluating the merits of participating in any transaction. The United States Government retains, and by accepting the article for publication, the publisher acknowledges that the United States Government retains, a nonexclusive, paid-up, irrevocable, worldwide license to publish or reproduce the published form of this work, or allow others to do so, for United States Government purposes.

\framebox{\parbox{.90\linewidth}{\scriptsize The submitted manuscript has been
created by UChicago Argonne, LLC, Operator of Argonne National Laboratory
(``Argonne''). Argonne, a U.S.\ Department of Energy Office of Science
laboratory, is operated under Contract No.\ DE-AC02-06CH11357.  The U.S.\
Government retains for itself, and others acting on its behalf, a paid-up
nonexclusive, irrevocable worldwide license in said article to reproduce,
prepare derivative works, distribute copies to the public, and perform publicly
and display publicly, by or on behalf of the Government.  The Department of
Energy will provide public access to these results of federally sponsored
research in accordance with the DOE Public Access Plan
\url{http://energy.gov/downloads/doe-public-access-plan}.}}

\end{bibunit}
\clearpage
\newpage
\starttocentries

\begin{bibunit}[apsrev4-2]

\title{Supplementary Information for: A Spin--Boson Mapping of the Quantum Approximate Optimization Algorithm}

\maketitle

\onecolumngrid
\tableofcontents
\newpage

\section{Computation of QAOA energy in infinite-size limit}
\label{sec:qaoa_energy_computation_with_spin_boson}
\subsection{Prior work}
Ref.~\cite{basso_et_al:LIPIcs.TQC.2022.7} proposed a classical method of evaluating the QAOA energy of \maxcut on high-girth $D$-regular graphs and the SK model. The work further showed these two energies are equal in the infinite degree limit ($D \to \infty$) of \maxcut and the infinite-size limit ($n\to\infty$) of SK, for appropriate normalizations of the problem Hamiltonians. In the following, we will refer to this common energy as the SK-QAOA energy, or simply the QAOA energy. Specifically, denote the average-case infinite-size-limit energy of a $p$-layer QAOA by
\begin{equation}\label{eq:nu_general_definition_SM}
    \nu_p(\bm\gamma,\bm\beta) = \lim_{n\rightarrow\infty}\mathbb{E}_J\bra{\bm\gamma, \bm\beta} C / n \ket{ \bm\gamma, \bm\beta},
\end{equation}
where the expectation is taken over the random choice of couplings $J$.

In the formalism of \cite{basso_et_al:LIPIcs.TQC.2022.7}, $\nu_p(\bm{\gamma},\bm{\beta})$ can be computed elementarily from a matrix $G^{(p)}\in\mathbb{C}^{(2p+1)\times (2p+1)}$ as
\begin{equation}\label{eq:nu1}
    \nu_p(\bm{\gamma},\bm{\beta}) = \frac{i}{2}\sum_{j=-p}^{p}\Gamma_j\left(G^{(p)}_{0,j}\right)^2.
\end{equation}
The matrix $G^{(p)}$ has the following index order
\[
\{1,\ldots,p,0,-p,\ldots,-1\},
\]
and is defined recursively through matrices $G^{(m)}\in\mathbb{C}^{(2p+1)\times (2p+1)}$, for $0\leq m \leq p$. This relationship is constructed as follows. First, a $(2p+1)$-dimensional vector $\bm{a}$ with entries $a_{i}\in\{+1,-1\}$ is introduced with the index order
\begin{equation}
    \bm{a} = (a_1,a_2, \ldots,a_p,a_0,a_{-p}\,\ldots,a_{-2},a_{-1}).
\end{equation}
We will also refer to this vector as a bitstring, where bits are understood to assume values in $\{1, -1\}$. Then the recursion relation is
\begin{align}
G^{(0)}_{j,k} &= \sum_{\bm{a}}f(\bm{a})a_{j}a_{k}, \label{eq:G0} \\
G^{(m)}_{j,k} &= \sum_{\bm{a}}f(\bm{a})a_{j}a_{k}\exp \left( -\frac{1}{2}\sum_{j',k'=-p}^{p}G_{j',k'}^{(m-1)}\Gamma_{j'}\Gamma_{k'}a_{j'}a_{k'} \right),\label{eq:Gm}
\end{align}
where
\begin{equation}
    \Gamma_j = 
    \begin{cases} 
        \gamma_j, & \text{if } j > 0 \\
        -\gamma_{-j}, & \text{if } j < 0 \\
        0, & \text{if } j = 0
    \end{cases},
\end{equation}
and
\begin{align}
     f\left(\bm{a}\right) & =  \frac{1}{2}\left(\prod_{l = 1}^{p - 1}\bra{a_l}e^{i\beta_lX}\ket{a_{l + 1}}\right)\bra{a_p}e^{i\beta_pX}\ket{a_0}\bra{a_0}e^{-i\beta_pX}\ket{a_{-p}}\left(\prod_{l = 1}^{p - 1}\bra{a_{-l - 1}}e^{-i\beta_lX}\ket{a_{-l}}\right).\label{eq:f_tensor_definition}
\end{align}
More explicitly, the $2 \times 2$ matrix elements occurring in $f\left(\bm{a}\right)$ read:
\begin{align}
    \bra{a}e^{i\beta X}\ket{b} = \left\{\begin{array}{cc}
         \cos\beta & \textrm{if } a = b\\
         i\sin\beta & \textrm{if } a \neq b
    \end{array}\right. \qquad \forall a, b \in \{1, -1\}.
\end{align}

Importantly, the matrices $G^{(m)}$ satisfy the following properties. First, the elements share the following symmetries:
\begin{align}\label{eq:Gsim}
    G_{j,k}^{(m)} &= G_{k,j}^{(m)} &  G_{0,r}^{(m)} &= G_{0,-r}^{(m)*} \nonumber \\
    G_{j,j}^{(m)} &= G_{j,-j}^{(m)}=1 &  G_{r,s}^{(m)} &= G_{r,-s}^{(m)}= G_{-r,-s}^{(m)*}= G_{-r,s}^{(m)*},
\end{align}
where the convention $1\leq r < s \leq p$ and $j,k\in\{1,\ldots,p,0,-p,\ldots,-1\}$ is used. Because of these symmetry relations, the whole $(2p+1)\times(2p+1)$ matrix can be fully described by any $(p+1)\times(p+1)$ corner of the matrix. Second, for $1\leq r < s \leq m$, $G^{(m)}_{r,s}$ only depends on $G^{(m - 1)}_{r',s'}$ where $1\leq r' < s' \leq m - 1$ (See Appendix A.4 in Ref.~\cite{basso_et_al:LIPIcs.TQC.2022.7}). 

Eqs.~\ref{eq:G0}-\ref{eq:Gm} form an iterative procedure that produces $G^{(p)}$, and hence $\nu_{p}(\bm\gamma,\bm\beta)$, when applied iteratively $p$ times to initial matrix $G^{(0)}$. This $G$-iteration is a finite sum over bitstrings, hence can be evaluated exactly (up to arithmetic error) in a finite number of steps. However, the number of terms in the sum scales as $\mathcal{O}(4^p)$, making the bruteforce summation approach impractical at large $p$. The current work introduces a mapping between this iterative procedure and a spin-boson system, reducing the cost of evaluating $\nu_{p}(\bm\gamma,\bm\beta)$.

\subsection{The spin-boson mapping}
\label{sec:spin_boson_mapping}

Our main theoretical result is an equivalence between the task of computing the SK-QAOA energy in the infinite-size limit averaged over random couplings $J$ and simulating a spin-1/2 particle coupled to $p$ bosonic modes. The central insight behind this result is that Eq.~\ref{eq:Gm} can be interpreted as a two-point correlation function of said spin-boson system. The partition function has a path integral measure
\begin{equation}
  \int  \mathcal{D}[\bm a] = \sum_{\bm a}
\end{equation}
and Feynman weight specified by an implicitly specified action $S$:
\begin{equation}
    e^{iS^{(m - 1)}[\bm a]/\hbar}=f\left(\bm{a}\right)\exp\left(-\frac{1}{2}\sum_{j', k' = -p}^pG^{(m - 1)}_{j', k'}\Gamma_{j'}\Gamma_{k'}a_{j'}a_{k'}\right).\label{eq:path_integral_measure_generalized}
\end{equation}

It was shown in Ref.~\cite{basso_et_al:LIPIcs.TQC.2022.7} that the partition function 
\begin{align}
    \mathcal{Z}&=\int  \mathcal{D}[\bm a]e^{iS^{(m - 1)}[\bm a]/\hbar}\\
    &=\sum_{\bm a}f\left(\bm{a}\right)\exp\left(-\frac{1}{2}\sum_{j', k' = -p}^pG^{(m - 1)}_{j', k'}\Gamma_{j'}\Gamma_{k'}a_{j'}a_{k'}\right) \\
    &= 1.
\end{align}
Then the entries $G^{(m)}_{j, k}$ are precisely the two-point correlation function
\begin{align}
    G^{(m)}_{j, k} &= C(a_j,a_k) = \frac{1}{\mathcal{Z}}\int  \mathcal{D}[\bm a]a_ja_ke^{iS^{(m - 1)}[\bm a]/\hbar}\label{eq:Gm_as_path_integral}
\end{align}

We now describe the construction of the spin-boson system. More specifically, we construct a spin-boson system where the spin's path integral measure is
\begin{align}
    f\left(\bm{a}\right)\exp\left(-\frac{1}{2}\sum_{j', k' = -p}^pG_{j', k'}\Gamma_{j'}\Gamma_{k'}a_{j'}a_{k'}\right),
\end{align}
where $G$ is an arbitrary complex symmetric matrix $\in \mathbb{C}^{(p + 1) \times (p + 1)}$. This construction can then be applied to iteration $m$ of SK-QAOA by letting $G := G^{(m - 1)}$. The Hilbert space of the physical system is given by:
\begin{align}
    \mathcal{H} = \mathbb{C}^2 \otimes \mathcal{F}_1\otimes\cdots\otimes \mathcal{F}_{p},
\end{align}
where $\mathcal{F}_i$ represents the infinite-dimensional Fock-space of the $i$-th bosonic mode. The states we construct will be obtained by applying unitary operations to the initial state
\begin{equation}\label{eq:psi0}
    \ket{+}\ket{\overline{0}_p}\equiv\ket{+}_{\mathbb{C}^2}\otimes\ket{\overline{0}}_{\mathcal{F}_1}\otimes\cdots\ket{\overline{0}}_{\mathcal{F}_{p}},
\end{equation}
where the bar is present to distinguish the bosonic zero mode from the qubit computational zero state. The evolution of the spin-bosons system is specified by $p$ unitary layers $U_1,\,\ldots,\,U_p$ applied to the initial state, Eq.~\ref{eq:psi0}:
\begin{align}\label{eq:psipure}
    \ket{\Psi} & = U_p \cdots U_1\ket{+}\ket{\overline{0}_{p}}.%
\end{align}
The unitary $U_{t}$ consists of two operators. The first is a product of displacement operators applied to each of the bosonic modes.  Given a vector of displacements $\bm{\alpha} = (\alpha_1,\ldots,\alpha_p)$, a multimode displacement operator has the form
\begin{equation}
    D\left(\bm{\alpha}\right)  = D_1(\alpha_1)\otimes\cdots\otimes D_{p}(\alpha_{p}),
\end{equation}
where, $D_{i}$ is the displacement operator on the $i$-th boson:
\begin{equation}\label{eq:Di}
    D_i(\alpha) = \exp\left({\alpha \hat{c}^\dagger_{i} - \alpha^* \hat{c}_i}\right).
\end{equation}
Here $\hat{c}_i$ denotes the annihilation operator of bosonic mode $i$. The multimode displacement operator term in $U_t$ is
\begin{equation}
    K(\bm\alpha) = \ket{0}\bra{0}\otimes D(\bm{\alpha}) + \ket{1}\bra{1}\otimes D(-\bm{\alpha}).
\end{equation}
In other words, the direction of the displacement vector is controlled by the spin. The second operator in $U_j$ is an $X$ rotation that acts only on the spin component. The final form of $U_t$ is
\begin{equation}\label{eq:Unit}
        U_t= \exp\left(-i\beta_tX\right)\,K\left(-i\gamma_t \bm{L_t}\right),
\end{equation}
where $\bm{L_t} = (L_{1,t},\ldots,L_{p,t})$ in the operator $K$ is the $t$-th column of the square matrix $L$ defined as a factorization of the $(p+1)$-dimensional Hermitian corner $G^{\mathrm{herm}}$ of $G$:
\begin{align}
    G^{\mathrm{herm}} & = L^{\dagger}L,\label{eq:Lfac}
\end{align}
where
\begin{align}
    G^{\mathrm{herm}}_{r, s} & = G_{r, -s}\quad\text{for } 0\leq r,s\leq p.
\end{align}
The Hermiticity of $G^{\mathrm{herm}}$ follows from the symmetries of SK-QAOA matrices $G^{(m)}$, which we assumed $G$ to satisfy. For Equation \ref{eq:Lfac} to make sense, we also need $G^\text{herm}$ to be positive semidefinite, which we will show below for the sequence of matrices $G^{(m)}$ arising from the SK-QAOA iteration.
In the statement of our main theoretical result (Proposition~\ref{spinBosonsStateDotProductPathIntegral}), the following variants of state \eqref{eq:psil}, with a single $Z$ operator inserted between two given consecutive unitaries, will play an important role. For $1 \leq l \leq p$, we define
\begin{align}\label{eq:psil}
    \ket{\Psi_l} &= U_p \cdots U_{l}\,Z\,  U_{l-1}\cdots U_1\ket{+}\ket{\overline{0}_{p}},%
\end{align}
and for index $0$, we let
\begin{align}
\label{eq:psi0z}
\ket{\Psi_0} &= Z\,U_p \cdots U_1\ket{+}\ket{\overline{0}_{p}},\nonumber \\
&=Z\ket{\Psi}
\end{align}

Given this system, the following result holds:

\begin{proposition}[Path integral representation of spin-bosons states overlap]
\label{spinBosonsStateDotProductPathIntegral}
Consider states $\ket{\Psi_l}$, $l \in \{1,\ldots,p,0\}$, defined in Eqs.~\ref{eq:psil}, \ref{eq:psi0z}. The dot products between any such two states can be expressed as
\begin{align}
    \left\langle \Psi_k|\Psi_j\right\rangle & = \sum_{\bm{a} \in \{1, -1\}^{2p + 1}}a_ka_{-j}f\left(\bm{a}\right)H\left(\bm{a}\right),\label{eq:spin_bosons_states_dot_product_main_text}
\end{align}
where
\begin{align}
    H\left(\bm{a}\right) & := \exp\left(-\frac{1}{2}\sum_{j', k' =-p}^pG_{j', k'}\Gamma_{j'}\Gamma_{k'}a_{j'}a_{k'}\right),
\end{align}
Hence, quasiprobability measure $f\left(\bm{a}\right)H\left(\bm{a}\right)$ can be regarded as the path integral measure generating the $Z$ time correlations of the spin in the spin-bosons system we constructed.
\end{proposition}

The proof of this Proposition is given in Appendix \ref{sec:sk_qaoa_iteration_spin_bosons}. Applying this Proposition recursively to the iteration \ref{eq:Gm}, we obtain the following Corollary used to compute QAOA energy:

\begin{corollary}[Evaluating the SK-QAOA iteration from correlation functions of a spin-bosons system]
\label{cor:evaluating_sk_qaoa_iteration_spin_bosons_system} The SK-QAOA energy in the infinite-size limit averaged over random choice of couplings $J$ (Eq.~\ref{eq:nu_general_definition_SM}) can be computed from overlaps of spin-boson states as follows
    \begin{equation}\label{eq:nu2}
         \nu_p(\bm{\gamma},\bm{\beta}) = \Im\left[\sum_{r=1}^{p}\gamma_r\Big(\braket{\Psi_0}{\Psi_r}\Big)^2\right].
    \end{equation}
\end{corollary}

A consequence of Proposition~\ref{spinBosonsStateDotProductPathIntegral} is that the quantity defined in Eq.~\ref{eq:spin_bosons_states_dot_product_main_text}:
\begin{align}
    \widetilde{G}^{\mathrm{herm}}_{j, k} & := \sum_{\bm{a} \in \{1, -1\}^{2p + 1}}a_ja_{-k}f\left(\bm{a}\right)H\left(\bm{a}\right)
\end{align}
is a Gram matrix, and hence, positive-semidefinite. It follows that $G^{\mathrm{herm}} \succeq 0$ automatically implies $\widetilde{G}^{\mathrm{herm}} \succeq 0$; applying this inductively to the SK-QAOA iterates $G^{(m)}$, rewriting initial iteration Eq.~\ref{eq:G0} as
\begin{align}
    G^{(0)}_{j, k} & = \sum_{\bm{a} \in \{1, -1\}^{2p + 1}}a_ja_kf\left(\bm{a}\right)H^{(-1)}\left(\bm{a}\right),\\
    H^{(-1)}\left(\bm{a}\right) & := \exp\left(-\frac{1}{2}\sum_{j', k' = -p}^pG^{(-1)}_{j', k'}\Gamma_{j'}\Gamma_{k'}a_{j'}a_{k'}\right)\nonumber\\
    G^{(-1)}_{j, k} & := 0
\end{align}
where evidently the Hermitian corner of $G^{(-1)}$ is nonnegative, we obtain that all iterates $G^{(m)}$ have nonnegative Hermitian corner. An earlier proof that all SK-QAOA iterates have non-negative Hermitian corner was provided in Ref.~\cite{leo2022Autocorrelation}, but using the interpretation of $G^{(m)}$ as the time autocorrelations of a Pauli $Z$ operator in $D$-regular, high-girth \maxcut QAOA as $D \to \infty$. In contrast, in the present framework, the non-negativity of $\widetilde{G}^{\mathrm{herm}}$ follows automatically from that of $G^{\mathrm{herm}}$, even if $G$ may not be realized as the time autocorrelation function of a QAOA circuit.

\subsection{Procedure}

We now describe the technique for computing the QAOA energy $\widetilde{\nu}_p(\bm\gamma, \bm\beta;d, \chi_\delta )$ as described in Eq.~\ref{eq:nu2}. 
We first note that this computation relies on evaluating the $(p+1)$-dimensional Hermitian matrix $G^\text{herm}$, the entries of which can be calculated as
\begin{align}
    G^\text{herm}_{j,k} & = \braket{\Psi_j}{\Psi_k}\nonumber\\
    & = \bra{+}\bra{\overline{0}_{p}}
    \Bigg(U_1^{\dagger}\cdots U_{j-1}^\dagger\Bigg)
    \,Z\,
    \Bigg(U_j^{\dagger}\cdots U_p^\dagger\Bigg)
    \Bigg( U_p\cdots U_k\Bigg)
    \,Z\,
    \Bigg(U_{k-1}\cdots U_1\Bigg)
    \ket{+}\ket{\overline{0}_{p}}\nonumber\\
    & = \bra{+}\bra{\overline{0}_{p}}
    \Bigg(U_1^{\dagger}\cdots U_{j-1}^\dagger\Bigg)
    \,Z\,
    \Bigg(U_{j - 1}\cdots U_k\Bigg)
    \,Z\,
    \Bigg(U_{k-1}\cdots U_1\Bigg)
    \ket{+}\ket{\overline{0}_{p}}\nonumber\\
\widetilde{G}^{\text{herm}}_{0,k} & = \braket{\Psi_0}{\Psi_k}\nonumber\\
 &= \bra{+}\bra{\overline{0}_{p}}
    \Bigg(U_1^{\dagger}\cdots U_{p}^\dagger\Bigg)
    \,Z\,
    \Bigg( U_p\cdots U_k\Bigg)
    \,Z\,
    \Bigg(U_{k-1}\cdots U_1\Bigg)
    \ket{+}\ket{\overline{0}_{p}}\label{eq:olap2}
\end{align}
Note that the elements $G^\text{herm}_{j,k}$ only depend on $j-1$ unitary layers. If we define the following states 
\begin{align}
\ket{\Psi^{(k)}} &= U_{k-1}\cdots U_1\ket{+}\ket{\overline{0}_p},\label{eq:psik}\\
\ket{\Psi_{j}^{(k)}} &= U_{k-1}\cdots U_{j}\,Z\,U_{j-1}\cdots U_1\ket{+}\ket{\overline{0}_p},\label{eq:psikj}
\end{align}
where the superscript in $\ket{\Psi}$ corresponds to the number of unitary layers applied, 
and the subscript corresponds to the location of the Pauli-$Z$ operator, then $G^\text{herm}_{j,k}$ for $1\leq k\leq j\leq p$ simplifies to
\begin{align}
    G^\text{herm}_{j,k} &= \braket{\Psi_{j}^{(j)}}{\Psi_{k}^{(j)}} \\
    G^\text{herm}_{0,k} &= \braket{\Psi_{p+1}^{(p+1)}}{\Psi_{k}^{(p+1)}}
\end{align}

We can construct $G^\text{herm}_{j,k}$ by iteratively evolving the states described in Eqs.~\ref{eq:psik} and \ref{eq:psikj}, which obey the recursion relations (assuming $1\leq k \leq j$)
\begin{align}
    \ket{\Psi^{(k)}} & = U_{k-1}\ket{\Psi^{(k-1)}}, \label{eq:prop1}\\
    \ket{\Psi^{(k)}_k} & = Z\ket{\Psi^{(k)}},\label{eq:prop2}\\
    \ket{\Psi^{(k)}_j} & = U_{k-1}\ket{\Psi_j^{(k-1)}} \label{eq:prop3}.
\end{align}
Suppose we have calculated $G^\text{herm}_{r,s}$ for $1\leq r,s \leq t$ using states 
\[\left\{\ket{\Psi^{(k)}},\, \ket{\Psi_{j}^{(k)}}\right\}_{1\leq j \leq k \leq t}.\] 
Then using Eqs.~\ref{eq:prop1}-\ref{eq:prop3}, we can use these computed states to prepare $G^\text{herm}_{t+1,s}$:
\begin{align}
G^\text{herm}_{t+1,s} &= \braket{\Psi_{t+1}^{(t+1)}}{\Psi_{s}^{(t+1)}} \label{eq:ghermolap}\\
&= \bra{\Psi^{(t)}}U^\dagger_t \,Z\,U_t\ket{\Psi_{s}^{(t)}}
\end{align}
Overall, we can iteratively grow the Hermitian matrix $G^\text{herm}$ from size $t$ to size $(t+1)$ from the states  $\ket{\Psi^{(t)}},\,\ket{\Psi_{1}^{(t)}},\ldots,\ket{\Psi_t^{(t)}}$, and operators $Z$ and $U_t$, producing a new set of states $\ket{\Psi^{(t+1)}},\,\ket{\Psi_{1}^{(t+1)}},\ldots,\ket{\Psi_t^{(t+1)}}$ for the next iteration. To obtain $U_t$, we require $\bm L_t$, the $t$-th column of the matrix $L$ obtained from the decomposition $G^\text{herm}=L^\dagger L$. At iteration $t$, where $G$ is size $t$, we perform the Cholesky decomposition to get a size $t$ matrix $L$, from which we get the $p$-dimensional vector $\bm L_t$ by padding the $t-$th column of $L$ with $p-t$ zeros.  

\subsection{The MPS algorithm}
Obtaining the QAOA energy in Eq.~\ref{eq:nu2} requires numerically computing and evolving the states in the form of Eq.~\ref{eq:psil}. This may appear challenging to compute given that the quantum states representing $p$ bosonic modes are \emph{infinite}-dimensional in the Fock basis. In practice, we find that using a small finite Fock dimension $d$ is sufficient to obtain energy values with high precision. In other words, we only consider the modes $\ket{\overline{0}},\ldots\ket{\overline{d-1}}$ for each boson. This truncates the displacement operator $D_i(\alpha)$ in Eq.~\ref{eq:Di} into a $d$-dimensional unitary. This finite-dimensional approximation makes the state $\ket{\Psi_l}$ representable as an MPS with $p$ boson sites of physical dimension $d$ and one spin site of physical dimension 2. In practice, we represent Eq.~\ref{eq:psil} as two MPS $\ket{\psi^{(l)}_{0}}$ and  $\ket{\psi^{(l)}_{1}}$ with $p$ boson sites, where
\begin{align}
    \ket{\psi^{(l)}_{0}} &= \Big(\bra{0}\otimes I^{\otimes p}\Big) \ket{\Psi_l} \\ 
    \ket{\psi^{(l)}_{1}} &= \Big(\bra{1}\otimes I^{\otimes p}\Big) \ket{\Psi_l}.
\end{align}
The states $\ket{\psi^{(l)}_{0}}$ and  $\ket{\psi^{(l)}_{1}}$ are projections of $\ket{\Psi_l}$ into the spin states $\ket{0}$ and $\ket{1}$, respectively, and live in the Hilbert space $\mathcal{F}_{1}\otimes\cdots\otimes\mathcal{F}_{p}$.

The most computationally costly part of the simulation is applying the QAOA layer operator $U_{t}$ to these states. The initial states correspond to a product state given by Eq.~\ref{eq:psi0} and have bond dimension $\chi=1$. 
The application of $K(\bm\alpha)$ is straightforward:
\begin{align}
    \ket{\psi_{0}} &\to D(-i\gamma_t\bm{L_t})\, \ket{\psi_{0}}\\
    \ket{\psi_{1}} &\to D(i\gamma_t\bm{L_t})\, \ket{\psi_{1}}.
\end{align}
This operation applies a product of single-site displacement operators $D_i$ to each site, thus making them local and keeping the MPS bond dimension unchanged. On the other hand, the mixing operator will grow the bond dimensions of the MPS. It transforms the MPS as
\begin{align}
    \ket{\psi_{0}} &\to \cos\beta_t \ket{\psi_{0}} -i\sin\beta_t \ket{\psi_{1}}\label{eq:mps_addition_psi0}\\
    \ket{\psi_{1}} &\to -i\sin\beta_t \ket{\psi_{0}} + \cos\beta_t \ket{\psi_{1}}\label{eq:mps_addition_psi1}.
\end{align}

The bond dimension of the transformed space is upper-bounded by the sum of the bond dimensions of the input state, so the bond dimension $\chi$ may grow exponentially with the number of applied unitaries $p$. However, we control the growth of the bond dimension by removing the singular values below a fixed cutoff. 

The overall procedure to compute the energy $\widetilde{\nu}_p(\bm\gamma,\bm\beta)$ is outlined in Alg.~\ref{alg:nu}, which describes the iterative procedure in growing the matrix $G^\text{herm}$. This method requires storing, evolving, and computing overlaps of spin-boson states which is detailed in Alg.~\ref{alg:apply}. We use the ITensor library to perform all our tensor network methods~\cite{ITensor} and deployed them on Perlmutter~\cite{perlmutter} and Polaris~\cite{polaris} supercomputers. To compute a single energy at QAOA depth $p$, we use $p$ Nvidia A100 GPUs. 

\begin{algorithm}[H]
\caption{$\texttt{AppLayer}(\vec{\Psi}_z, \vec{\alpha}, \beta)$}\label{alg:apply}
\KwIn{$\Psi_z[1,\ldots,l+1]$, $\alpha[1,\ldots,p]$, $\beta$}
\KwOut{$G_l[1,\ldots,l]$, $\Psi_z[1,\ldots,l+2]$}
\DontPrintSemicolon
\;
\If{$l+1 = 0$}{
    $\ket{\Psi^{(1)}} \gets \left(\frac{1}{\sqrt{2}}\ket{\overline{0}_p}, \frac{1}{\sqrt{2}}\ket{\overline{0}_p}\right)$\;
    $\ket{\Psi^{(1)}_{1}} \gets \left(\frac{1}{\sqrt{2}}\ket{\overline{0}_p},\,-\frac{1}{\sqrt{2}}\ket{\overline{0}_p}\right)$\;
    $\Psi_z \gets \Big[\ket{\Psi^{(1)}_{1}}, \ket{\Psi^{(1)}}\Big]$\;
    $l \gets 1$\;
}
\;

\For{$i=1$ \KwTo $l+1$}{
    $(\ket{\psi_0}, \ket{\psi_1}) \gets \Psi_z[i]$\;
    $\ket{\psi_0} \gets \hat{D}(\alpha[1,\ldots,p]) \ket{\psi_0}$\;
    $\ket{\psi_1} \gets \hat{D}(-\alpha[1,\ldots,p]) \ket{\psi_1}$\;
    \;
    $
    \begin{pmatrix}
        \ket{\psi_0} \\
        \ket{\psi_1}
    \end{pmatrix}
    \gets
    \begin{pmatrix}
        \cos\beta & -i\sin\beta \\
        -i\sin\beta & \cos\beta
    \end{pmatrix}
    \begin{pmatrix}
        \ket{\psi_0} \\
        \ket{\psi_1}
    \end{pmatrix}
    $
    \;
    \;
    $\Psi_z[i] \gets \left(\ket{\psi_0}, \ket{\psi_1}\right)$\;
}
\;
$\ket{\Psi^{(l+1)}} \gets \Psi_z[l+1]$\;
$\ket{\Psi^{(l+1)}_{l+1}} \gets \hat{Z} \ket{\Psi^{(l+1)}}$\;
$\Psi_z[l+1] \gets \ket{\Psi^{(l+1)}_{l+1}}$\;
\texttt{Append}$(\Psi_z, \ket{\Psi^{(l+1)}})$\;
\;
$G_l[1,\ldots,l] \gets [0,\ldots,0]$\;
\For{$i=1$ \KwTo $l$}{
    $\ket{\Psi^{(l+1)}_{i}} \gets \Psi_z[i]$\;
    $G_l[i] \gets \braket{\Psi^{(l+1)}_{i}}{\Psi^{(l+1)}_{l+1}}$\;
}
\end{algorithm}

\begin{algorithm}[H]
\caption{Compute $\nu_{p}(\vec{\gamma}, \vec{\beta})$}\label{alg:nu}
\KwIn{$\gamma[1,\ldots,p]$, $\beta[1,\ldots,p]$}
\KwOut{$\nu_p$}
\DontPrintSemicolon

Initialize $\Psi_z \gets$ empty vector\;
Initialize $L \gets$ zero matrix of size $(p+1) \times (p+1)$\;
\;
$L[1,1] \gets 1$\;

\For{$l = 1$ \KwTo $p$}{
    $\alpha \gets$ zero vector of length $p$\;
    $\alpha[1,\ldots,l] \gets -i \cdot \gamma[l] \cdot L[1,\ldots,l,l]$\;
    $(G_l[1,\ldots,l], \Psi_z) \gets \texttt{AppLayer}(\Psi_z, \alpha, \beta[l])$\;
    $L_l[1,\ldots,l] \gets (L[1:l,1:l]^\dagger)^{-1} \cdot G_l$\;
    $L[1,\ldots,l,l+1] \gets L_l[1,\ldots,l]$\;
    $L[l+1,l+1] \gets \sqrt{1 - \sum_{i=1}^l L_l^*[i] \cdot L_l[i]}$\;
}
\;
$G \gets L^\dagger L$\;
$\nu_p \gets \sum_{i=1}^p \text{Im}\big(\gamma[i]\,G[p+1,i]\big)$\;
\end{algorithm}

\subsection{MPS performance}

By truncating both the Fock-space dimension to $d$ and the bond dimension to $\chi$, we compute a value $\widetilde{\nu}_{p}(\bm{\gamma}, \bm{\beta}; d, \chi)$ that approximates the true QAOA energy $\nu_{p}(\bm{\gamma}, \bm{\beta})$. Then $d$ and $\chi$ are tunable parameters that determine the absolute error $\Delta\nu_p(d,\chi)$ of QAOA energy computed via MPS:
\begin{equation}\label{eq:approx}
    \left|\widetilde{\nu}_{p}(\bm{\gamma}, \bm{\beta}; d, \chi) - \nu_{p}(\bm{\gamma},\bm{\beta})\right| \leq \Delta\nu_p(d,\chi),
\end{equation}
with $\Delta\nu_p(d,\chi)\to0$ as $d\to\infty$ and $\chi\to2^p$. The truncation of the bond dimension between two MPS sites is performed by fixing a singular value decomposition (SVD) cutoff parameter $\delta$, so the bond dimension is determined by the cutoff parameter $\chi = \chi_\delta$. Empirically, we find that the absolute error $\Delta\nu_p(d,\chi_\delta)$ decays exponentially with $d$ and polynomially with $\log\left(1/\delta\right)$ (see Fig.~\ref{fig:gt} below).

\begin{figure*}[t]
\centering
\includegraphics[width=\linewidth]{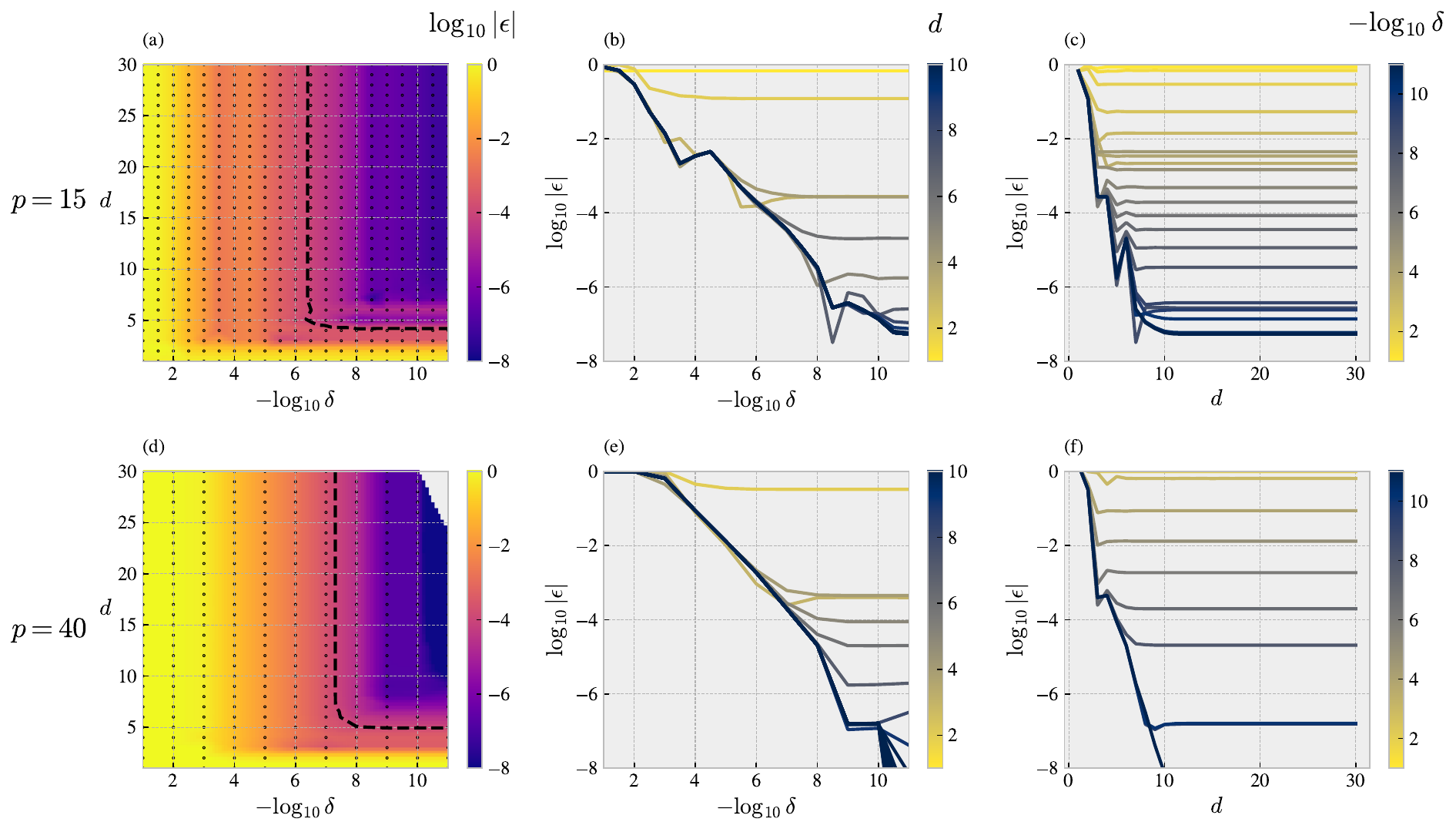}
\caption{
Convergence of the approximate tensor network simulation methods.
The top row (a-c) shows convergence for $p = 15$, and the bottom row (d-f) corresponds to $p = 40$. (a, d) Contour plots of the logarithmic relative error $\epsilon = 1-\frac{\widetilde{\nu}_p }{\nu_p}$ as a function of the Fock-space dimension $d$ and SVD cutoff parameter $\delta$ . The dashed line marks where the absolute error is $0.01\%$. (b, e) Horizontal slices of (a, d) showing convergence of error to zero with decreasing SVD cutoff $\delta$ at various Fock-space dimensions $d$. Clear $d$-dependent floor on relative error is observed as expected. (c, f) Vertical slices of (a, d) showing convergence of error to zero with growing Fock-space dimension $d$ at various SVD cutoff parameters $\delta$. For $p=15$, the error is computed relative to the exact QAOA energy obtained from separate methods. for $p=40$, the error is computed relative to the energy obtained at $d=25$ and $\delta=10^{-11}$.}
\label{fig:gt}
\end{figure*}

The cost of computing QAOA energy $\widetilde{\nu}_p$ is dominated by the cost of applying a unitary layer $U_t$ to the spin-boson MPS. Given a fixed bond dimension $\chi$, the time it takes to apply $U_t$ on one $p$-site state is $\mathcal{O}(pd\chi^3)$, using SVD and truncation schemes. The singular value decomposition (SVD) scaling cost is notable here, as it involves operations that scale as $\mathcal{O}(d\chi^3)$. Since there are $(p+1)$ MPS in Eq.~\ref{eq:nu2}, each requiring $\mathcal{O}(p)$ unitary applications to the initial product state, the total time to compute $\widetilde{\nu}_p$ is $\mathcal{O}\left(p^3d\chi^3\right)$. By parallelizing the application of a single unitary $U_t$ to each of the $p+1$ MPS, we reduce the time complexity to $\mathcal{O}\left(p^2 d \chi^3\right)$. We summarize the application of $U_t$ in Alg.~\ref{alg:apply}.

\begin{figure*}
\centering
\includegraphics[width=\linewidth]{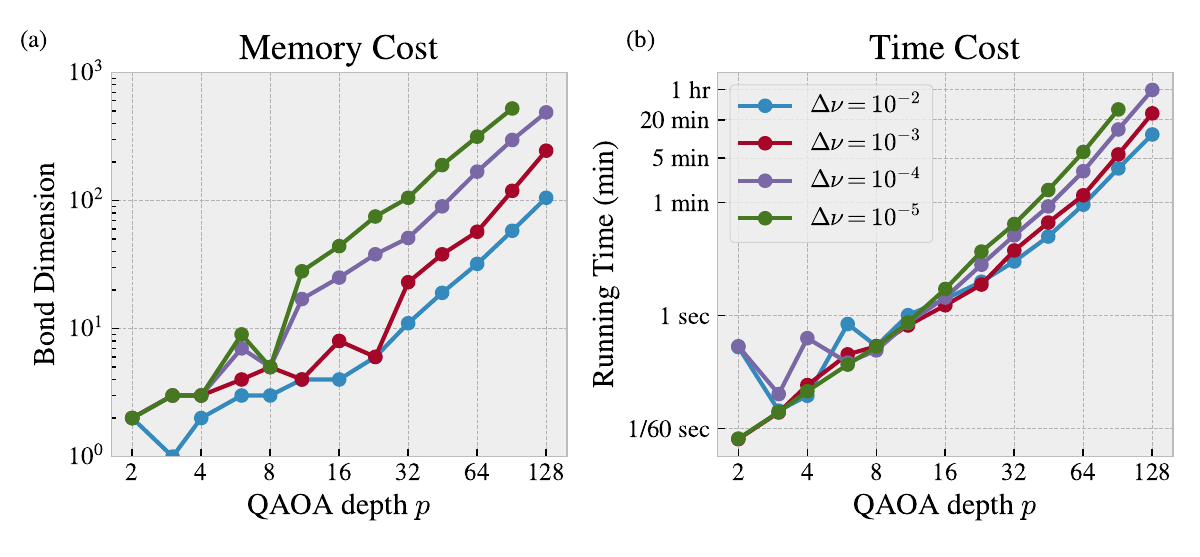}
\caption{
Cost to compute approximate QAOA energy $\widetilde{\nu}_p(\bm{\gamma},\bm{\beta};d,\chi)$ (a)  Growth of bond dimension required to achieve fixed precision for varying QAOA depth $p$. For a given QAOA depth $p$, we use $p$ A100 GPUs to compute the QAOA energy at a given precision. (b) Log-log plot of walltime (in minutes) versus QAOA depth $p$ for different precision levels and fixed Fock-space dimension $d=10$. Total number of GPU hours is $p$ times the walltime.
}
\label{fig:scalings}
\end{figure*}

We now show empirically that the approximation error $\Delta\nu_p(d,\chi_\delta)$ converges to zero as a function of truncation parameters $d$ and $\delta$, allowing us to compute QAOA energy $\nu_{p}(\bm{\gamma},\bm{\beta})$ up to any desired level of precision. We validate the convergence as follows. We start by taking the $p=15$ case and compare $\widetilde{\nu}_{p}(\bm{\gamma}, \bm{\beta}; d, \chi)$ to the exact value. The exact value is computed to 80-bit floating point precision by brute-force summation of the $G$-matrix iteration in Eqs.~\ref{eq:G0} and \ref{eq:Gm} using the implementation \cite{large_girth_maxcut_qaoa_github} provided by Ref.~\cite{basso_et_al:LIPIcs.TQC.2022.7}. We estimate the uncertainty on the energy to be less than $10^{-11}$. Fig~\ref{fig:gt}(a-c) shows the relative energy error as we tune both $d$ and $\delta$. The exact energy value is computed up to 11 digits. We find that the energy error vanishes polynomially with the truncation cutoff $\delta$ as expected~\cite{Verstraete_2006}, but, surprisingly, vanishes \emph{exponentially} with the Fock space dimension $d$. For instance, a Fock-space dimension $d=5$ is enough to have a relative energy error of $0.01\%$ for $p\lesssim 40$. We indeed find that as we tune both the Fock-space dimension and the cutoff, the energy we compute converges to the exact value. 

For large values of $p$, we cannot obtain an exact energy because of the exponential scaling $\widetilde{\mathcal{O}}\left(4^p\right)$ of the brute-force computation, and therefore cannot verify that the computed energy converges to the exact value as we tune our parameters $d$ and $\delta$. Instead, we validate that the convergence with $d$ and $\chi_b$ observed at $p=15$ persists for higher $p$. Fig.~\ref{fig:gt}(d-f) shows the relative errors for $p=40$ energies using the highest $d$ and smallest $\delta$ computable on Nvidia A100 GPUs, namely $d=25$, $\delta= 10^{-11}$. For $p=40$, the highest $d$ and smallest $\delta$ gives the best proxy for the exact energy. We observe the same monotonic decay in the error at $p=40$ as at $p=15$, providing strong evidence that the energy converges to the exact value $\nu_p$ as we tune $d\to\infty$ and $\delta\to0$.  In both the $p=15$ and $p=40$ cases, we see a nearly identical trend, except that for the $p=40$ case, we require a smaller cutoff parameter to yield the same relative energy error as the $p=15$ case.

In addition to observing that our method can compute QAOA energies up to very high precision, we also find that modest values of both the bond dimension and the Fock-space dimension are sufficient to compute QAOA energies at a high precision. Fig.~\ref{fig:scalings}(a) shows the growth of bond dimension required for different QAOA depths and precision levels. Surprisingly, the bond dimension empirically scales polynomially with $p$ for a given desired level of precision. Consequently, the total simulation time also scales mildly with $p$, as shown in Fig.~\ref{fig:scalings}(b). Here, we use $p$ Nvidia A100 GPUs to compute $\widetilde{\nu}_{p}$.

\section{QAOA parameters optimized in the infinite-size limit}\label{app:parameters}

We describe how we find optimized QAOA parameters. Since evaluating $\nu_p(\bm\gamma,\bm\beta)$ exactly requires an exponential amount of resources with respect to $p$, in practice, we optimize the parameters to maximize an approximation to Eq.~\ref{eq:nu_general_definition_SM}, namely $\widetilde{\nu}_{p}(\bm{\gamma}, \bm{\beta}; d, \chi_\delta)$ in Eq.~\ref{eq:approx} where $\delta$ is the MPS cutoff parameter and $d$ is the truncated Fock-space dimension. For all our simulations, we find $d=10$ suffices. The bond dimension $\chi_\delta$ is determined by the desired level of precision $\Delta\nu_p$ based on Fig.~\ref{fig:gt}. We set $\Delta\nu_p= 10^{-6}$ for $p < 100$, we set $\Delta\nu_p= 10^{-4}$ and for $p>100$. 

We must numerically optimize $\widetilde{\nu}_{p}\left(\bm{\gamma}, \bm{\beta}; d, \chi_\delta\right)$ but since our simulation does not provide access to gradients of $\widetilde{\nu}_p$ with respect to $\bm{\gamma}$ or $\bm{\beta}$, we resort to using derivative-free methods. However, we do note that there is considerable structure in the calculation of the energy: the simulation is not a pure single-output black-box returning only $\widetilde{\nu}_{p}$ values, but is rather a known function of multiple simulation quantities. Rather than modeling the energy as a function of $(\bm{\gamma},\bm{\beta}$), we can instead model the $p$ elements $\left\{G_{p,i}(\bm{\gamma},\bm{\beta})\right\}_{i=1}^{p-1}$ and combine these models using the known form of the objective. Towards this end, we use methods in the IBCDFO library for interpolation-based composite derivative-free optimization~\cite{ibcdfo}. Exploiting such objective information can often decrease the number of objective queries needed to achieve a solution quality (as compared to general-purpose approaches)~\cite{Larson2022}. 

We can slightly simplify the definition of $\widetilde{\nu}$ because $\bm{\gamma}$ is pure real:
\begin{align*}
  \widetilde{\nu}_p(\bm{\gamma},\bm{\beta}) &= \text{Im}\left[\sum_{i=1}^{p}\gamma_{i}G_{p,i-1}^{2}\left(\bm{\gamma},\bm{\beta}\right)\right]     \\
  &=\sum_{i=1}^{p}\gamma_{i}\,\text{Im}\left[G_{p,i-1}^{2}\left(\bm{\gamma},\bm{\beta}\right)\right]\\
  &= 2 \sum_{i=1}^p \gamma_i \,\text{Im}\left[G_{p,i-1}(\bm{\gamma},\bm{\beta})\right] \, \text{Re}\left[G_{p,i-1}(\bm{\gamma},\bm{\beta})\right].
\end{align*}

We find that the $p$ components of $G$ are sufficiently smooth in $(\bm{\gamma},\bm{\beta})$ to allow us to model each term in the sum using quadratic models. We combine those models using the above form of $\widetilde{\nu}$ above, to get a more accurate (in practice) model of $\widetilde{\nu}$ than building a single quadratic model of $\widetilde{\nu}$. Within the IBCDFO library we use the Practical Optimization Using No Derivatives for sums of Squares (POUNDerS) method, which we modify to explicitly to account for this known functional mapping (instead of traditional use for least-squares mappings)~\cite{SWCHAP14}. We use this optimizer to find optimal SK-QAOA angles.

Parameter initialization is critical in both limiting computational expense for our simulations as well as finding good minima. In QAOA, finding high-quality initial parameters for depth $p$ can be found by extrapolating from a smaller depth $p'<p$  by expanding in the Fourier basis~\cite{Zhou2020}, although other methods have been studied as well~\cite{2504.01694,hao2024endtoendprotocolhighqualityqaoa,Shaydulin2019}.  For initial parameters, we use have optimized parameters for a given QAOA depth $p$ and extrapolate them using a Fourier expansion to obtain initial parameters for QAOA depth $p+\Delta p$. 

We limit our numerical optimization to use at most $20p$ function evaluations for each SK-QAOA optimization. In employing the POUNDerS optimization method, many of these function evaluations can be done concurrently. This is due to to the optimizer requesting geometry (model-building) points, and optimization points. The former type of points are produced in a batch and can be evaluated independently as needed, and the vast majority of points requested by the method are such model-building points. 

\begin{figure*}[!h]
    \centering
    \includegraphics[width=\linewidth]{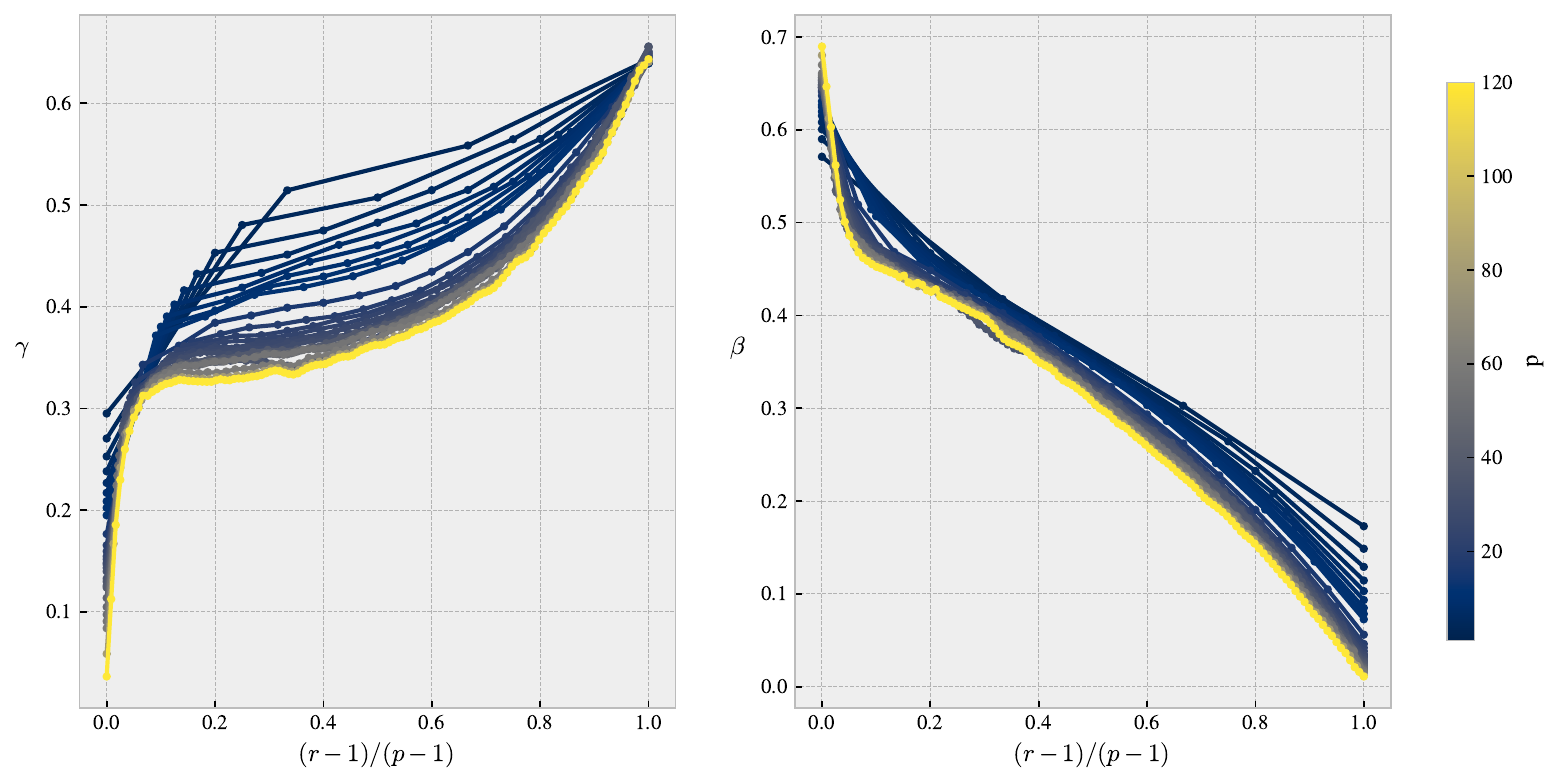}
    \caption{Parameter values for $\bm\gamma$ (left) and $\bm\beta$ (right) for each QAOA depth $p$ where the energy was evaluated. The parameter index $r$ is normalized by the QAOA depth $p$.}
    \label{fig:qaoa_parameters}
\end{figure*}

Figure~\ref{fig:qaoa_parameters} shows the QAOA parameters we obtained by optimizing $\nu_p(\bm\gamma,\bm\beta)$. Table~\ref{table:extended_values} shows QAOA energy with the optimized parameters.

\begin{table}[ht]
\centering
\begin{tabular}{cccccccccc}
\hline
\multicolumn{1}{|c|}{$p$} & 1 & 2 & 3 & 4 & 5 & 6 & 7 & 8 & \multicolumn{1}{c|}{9} \\ \hline
\multicolumn{1}{|c|}{$\widetilde{\nu}_p$} & $0.3033$ & $0.4075$ & $0.4726$ & $0.5157$ & $0.5476$ & $0.5721$ & $0.5915$ & $0.6073$ & \multicolumn{1}{c|}{$0.6203$} \\ \hline 
\\
\cline{1-10}
\multicolumn{1}{|c|}{$p$} & 10 & 11 & 12 & 13 & 14 & 15 & 16 & 17 & \multicolumn{1}{c|}{18} \\ \cline{1-10}
\multicolumn{1}{|c|}{$\widetilde{\nu}_p$} &  $0.6314$ & $0.6408$ & $0.6489$ & $0.6561$ & $0.6623$ & $0.6678$ & $0.6729$ & $0.6772$ & \multicolumn{1}{c|}{$0.6814$} \\ \cline{1-10}
\\
\cline{1-10}
\multicolumn{1}{|c|}{$p$} & 19 & 20 & 22 & 24 & 26 & 27 & 28 & 30 & \multicolumn{1}{c|}{31} \\ \cline{1-10}
\multicolumn{1}{|c|}{$\widetilde{\nu}_p$}  & $0.6850$ & $0.6884$ & $0.6942$ & $0.6992$ & $0.7035$ & $0.7055$ & $0.7073$ & $0.7106$ & \multicolumn{1}{c|}{$0.7115$} \\ \cline{1-10}
\\
\cline{1-10}
\multicolumn{1}{|c|}{$p$} & 32 & 34 & 36 & 38 & 40 & 48 & 60  & 80 &\multicolumn{1}{c|}{120} \\ \cline{1-10}
\multicolumn{1}{|c|}{$\widetilde{\nu}_p$}  & $0.7133$ & $0.7159$ & $0.7182$ & $0.7202$ & $0.7222$ & $0.7276$ & $0.7337$ & $0.7397$ & \multicolumn{1}{c|}{$0.7447$} \\ \cline{1-10}
\\
\cline{1-2}
\multicolumn{1}{|c|}{$p$} &\multicolumn{1}{c|}{160} \\ \cline{1-2}
\multicolumn{1}{|c|}{$\widetilde{\nu}_p$} & \multicolumn{1}{c|}{$0.7461$} \\ \cline{1-2}
\end{tabular}
\caption{Best infinite-size-limit QAOA energies $\widetilde{\nu}_p(\bm\gamma,\bm\beta)$ obtained for QAOA depth $p$. Further optimizations might be possible for $p\leq 80$, whereas for $p>80$, we report an unoptimized lower bound for the QAOA energy obtained through extrapolating from the $p=80$ optimized parameters. The optimal value is $P_*\approx0.763166$.}
\label{table:extended_values}
\end{table}

\section{Alternative parameter fits}
\label{app:alt_fittings}
\begin{figure}
    \centering
    \includegraphics[width=\linewidth]{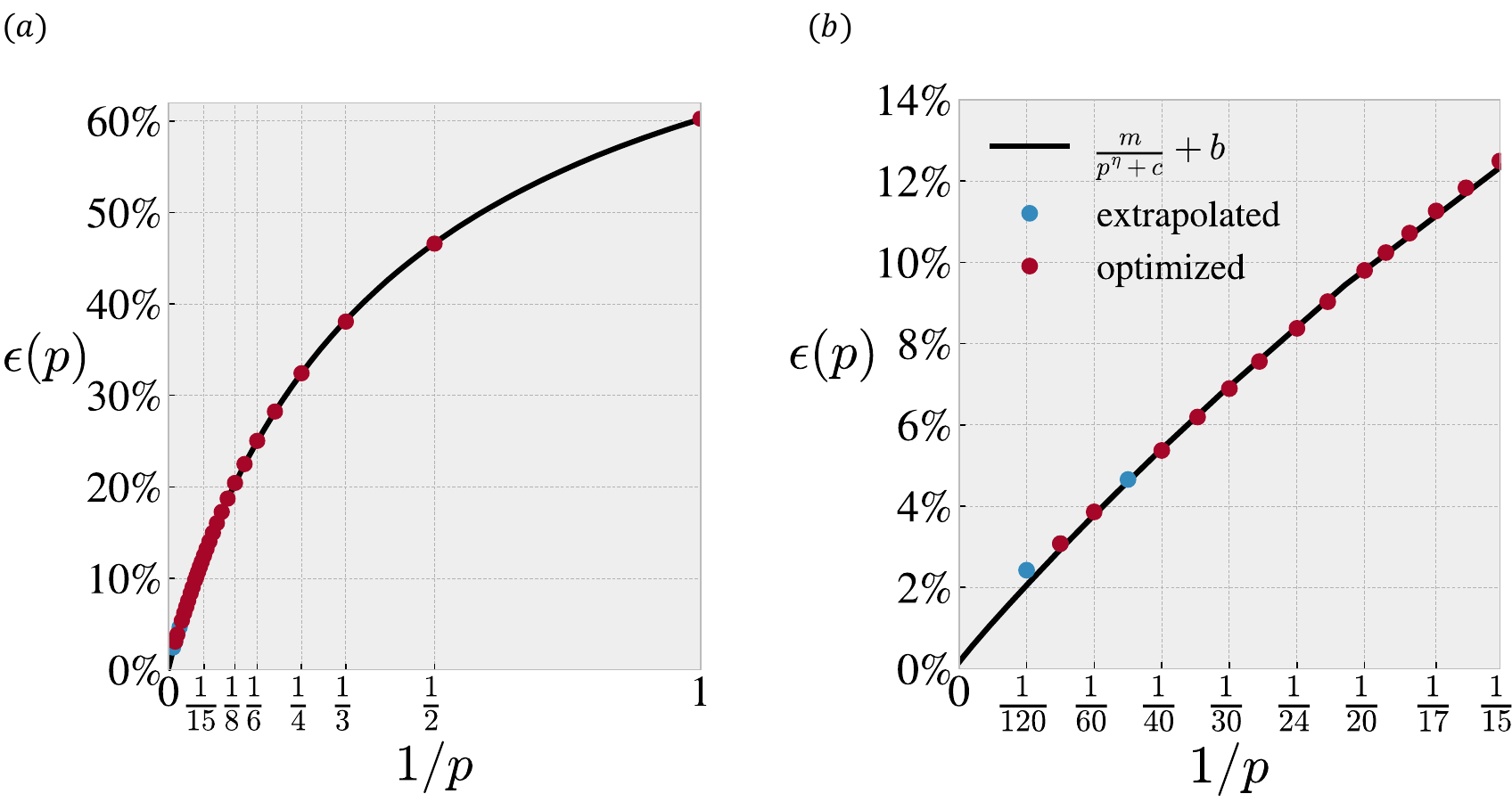}
    \caption{Fitting the data points $(p, \nu_p)$ for depth $p$ QAOA energies $\nu_p$  in the infinite size limit to Eq.~\ref{eq:altfit}.  Data points are fitted using all optimized points from $p=1$ to $p=80$. The entire fit is shown in $(a)$, while, a zoomed-in version is shown in $(b)$.
    }
    \label{fig:alt_fittings}
\end{figure}

Here, we follow Ref.~\cite{leoTQC2022talk} and fit the obtained depth $p$ QAOA energies $\nu_p$ at optimized angles to the function
\begin{equation}\label{eq:altfit}
    \epsilon(p) \equiv 1 - \frac{\nu_p}{P_*} = \frac{m}{p^\eta + c} + b
\end{equation}
which has parameters $(m, \eta, c, b)$. We fit  the dataset $(p,\nu_p)$  $p=1$ to $p=80$. This yields fitting parameters
\begin{align}
m   &= 1.9540  \pm 0.05160\\
\eta &= 0.9635\pm 0.01617\\
c   &= 2.2540  \pm  0.1001\\
b &= 0.001610 \pm 0.002252,
\end{align}
with a goodness of fit $1-R^2 = 1.805\times10^{-5}$.  This fit is shown in Fig.~\ref{fig:alt_fittings}

\section{Generalized SK-QAOA iteration from spin-bosons system (proof of Proposition \ref{spinBosonsStateDotProductPathIntegral})}
\label{sec:sk_qaoa_iteration_spin_bosons}

This appendix establishes the main technical result of this work (Proposition~\ref{spinBosonsStateDotProductPathIntegral}), stating that
\begin{align}
    f\left(\bm{a}\right)H\left(\bm{a}\right),\label{eq:path_integral_weight_reminder}
\end{align}
where
\begin{align}
    f\left(\bm{a}\right) & = \frac{1}{2}\bra{a_p}e^{i\beta_pX}\ket{a_0}\bra{a_0}e^{-i\beta_pX}\ket{a_{-p}}\prod_{1 \leq t \leq p - 1}\bra{a_t}e^{i\beta_tX}\ket{a_{t + 1}}\bra{a_{-t - 1}}e^{-i\beta_tX}\ket{a_{-t}},\\
    H\left(\bm{a}\right) & = \exp\left(-\frac{1}{2}\sum_{j', k' = -p}^pG_{j', k'}\Gamma_{j'}\Gamma_{k'}a_{j'}a_{k'}\right),\\
    \bm{a} & = \left(a_1, a_2, \ldots, a_{p - 1}, a_p, a_0, a_{-p}, a_{-(p - 1)}, \ldots, a_{-2}, a_{-1}\right) \in \{1, -1\}^{2p + 1},
\end{align}
can be interpreted as the path integral weight of a spin in a spin-bosons system. More specifically, the spin-bosons system's evolution can be written in the form:
\begin{align}
    \overleftarrow{\prod_{t = 1}^p}\exp\left(-i\beta_tX\right)D\left(-i\gamma_tZ\bm{L}_{:,\,\bm{a}}\right)\ket{+} \otimes \ket{\overline{0}},
\end{align}
The intuition can be summarized as follows. Using an original idea introduced in the context of QAOA by Refs.~\cite{Farhi_2022,basso_et_al:LIPIcs.TQC.2022.7}, we proceed to expand this state by summing over computational basis paths
\begin{align}
    \left(a_{p + 1}, a_{p}, \ldots, a_{2}, a_{1}\right) \in \{1, -1\}^{p + 1}
\end{align}
of the spin. Note such a spin's path has $(p + 1)$ steps as opposed to bitstrings $\bm{a}$ occurring in $f\left(\bm{a}\right)H\left(\bm{a}\right)$, which have $(2p + 1)$ bits. This is because the latter path integral measure computes expectations rather than state amplitudes, hence the double number of path steps. From this computational basis path expansion, multiplicative contribution $f\left(\bm{a}\right)$ in Eq.~\ref{eq:path_integral_weight_reminder} arises from the coupling of path steps induced by $X$ rotations $e^{-i\beta_tX}$. The origin of contribution $H\left(\bm{a}\right)$ may be less evident. As will be shown in detail in the proofs, this contribution essentially arises from the multiplication identity for displacement operators (considering single-mode operators here for simplicity):
\begin{align}
    D\left(\eta\right)D\left(\theta\right) & = \exp\left(\left(\eta\overline{\theta} - \overline{\eta}\theta\right)/2\right)D\left(\eta + \theta\right)\nonumber\\
    & = \exp\left(i\Im\left(\eta\overline{\theta}\right)\right)D\left(\eta + \theta\right),\\
    \eta, \theta & \in \mathbb{C}.
\end{align}
The prefactor in the last expression is the exponential of a quadratic function of the displacement operators' arguments. This is intuitively how $H\left(\bm{a}\right)$, the exponential of a quadratic function of $\bm{a}$, arises in the path integral weight. Besides, we will see that the symmetries of the $\bm{G}$ matrix (Eqns.~\ref{eq:Gsim}), relating 
real and imaginary parts of different corners of $\bm{G}$, naturally emerge from the same multiplication identity for displacement operators.

We now make this sketch of derivation more precise. Consider a matrix
\begin{align}
    \bm{L} \in \mathbb{C}^{p' \times p},\label{eq:l_matrix}
\end{align}
which we regard as a collection of column vectors. Let us then introduce the Gram matrix of these column vectors, denoted by $\bm{G}^{(\mathrm{herm})}$: 

\begin{definition}[Gram matrix of vectors]
\label{def:gram_matrix}
Given a matrix $\bm{L}$ of column vectors defined in equation \ref{eq:l_matrix}, we denote the Gram matrix of these column vectors as:
\begin{align}
    \bm{G}^{(\mathrm{herm})} & := \bm{L}^{\dagger}\bm{L}.
\end{align}
\end{definition}

The relation between $\bm{L}$ and $\bm{G}^{\left(\mathrm{herm}\right)}$ in Definition~\ref{def:gram_matrix} is precisely Eqn.~\ref{eq:Lfac}. When describing the spin-boson mapping of the SK-QAOA $G$ iteration (Section~\ref{sec:spin_boson_mapping}), given $\bm{G}^{\left(\mathrm{herm}\right)}$, we introduced $\bm{L}$ as a factorization of this Hermitian matrix ---among many possible. In this Section, it will be more natural to start with $\bm{L}$ and define $\bm{G}^{\left(\mathrm{herm}\right)} := \bm{L}^{\dagger}\bm{L}$. Indeed, as we will show, the spin-bosons system's construction depends explicitly on $\bm{L}$, as opposed to the spin's path-integral measure (Eqn.~\ref{eq:path_integral_measure_generalized}) which only depends on $\bm{G}^{\left(\mathrm{herm}\right)}$. Given $\bm{G}^{\left(\mathrm{herm}\right)}$ defined from $\bm{L}$, it will be convenient to introduce a symmetric counterpart of this matrix ---defined by same lower triangle, but symmetric rather than Hermitian:

\begin{definition}[Symmetric counterpart of Gram matrix of vectors]
\label{def:gram_matrix_symmetric}
From the Hermitian non-negative Gram matrix introduced in Definition \ref{def:gram_matrix}, we define its symmetric counterpart:
\begin{align}
    \bm{G}^{(\mathrm{sym})},
\end{align}
where
\begin{align}
    G^{(\mathrm{sym})}_{j, k} & := G^{(\mathrm{herm})}_{k, j} \qquad \forall 1 \leq j \leq k \leq p.
\end{align}
\end{definition}

\begin{definition}[Generalized QAOA state]
\label{def:generalized_qaoa_state}
The generalized QAOA state lives in the tensor product of $p'$ copies of $L^2\left(\mathbf{R}, \mathrm{d}x\right)$ (boson space) and $\mathbb{C}^2$ (spin space). We denote by
\begin{align}
    \hat{c}_1, \hat{c}_2, \ldots, \hat{c}_{p - 1}, \hat{c}_{p'}
\end{align}
the annihilation operators associated to the $p'$ copies of $L^2\left(\mathbf{R}, \mathrm{d}x\right)$. We consistently denote by
\begin{align}
    D_1\left(\alpha_1\right) & := \exp\left(\alpha_1\hat{c}_1^{\dagger} - \overline{\alpha_1}\hat{c}_1\right), \qquad \alpha_1 \in \mathbb{C},\\
    \ldots\\
    D_{p'}\left(\alpha_{p'}\right) & := \exp\left(\alpha_1\hat{c}_{p'}^{\dagger} - \overline{\alpha_1}\hat{c}_{p'}\right), \qquad \alpha_{p'} \in \mathbb{C}.
\end{align}
Also, for all displacement vectors
\begin{align}
    \bm{\alpha} = \left(\alpha_1, \ldots, \alpha_{p'}\right) \in \mathbb{C}^{p'},
\end{align}
we denote by
\begin{align}
    D(\bm{\alpha}) & := \bigotimes_{l = 1}^{p'}D_l\left(\alpha_l\right)
\end{align}
the tensor product of displacement operators acting independently on each mode. Finally, we write
\begin{align}
    X, Y, Z, I
\end{align}
for the Pauli matrices acting on $\mathbb{C}^2$. Then, the generalized QAOA state of angles
\begin{align}
    \bm{\gamma} & := \left(\gamma_1, \ldots, \gamma_p\right) \in \mathbf{R}^p,\\
    \bm{\beta} & := \left(\beta_1, \ldots, \beta_p\right) \in \mathbf{R}^p,
\end{align}
is defined as:
\begin{align}
    \ket{\Psi} & := \left(\overleftarrow{\prod_{t = 1}^p}\exp\left(-i\beta_tX\right)D\left(-i\gamma_tZ\bm{L}_{:,\,t}\right)\right)\ket{+}_{\mathbb{C}^2} \otimes \ket{0}_{L^2\left(\mathbf{R},\,\mathrm{d}x\right)}^{\otimes p'}.\label{eq:generalized_qaoa_state}
\end{align}
In the above expression, each bosonic mode is initialized to the ground state of the quantum harmonic oscillator, denoted by $\ket{0}_{L^2\left(\mathbf{R},\,\mathrm{d}x\right)}$. In the following, we will drop indices on kets to lighten the notation, keeping in mind the ordering of the spaces: the spin one first, and the $p'$ bosonic modes next. Note the displacement operators in the above formula entangle the spin space and the bosonic spaces due to the occurrence of spin operator $Z$ in the argument the displacement vector.
\end{definition}

It is ``easy" to express the coordinates of the generalized QAOA state in the coherent states basis as a path integral over the spin space. For this purpose, we will first need the following lemma that will be used in several places:

\begin{lemma}[Product of displacement operators conditioned on $Z$ values]
\label{lemma:product_displacement_operators_spin_path}
Consider the product of displacement operators appearing in Definition \ref{def:generalized_qaoa_state} of the generalized QAOA state, but with all $Z$ replaced by given spin values
\begin{align}
    \bm{a} & = \left(a_1, \ldots, a_p\right) \in \{1, -1\}^p.
\end{align}
Then, this product obeys identity:
\begin{align}
    \overleftarrow{\prod_{t = 1}^p}D\left(-i\gamma_ta_t\bm{L}_{:,\,t}\right) & = \exp\left(-\frac{1}{2}\sum_{1 \leq r, s \leq p}i\Im G^{(\mathrm{sym})}_{r, s}\gamma_r\gamma_sa_ra_s\right)D\left(-i\sum_{1 \leq t \leq p}\gamma_ta_t\bm{L}_{:,\,t}\right),\label{eq:displacement_operator_spin_path_integral_action_g_sym_imag}
\end{align}
where $\bm{G}^{(\mathrm{sym})}$, the symmetric version of $\bm{G}^{\left(\mathrm{herm}\right)} := \bm{L}^{\dagger}\bm{L}$, was introduced in Definition \ref{def:gram_matrix_symmetric}. In particular, applying this identity to the ground state $\ket{0}^{\otimes p'}$ of the $p'$ bosonic modes,
\begin{align}
    \overleftarrow{\prod_{t = 1}^p}D\left(-i\gamma_ta_t\bm{L}_{:,\,t}\right)\ket{0}^{\otimes p'} & = \exp\left(-\frac{1}{2}\sum_{1 \leq r, s \leq p}i\Im G^{(\mathrm{sym})}_{r, s}\gamma_r\gamma_ra_ra_s\right)\Bigg|-i\sum_{1 \leq t \leq p}\gamma_ta_t\bm{L}_{:,\,t}\Bigg\rangle\label{eq:displacement_operator_spin_path_integral_vacuum_action_g_sym_imag}\\
    & = \exp\left(-\frac{1}{2}\sum_{1 \leq r, s \leq p}G^{(\mathrm{sym})}_{r, s}\gamma_r\gamma_sa_ra_s\right)\bigotimes_{l = 1}^{p'}\exp\left(-i\hat{c}_l^{\dagger}\sum_{1 \leq t \leq p}L_{l, t}\gamma_ta_t\right)\ket{0}^{\otimes p'}\label{eq:displacement_operator_spin_path_integral_vacuum_action_g_sym}
\end{align}
\begin{proof}
We prove by induction on $q = 1, \ldots, p$ that
\begin{align}
    \overleftarrow{\prod_{t = 1}^q}D\left(-i\gamma_ta_t\bm{L}_{:, t}\right) & = \exp\left(-\frac{1}{2}\sum_{1 \leq r, s \leq q }i\Im G^{(\mathrm{sym})}_{r, s}\gamma_r\gamma_ra_ra_s\right)D\left(-i\sum_{1 \leq t \leq q}\gamma_ta_t\bm{L}_{:,\,t}\right).
\end{align}
For the initialization $q = 1$, we note that 
\begin{align*}
    \Im G^{(\mathrm{sym})}_{1, 1} & = \Im G^{(\mathrm{herm})}_{1, 1}\\
    & = \Im\left\langle \bm{L}_{:, 1}, \bm{L}_{:, 1} \right\rangle\\
    & = \Im\left\lVert \bm{L}_{:, 1} \right\rVert^2_2\\
    & = 0,
\end{align*}
so the exponential evaluates to $1$ and the equality holds trivially.

Now, assume the identity holds up to level $q$ and let us prove it at level $q + 1$. Then,
\begin{align}
    \overleftarrow{\prod_{t = 1}^{q + 1}}D\left(-i\gamma_ta_t\bm{L}_{:, t}\right) & = D\left(-i\gamma_{q + 1}a_{q + 1}\bm{L}_{:, q + 1}\right)\overleftarrow{\prod_{t = 1}^{q}}D\left(-i\gamma_ta_t\bm{L}_{:, t}\right)\\
    & = \exp\left(-\frac{1}{2}\sum_{1 \leq r, s \leq q }i\Im G^{(\mathrm{sym})}_{r, s}\gamma_r\gamma_ra_ra_s\right)D\left(-i\gamma_{q + 1}a_{q + 1}\bm{L}_{:, q + 1}\right)D\left(-i\sum_{1 \leq t \leq q}\gamma_ta_t\bm{L}_{:, t}\right)\label{eq:product_displacement_operators_spin_path_apply_induction step}
\end{align}
We now compute the product of displacement operators:
\begin{align}
    & D\left(-i\gamma_{q + 1}a_{q + 1}\bm{L}_{:, q + 1}\right)D\left(-i\sum_{1 \leq t \leq q}\gamma_ta_t\bm{L}_{:, t}\right)\\
    & = \bigotimes_{l = 1}^{p'}D_l\left(-i\gamma_{q + 1}a_{q + 1}L_{l, q + 1}\right)D_l\left(-i\sum_{1 \leq t \leq q}\gamma_ta_tL_{l, t}\right)\\
    & = \bigotimes_{l = 1}^{p'}\exp\left(-i\Im\left\{\overline{\left(-i\gamma_{q + 1}a_{q + 1}L_{l, q + 1}\right)}\left(-i\sum_{1 \leq t \leq q}\gamma_ta_tL_{l, t}\right)\right\}\right)D\left(-i\sum_{1 \leq t \leq q + 1}\gamma_ta_tL_{l, t}\right)\\
    & = \exp\left(-i\Im\left\{\sum_{1 \leq l \leq p'}\sum_{1 \leq t \leq q}\overline{L}_{l, q + 1}L_{l, t}\gamma_{q + 1}\gamma_ta_{q + 1}a_t\right\}\right)D\left(-i\sum_{1 \leq t \leq q + 1}\gamma_ta_t\bm{L}_{:,\,t}\right)
\end{align}
Noting that
\begin{align}
    \sum_{1 \leq l \leq p'}\overline{L_{l, q + 1}}L_{l, t} & = G^{(\mathrm{herm})}_{q + 1, t}\\
    & = G^{(\mathrm{sym})}_{q + 1, t}
\end{align}
(the latter because $1 \leq t \leq q + 1$), the product of displacement operators has been recast to:
\begin{align}
    D\left(-i\gamma_{q + 1}a_{q + 1}\bm{L}_{:, q + 1}\right)D\left(-i\sum_{1 \leq t \leq q}\gamma_ta_t\bm{L}_{:, t}\right) & = \exp\left(-i\sum_{1 \leq t \leq q}\Im G^{(\mathrm{sym})}_{q + 1, t}\gamma_{q + 1}\gamma_ta_{q + 1}a_t\right)
\end{align}
Plugging this into equation \ref{eq:product_displacement_operators_spin_path_apply_induction step} completes the induction step and the proof.

We now prove auxiliary equations \ref{eq:displacement_operator_spin_path_integral_vacuum_action_g_sym_imag}, \ref{eq:displacement_operator_spin_path_integral_vacuum_action_g_sym}, rephrasing operator equality \ref{eq:displacement_operator_spin_path_integral_action_g_sym_imag} in terms of action on the ground state. Equation \ref{eq:displacement_operator_spin_path_integral_vacuum_action_g_sym_imag} is immediate, resulting from identity
\begin{align}
    \ket{\bm{\alpha}} & = D(\bm{\alpha})\ket{0}^{\otimes p'}.
\end{align}
As for equation \ref{eq:displacement_operator_spin_path_integral_vacuum_action_g_sym}, it follows by using representation
\begin{align}
    D(\eta) & = e^{-|\eta|^2/2}e^{\eta\hat{c}^{\dagger}}e^{\overline{\eta}\hat{c}}
\end{align}
for a single-mode displacement operator. Applying that to the multimode displacement operator,
\begin{align*}
    \Bigg|-i\sum_{1 \leq t \leq p}\gamma_ta_t\bm{L}_{:,\,t}\Bigg\rangle & = D\left(-i\sum_{1 \leq t \leq p}\gamma_ta_t\bm{L}_{:,\,t}\right)\ket{0}^{\otimes p'}\\
    & = \exp\left(-\frac{1}{2}\left|-i\sum_{1 \leq t \leq p}\gamma_ta_t\bm{L}_{:, t}\right|^2\right)\bigotimes_{l = 1}^{p'}\exp\left(-i\sum_{1 \leq t \leq p}\gamma_ta_tL_{l, t}\hat{c}^{\dagger}_l\right)\ket{0}^{\otimes p'}
\end{align*}
It remains to compute the norm in the exponential. This is:
\begin{align*}
    \left|-i\sum_{1 \leq t \leq p}\gamma_ta_t\bm{L}_{:, t}\right|^2 & = \sum_{1 \leq l \leq p'}\overline{\left(-i\sum_{1 \leq t \leq p}\gamma_ta_tL_{l, t}\right)}\left(-i\sum_{1 \leq t \leq p}\gamma_ta_tL_{:, t}\right),\\
    & = \sum_{1 \leq r, s \leq p}\gamma_r\gamma_sa_ra_s\sum_{1 \leq l \leq p'}\overline{L_{l, r}}L_{l, s}\\
    & = \sum_{1 \leq r, s \leq p}\gamma_r\gamma_sa_ra_sG^{(\mathrm{herm})}_{r, s}\\
    & = \sum_{1 \leq r, s \leq p}\gamma_r\gamma_sa_ra_s\left(\frac{G^{(\mathrm{herm})} + G^{(\mathrm{herm})T}}{2}\right)_{r, s}\\
    & = \sum_{1 \leq r, s \leq p}\gamma_r\gamma_sa_ra_s\Re G^{(\mathrm{herm})}_{r, s}\\
    & = \sum_{1 \leq r, s \leq p}\gamma_r\gamma_sa_ra_s\Re G^{(\mathrm{sym})}_{r, s}.
\end{align*}
Plugging this expression of the norm squared into the exponential, equation \ref{eq:displacement_operator_spin_path_integral_vacuum_action_g_sym} follows.
\end{proof}
\end{lemma}

We can now apply Lemma \ref{lemma:product_displacement_operators_spin_path} to obtain a representation of the state produced by unitary evolution \ref{eq:generalized_qaoa_state} as a path integral over the spin's computational basis state. This path integral is a finite sum over $p$ bits (the computational state of the spin after each layer), which can be viewed as a ``square root" of the usual $G$ matrix iteration sum over $2p + 1$ bits.

\begin{proposition}[Spin path integral representation of the generalized QAOA state]
\label{prop:generalized_qaoa_state_path_integral}
Consider the generalized QAOA state produced by the unitary dynamics from equation \ref{eq:generalized_qaoa_state}, i.e.:
\begin{align}
    \ket{\Psi} & := \left(\overleftarrow{\prod_{t = 1}^p}\exp\left(-i\beta_tX\right)D\left(-i\gamma_tZ\bm{L}_{:,\,t}\right)\right)\ket{0}_{\mathbb{C}^2} \otimes \ket{0}_{\mathbf{L}_2\left(\mathbf{R}\right)}^{\otimes p'}.
\end{align}
This state the following representation as a path integral over the spin's computational basis state:
\begin{align}
    \ket{\Psi} & = \sum_{\bm{a}_{1:p + 1} \in \{1, -1\}^{p + 1}}f_{1:p + 1}\left(\bm{a}_{1:p + 1}\right)\exp\left(-\frac{i}{2}\sum_{1 \leq r, s \leq p}\Im G_{r, s}\gamma_r\gamma_sa_ra_s\right)\ket{a_{p + 1}} \otimes \Bigg|-i\sum_{1 \leq t \leq p}\gamma_ta_t\bm{L}_{:,\,t}\Bigg\rangle\\
    & = \sum_{\bm{a}_{1:p + 1} \in \{1, -1\}^{p + 1}}f_{1:p + 1}\left(\bm{a}_{1:p + 1}\right)\exp\left(-\frac{1}{2}\sum_{1 \leq r, s \leq p}G_{r, s}\gamma_r\gamma_sa_ra_s\right)\ket{a_{p + 1}} \otimes \bigotimes_{l = 1}^p\exp\left(-i\hat{c}_l^{\dagger}\sum_{1 \leq t \leq p}\gamma_ta_tL_{l, t}\right)\ket{0}^{\otimes p'},
\end{align}
where $f_{1;p + 1}$ is defined as: 
\begin{align}
    f_{1:p + 1}\left(\bm{a}_{1:p + 1}\right) & := \prod_{t = 1}^p\bra{a_{t + 1}}\exp\left(-i\beta_tX\right)\ket{a_t}.
\end{align}
\end{proposition}

Besides the generalized QAOA state introduced in Definition \ref{def:generalized_qaoa_state}, and expressed as a path integral in Proposition \ref{prop:generalized_qaoa_state_path_integral}, it will be useful to introduce states generated by the same sequence of unitaries (equation \ref{eq:generalized_qaoa_state}), but with a single $Z$ inserted after a given unitary layer:

\begin{definition}[Generalized QAOA state with single $Z$ insertion]
\label{def:generalized_qaoa_state_single_z_insertion}
The generalized QAOA state with a single $Z$ insertion before layer $r$ is defined as follows:
\begin{align}
    \ket{\Psi_r} & := \left(\overleftarrow{\prod_{t = r}^p}\exp\left(-i\beta_tX\right)D\left(-i\gamma_tZ\bm{L}_{:,\,t}\right)\right)Z\left(\overleftarrow{\prod_{t = 1}^{r - 1}}\exp\left(-i\beta_tX\right)D\left(-i\gamma_tZ\bm{L}_{:,\,s}\right)\right)\ket{0}_{\mathbb{C}^2}\ket{0}_{\mathbf{L}_2\left(\mathbf{R}\right)}^{\otimes p'}
\end{align}
Note this is also well-defined for $r = p + 1$, in which case
\begin{align}
    \ket{\Psi_{p + 1}} & = Z\left(\overleftarrow{\prod_{t = 1}^p}\exp\left(-i\beta_tX\right)D\left(-i\gamma_tZ\bm{L}_{:,\,t}\right)\right)\ket{0}_{\mathbb{C}^2}\otimes\ket{0}_{\mathbf{L}_2\left(\mathbf{R}\right)}^{\otimes p'}\nonumber\\
    & = Z\ket{\Psi}.
\end{align}
Similar to Proposition \ref{prop:generalized_qaoa_state_path_integral}, these states can be written as a path integral over the spin's computational basis state:
\begin{align}
    \ket{\Psi_t} & = \sum_{\bm{a}_{1:p + 1} \in \{1, -1\}^{p + 1}}a_tf_{1:p + 1}\left(\bm{a}_{1:p + 1}\right)\exp\left(-\frac{i}{2}\sum_{1 \leq r, s \leq p}\Im G_{r, s}\gamma_r\gamma_sa_ra_s\right)\ket{a_{p + 1}} \otimes \Bigg|-i\sum_{1 \leq t \leq p}\gamma_ta_t\bm{L}_{:,\,t}\Bigg\rangle\\
    & = \sum_{\bm{a}_{1:p + 1} \in \{1, -1\}^{p + 1}}a_tf_{1:p + 1}\left(\bm{a}_{1:p + 1}\right)\exp\left(-\frac{1}{2}\sum_{1 \leq r, s \leq p}G_{r, s}\gamma_r\gamma_sa_ra_s\right)\ket{a_{p + 1}} \otimes \bigotimes_{l = 1}^p\exp\left(-i\hat{c}_l^{\dagger}\sum_{1 \leq t \leq p}\gamma_ta_tL_{l, t}\right)\ket{0}^{\otimes p'}.
\end{align}
\end{definition}

Applying Lemma \ref{lemma:product_displacement_operators_spin_path}, we can explicitly recover the QAOA iteration (sum over $2p + 1$ spins) in equation \ref{eq:Gm}. More specifically, the sum in the right-hand side of the iteration in equation \ref{eq:Gm} can be interpreted as a time autocorrelation of the $Z$ operator applied to the spin, similar to the the result established for $D$-regular MAXCUT QAOA in earlier work Ref.~\cite{leo2022Autocorrelation}.

\begin{repproposition}{spinBosonsStateDotProductPathIntegral}[Path integral representation of spin-bosons states overlap, restated]
Let matrix
\begin{align}
    \bm{G} & = \left(G_{r, s}\right)_{r, s \in \pm [p]}
\end{align}
be defined from $\bm{G}^{\left(\mathrm{herm}\right)}$, and related matrix $\bm{G}^{\left(\mathrm{sym}\right)}$, by the 4 following blocks:
\begin{align}
    G_{r, s} & := \overline{G^{\left(\mathrm{sym}\right)}_{r, s}},\label{eq:g_from_gherm_plus_plus}\\
    G_{-r, -s} & := G^{\left(\mathrm{sym}\right)}_{r, s},\label{eq:g_from_gherm_minus_minus}\\
    G_{-r, s} & := G^{\left(\mathrm{herm}\right)}_{s, r}\label{eq:g_from_gherm_minus_plus},\\
    G_{s, -r} & := G^{\left(\mathrm{herm}\right)}_{s, r},\label{eq:g_from_gherm_plus_minus}
\end{align}
where $r, s \in [p]$ in the 4 equations. Note $\bm{G}$ is symmetric by construction. Then, for all $1 \leq j, k \leq p + 1$, the dot product between $Z$-inserted states $\ket{\Psi_j}$ and $\ket{\Psi_k}$ admits the following path integral representation
\begin{align}
    \braket{\Psi_k}{\Psi_j} & = \sum_{\bm{a} \in \{1, -1\}^{2p + 1}}a_ka_{-j}f\left(\bm{a}\right)\exp\left(-\frac{1}{2}\sum_{r, s \in \pm [p]}G_{r, s}\Gamma_r\Gamma_sa_ra_s\right),\label{eq:generalized_qaoa_state_single_z_insertion_dot_product_path_integral}
\end{align}
where the sum is taken over $(2p + 1)$-bit bitstrings
\begin{align}
    \bm{a} & = \left(a_1, a_2, \ldots, a_{p - 1}, a_p, a_0, a_{-p}, a_{-(p - 1)}, \ldots, a_{-2}, a_{-1}\right)\\
    & \in \{1, -1\}^{2p + 1}
\end{align}
and $f\left(\bm{a}\right)$ is the two-sided $f$ tensor (Definition \ref{def:f_tensor_one_sided}) evaluated at bitstring $\bm{a}$:
\begin{align}
    f\left(\bm{a}\right) & = \frac{1}{2}\bra{a_p}e^{i\beta_pX}\ket{a_0}\bra{a_0}e^{-i\beta_pX}\ket{a_{-p}}\prod_{1 \leq t < p}\bra{a_t}e^{i\beta_tX}\ket{a_{t + 1}}\bra{a_{-t - 1}}e^{-i\beta_tX}\ket{a_{-t}}.
\end{align}
\begin{proof}
In order to apply Lemma \ref{lemma:product_displacement_operators_spin_path}, more specifically equation \ref{eq:displacement_operator_spin_path_integral_vacuum_action_g_sym}, we rewrite the time autocorrelations between times $j, k$ ($1 \leq j \leq k \leq p$) as the dot product of two vectors:
\begin{align*}
    & \bra{+}\bra{0}^{\otimes p'}\left\{\overrightarrow{\prod_{t = 1}^p}D\left(i\gamma_t\bm{L}_{:,\,t}Z\right)\exp\left(i\beta_tX\right)\right\}\left\{\overleftarrow{\prod_{t = k}^{p}}\exp\left(-i\beta_tX\right)D\left(-i\gamma_t\bm{L}_{:,\,t}Z\right)\right\}\\
    & \hspace*{45px} Z\left\{\overleftarrow{\prod_{t = j}^{k - 1}}\exp\left(-i\beta_tX\right)D\left(-i\gamma_t\bm{L}_{:,\,t}Z\right)\right\}Z\left\{\overleftarrow{\prod_{t = 1}^{j - 1}}\exp\left(-i\beta_tX\right)D\left(-i\gamma_t\bm{L}_{:,\,t}Z\right)\right\}\ket{+}\ket{0}^{\otimes p'}\\
    & = \left(\overleftarrow{\prod_{t = 1}^p}\exp\left(-i\beta_tX\right)D\left(-i\gamma_t\bm{L}_{:,\,t}Z\right)\ket{+}\ket{0}^{\otimes p'}\right)^{\dagger}\left\{\overleftarrow{\prod_{t = k}^{p}}\exp\left(-i\beta_tX\right)D\left(-i\gamma_t\bm{L}_{:,\,t}Z\right)\right\}\nonumber\\
    & \hspace*{40px} \times Z\left\{\overleftarrow{\prod_{t = j}^{k - 1}}\exp\left(-i\beta_tX\right)D\left(-i\gamma_t\bm{L}_{:,\,t}Z\right)\right\}Z\left\{\overleftarrow{\prod_{t = 1}^{j - 1}}\exp\left(-i\beta_tX\right)D\left(-i\gamma_t\bm{L}_{:,\,t}Z\right)\right\}\ket{+}\ket{0}^{\otimes p'}
\end{align*}
Note we could have truncated the product to layer $k$ rather than $p$. We now decompose each of the statevectors as a path integral. For the daggered vector:
\begin{align*}
    & \overleftarrow{\prod_{t = 1}^p}\exp\left(-i\beta_tX\right)D\left(-i\gamma_t\bm{L}_{:,\,t}Z\right)\ket{+}\ket{0}^{\otimes p'}\\
    & = \sum_{\substack{\bm{a}_{1:p + 1} \in \{1, -1\}^{p + 1}}}\ket{a_{p + 1}}\bra{a_{p + 1}}\left\{\overleftarrow{\prod_{t = 1}^p}\exp\left(-i\beta_tX\right)D\left(-i\gamma_t\bm{L}_{:,\,t}Z\right)\ket{a_{t}}\bra{a_{t}}\right\}\ket{+}\ket{0}^{\otimes p'}\\
    & = \sum_{\bm{a}_{1:p + 1} \in \{1, -1\}^{p + 1}}\underbrace{\frac{1}{\sqrt{2}}\left(\prod_{t = 1}^p\bra{a_{t + 1}}\exp\left(-i\beta_tX\right)\ket{a_{t}}\right)}_{= f_{1:p + 1}(\bm{a}_{-1:-p - 1})}\ket{a_{p + 1}} \otimes \left(\overleftarrow{\prod_{t = 1}^p}D\left(-i\gamma_t\bm{L}_{:,\,t}a_{t}\right)\right)\ket{0}^{\otimes p'}\\
    & = \sum_{\bm{a}_{1:p + 1} \in \{1, -1\}^{p + 1}}f_{1:p + 1}(\bm{a}_{1:p + 1})\exp\left(-\frac{1}{2}\sum_{1 \leq r, s \leq p}G^{(\mathrm{sym})}_{r, s}\gamma_r\gamma_ra_{r}a_{s}\right)\ket{a_{p + 1}} \otimes \bigotimes_{l = 1}^{p'}\left\{\exp\left(-i\gamma_t\hat{c}_l^{\dagger}\sum\limits_{1 \leq t \leq p}L_{l, t}\gamma_ta_t\right)\ket{0}\right\},
\end{align*}
where in the last step we used equation \ref{eq:displacement_operator_spin_path_integral_vacuum_action_g_sym} from Lemma \ref{lemma:product_displacement_operators_spin_path}. The path integral expansion of the non-daggered vector is very similar ---the integrand only picking an extra factor $a_{-j}a_{-k}$ due to the inserted $Z$ operators:
\begin{align*}
    & \left\{\overleftarrow{\prod_{t = k}^{p}}\exp\left(-i\beta_tX\right)D\left(-i\gamma_t\bm{L}_{:,\,t}Z\right)\right\}Z\left\{\overleftarrow{\prod_{t = j}^{k - 1}}\exp\left(-i\beta_tX\right)D\left(-i\gamma_t\bm{L}_{:,\,t}Z\right)\right\}Z\left\{\overleftarrow{\prod_{t = 1}^{j - 1}}\exp\left(-i\beta_tX\right)D\left(-i\gamma_t\bm{L}_{:,\,t}Z\right)\right\}\ket{+}\ket{0}^{\otimes p'}\\
    & = \sum_{\bm{a}_{-1:-p - 1} \in \{1, -1\}^{p + 1}}a_{-j}a_{-k}f_{1:p + 1}(\bm{a}_{-1:-p - 1})\exp\left(-\frac{1}{2}\sum_{1 \leq r, s \leq p}G^{(\mathrm{sym})}_{r, s}\gamma_r\gamma_ra_{-r}a_{-s}\right)\nonumber\\
    & \hspace*{130px} \times \ket{a_{-p - 1}} \otimes \bigotimes_{l = 1}^{p'}\left\{\exp\left(-i\gamma_t\hat{c}_l^{\dagger}\sum\limits_{1 \leq t \leq p}L_{l, t}\gamma_ta_{-t}\right)\ket{0}\right\}
\end{align*}
One can now express the desired dot product as follows from the path integrals expansion:
\begin{align*}
    \braket{\Psi_k}{\Psi_j} & = \sum_{\substack{\bm{a}_{-1:-p - 1} \in \{1, -1\}^{p + 1}\\\bm{a}_{1:p + 1} \in \{1, -1\}^{p + 1}}}\hspace*{-10px}a_{-j}a_{-k}f_{1:p + 1}\left(\bm{a}_{-1:-p - 1}\right)\exp\left(-\frac{1}{2}\sum_{1 \leq r, s \leq p}G^{(\mathrm{sym})}_{r, s}\gamma_r\gamma_sa_{-r}a_{-s}\right)\nonumber\\
    & \hspace*{100px} \times \overline{f_{1:p + 1}(\bm{a}_{1:p + 1})}\exp\left(-\frac{1}{2}\sum_{1 \leq r, s \leq p}\overline{G^{(\mathrm{sym})}_{r, s}}\gamma_r\gamma_sa_ra_s\right)\\
    & \hspace*{100px} \times \mathbf{1}\left[a_{-p - 1} = a_{p + 1}\right]\prod_{1 \leq l \leq p'}\bra{0}\exp\left(i\hat{c}_l\sum_{1 \leq t \leq p}\overline{L_{l, t}}\gamma_ta_t\right)\exp\left(-i\hat{c}^{\dagger}_l\sum_{1 \leq t \leq p}L_{l, t}\gamma_ta_{-t}\right)\ket{0}
\end{align*}
Using constraint $a_{-p - 1} = a_{p + 1}$, one may let $a_0 := a_{p + 1} = a_{-p - 1}$ and replace the summation over $\bm{a}_{1:p + 1}, \bm{a}_{-1:-p - 1} \in \{1, -1\}^{p + 1}$ by one over
\begin{align}
    \bm{a} & := \left(a_1, a_2, \ldots, a_{p - 1}, a_p, a_0, a_{-p}, a_{-(p - 1)}, \ldots, a_{-2}, a_{-1}\right)  \in \{1, -1\}^{2p + 1}.
\end{align}
From this notation, the two one-sided $f$ tensors combine to a double-sided $f$ tensor (Definition \ref{def:f_tensor_one_sided}):
\begin{align}
    \overline{f_{1:p + 1}\left(\bm{a}_{1:p + 1}\right)}f_{1:p + 1}\left(\bm{a}_{-1:-p - 1}\right) & = f\left(\bm{a}\right).
\end{align}
Next, we evaluate the matrix elements of exponentiated creation operators:
\begin{align}
    & \prod_{1 \leq l \leq p'}\bra{0}\exp\left(i\hat{c}_l\sum_{1 \leq t \leq p}\overline{L_{l, t}}\gamma_ta_t\right)\exp\left(-i\hat{c}_l^{\dagger}\sum_{1 \leq t \leq p}L_{l, t}\gamma_ta_{-t}\right)\ket{0}\nonumber\\
    & = \prod_{1 \leq l \leq p'}\exp\left(\left(i\sum_{1 \leq t \leq p}\overline{L_{l, t}}\gamma_ta_t\right)\left(-i\sum_{1 \leq t \leq p}L_{l, t}\gamma_ta_{-t}\right)\right)\nonumber\\
    & = \exp\left(\sum_{1 \leq r, s \leq p}\left(\underbrace{\sum_{1 \leq l \leq p'}\overline{L_{l ,r}}L_{l, s}}_{= G^{(\mathrm{herm})}_{r, s}}\right)\gamma_r\gamma_sa_ra_{-s}\right)\nonumber\\
    & = \exp\left(\sum_{1 \leq r, s \leq p}G^{(\mathrm{herm})}_{r, s}\gamma_r\gamma_sa_ra_{-s}\right).
\end{align}
It follows:
\begin{align}
    \braket{\Psi_k}{\Psi_j} & = \sum_{\bm{a} \in \{1, -1\}^{p + 1}}a_{-k}a_{-j}f\left(\bm{a}\right)\exp\left(-\frac{1}{2}\sum_{1 \leq r, s \leq p}G^{\left(\mathrm{sym}\right)}_{r, s}\gamma_r\gamma_sa_{-r}a_{-s} - \frac{1}{2}\sum_{1 \leq r, s \leq p}\overline{G^{\left(\mathrm{sym}\right)}_{r, s}}\gamma_r\gamma_sa_ra_s\right)\nonumber\\
    & \hspace*{105px} \times \exp\left(\sum_{1 \leq r, s \leq p}G^{\left(\mathrm{herm}\right)}_{r, s}\gamma_r\gamma_sa_ra_{-s}\right)
\end{align}
Let us now define a matrix
\begin{align}
    \bm{G} & = \left(G_{r, s}\right)_{r, s \in \pm [p]}
\end{align}
by the following 4 blocks:
\begin{align}
    G_{r, s} & := \overline{G^{\left(\mathrm{sym}\right)}_{r, s}},\\
    G_{-r, -s} & := G^{\left(\mathrm{sym}\right)}_{r, s},\\
    G_{-r, s} & := G^{\left(\mathrm{herm}\right)}_{s, r},\\
    G_{s, -r} & := G^{\left(\mathrm{herm}\right)}_{s, r},
\end{align}
where $r, s \in [p]$ in the 4 equations. Note $\bm{G}$ is symmetric by construction. Finally defining 
\begin{align}
    \bm{\Gamma} & := \left(\Gamma_1, \Gamma_1, \ldots, \Gamma_{p - 1}, \Gamma_p, \Gamma_{-p}, \Gamma_{-(p - 1)}, \ldots, \Gamma_{-2}, \Gamma_{-1}\right)\\
    & := \left(\gamma_1, \gamma_2, \ldots, \gamma_{p - 1}, \gamma_p, -\gamma_p, -\gamma_{p - 1}, \ldots, -\gamma_2, -\gamma_1\right),
\end{align}
the product of exponentials in the path integral expansion of $\braket{\Psi_k}{\Psi_j}$ can be rewritten:
\begin{align}
    & \exp\left(-\frac{1}{2}\sum_{1 \leq r, s \leq p}G^{\left(\mathrm{sym}\right)}_{r, s}\gamma_r\gamma_sa_{-r}a_{-s} - \frac{1}{2}\sum_{1 \leq r, s \leq p}\overline{G^{\left(\mathrm{sym}\right)}_{r, s}}\gamma_r\gamma_sa_ra_s + \sum_{1 \leq r, s \leq p}G^{\left(\mathrm{herm}\right)}_{r, s}\gamma_r\gamma_sa_ra_{-s}\right)\nonumber\\
    & = \exp\left(-\frac{1}{2}\sum_{r, s \in \pm [p]}G_{r, s}\Gamma_r\Gamma_sa_ra_s\right).
\end{align}
Hence, 
\begin{align}
    \braket{\Psi_k}{\Psi_j} & = \sum_{\bm{a} \in \{1, -1\}^{2p + 1}}a_{-j}a_{-k}f\left(\bm{a}\right)\exp\left(-\frac{1}{2}\sum_{r, s \in \pm [p]}G_{r, s}\Gamma_r\Gamma_sa_ra_s\right).
\end{align}
\end{proof}
\end{repproposition}

Summarizing, Proposition \ref{spinBosonsStateDotProductPathIntegral} expresses the dot product of states $\ket{\Psi_j}$ as a sum over bitstrings (equation \ref{eq:generalized_qaoa_state_single_z_insertion_dot_product_path_integral}) identical to the SK-QAOA $G$ matrix iteration (equation \ref{eq:Gm}). The difference with the SK-QAOA iteration is, in this section, the $\bm{G}$ matrix on the right-hand side of equation \ref{eq:generalized_qaoa_state_single_z_insertion_dot_product_path_integral} is not assumed to proceed from a previous SK-QAOA iteration (i.e., to be some recursively computed $\bm{G}^{(m)}$), but may be any matrix satisfying the symmetries of the SK-QAOA $G$ matrix (equation \ref{eq:Gsim}) and specified by a non-negative Hermitian corner $\bm{G}^{\mathrm{herm}}$ as per equations 
\ref{eq:g_from_gherm_plus_plus}--\ref{eq:g_from_gherm_plus_minus}.

\section{An analytic estimate of the truncated Fock space dimension}
\label{sec:fock_space_truncation_analytic}

Approximate numerical simulation of the spin-bosons system constructed in Appendix~\ref{sec:sk_qaoa_iteration_spin_bosons} is enabled by a truncation of the Fock space to a modest dimension. In the current Appendix, we provide analytic estimates of the required Fock space dimension. We leave the practical tightness of these estimates to future numerical explorations. Besides, we note the current section's analysis fall short of establishing the system has low entanglement, another crucial ingredient enabling the MPS simulation.

\subsection{Proof of analytic estimate of truncation threshold}

In this Section, we establish simple estimates on the infidelity of the spin-boson states $\ket{\Phi}, \ket{\Phi_t}$ constructed in Proposition~\ref{prop:generalized_qaoa_state_path_integral} resulting from truncation of the Fock space. We show the infidelity to decrease exponentially above a Fock space dimension threshold $\mathcal{O}\left(\gamma_{\mathrm{max}}^2p^2\right)$, where $\gamma_{\mathrm{max}}$ is the maximum $\gamma$ angles. The basic idea is that $\ket{\Phi}$ or $\ket{\Phi_t}$ can be expressed as a sum over spin paths $\bm{a}_{1:p + 1} = \left(a_1, \ldots, a_{p + 1}\right)$ of coherent states of the form
\begin{align}
    \bigotimes_{l = 1}^{p'}\ket{-i\sum_{1 \leq t \leq p}L_{l, t}\gamma_ta_t}.
\end{align}
We then use the observation that for a single-mode coherent state $\ket{\alpha}$, the number distribution decreases exponentially above threshold number $|\alpha|^2$, and compound this result over $p'$ modes.

We start by proving a truncation result for a single mode (Proposition~\ref{prop:single_mode_fock_space_truncation_error}), before extending it to several modes (Proposition~\ref{prop:multi_mode_fock_space_truncation_error}). The following Lemma allows to simply relate the single and multi mode cases.

\begin{lemma}[Incoherent norm bound for application of commuting projectors]
\label{lemma:commuting_projection_incoherent_error_bound}
Let $\Pi_1, \ldots, \Pi_m$ a sequence of commuting orthogonal projectors in a Hilbert space $\mathcal{H}$. Then, for any $\ket{\psi} \in \mathcal{H}$,
\begin{align}
    \left\lVert \ket{\psi}- \Pi_1\ldots\Pi_m\ket{\psi} \right\rVert_2 & \leq \sqrt{\sum_{1 \leq j \leq m}\left\lVert \left(I - \Pi_j\right)\ket{\psi} \right\rVert_2^2}.
\end{align}
\begin{proof}
To estimate $\left\lVert \ket{\psi} - \Pi_1\ldots\Pi_m\ket{\psi} \right\rVert_2 = \left\lVert \left(I - \Pi_1\ldots\Pi_m\right)\ket{\psi} \right\rVert_2$, we express the difference between the identity and the product of projectors as a telescopic sum, replacing one projector by identity at a time:
\begin{align}
    I - \Pi_1\ldots\Pi_m & = \left(I - \Pi_1\right) + \left(\Pi_1 - \Pi_1\Pi_2\right) + \left(\Pi_1\Pi_2 - \Pi_1\Pi_2\Pi_3\right) + \left(\Pi_1\Pi_2\Pi_3 - \Pi_1\Pi_2\Pi_3\Pi_4\right) + \ldots\nonumber\\
    & = \left(I - \Pi_1\right) + \Pi_1\left(I - \Pi_2\right) + \Pi_1\Pi_2\left(I - \Pi_3\right) + \Pi_1\Pi_2\Pi_3\left(I - \Pi_4\right) + \ldots\nonumber\\
    & = \sum_{1 \leq j \leq m}\left(I - \Pi_j\right)\prod_{1 \leq k < j}\Pi_k.
\end{align}
Applying this additive decomposition to a vector $\ket{\psi}$ provides an orthogonal decomposition of $\ket{\psi}$. Hence, the terms add incoherently:
\begin{align}
    \left\lVert \left(I - \Pi_1\ldots\Pi_m\right)\ket{\psi} \right\rVert_2 & = \left\lVert \sum_{1 \leq j \leq m}\left(\prod_{1 \leq k < j}\Pi_k\right)\left(I - \Pi_j\right)\ket{\psi} \right\rVert_2\nonumber\\
    & \leq \sqrt{\sum_{1 \leq j \leq m}\left\lVert \left(\prod_{1 \leq k < j}\Pi_k\right)\left(I - \Pi_j\right)\ket{\psi} \right\rVert_2^2}\nonumber\\
    & \leq \sqrt{\sum_{1 \leq j \leq m}\left\lVert \left(I - \Pi_j\right)\ket{\psi} \right\rVert_2^2},
\end{align}
where the last inequality uses that orthogonal projectors reduce the norm.
\end{proof}
\end{lemma}

We now show an explicit truncation bound for the single-mode case:

\begin{proposition}[Error for single-mode Fock space truncation]
\label{prop:single_mode_fock_space_truncation_error}
For any mode index $l \in [p']$, and any integer $n \geq 0$, let $\Pi_{l, n}$ orthogonally project mode $l$ onto number state $n$. Besides, given some integer $d$, let $\Pi_{l,\,\leq d}$ orthogonally project mode $l$ onto the states of number at most $d$. Then for any integer $d_l \geq 0$ and truncation number 
\begin{align}
    d \geq ed_{*, l} + d_l,
\end{align}
where
\begin{align}
    d_{*, l} := \left(\sum_{1 \leq t \leq p}|L_{l, t}||\gamma_t|\right)^2,
\end{align}
the following Fock space truncation bound holds for the generalized QAOA state $\ket{\Phi}$:
\begin{align}
    \left\lVert \ket{\Phi} - \Pi_{l, \leq d}\ket{\Phi} \right\rVert_2 & \leq 2e^{-d_l}. 
\end{align}
The same error bound holds for the states with a single $Z$ insertion:
\begin{align}
    \left\lVert \ket{\Phi_t} - \Pi_{l, \leq d}\ket{\Phi_t} \right\rVert_2 & \leq 2e^{-d_l}.
\end{align}
\begin{proof}
We decompose the error by summing over all boson numbers $n > d$:
\begin{align}
    \left\lVert \ket{\Phi} - \Pi_{l, \leq d}\ket{\Phi} \right\rVert_2^2 & = \sum_{n > d}\left\lVert \Pi_{l, n}\ket{\Phi} \right\rVert_2^2.
\end{align}
Recalling the path integral representation of $\Phi$ (Proposition~\ref{prop:generalized_qaoa_state_path_integral}):
\begin{align}
    \ket{\Phi} & = \sum_{\bm{a}_{1:p + 1} \in \{1, -1\}^{p + 1}}f_{-1:-p - 1}\left(\bm{a}_{-1:p - 1}\right)\exp\left(-\frac{1}{2}\sum_{1 \leq r, s \leq p}G^{\mathrm{(sym)}}_{r, s}\Gamma_r\Gamma_sa_{-r}a_{-s}\right)\ket{a_{-p - 1}}_{\mathbb{C}^2} \otimes \bigotimes_{1 \leq m \leq p'}\left|-i\sum_{1 \leq t \leq p}L_{m, t}\gamma_ta_{-t}\right\rangle
\end{align}
and identity
\begin{align}
    \bra{n}e^{\alpha\hat{c}^{\dagger}}\ket{0} & = \frac{\alpha^n}{\sqrt{n!}}
\end{align}
for a single-mode creation operator $\hat{c}^{\dagger}$ and $\alpha \in \mathbb{C}$, we find:
\begin{align}
    {}_l\braket{n}{\Phi} & = \sum_{\bm{a} \in \{1, -1\}^{p + 1}}f_{1:p + 1}\left(\bm{a}\right)\exp\left(-\frac{1}{2}\sum_{1 \leq r, s \leq p}G^{\mathrm{(sym)}}_{r, s}\Gamma_r\Gamma_sa_{-r}a_{-s}\right)\frac{1}{\sqrt{n!}}\left(-i\sum_{1 \leq t \leq p}L_{l, t}\gamma_ta_{-t}\right)^n\nonumber\\
    & \hspace*{70px} \times \ket{a_{-p - 1}}_{\mathbb{C}^2} \otimes \bigotimes_{\substack{1 \leq m \leq p'\\m \neq l}}\left|-i\sum_{1 \leq t \leq p}L_{m, t}\gamma_ta_{-t}\right\rangle,
\end{align}
where we denoted by $\ket{n}_l$ the state of number $n$ for bosonic mode $l$. Squaring this and reasoning as in Proposition~\ref{spinBosonsStateDotProductPathIntegral} yields
\begin{align}
    \left\lVert \Pi_{l, n}\ket{\Phi} \right\rVert_2^2 & = \left\lVert {}_l\braket{n}{\Phi} \right\rVert_2^2\nonumber\\
    & = \sum_{\bm{a} \in \{1, -1\}^{2p + 1}}f\left(\bm{a}\right)\exp\left(-\frac{1}{2}\sum_{r, s \in \pm [p]}G_{r, s}\Gamma_r\Gamma_sa_ra_s\right)\frac{1}{n!}\left(i\sum_{1 \leq t \leq p}\overline{L_{l, t}}\gamma_ta_t\right)^n\left(-i\sum_{1 \leq t \leq p}L_{l, t}\gamma_ta_{-t}\right)^n,
\end{align}
with $\bm{G} = \left(G_{r, s}\right)_{r, s \in \pm [p]}$ defined by Eqns.~\ref{eq:g_from_gherm_plus_plus}-\ref{eq:g_from_gherm_plus_minus} from the Proposition.

To bound this, we imagine the functions of bits $\bm{a}_t$ ($t \in \pm [p]$) raised to the power $n$ as a Boolean polynomial in these $\pm 1$-valued variables:
\begin{align}
    \frac{1}{n!}\left(i\sum_{1 \leq t \leq p}\overline{L_{l, t}}\gamma_ta_t\right)^n\left(-i\sum_{1 \leq t \leq p}L_{l, t}\gamma_ta_{-t}\right)^n & = \sum_{S \subset \pm [p]}\hat{c}_S\prod_{t \in S}a_t,
\end{align}
where the precise expression of coefficients $\hat{c}_S$ is unimportant. By linearity, its then sufficient to bound
\begin{align}
    \sum_{\bm{a} \in \{1, -1\}^{2p + 1}}f\left(\bm{a}\right)\exp\left(-\frac{1}{2}\sum_{r, s \in \pm [p]}G_{r, s}\Gamma_r\Gamma_sa_ra_s\right)\prod_{t \in S}a_t
\end{align}
for all $S \subset \pm [p]$. But from the interpretation of 
\begin{align}
    f\left(\bm{a}\right)\exp\left(-\frac{1}{2}\sum_{r, s \in \pm [p]}G_{r, s}\Gamma_r\Gamma_sa_ra_s\right)
\end{align}
as the path integral measure of the spin-bosons system (Proposition~\ref{spinBosonsStateDotProductPathIntegral}), the above can be interpreted as a time autocorrelation of the $Z$ operator. That is, a matrix element of the spin-bosons time evolution with $Z$ insertions. For instance, taking $p = 3$ and $S = \{1, 2, -2\}$ for definiteness,
\begin{align}
    \sum_{\bm{a} \in \{1, -1\}^{7}}f\left(\bm{a}\right)\exp\left(-\frac{1}{2}\sum_{r, s \in \pm [3]}G_{r, s}\Gamma_r\Gamma_sa_ra_s\right)a_1a_2a_{-1} & = \bra{0}_{\mathbb{C}^2}\bra{0}_{\mathbf{L}_2\left(\mathbf{R}\right)}^{\otimes p'}U_1^{\dagger}ZU_2^{\dagger}U_3^{\dagger}U_3U_2ZU_1Z\ket{0}_{\mathbb{C}^2}\ket{0}_{\mathbf{L}_2\left(\mathbf{R}\right)}^{\otimes p'},
\end{align}
where we denoted spin-bosons unitaries by $U_t := e^{-i\beta_tX}D\left(-i\gamma_tZ\bm{L}_{:,\,t}\right)$. Since $\left\lVert Z \right\rVert \leq 1$, any such sum is then upper-bounded by $1$. As a result, by the triangle inequality:
\begin{align}
    & \left|\sum_{\bm{a} \in \{1, -1\}^{2p + 1}}f\left(\bm{a}\right)\exp\left(-\frac{1}{2}\sum_{r, s \in \pm [p]}G_{r, s}\Gamma_r\Gamma_sa_ra_s\right)\frac{1}{n!}\left(i\sum_{1 \leq t \leq p}\overline{L_{l, t}}\gamma_ta_t\right)^n\left(-i\sum_{1 \leq t \leq p}L_{l, t}\gamma_ta_{-t}\right)^n\right|\nonumber\\
    & = \left|\sum_{S \subset \pm [p]}\hat{c}_S\sum_{\bm{a} \in \{1, -1\}^{2p + 1}}f\left(\bm{a}\right)\exp\left(-\frac{1}{2}\sum_{r, s \in \pm [p]}G_{r, s}\Gamma_r\Gamma_sa_ra_s\right)\prod_{t \in S}a_t\right|\nonumber\\
    & \leq \sum_{S \subset \pm [p]}\left|\hat{c}_S\right|\left|\sum_{\bm{a} \in \{1, -1\}^{2p + 1}}f\left(\bm{a}\right)\exp\left(-\frac{1}{2}\sum_{r, s \in \pm [p]}G_{r, s}\Gamma_r\Gamma_sa_ra_s\right)\prod_{t \in S}a_t\right|\nonumber\\
    & \leq \sum_{S \subset \pm [p]}\left|\hat{c}_S\right|.
\end{align}
Next, it is easily seen that for all $S \subset \pm [p]$, $\left|\hat{c}_S\right|$ is upper-bounded by Boolean coefficient $S$ of function
\begin{align}
    \frac{1}{n!}\left(\sum_{1 \leq t \leq p}\left|L_{l, t}\right||\gamma_t|a_t\right)^n\left(\sum_{1 \leq t \leq p}\left|L_{l, t}\right||\gamma_t|a_{-t}\right)^n.
\end{align}
Hence, $\sum_S\left|\hat{c}_S\right|$ is upper-bounded by the above function evaluated at $a_t := 1$ for all $t \in \pm [p]$, i.e.
\begin{align}
    \frac{1}{n!}\left(\sum_{1 \leq t \leq p}|L_{l, t}||\gamma_t|\right)^{2n}.
\end{align}
All in all, we showed 
\begin{align}
    \left\lVert \Pi_{l, n}\ket{\Phi} \right\rVert_2^2 & \leq \frac{1}{n!}\left(\sum_{1 \leq t \leq p}|L_{l, t}||\gamma_t|\right)^{2n}.
\end{align}
Letting
\begin{align}
    d_{*, l} & := \left(\sum_{1 \leq t \leq p}|L_{l, t}||\gamma_t|\right)^2,
\end{align}
one can sum this bound over
\begin{align}
    n \geq ed_{*, l} + d'_l
\end{align}
with the help of Lemma \ref{lemma:exp_taylor_coefficients_bound}. This gives:
\begin{align}
    \left\lVert \Pi_l\ket{\Phi} \right\rVert_2^2 & = \sum_{n \geq ed_{*, l} + d_l}\left\lVert \Pi_{l, n}\ket{\Phi} \right\rVert_2^2\nonumber\\
    & \leq \sum_{n \geq ed_{*, l} + d'_l}\frac{d_{*, l}^n}{n!}\nonumber\\
    & = e^{-d'_l}\sum_{n \geq ed_{*, l} + d'_l}e^{-(n - ed_{*, l} - d'_l)}\nonumber\\
    & \leq \frac{e^{-d'_l}}{1 - 1/e}\nonumber\\
    & \leq 2e^{-d'_l}.
\end{align}
The same proof carries over for states with a single $Z$ insertion $\ket{\Phi_t}$; indeed, the additional $Z$ insertion translates as a factor $a_t$ in the path integral measure, allowing the same reasoning in terms of autocorrelations.
\end{proof}
\end{proposition}
Proposition~\ref{prop:single_mode_fock_space_truncation_error} shows the infidelity decreases exponentially with the Fock space truncation dimension, provided the latter exceeds a certain threshold $d_{*, l}$. Since the columns of $\bm{L}$ are normalized, simple bound $\left|L_{l, t}\right| \leq 1$, gives 
\begin{align}
    d_{*, l} = \left(\sum_{1 \leq t \leq p}\left|L_{l, t}\right||\gamma_t|\right)^2 \leq p^2\gamma_{\mathrm{max}}^2,
\end{align}
with $\gamma_{\mathrm{max}} := \max_{1 \leq t \leq p}\left|\gamma_t\right|$ the largest $\gamma$ angle. In particular, the truncation threshold is polynomial in $p$. We leave the tightness of this bound for future investigations.

Proposition~\ref{prop:single_mode_fock_space_truncation_error}, bounding the error from truncation of a single mode, can be extended to all $p$ modes with the help of Lemma~\ref{lemma:commuting_projection_incoherent_error_bound}:

\begin{proposition}[Error for multi-modes Fock space truncation]
\label{prop:multi_mode_fock_space_truncation_error}
Let $\Pi_{\leq d} := \Pi_{1, \leq d} \ldots \Pi_{p' ,\leq d}$ project onto the number states of number at most $d$ for all modes. Then, provided $d$ satisfies:
\begin{align}
    d \geq p^2\gamma_{\mathrm{max}}^2 + d',
\end{align}
where $d' \geq 0$, the $2$-norm error from Fock space truncation can be bounded as:
\begin{align}
    \left\lVert \ket{\Phi} - \Pi_{\leq d}\ket{\Phi} \right\rVert_2 & \leq 2\sqrt{p'}e^{-d'}.
\end{align}
\begin{proof}
Invoking Lemma~\ref{lemma:commuting_projection_incoherent_error_bound},
\begin{align}
    \left\lVert \ket{\Phi} - \Pi_{\leq d}\ket{\Phi} \right\rVert_2 & \leq \left\lVert \ket{\Phi} - \Pi_{1, \leq d} \ldots \Pi_{p', \leq d}\ket{\Phi} \right\rVert_2\nonumber\\
    & \leq \sqrt{\sum_{1 \leq l \leq p'}\left\lVert \left(I - \Pi_{l, \leq d}\right)\ket{\Phi} \right\rVert_2^2}.
\end{align}
We now apply Proposition \ref{prop:single_mode_fock_space_truncation_error} using $d \geq p^2\gamma_{\mathrm{max}}^2 + d' \geq ed_{*, l} + d'$ to bound each single-mode projection error. This gives:
\begin{align}
    \left\lVert \ket{\Phi} - \Pi_{\leq d}\ket{\Phi} \right\rVert_2 & \sqrt{\sum_{1 \leq l \leq p'}\left(2e^{-d'}\right)^2}\nonumber\\
    & \leq 2\sqrt{p'}e^{-d'}.
\end{align}
\end{proof}
\end{proposition}

\subsection{Auxiliary technical results}

In this Section, we collect auxiliary technical results used in establishing the main truncation bounds Propositions~\ref{prop:single_mode_fock_space_truncation_error} and \ref{prop:multi_mode_fock_space_truncation_error}.

\begin{lemma}[Crude Stirling's formula]
\label{lemma:crude_stirling_inequality}
    For all integer $n \geq 1$,
    \begin{align}
        n! & \geq n^ne^{-n}
    \end{align}
\end{lemma}

\begin{lemma}[Simple bound on the Taylor coefficients of the exponential]
\label{lemma:exp_taylor_coefficients_bound}
    Let $x \geq 0$ an arbitrary non-negative number. Then for all integer $n$ and real $y$ satisfying
    \begin{align}
        n & \geq ex + y,\\
        y & \geq 0,
    \end{align}
    the following inequality holds:
    \begin{align}
        \frac{x^n}{n!} \leq e^{-y}.
    \end{align}
\begin{proof}
According to the simplified Stirling's formula in proposition \ref{lemma:crude_stirling_inequality},
\begin{align*}
    \frac{x^n}{n!} & \leq \left(\frac{ex}{n}\right)^n\\
    & = \exp\left(n\log\left(\frac{ex}{n}\right)\right)\\
    & = \exp\left(n + n\log x - n\log n\right)\\
    & = \exp\left(\varphi(n)\right),
\end{align*}
where we introduced function
\begin{align}
    \varphi(u) & := u + u\log x - u \log u
\end{align}
defined over the positive reals. It is strictly concave with critical point
\begin{align}
    u^* & = x.
\end{align}
This means $\varphi$ is strictly decreasing on $\left[u^*, +\infty\right)$, hence in particular on $\left[eu^*, +\infty\right)$, and further satisfies inequality
\begin{align}
    \varphi\left(eu^* + v\right) & \leq \varphi\left(eu^*\right) + v\varphi'\left(eu^*\right) \qquad \forall v \geq 0.
\end{align}
Substituting the expression of $\varphi$ and $\varphi'$, the above inequality reads:
\begin{align}
    \varphi\left(eu^* + v\right) & \leq -v
\end{align}
Hence, coming back to the bound on $x^n/n!$,
\begin{align*}
    \frac{x^n}{n!} & \leq \exp\left(\varphi(n)\right)\\
    & \leq \exp\left(\varphi(ex + y)\right)\\
    & \leq \exp\left(-y\right)
\end{align*}
as claimed.
\end{proof}
\end{lemma}

\section{Interpretation of mapping in $D$-regular MAXCUT-QAOA}
\label{sec:maxcut_mapping_interpretation}

In Appendix \ref{sec:sk_qaoa_iteration_spin_bosons}, we proved the main theoretical result of the work (Proposition \ref{spinBosonsStateDotProductPathIntegral}) by ``direct calculation", showing the spin-bosons system constructed there had $Z$ time correlations given by a sum over bitstrings generalizing the one from the $\bm{G}$ matrix iteration. This analysis was sufficient to derive the theorem as well as justify our numerical simulation techniques, indeed directly simulating the spin-bosons system without reference to the original QAOA qubits system. In this section, of exclusively fundamental interest, we show how the spin-bosons mapping naturally emerges in the $D$-regular, high-girth \maxcut-QAOA. More specifically, we construct an explicit isometry between the state produced by \maxcut-QAOA on a high-girth $D$-regular graph---or more specifically the local state produced on a tree thanks to the high girth assumption---and the spin-bosons systems.

\subsection{Basic definitions and $\mathbf{G}$ matrix as time correlations}

We begin by defining the tree on which the state lives according to our definition of \maxcut-QAOA. We name this tree the ``alternative Bethe lattice" in reference to the standard Bethe lattice ---an infinite regular tree. The difference with the Bethe lattice is that the tree has finite depth $\Delta$ in this work, and the root further has degree one less than the other non-leaf nodes: 

\begin{definition}[Alternative Bethe lattice]
\label{def:bethe_lattice}
The alternative Bethe lattice of degree $D$ and depth $\Delta$ is the depth-$\Delta$ tree with vertex set:
\begin{itemize}
    \item $\varnothing$ (root vertex);
    \item $v \in [D]^{\Delta'}$ for all $1 \leq \Delta' \leq \Delta$ (vertices of depth $\Delta'$).
\end{itemize}
We regard $\varnothing$ as the zero-length tuple and a single integer $j \in [D]$ as a length-$1$ tuple. Hence, vertices are labelled by tuple from length $0$ to $\Delta$ (inclusive) of integers $\in [D]$. The directed edge set of the alternative Bethe lattice of depth $\Delta$ is defined as follows:
\begin{itemize}
    \item the root $\varnothing$ has children $j$ for $j \in [D]$;
    \item for all $1 \leq \Delta' < \Delta$, $v \in [D]^{\Delta'}$ has children $(v, j) \in [D]^{\Delta' + 1}$ for all $j \in [D]$.
\end{itemize}
In the last statement, $(v, j)$ refers to the concatenation of tuple $u$ with $j$. The alternative Bethe lattice of depth $\Delta$ and degree $D$, defined by the previous edge set, is denoted $\mathcal{T}^{\left(\Delta,\,D\right)} := \left(\mathcal{V}^{\left(\Delta,\,D\right)}, \mathcal{E}^{\left(\Delta,\,D\right)}\right)$, with $\mathcal{V}^{\left(\Delta,\,D\right)}$ referring to the vertex set and $\mathcal{E}^{\left(\Delta,\,D\right)}$ to the edge set.
\end{definition}

An advantage of the tuple notation for vertices $\mathcal{V}^{\left(\Delta,\,D\right)}$ of tree $\mathcal{T}^{\left(\Delta,\,D\right)}$ is, parent-child relationships can be captured by the following simple partial order on tuples:

\begin{definition}[Partial ordering of variable-length tuples]
Given variable-length tuples $u, v$, we denote $u \prec v$ to signify $v$ is the concatenation of $u$ and an arbitrary number of elements (at least one). We also denote $u \preceq v$ to signify $u \prec v$ or $u = v$. Naturally, we also let $u \succ v$ to mean $v \prec u$ and $u \succeq v$ to mean $v \preceq u$. $\prec$ and $\succ$ (resp. $\preceq$ and $\succeq$) thus specified define partial orderings (resp. strict partial orderings) of variable-length tuples.
\end{definition}

When applied to the variable-length tuples $u, v \in \mathcal{V}^{\left(\Delta,\,D\right)}$ labeling nodes of $\mathcal{T}^{\left(\Delta,\,D\right)}$ (definition \ref{def:bethe_lattice}), $u \prec v$ means $v$ is a descendant of $u$ in $\mathcal{T}^{\left(\Delta,\,D\right)}$. Next, we introduce further notation for subtrees of $\mathcal{T}^{\left(\Delta,\,D\right)}$, which we will need to consider when rephrasing iterate $\bm{G}^{(m)}$ (equation \ref{eq:Gm}) as the correlation of some quantum operator on a tree lattice of depth $m$: 

\begin{definition}[Subtrees of the alternative Bethe lattice]
\label{def:bethe_lattice_subtree}
Consider definition \ref{def:bethe_lattice} of the alternative Bethe lattice $\mathcal{T}^{\left(\Delta,\,D\right)}$ of depth $\Delta$ and degree $D$. For any node of this tree $t \in \mathcal{V}^{\left(\Delta,\,D\right)}$, we define $\mathcal{T}^{\left(\Delta,\,D\right)}_{\succeq t}$ as the subtree of $\mathcal{T}^{\left(\Delta,\,D\right)}$ rooted at $t$, including $t$ and its descendants. We denote by $\mathcal{V}^{\left(\Delta,\,D\right)}_{\succeq t}$ and $\mathcal{E}^{\left(\Delta,\,D\right)}_{\succeq t}$ the corresponding vertex and directed edge set of this subgraph.
\end{definition}

Having introduced the tree lattice on which the QAOA state lives, let us then define the QAOA state itself. This state can be considered a faithful proxy for understanding the energy (or any other local observable) of QAOA on high-girth regular graphs due to the finite range of QAOA correlations (see e.g., \cite{basso_et_al:LIPIcs.TQC.2022.7}).

\begin{definition}[MAXCUT-QAOA on depth-$\Delta$ tree]
\label{def:qaoa_state}
The cost Hamiltonian is:
\begin{align}
    C^{(\Delta,\,D)} & := \frac{1}{\sqrt{D}}\sum_{\left(v,\,(v,\,j)\right) \in \mathcal{E}^{\left(\Delta,\,D\right)}}Z_vZ_{(v,\,j)}.
\end{align}
The mixer Hamiltonian is:
\begin{align}
    B^{(\Delta,\,D)} & := \sum_{u \in \mathcal{V}^{\left(\Delta,\,D\right)}}X_u.
\end{align}
The QAOA state with angles
\begin{align}
    \bm\gamma \in \mathbf{R}^p, \quad \bm\beta \in \mathbf{R}^p,
\end{align}
is
\begin{align}
\label{eq:qaoa_state}
    \ket{\Psi^{(\Delta,\,D)}(\bm\gamma, \bm\beta)} & := \left(\overleftarrow{\prod_{t = 1}^p}e^{-i\beta_tB^{\left(\Delta,\,D\right)}}e^{-i\gamma_tC^{\left(\Delta,\,D\right)}}\right)\ket{+}_{\mathcal{V}^{\left(\Delta,\,D\right)}}
\end{align}
Consistent with the notation from definition \ref{def:bethe_lattice_subtree}, we denote by 
\begin{align}
    C^{\left(\Delta,\,D\right)}_{\succeq u} & := \frac{1}{\sqrt{D}}\sum_{\left(v,\,(v,\,j)\right) \in \mathcal{E}^{\left(\Delta,\,D\right)}_{\succeq u}}Z_vZ_{(v,\,j)},\\
    B^{\left(\Delta,\,D\right)}_{\succeq u} & := \sum_{v \in \mathcal{V}^{\left(\Delta,\,D\right)}_{\succeq u}}X_v
\end{align}
the restrictions of the cost and mixer Hamiltonians to the subtree $\mathcal{T}^{\left(\Delta,\,D\right)}_{\succeq u}$ of $\mathcal{T}^{\left(\Delta,\,D\right)}$ rooted at $u$.
\end{definition}

From the QAOA cost and mixer Hamiltonians introduced in definition \ref{def:qaoa_state}, we also define a variant of the QAOA state with a single insertion of a $Z$ operator acting on the lattice root. The definition is similar to the spin-bosons states with a single $Z$ insertion introduced in equation \ref{eq:qaoa_state_z_inserted}. We will show these are indeed the same state under the isometry constructed in this appendix:

\begin{definition}[QAOA states with $Z_{\varnothing}$ inserted]
\label{def:qaoa_state_z_inserted}
We define
\begin{align}
\label{eq:qaoa_state_z_inserted}
    \ket{\Psi^{(\Delta,\,D)}_l(\bm\gamma, \bm\beta)} & := \left(\overleftarrow{\prod_{t = l}^p}e^{-i\beta_tB^{\left(\Delta,\,D\right)}}e^{-i\gamma_tC^{\left(\Delta,\,D\right)}}\right)Z_{\varnothing}\left(\overleftarrow{\prod_{t = 1}^{l - 1}}e^{-i\beta_tB^{\left(\Delta,\,D\right)}}e^{-i\gamma_tC^{\left(\Delta,\,D\right)}}\right)\ket{+}
\end{align}
the QAOA variational state modified with a single $Z_{\varnothing}$ inserted before layer $l$ of the ansatz.
\end{definition}

We will then need to introduce a variant of the $f\left(\bm{a}\right)$ ($\bm{a} \in \{1, -1\}^{2p + 1}$) tensor occurring in the SK-QAOA iteration (equation \ref{eq:Gm}). We call this tensor $f_{1:p}\left(\bm{a}_{1:p + 1}\right)$ ($\bm{a}_{1:p + 1} \in \{1, -1\}^{p + 1}$) the \textit{one-sided} $f\left(\bm{a}\right)$ tensor as the formula only involves one half of that of $f\left(\bm{a}\right)$. The intuition for this object is that $f\left(\bm{a}\right)$ occurs in the evaluation of a quantum expectation (the QAOA energy evaluated by the iteration), whereas $f_{1:p + 1}\left(\bm{a}\right)$ occurs in the expression of states, hence the half tensor degree:

\begin{definition}[$f$ tensor]
\label{def:f_tensor_one_sided}
For all $(p + 1)$-bits bitstring
\begin{align}
    \bm{a} = \bm{a}_{1:p + 1} = \left(a_1, \ldots, a_{p + 1}\right) & \in \{1, -1\}^{p + 1},
\end{align}
we define the \textit{one-sided $f$ tensor} as:
\begin{align}
    f_{1:p + 1}(\bm{a}_{1:p + 1}) & := f_{1:p + 1}\left(a_1, a_2, \ldots, a_p, a_{p + 1}\right)\\
    & := \frac{1}{\sqrt{2}}\prod_{1 \leq t \leq p}\bra{a_{t + 1}}e^{-i\beta_tX}\ket{a_t}\label{eq:f_tensor_one_sided}
\end{align}
For all $(2p + 1)$-bits bitstring
\begin{align}
    \bm{a} = \frac{1}{2}\left(a_1, a_2, \ldots, a_{p - 1}, a_p, a_0, a_{-p}, a_{-(p - 1)}, \ldots, a_{-2}, a_{-1}\right) \in \{1, -1\}^{2p + 1},
\end{align}
we define the \textit{two-sided $f$ tensor} by:
\begin{align}
    f\left(\bm{a}\right) & = \bra{a_p}e^{i\beta_pX}\ket{a_0}\bra{a_0}e^{-i\beta_pX}\ket{a_{-p}}\prod_{1 \leq t \leq p - 1}\bra{a_t}e^{i\beta_tX}\ket{a_{t + 1}}\bra{a_{-t - 1}}e^{-i\beta_tX}\ket{a_{-t}}\\
    & = f_{1:p + 1}\left(a_{-1}, a_{-2}, \ldots, a_{-(p - 1)}, a_{-p}, a_0\right)\overline{f_{1:p + 1}\left(a_1, a_2, \ldots, a_{p - 1}, a_p, a_0\right)}.\label{eq:f_tensor_two_sided}
\end{align}
\end{definition}

The next two definitions introduce matrices named ``$\bm{G}$ matrices" (the symmetric and Hermitian $\bm{G}$ matrices respectively). Importantly, unlike in equation \ref{eq:Gm}, we do not define these as arising from the SK-QAOA iteration in the $D \to \infty$ limit, but rather as the time autocorrelations of some operator at all finite $D$. A first result from this appendix will then show they converge to the usual $\bm{G}$ matrices in the limit $D \to \infty$. This result can be compared to earlier work Ref.~\cite{leo2022Autocorrelation}, which proved SK-QAOA iterates $G^{(m)}$ identified to time autocorrelations in the $D \to \infty$ limit. The present proof can be seen as taking the opposite direction, showing that time autocorrelations at finite $D$ converge to the SK-QAOA iterates in the $D \to \infty$ limit. 

\begin{definition}[Symmetric $G$ matrix $\bm{G}^{(\mathrm{sym},\,\Delta)}$]
\label{def:symmetric_g}
We let $\bm{G}^{(\mathrm{sym},\,\Delta,\,D)}$ the symmetric matrix with entry $(j, k)$, $j \leq k$, defined by the two-times autocorrelation of $Z_{\varnothing}$ on the alternative Bethe lattice of depth $\Delta$, where $Z_{\varnothing}$ is inserted at positions $j, k$ of unitary evolution, i.e.:
\begin{align}
    G^{(\mathrm{sym},\,\Delta,\,D)}_{j, k} & := \bra{+}\left(\overrightarrow{\prod_{l = 1}^p}e^{i\gamma_lC^{\left(\Delta,\,D\right)}}e^{i\beta_lB^{\left(\Delta,\,D\right)}}\right)\left(\overleftarrow{\prod_{l = k}^p}e^{-i\beta_lB^{\left(\Delta,\,D\right)}}e^{-i\gamma_lC^{\left(\Delta,\,D\right)}}\right)Z_{\varnothing}\nonumber\\
    & \hspace*{30px} \times \left(\overleftarrow{\prod_{l = j}^{k - 1}}e^{-i\beta_lB^{\left(\Delta,\,D\right)}}e^{-i\gamma_lC^{\left(\Delta,\,D\right)}}\right)Z_{\varnothing}\left(\prod_{l = 1}^{j - 1}e^{-i\beta_lB^{\left(\Delta,\,D\right)}}e^{-i\gamma_lC^{\left(\Delta,\,D\right)}}\right)\ket{+}
\end{align}
\end{definition}

\begin{definition}[Hermitian $G$ matrix $\bm{G}^{(\mathrm{herm},\,\Delta)}$]
\label{def:hermitian_g}
We let $\bm{G}^{(\mathrm{herm},\,\Delta)}$ the Gram matrix of the family of vectors introduced in definition \ref{def:qaoa_state_z_inserted}, i.e.
\begin{align}
    G^{(\mathrm{herm},\,\Delta,\,D)}_{j, k} & := \braket{\Psi_j^{(\Delta,\,D)}(\bm\gamma, \bm\beta)}{\Psi_k^{(\Delta,\,D)}(\bm\gamma, \bm\beta)}
\end{align}
Plugging in the explicit expressions of $\ket{\Psi_j^{(\Delta,\,D)}\left(\bm\gamma, \bm\beta\right)}, \ket{\Psi_k^{(\Delta,\,D)}\left(\bm\gamma, \bm\beta\right)}$ (definition \ref{def:qaoa_state_z_inserted}) and comparing with definition \ref{def:symmetric_g} of $\bm{G}^{(\mathrm{sym},\,\Delta,\,D)}$, we see:
\begin{align}
    G^{(\mathrm{herm},\,\Delta,\,D)}_{j, k} & = \left\{\begin{array}{cc}
        G^{(\mathrm{sym},\,\Delta,\,D)}_{j, k} & \textrm{if } j \geq k\\
        \overline{G^{(\mathrm{sym},\,\Delta,\,D)}_{j, k}} &  \textrm{otherwise}
    \end{array}\right.
\end{align}
\end{definition}

Note it follows automatically from definitions \ref{def:symmetric_g}, \ref{def:hermitian_g}, that the diagonal elements of $\bm{G}^{\left(\mathrm{sym},\,\Delta,\,D\right)}, \bm{G}^{\left(\mathrm{herm},\,\Delta,\,D\right)}$ are $1$. A much less trivial fact is that in the limit $D \to \infty$, the off-diagonal elements can also be exactly computed, coinciding with the $\bm{G}$ matrix introduced in \cite{basso_et_al:LIPIcs.TQC.2022.7}. This is the object of the following proposition:

\begin{proposition}[Relation between $\bm{G}^{\left(\mathrm{sym},\,\Delta,\,D\right)}$ matrix and $\bm{G}$ matrix]
\label{prop:relation_g_matrix}
In the infinite-degree limit $D \to\ \infty$, $\bm{G}^{(\mathrm{sym},\,\Delta,\,D)}$ and $\bm{G}^{(\mathrm{herm},\,\Delta,\,D)}$ converge to $\bm{G}$ defined in \cite{basso_et_al:LIPIcs.TQC.2022.7}. More precisely, as the degree goes to infinity (with angles $\bm\gamma, \bm\beta$ kept constant), these matrices admit limits:
\begin{align}
    \bm{G}^{(\mathrm{sym},\,\Delta,\,D)} & \xrightarrow[D \to \infty]{} \bm{G}^{(\mathrm{sym},\,\Delta)},\\
    \bm{G}^{(\mathrm{herm},\,\Delta,\,D)} & \xrightarrow[D \to \infty]{} \bm{G}^{(\mathrm{herm},\,\Delta)},
\end{align}
and these limits are related to $G^{(\Delta)}$ (the $G$ matrix computed at iteration $\Delta \geq 0$) as follows:
\begin{align}
    \bm{G}^{(\mathrm{sym},\,\Delta)}_{j, k} & = \bm{G}^{(\Delta)}_{-j, -k} && \forall 1 \leq j, k \leq p,\\
    \bm{G}^{(\mathrm{herm},\,\Delta)}_{j, k} & = \bm{G}^{(\Delta)}_{j, -k} && \forall 1 \leq j, k \leq p.
\end{align}
\end{proposition}

The rest of the present section is dedicated to the proof of proposition \ref{prop:relation_g_matrix}.

Given the correspondence between $\bm{G}^{\left(\mathrm{sym},\,\Delta,\,D\right)}$ and $\bm{G}^{\left(\mathrm{herm},\,\Delta,\,D\right)}$ (definition \ref{def:hermitian_g}), it suffices to prove the convergence result for $\bm{G}^{\left(\mathrm{sym}\right)}$. For that purpose, we proceed by induction on the depth of the lattice $\Delta \geq 0$.

\paragraph{Base induction case: $\Delta = 0$}\mbox{}

Let us start with the base case $\Delta = 0$. In this case, the lattice consists of a single vertex and Hamiltonians act on a single qubit. For $1 \leq j < k \leq p$,
\begin{align}
    G^{\left(\mathrm{sym},\,0,\,D\right)}_{j, k} & = \bra{+}\left(\overrightarrow{\prod_{t = 1}^p}e^{i\gamma_tC^{\left(0,\,D\right)}}e^{i\beta_tB^{\left(0,\,D\right)}}\right)\left(\overleftarrow{\prod_{t = k}^p}e^{-i\beta_tB^{\left(0,\,D\right)}}e^{-i\gamma_tC^{\left(0,\,D\right)}}\right)Z_{\varnothing}\nonumber\\
    & \hspace*{30px} \times \left(\overleftarrow{\prod_{t = j}^{k - 1}}e^{-i\beta_tB^{\left(0,\,D\right)}}e^{-i\gamma_tC^{\left(0,\,D\right)}}\right)Z_{\varnothing}\left(\overleftarrow{\prod_{t = 1}^{j - 1}}e^{-i\beta_tB^{\left(0,\,D\right)}}e^{-i\gamma_tC^{\left(0,\,D\right)}}\right)\ket{+}
\end{align}
Now, at depth $\Delta = 0$, the cost and mixer Hamiltonians simplify as follows:
\begin{align}
    C^{\left(0,\,D\right)} & = 0 \qquad \textrm{(no edge in the tree)}\\
    B^{\left(0,\,D\right)} & = X_{\varnothing}
\end{align}
Hence,
\begin{align}
    G^{\left(\mathrm{sym},\,0,\,D\right)}  & = \bra{+}\left(\overrightarrow{\prod_{t = 1}^p}e^{i\beta_tX_{\varnothing}}\right)\left(\overleftarrow{\prod_{t = k}^p}e^{-i\beta_tX_{\varnothing}}\right)Z_{\varnothing}\left(\overleftarrow{\prod_{t = j}^{k - 1}}e^{-i\beta_tX_{\varnothing}}\right)Z_{\varnothing}\left(\overleftarrow{\prod_{t = 1}^{j - 1}}e^{-i\beta_tX_{\varnothing}}\right)\ket{+}\label{eq:depth_0_time_autocorrelation_braket_form}.
\end{align}
This can be simply evaluated using identities
\begin{align}
    e^{i\theta X}\ket{\pm} & = e^{\pm i\theta}\ket{\pm},\\
    Z\ket{\pm} & = \ket{\mp},
\end{align}
yielding
\begin{align}
    G^{\left(\mathrm{sym},\,1,\,D\right)}_{j,\,k} & = \exp\left(2i\sum_{j \leq t < k}\beta_t\right)
\end{align}
which is independent of $D$, hence trivially admits a limit 
\begin{align}
    G^{\left(\mathrm{sym},\,D\right)}_{j,\,k} & := \exp\left(2i\sum_{j \leq t < k}\beta_t\right)
\end{align}
as $D \to \infty$. Alternatively, one could have inserted $(2p + 1)$ computational basis completeness relations:
\begin{align}
    \sum_{z^{[l]}_{\varnothing} \in \{\pm 1\}}\ket{z^{[l]}_{\varnothing}}\bra{z^{[l]}_{\varnothing}} & = \bm{I}_{\mathbb{C}^2} \qquad \forall -p \leq l \leq p
\end{align}
inside equation \ref{eq:depth_0_time_autocorrelation_braket_form} to obtain as path integral representation of the same quantity:
\begin{align}
    & G^{\left(\mathrm{sym},\,0,\,D\right)}\nonumber\\
    & = \sum_{\substack{z^{[l]}_{\varnothing} \in \{\pm 1\}\\\forall -p \leq l \leq p}}\bra{+}\left(\overrightarrow{\prod_{t = 1}^p}\ket{z^{[t]}_{\varnothing}}\bra{z^{[t]}_{\varnothing}}e^{i\beta_t X_{\varnothing}}\right)\ket{z^{[0]}_{\varnothing}}\bra{z^{[0]}_{\varnothing}}\left(\overleftarrow{\prod_{t = k}^p}e^{-i\beta_tX_{\varnothing}}\ket{z^{[-t]}_{\varnothing}}\bra{z^{[-t]}_{\varnothing}}\right)Z_{\varnothing}\nonumber\\
    & \hspace*{80px} \times \left(\overleftarrow{\prod_{t = j}^{k - 1}}e^{-i\beta_tX_{\varnothing}}\ket{z^{[-t]}_{\varnothing}}\bra{z^{[-t]}_{\varnothing}}\right)Z_{\varnothing}\left(\overleftarrow{\prod_{t = 1}^{j - 1}}e^{-i\beta_tX}\ket{z^{[-t]}_{\varnothing}}\bra{z^{[-t]}_{\varnothing}}\right)\ket{+}\nonumber\\
    & = \sum_{\substack{z^{[l]}_{\varnothing} \in \{\pm 1\}\\\forall -p \leq l \leq p}}z^{[j]}_{\varnothing}z^{[k]}_{\varnothing}\braket{+}{z^{[1]}_{\varnothing}}\braket{z^{[-1]}_{\varnothing}}{+}\bra{z^{[p]}_{\varnothing}}e^{i\beta_pX}\ket{z^{[0]}_{\varnothing}}\bra{z^{[0]}_{\varnothing}}e^{-i\beta_pX}\ket{z^{[-p]}_{\varnothing}}\prod_{t = 1}^{p - 1}\bra{z^{[t]}_{\varnothing}}e^{i\beta_tX}\ket{z^{[t + 1]}_{\varnothing}}\bra{z^{[-t - 1]}_{\varnothing}}e^{-i\beta_tX}\ket{z^{[-t]}_{\varnothing}}\nonumber\\
    & = \sum_{\substack{z^{[l]}_{\varnothing} \in \{\pm 1\}\\\forall -p \leq l \leq p}}z^{[j]}_{\varnothing}z^{[k]}_{\varnothing}\frac{1}{2}\bra{z^{[p]}_{\varnothing}}e^{i\beta_pX}\ket{z^{[0]}_{\varnothing}}\bra{z^{[0]}_{\varnothing}}e^{-i\beta_pX}\ket{z^{[-p]}_{\varnothing}}\prod_{t = 1}^{p - 1}\bra{z^{[t]}_{\varnothing}}e^{i\beta_tX}\ket{z^{[t + 1]}_{\varnothing}}\bra{z^{[-t - 1]}_{\varnothing}}e^{-i\beta_tX}\ket{z^{[-t]}_{\varnothing}}
\end{align}
Introducing notation
\begin{align}
    \bm{z}_{\varnothing} & := \left(z^{[1]}_{\varnothing}, z^{[2]}_{\varnothing}, \ldots, z^{[p - 1]}_{\varnothing}, z^{[p]}_{\varnothing}, z^{[0]}_{\varnothing}, z^{[-p]}_{\varnothing}, z^{[-(p - 1)]}_{\varnothing}, \ldots, z^{[-2]}_{\varnothing}, z^{[-1]}_{\varnothing}\right) \in \{1, -1\}^{2p + 1}
\end{align}
for the bitstring formed by the path integral expansion bits, we recognize:
\begin{align}
    \frac{1}{2}\bra{z^{[p]}_{\varnothing}}e^{i\beta_pX}\ket{z^{[0]}_{\varnothing}}\bra{z^{[0]}_{\varnothing}}e^{-i\beta_pX}\ket{z^{[-p]}_{\varnothing}}\prod_{t = 1}^{p - 1}\bra{z^{[t]}_{\varnothing}}e^{i\beta_tX}\ket{z^{[t + 1]}_{\varnothing}}\bra{z^{[-t - 1]}_{\varnothing}}e^{-i\beta_tX}\ket{z^{[-t]}_{\varnothing}} & = f\left(\bm{z}_{\varnothing}\right),
\end{align}
where $f\left(\bm{a}\right), \bm{a} \in \{1, -1\}^{2p + 1}$ was defined in equation \ref{eq:f_tensor_definition}. Hence,
\begin{align}
    G^{\left(\mathrm{sym},\,0,\,D\right)}_{j, k} & = \sum_{\bm{z}_{\varnothing} \in \{1, -1\}^{2p + 1}}z^{[j]}_{\varnothing}z^{[k]}_{\varnothing}f\left(\bm{z}_{\varnothing}\right),
\end{align}
which coincides with the expression for $G^{(0)}_{j, k}$ given in equation \ref{eq:G0}. This concludes the initial induction step.

\paragraph{General induction step}\mbox{}

We now assume the induction statement was proven up to tree depth $\Delta$ (inclusive) and consider proving it at depth $\Delta + 1$. Let $1 \leq j < k \leq p$ and consider the following autocorrelation of $Z_{\varnothing}$:
\begin{align}
    & G^{\left(\mathrm{sym},\,\Delta + 1,\,D\right)}_{j, k}\nonumber\\
    & = \bra{+}_{\mathcal{V}^{\left(\Delta + 1,\,D\right)}}\left(\overrightarrow{\prod_{t = 1}^p}e^{i\gamma_tC^{\left(\Delta + 1,\,D\right)}}e^{i\beta_tB^{\left(\Delta + 1,\,D\right)}}\right)\left(\overleftarrow{\prod_{t = k}^p}e^{-i\beta_tB^{\left(\Delta + 1,\,D\right)}}e^{-i\gamma_tC^{\left(\Delta + 1,\,D\right)}}\right)Z_{\varnothing}\nonumber\\
    & \hspace*{60px} \times \left(\overleftarrow{\prod_{t = j}^{k - 1}}e^{-i\beta B^{\left(\Delta + 1,\,D\right)}}e^{-i\gamma_tC^{\left(\Delta + 1,\,D\right)}}\right)Z_{\varnothing}\left(\overleftarrow{\prod_{t = 1}^{j - 1}}e^{-i\beta_tB^{\left(\Delta + 1,\,D\right)}}e^{-i\gamma_tC^{\left(\Delta + 1,\,D\right)}}\right)\ket{+}_{\mathcal{V}^{\left(\Delta + 1,\,D\right)}}.
\end{align}
In the above equation, we indexed the $+$ bra and ket by $\mathcal{V}^{\left(\Delta + 1,\,D\right)}$ to indicate the $+$ state over qubits indexed by $\mathcal{V}^{\left(\Delta + 1,\,D\right)}$. In other words,
\begin{align}
    \ket{+}_{\mathcal{V}^{\left(\Delta + 1,\,D\right)}} & \simeq \ket{+}^{\otimes\left|\mathcal{V}^{\left(\Delta + 1,\,D\right)}\right|}.
\end{align}
We will prefer the left-hand side notation to the right-hand side one, as it will be more convenient to label qubits by tree nodes rather than integers. Similar to the base induction case, let us insert $(2p + 1)$ completeness relations over the computational basis state of the root vertex:
\begin{align}
    \sum_{z^{[l]}_{\varnothing} \in \{1, -1\}^{2p + 1}}\ket{z^{[l]}_{\varnothing}}\bra{z^{[l]}_{\varnothing}}_{\varnothing} \otimes \bm{I}_{\mathcal{V}^{\left(\Delta + 1,\,D\right)} - \{\varnothing\}} & = \bm{I}_{\mathcal{V}^{\left(\Delta + 1,\,D\right)}}, \qquad -p \leq l \leq p.
\end{align}
Similar to the induction's initialization, joint summation over $z_{\varnothing}^{[l]} \in \{1, -1\}$ for $-p \leq l \leq p$ will also be denoted as summation over $\bm{z}_{\varnothing} \in \{1, -1\}^{2p + 1}$, where
\begin{align}
    \bm{z}_{\varnothing} & := \left(z^{[1]}_{\varnothing}, z^{[2]}_{\varnothing}, \ldots, z^{[p - 1]}_{\varnothing}, z^{[p]}_{\varnothing}, z^{[0]}_{\varnothing}, z^{[-p]}_{\varnothing}, z^{[-(p - 1)]}_{\varnothing}, \ldots, z^{[-2]}_{\varnothing}, z^{[-1]}_{\varnothing}\right).
\end{align}
To lighten the notation, we will not write the identity matrices anymore in the following formulae. With these completeness relations inserted, the correlation of interest can be rewritten
\begin{align}
    G^{\left(\Delta + 1,\,D\right)}_{j, k} & = \sum_{\substack{z^{[t]}_{\varnothing} \in \{1, -1\}\\\forall -p \leq t \leq p}}\bra{+}_{\mathcal{V}^{\left(\Delta + 1,\,D\right)}}\nonumber\\
    & \hspace*{70px} \left(\overrightarrow{\prod_{t = 1}^p}\ket{z^{[t]}_{\varnothing}}\bra{z^{[t]}_{\varnothing}}_{\varnothing}e^{i\gamma_tC^{\left(\Delta + 1,\,D\right)}}e^{i\beta_tB^{\left(\Delta + 1,\,D\right)}}\right)\ket{z^{[0]}_{\varnothing}}\bra{z^{[0]}_{\varnothing}}_{\varnothing}\nonumber\\
    & \hspace*{70px} \times \left(\overleftarrow{\prod_{t = k}^p}e^{-i\beta_tB^{\left(\Delta + 1,\,D\right)}}e^{-i\gamma_tC^{\left(\Delta + 1,\,D\right)}}\ket{z^{[-t]}_{\varnothing}}\bra{z^{[-t]}_{\varnothing}}_{\varnothing}\right)\nonumber\\
    & \hspace*{70px} \times Z_{\varnothing}\left(\overleftarrow{\prod_{t = j}^{k - 1}}e^{-i\beta_tB^{\left(\Delta + 1,\,D\right)}}e^{-i\gamma_tC^{\left(\Delta + 1,\,D\right)}}\ket{z^{[-t]}_{\varnothing}}\bra{z^{[-t]}_{\varnothing}}_{\varnothing}\right)Z_{\varnothing}\nonumber\\
    & \hspace*{80px} \times \left(\overleftarrow{\prod_{t = 1}^{j - 1}}e^{-i\beta_tB^{\left(\Delta + 1,\,D\right)}}e^{-i\gamma_tC^{\left(\Delta + 1,\,D\right)}}\ket{z^{[-t]}_{\varnothing}}\bra{z^{[-t]}_{\varnothing}}_{\varnothing}\right)\nonumber\\
    & \hspace*{55px} \ket{+}_{\mathcal{V}^{\left(\Delta + 1,\,D\right)}}\nonumber\\
    & = \sum_{\substack{z^{[t]}_{\varnothing} \in \{1, -1\}\\\forall -p \leq t \leq p}}\braket{+}{z^{[1]}_{\varnothing}}\braket{z^{[-1]}_{\varnothing}}{+}z^{[-k]}_{\varnothing}z^{[-j]}_{\varnothing}\nonumber\\
    & \hspace*{60px} \times \bra{+}_{\mathcal{V}^{\left(\Delta + 1,\,D\right)} - \{\varnothing\}}\nonumber\\
    & \hspace*{90px} \left(\overrightarrow{\prod_{t = 1}^{p - 1}}\bra{z^{[t]}_{\varnothing}}_{\varnothing}e^{i\gamma_tC^{\left(\Delta + 1,\,D\right)}}e^{i\beta_tB^{\left(\Delta + 1,\,D\right)}}\ket{z^{[t + 1]}_{\varnothing}}_{\varnothing}\right)\nonumber\\
    & \hspace*{90px} \times \bra{z^{[p]}_{\varnothing}}_{\varnothing}e^{i\gamma_pC^{\left(\Delta + 1,\,D\right)}}e^{i\beta_tB^{\left(\Delta + 1,\,D\right)}}\ket{z^{[0]}_{\varnothing}}_{\varnothing}\nonumber\\
    & \hspace*{90px} \times  \bra{z^{[0]}_{\varnothing}}_{\varnothing}e^{-i\beta_pB^{\left(\Delta + 1,\,D\right)}}e^{-i\gamma_pC^{\left(\Delta + 1,\,D\right)}}\ket{z^{[-p]}_{\varnothing}}_{\varnothing}\nonumber\\
    & \hspace*{90px} \times \left(\overleftarrow{\prod_{t = 1}^{p - 1}}\bra{z^{[-t - 1]}_{\varnothing}}_{\varnothing}e^{-i\beta_tB^{\left(\Delta + 1,\,D\right)}}e^{-i\gamma_tC^{\left(\Delta + 1,\,D\right)}}\ket{z^{[-t]}_{\varnothing}}_{\varnothing}\right)\nonumber\\
    & \hspace*{70px} \ket{+}_{\mathcal{V}^{\left(\Delta + 1,\,D\right)} - \{\varnothing\}}
\end{align}
In the last step, we used that the $Z_{\varnothing}$ operators acts as a scalar on the inserted computational basis projectors, i.e.
\begin{align}
    Z_{\varnothing}\ket{z^{[t]}_{\varnothing}}\bra{z^{[t]}_{\varnothing}}_{\varnothing} & = z^{[t]}_{\varnothing}\ket{z^{[t]}_{\varnothing}}\bra{z^{[t]}_{\varnothing}}_{\varnothing} \qquad t = j, k.
\end{align}
We now need to evaluate the ``matrix elements" (more accurately, the partial traces) between inserted computational basis states and QAOA unitaries. Without loss of generality, let us consider
\begin{align}
    \bra{z^{[t]}_{\varnothing}}_{\varnothing}e^{i\gamma_tC^{\left(\Delta + 1,\,D\right)}}e^{i\beta_tB^{\left(\Delta + 1,\,D\right)}}\ket{z^{[t + 1]}_{\varnothing}}_{\varnothing}
\end{align}
for some $1 \leq t \leq p - 1$. To commute the left-hand side bra through the phase separator unitary
\begin{align}
    e^{i\gamma_tC^{\left(\Delta + 1,\,D\right)}},
\end{align}
we decompose the cost Hamiltonian by singling out edges involving root vertex $\varnothing$:
\begin{align}
    C^{\left(\Delta + 1,\,D\right)} & = \frac{1}{\sqrt{D}}\sum_{\substack{\left(v,\,(v,\,j)\right) \in \mathcal{E}^{\left(\Delta + 1,\,D\right)}}}Z_vZ_{(v,\,j)}\nonumber\\
    & = \frac{1}{\sqrt{D}}\sum_{j \in [D]}Z_{\varnothing}Z_j + \frac{1}{\sqrt{D}}\sum_{\substack{\left(v,\,(v,\,j)\right) \in \mathcal{E}^{\left(\Delta + 1,\,D\right)}\\v \neq \varnothing}}Z_vZ_{\left(v,\,j\right)}\nonumber\\
    & = \frac{1}{\sqrt{D}}\sum_{j \in [D]}Z_{\varnothing}Z_j + \frac{1}{\sqrt{D}}\sum_{j \in [D]}\sum_{\left(v,\,(v,\,k)\right) \in \mathcal{E}^{\left(\Delta + 1,\,D\right)}_{\succeq j}}Z_vZ_{(v,\,k)}\nonumber\\
    & = \frac{1}{\sqrt{D}}\sum_{j \in [D]}\left(Z_{\varnothing}Z_j + \sum_{\left(v,\,(v,\,k)\right) \in \mathcal{E}^{\left(\Delta + 1,\,D\right)}_{\succeq j}}Z_vZ_{(v,\,k)}\right),
\end{align}
where in the third line, we enumerated edges of $\mathcal{T}^{\left(\Delta + 1,\,D\right)}$ not involving the root as edges of subtrees $\mathcal{T}^{\left(\Delta + 1,\,D\right)}_{\succeq j}$. Note the second additive contribution of the Hamiltonian is supported out of the root vertex $\varnothing$. From this rewriting, the computational basis bra can be commmuted through the phase separator unitary as follows:
\begin{align}
    \bra{z^{[t]}_{\varnothing}}_{\varnothing}\exp\left(i\gamma_tC^{\left(\Delta + 1,\,D\right)}\right) & = \bra{z^{[t]}_{\varnothing}}_{\varnothing}\exp\left(\frac{i\gamma_t}{\sqrt{D}}\sum_{j \in [D]}\left(Z_{\varnothing}Z_j + \sum_{\left(v,\,(v,\,k)\right) \in \mathcal{E}_{\succeq j}^{\left(\Delta + 1,\,D\right)}}Z_vZ_{(v,\,k)}\right)\right)\nonumber\\
    & = \exp\left(\frac{i\gamma_t}{\sqrt{D}}\sum_{j \in [D]}\left(z^{[t]}_{\varnothing}Z_j + \sum_{\left(v,\,(v,\,k)\right) \in \mathcal{E}_{\succeq j}^{\left(\Delta + 1,\,D\right)}}Z_vZ_{(v,\,k)}\right)\right)\bra{z^{[t]}_{\varnothing}}_{\varnothing}
\end{align}
The treatment of the mixer is trivial due to its product form. All in all, the desired ``matrix element" reads:
\begin{align}
    & \bra{z^{[t]}_{\varnothing}}_{\varnothing}\exp\left(\frac{i\gamma_t}{\sqrt{D}}C^{\left(\Delta + 1,\,D\right)}\right)\exp\left(i\beta_tB^{\left(\Delta + 1,\,D\right)}\right)\ket{z^{[t + 1]}_{\varnothing}}\nonumber\\
    & = \exp\left(\frac{i\gamma_t}{\sqrt{D}}\sum_{j \in [D]}\left(z^{[t]}_{\varnothing}Z_j + \sum_{\left(v,\,(v,\,k)\right) \in \mathcal{E}_{\succeq j}^{\left(\Delta + 1,\,D\right)}}Z_vZ_{(v,\,k)}\right)\right)\exp\left(i\beta_t\sum_{j \in [D]}\sum_{v \in \mathcal{V}_{\succeq j}^{\left(\Delta + 1,\,D\right)}}X_v\right)\bra{z^{[t]}_{\varnothing}}e^{i\beta_tX}\ket{z^{[t + 1]}_{\varnothing}}\nonumber\\
    & = \bra{z^{[t]}_{\varnothing}}e^{i\beta_tX}\ket{z^{[t + 1]}_{\varnothing}}\prod_{j \in [D]}\exp\left(\frac{i\gamma_t}{\sqrt{D}}\left(z^{[t]}_{\varnothing}Z_j + \sum_{\left(v,\,(v,\,k)\right) \in \mathcal{E}_{\succeq j}^{\left(\Delta + 1,\,D\right)}}Z_vZ_{(v,\,k)}\right)\right)\exp\left(i\beta_t\sum_{v \in \mathcal{V}^{\left(\Delta + 1\right)}_{\succeq j}}X_v\right)\label{eq:maxcut_qaoa_tree_induction_matrix_element}
\end{align}
We stress the braket in the last two equations, involving a single-qubit $X$ rotation, is a scalar rather than an operator acting over $\mathcal{V}^{\left(\Delta + 1,\,D\right)} - \{\varnothing\}$. We now observe the above operator acts independently and identically on all subsystems $\mathcal{V}^{\left(\Delta + 1,\,D\right)}_{\succeq j}, j \in [D]$. Said differently, the product over $j \in [D]$ in the final equation can effectively be regarded as a tensor product. To reflect this in the notation, we choose to order qubits according to sequence $\mathcal{V}^{\left(\Delta + 1,\,D\right)}_{\succeq 1}, \ldots, \mathcal{V}^{\left(\Delta + 1,\,D\right)}_{\succeq D}$. The order of qubits within each tree does not matter, but should for be chosen identical (up to isomorphism) between the trees. Besides, we use tree isomorphism:
\begin{align}
    \mathcal{T}^{\left(\Delta + 1,\,D\right)}_{\succeq j} \simeq \mathcal{T}^{\left(\Delta,\,D\right)} \implies \left(\mathbb{C}^2\right)^{\otimes \mathcal{V}^{\left(\Delta + 1,\,D\right)}_{\succeq j}} \simeq \left(\mathbb{C}^2\right)^{\otimes \mathcal{V}^{\left(\Delta,\,D\right)}}.
\end{align}
Following our choice of tensor product ordering and plugging in the above isomorphism, the ``matrix element" in equation \ref{eq:maxcut_qaoa_tree_induction_matrix_element} reads:
\begin{align}
    & \bra{z^{[t]}_{\varnothing}}_{\varnothing}\exp\left(i\gamma_tC^{\left(\Delta + 1,\,D\right)}\right)\exp\left(i\beta_tB^{\left(\Delta + 1,\,D\right)}\right)\ket{z^{[t + 1]}_{\varnothing}}\nonumber\\
    & \simeq \bra{z^{[t]}_{\varnothing}}e^{i\beta_tX}\ket{z_{\varnothing}^{[t + 1]}}\bigotimes_{j \in [D]}\exp\left(\frac{i\gamma_t}{\sqrt{D}}z_{\varnothing}^{[t]}Z_{\varnothing} + \frac{i\gamma_t}{\sqrt{D}}\sum_{\left(v,\,(v,\,k)\right) \in \mathcal{E}^{\left(\Delta,\,D\right)}}Z_vZ_{(v,\,k)}\right)\exp\left(i\beta_t\sum_{v \in \mathcal{V}^{\left(\Delta,\,D\right)}}X_v\right)\nonumber\\
    & \simeq \bra{z^{[t]}_{\varnothing}}e^{i\beta_tX}\ket{z_{\varnothing}^{[t + 1]}}\left(\exp\left(\frac{i\gamma_t}{\sqrt{D}}z_{\varnothing}^{[t]}Z_{\varnothing} + \frac{i\gamma_t}{\sqrt{D}}\sum_{\left(v,\,(v,\,k)\right) \in \mathcal{E}^{\left(\Delta,\,D\right)}}Z_vZ_{(v,\,k)}\right)\exp\left(i\beta_t\sum_{v \in \mathcal{V}^{\left(\Delta,\,D\right)}}X_v\right)\right)^{\otimes D}\label{eq:maxcut_qaoa_tree_induction_matrix_element_as_tensor_product}
\end{align}
To stress again the principle of the reorganization between equations \ref{eq:maxcut_qaoa_tree_induction_matrix_element} and \ref{eq:maxcut_qaoa_tree_induction_matrix_element_as_tensor_product}, the new system, isomorphic to the original one, now consists of $D$ independent copy of the depth-$\Delta$ tree of degree $D$, namely $\mathcal{T}^{\left(\Delta,\,D\right)} = \left(\mathcal{V}^{\left(\Delta,\,D\right)}, \mathcal{E}^{\left(\Delta,\,D\right)}\right)$. Proceeding similarly for the other ``matrix elements", we finally get the following expression for the desired two-times correlation of $Z_{\varnothing}$:
\begin{align}
    & G^{\left(\mathrm{sym},\,\Delta + 1,\,D\right)}_{j,\,k}\nonumber\\
    & = \sum_{\substack{z^{[t]}_{\varnothing} \in \{\pm 1\}\\\forall -p \leq t \leq p}}z^{[-k]}_{\varnothing}z^{[-j]}_{\varnothing}\bra{z^{[p]}_{\varnothing}}e^{i\beta_pX}\ket{z^{[0]}_{\varnothing}}\bra{z^{[0]}_{\varnothing}}e^{-i\beta_pX}\ket{z^{[-p]}_{\varnothing}}\prod_{1 \leq t \leq p - 1}\bra{z^{[t]}_{\varnothing}}e^{i\beta_tX}\ket{z^{[t + 1]}_{\varnothing}}\bra{z^{[-t - 1]}_{\varnothing}}e^{-i\beta_tX}\ket{z^{[-t]}_{\varnothing}}\nonumber\\
    & \hspace*{50px} \times \bra{+}_{\mathcal{V}^{\left(\Delta,\,D\right)}}^{\otimes D}\nonumber\\
    & \hspace*{75px} \overrightarrow{\prod_{t = 1}^p}\left(\exp\left(\frac{i\gamma_t}{\sqrt{D}}z_{\varnothing}^{[t]}Z_{\varnothing} + \frac{i\gamma_t}{\sqrt{D}}\sum_{\left(v,\,(v,\,k)\right) \in \mathcal{E}^{\left(\Delta,\,D\right)}}Z_vZ_{(v,\,k)}\right)\exp\left(i\beta_t\sum_{v \in \mathcal{V}^{\left(\Delta,\,D\right)}}X_v\right)\right)^{\otimes D}\nonumber\\
    & \hspace*{75px} \overleftarrow{\prod_{t = 1}^p}\left(\exp\left(-i\beta_t\sum_{v \in \mathcal{V}^{\left(\Delta,\,D\right)}}X_v\right)\exp\left(-\frac{i\gamma_t}{\sqrt{D}}z_{\varnothing}^{[t]}Z_{\varnothing} - \frac{i\gamma_t}{\sqrt{D}}\sum_{\left(v,\,(v,\,k)\right) \in \mathcal{E}^{\left(\Delta,\,D\right)}}Z_vZ_{(v,\,k)}\right)\right)^{\otimes D}\nonumber\\
    & \hspace*{60px} \ket{+}_{\mathcal{V}^{\left(\Delta,\,D\right)}}^{\otimes D}\label{eq:tree_root_correlation_cavity_expression}
\end{align}
In the above expression, we separated the scalar contributions ---involving $Z_{\varnothing}$ operator insertions at position $j, k$, as well as the matrix elements of $X$ rotations with bits of the computational basis path $\bm{z}_{\varnothing}$--- and the remaining product of bras, kets and operators in the tensor product space
\begin{align}
    \left(\left(\mathbb{C}^2\right)^{\otimes \mathcal{V}^{\left(\Delta,\,D\right)}}\right)^{\otimes D}.
\end{align}
We recognize the scalar contribution to be:
\begin{align}
    & z^{[-k]}_{\varnothing}z^{[-j]}_{\varnothing}\bra{z^{[p]}_{\varnothing}}e^{i\beta_pX}\ket{z^{[0]}_{\varnothing}}\bra{z^{[0]}_{\varnothing}}e^{-i\beta_pX}\ket{z^{[-p]}_{\varnothing}}\prod_{1 \leq t \leq p - 1}\bra{z^{[t]}_{\varnothing}}e^{i\beta_tX}\ket{z^{[t + 1]}_{\varnothing}}\bra{z^{[-t - 1]}_{\varnothing}}e^{-i\beta_tX}\ket{z^{[-t]}_{\varnothing}}\nonumber\\
    & = z^{[-k]}_{\varnothing}z^{[-j]}_{\varnothing}f\left(\bm{z}_{\varnothing}\right).\label{eq:tree_root_correlation_cavity_expression_root_expectation}
\end{align}
It therefore remains to estimate the dot product in the tensor product space. In doing so, it will help to keep in mind the summation over the computational basis path $\bm{z}_{\varnothing} \in \{1, -1\}^{2p + 1}$ has a number of terms independent of $D$. First, thanks to the tensor product structure of states, the dot product factors into a matrix element over a single tree copy $\mathcal{T}^{\left(\Delta,\,D\right)}$, raised to the power $D$:
\begin{align}
    & \bra{+}_{\mathcal{V}^{\left(\Delta,\,D\right)}}^{\otimes D}\nonumber\\
    & \hspace*{15px} \overrightarrow{\prod_{t = 1}^p}\left(\exp\left(\frac{i\gamma_t}{\sqrt{D}}z_{\varnothing}^{[t]}Z_{\varnothing} + \frac{i\gamma_t}{\sqrt{D}}\sum_{\left(v,\,(v,\,k)\right) \in \mathcal{E}^{\left(\Delta,\,D\right)}}Z_vZ_{(v,\,k)}\right)\exp\left(i\beta_t\sum_{v \in \mathcal{V}^{\left(\Delta,\,D\right)}}X_v\right)\right)^{\otimes D}\nonumber\\
    & \hspace*{15px} \overleftarrow{\prod_{t = 1}^p}\left(\exp\left(-i\beta_t\sum_{v \in \mathcal{V}^{\left(\Delta,\,D\right)}}X_v\right)\exp\left(-\frac{i\gamma_t}{\sqrt{D}}z_{\varnothing}^{[t]}Z_{\varnothing} - \frac{i\gamma_t}{\sqrt{D}}\sum_{\left(v,\,(v,\,k)\right) \in \mathcal{E}^{\left(\Delta,\,D\right)}}Z_vZ_{(v,\,k)}\right)\right)^{\otimes D}\nonumber\\
    & \ket{+}_{\mathcal{V}^{\left(\Delta,\,D\right)}}^{\otimes D}\nonumber\\
    & = \left(\bra{+}\overrightarrow{\prod_{t = 1}^p}\exp\left(\frac{i\gamma_t}{\sqrt{D}}z_{\varnothing}^{[t]}Z_{\varnothing} + \frac{i\gamma_t}{\sqrt{D}}\sum_{\left(v,\,(v,\,k)\right) \in \mathcal{E}^{\left(\Delta,\,D\right)}}Z_vZ_{(v,\,k)}\right)\exp\left(i\beta_t\sum_{v \in \mathcal{V}^{\left(\Delta,\,D\right)}}X_v\right)\right.\nonumber\\
    & \hspace*{35px} \left.\overleftarrow{\prod_{t = 1}^p}\exp\left(-i\beta_t\sum_{v \in \mathcal{V}^{\left(\Delta,\,D\right)}}X_v\right)\exp\left(-\frac{i\gamma_t}{\sqrt{D}}z_{\varnothing}^{[t]}Z_{\varnothing} - \frac{i\gamma_t}{\sqrt{D}}\sum_{\left(v,\,(v,\,k)\right) \in \mathcal{E}^{\left(\Delta,\,D\right)}}Z_vZ_{(v,\,k)}\right)\ket{+}\right)^D\label{eq:tree_root_correlation_cavity_expression_subtrees_expectation}
\end{align}
The problem then reduces to estimating the matrix element over a single tree $\mathcal{T}^{\left(\Delta,\,D\right)}$. For that purpose, we will apply the following lemma, formulated for any sequence of unitaries, to the special case of the QAOA unitaries:

\begin{lemma}[Second-order expansion of time correlations of near-identity operators]
\label{lemma:second_order_expansion_time_correlations}
Consider a sequence of $q$ unitaries
\begin{align}
    V_1, V_2, \ldots, V_{q - 1}, V_q
\end{align}
acting on a finite-dimensional Hilbert space $\mathcal{H}$, and let
\begin{align}
    \ket{\psi} \in \mathcal{H}
\end{align}
an arbitrary state from the space. Consider Hermitian operators
\begin{align}
    O_1, O_2, \ldots, O_q, O_{q + 1}
\end{align}
and consider expectation
\begin{align}
    & \bra{\psi}e^{iO_{q + 1}/\sqrt{D}}\left(\overleftarrow{\prod_{t = 1}^{q}}V_te^{iO_t/\sqrt{D}}\right)\ket{\psi}\label{eq:general_time_correlations_form_1}\\
    & = \bra{\psi}\left(\overleftarrow{\prod_{t = 1}^{q}}e^{iO_{t + 1}/\sqrt{D}}V_t\right)e^{iO_{1}/\sqrt{D}}\ket{\psi}\label{eq:general_time_correlations_form_2}\\
    & = \bra{\psi}e^{iO_{q + 1}/\sqrt{D}}V_qe^{iO_{q}/\sqrt{D}}V_{q - 1}\ldots e^{iO_3/\sqrt{D}}V_2e^{iO_2/\sqrt{D}}V_1e^{iO_1/\sqrt{D}}\ket{\psi}\label{eq:general_time_correlations_form_3}
\end{align}
We are interested in the behavior of this expectation in the limit $D \to \infty$, where all inserted operators $e^{iO_t/\sqrt{D}}$ are close to identity but the number of unitaries $q$ does not grow with $D$. Define the ``zero-order moment" as the expectation without inserted operators:
\begin{align}
    \mu^{(0)} & := \bra{\psi}\overleftarrow{\prod_{t = 1}^p}V_t\ket{\psi}\\
    & = \bra{\psi}V_qV_{q - 1}\ldots V_2V_1\ket{\psi}.
\end{align}
Next, define the ``first-order moments" of operators $\left(O_t\right)_{t \in [q + 1]}$ as the expectation of the product of operators $V_1, V_2, \ldots, V_{q - 1}, V_q$ with a single $O_l$ inserted at position $l$:
\begin{align}
    \bm{\mu}^{(1)} & := \left(\mu^{(1)}_l\right)_{l \in [q + 1]},\\
    \mu^{(1)}_l & := \bra{\psi}\left(\overleftarrow{\prod_{t = l}^q}V_t\right)O_l\left(\overleftarrow{\prod_{t = 1}^{l - 1}}V_t\right)\ket{\psi}\\
    & = \bra{\psi}V_qV_{q - 1}\ldots V_{l + 1}V_lO_lV_{l - 1}V_{l - 2}\ldots V_2V_1\ket{\psi}
\end{align}
Similarly, define the ``second-order moments" of operators $\left(O_t\right)_{t \in [q + 1]}$ as the expectation of the product of operator $V_1, V_2, \ldots, V_{q - 1}, V_q$ with a pair of operators $O_l, O_m$ inserted at positions $l, m$:
\begin{align}
    \bm{\mu}^{(2)} & := \left(\mu^{(2)}_{l,\,m}\right)_{l, m \in [q + 1]},\\
    \mu^{(2)}_{l,\,m} & := \bra{\psi}\left(\overleftarrow{\prod_{t = m}^q}V_t\right)O_m\left(\overleftarrow{\prod_{t = l}^{m - 1}}V_t\right)O_l\left(\overleftarrow{\prod_{t = 1}^{l - 1}}V_t\right)\ket{\psi} \qquad l \leq m\\
    & = \bra{\psi}V_qV_{q - 1}\ldots V_{m + 1}V_mO_mV_{m - 1}V_{m - 2}\ldots V_{l + 1}V_lO_lV_{l - 1}V_{l - 2}\ldots V_2V_1\ket{\psi}
\end{align}
The above $\bm{\mu}^{(2)}$ matrix, defined only in the upper triangle by the previous equation, is completed by symmetry to the lower triangle. Under the previous definitions of $\mu^{(0)}, \bm{\mu}^{(1)}, \bm{\mu}^{(2)}$, the following second-order expansion in $1/\sqrt{D}$ of expectation value \ref{eq:general_time_correlations_form_1} holds:
\begin{align}
    \bra{\psi}e^{iO_{q + 1}/\sqrt{D}}\left(\overleftarrow{\prod_{t = 1}^{q}}V_te^{iO_t/\sqrt{D}}\right)\ket{\psi} & = \mu^{(0)} + \frac{i}{\sqrt{D}}\sum_{1 \leq l \leq q + 1}\mu^{(1)}_l - \frac{1}{2D}\sum_{1 \leq l, m \leq q + 1}\mu^{(2)}_{l,\,m} + \mathcal{O}\left(D^{-3/2}\right),\label{eq:second_order_expansion_time_correlations}
\end{align}
where the constant implied in the $\mathcal{O}$ depends only on $q$ and operator norms
\begin{align}
    \left\lVert O_1 \right\rVert, \left\lVert O_2 \right\rVert, \ldots, \left\lVert O_{q - 1} \right\rVert, \left\lVert O_{q + 1} \right\rVert
\end{align}
(and not for instance on the dimension of the Hilbert space $\mathcal{H}$ or unitaries $V_1, V_2, \ldots, V_{q - 1}, V_q$).
\begin{proof}
For convenience, let us pad the sequence of unitaries with a final dummy identity operator:
\begin{align}
    V_{q + 1} & := I_{\mathcal{H}},
\end{align}
so we may write
\begin{align}
    \bra{\psi}e^{iO_{q + 1}/\sqrt{D}}\left(\overleftarrow{\prod_{t = 1}^q}V_te^{iO_t/\sqrt{D}}\right)\ket{\psi} & = \bra{\psi}\left(\overleftarrow{\prod_{t = 1}^{q + 1}}V_te^{iO_t/\sqrt{D}}\right)\ket{\psi}.
\end{align}
We first show that each near-identity unitary can be replaced by its expansion up to second order in $1/\sqrt{D}$ for an error not exceeding $\mathcal{O}\left(D^{-3/2}\right)$, where the implicit constant only depends on  $q$ and the operator norms of the $O_t$. The expansion of a near-identity unitaries up to second order reads:
\begin{align}
    e^{iO_t/\sqrt{D}} & = o_t + e_t,
\end{align}
where the main term of the expansion is
\begin{align}
    o_t & := I_{\mathcal{H}} + \frac{iO_t}{\sqrt{D}} - \frac{1}{2D}O_t^2,
\end{align}
and the error term is
\begin{align}
    e_t & := e^{iO_t/\sqrt{D}} - o_t,\\
    \left\lVert e_t \right\rVert & \leq \frac{\left\lVert O_t \right\rVert^3}{6D^{-3/2}}.
\end{align}
The latter bound on the error follows from inequality
\begin{align}
    \left|e^{ix} - 1 - ix + \frac{x^2}{2}\right| & \leq \frac{x^3}{6}, \qquad x \in \mathbf{R},
\end{align}
holding for all real number $x$, and promoted to Hermitian operators. We perform the replacement iteratively. Namely, we start with $e^{iO_{q + 1}/\sqrt{D}}$, the following bound holds on the state error:
\begin{align}
    \left\lVert \left(\overleftarrow{\prod_{t = 1}^{q + 1}}V_te^{iO_t/\sqrt{D}}\right)\ket{\psi} - \left(\overleftarrow{\prod_{t = 2}^{q + 1}}V_te^{iO_t/\sqrt{D}}\right)V_1o_1\ket{\psi}\right\rVert_2 & = \left\lVert \left(\overleftarrow{\prod_{t = 2}^{q + 1}}V_te^{iO_t/\sqrt{D}}\right)V_1\left(e^{iO_1/\sqrt{D}} - o_1\right)\ket{\psi} \right\rVert_2\nonumber\\
    & = \left\lVert \left(\overleftarrow{\prod_{t = 2}^{q + 1}}V_te^{iO_t/\sqrt{D}}\right)V_1e_1\ket{\psi} \right\rVert_2\nonumber\\
    & \leq \prod_{t = 2}^{q + 1}\left\lVert V_te^{iO_t/\sqrt{D}} \right\rVert \times \lVert e_1 \rVert \times \lVert \ket{\psi} \rVert_2\nonumber\\
    & \leq \left\lVert e_1 \right\rVert\nonumber\\
    & \leq \frac{\lVert O_1 \rVert}{6D^{-3/2}}\nonumber\\
    & = \mathcal{O}\left(D^{-3/2}\right).
\end{align}
Hence, we approximated statevector
\begin{align}
    \left(\overleftarrow{\prod_{t = 1}^{q + 1}}V_te^{iO_t/\sqrt{D}}\right)\ket{\psi}
\end{align}
by (unnnormalized) vector
\begin{align}
    \left(\prod_{t = 2}^{q + 1}V_te^{iO_t/\sqrt{D}}\right)V_1o_1\ket{\psi}
\end{align}
up to the desired additive error. We now approximate the latter vector by replacing near-identity unitary $e^{iO_2/\sqrt{D}}$ by its second-order expansion $o_2$ in the second layer. This gives:
\begin{align}
    \left\lVert \left(\overleftarrow{\prod_{t = 2}^{q + 1}}V_te^{iO_t/\sqrt{D}}\right)V_1o_1\ket{\psi} - \left(\overleftarrow{\prod_{t = 3}^{q + 1}}V_te^{iO_t/\sqrt{D}}\right)V_2o_2V_1o_1\ket{\psi} \right\rVert_2  & = \left\lVert \left(\overleftarrow{\prod_{t = 3}^{q + 1}}V_te^{iO_t/\sqrt{D}}\right)V_2\left(e^{iO_2/\sqrt{D}} - o_2\right)V_1o_1\ket{\psi} \right\rVert_2\nonumber\\
    & = \left\lVert \left(\overleftarrow{\prod_{t = 3}^{q + 1}}V_te^{iO_t/\sqrt{D}}\right)V_2e_2V_1o_1\ket{\psi} \right\rVert_2\nonumber\\
    & \leq \left\lVert e_2 \right\rVert \left\lVert o_1 \right\rVert\nonumber\\
    & = \left\lVert e_2 \right\rVert \left\lVert e^{iO_1/\sqrt{D}} - e_1 \right\rVert\nonumber\\
    & \leq \left\lVert e_2 \right\rVert \left(\left\lVert e^{iO_1/\sqrt{D}} \right\rVert + \left\lVert e_1 \right\rVert\right)\nonumber\\
    & \leq \left\lVert e_2 \right\rVert \left(1 + \left\lVert e_1 \right\rVert\right)\nonumber\\
    & \leq \frac{\left\lVert O_2 \right\rVert}{6D^{-3/2}}\left(1 + \frac{\left\lVert O_2 \right\rVert}{6D^{-3/2}}\right)\\
    & = \mathcal{O}\left(D^{-3/2}\right).
\end{align}
Generalizing the previous calculation,
\begin{align}
    \left\lVert \left(\overleftarrow{\prod_{t = q'}^{q + 1}}V_te^{iO_t/\sqrt{D}}\right)\left(\overleftarrow{\prod_{t = 1}^{q' - 1}}V_to_t\right)\ket{\psi} - \left(\overleftarrow{\prod_{t = q' + 1}^{q + 1}}V_te^{iO_t/\sqrt{D}}\right)\left(\overleftarrow{\prod_{t = 1}^{q'}}V_to_t\right)\ket{\psi} \right\rVert_2  & \leq \frac{\left\lVert O_{q'} \right\rVert}{6D^{-3/2}}\prod_{1 \leq t \leq q' - 1}\left(1 + \frac{\left\lVert O_t \right\rVert}{6D^{-3/2}}\right)\nonumber\\
    & = \mathcal{O}\left(D^{-3/2}\right).
\end{align}
Adding up all the inequalities yields
\begin{align}
    \left\lVert \left(\overleftarrow{\prod_{t = 1}^{q + 1}}V_te^{iO_t/\sqrt{D}}\right)\ket{\psi} - \left(\overleftarrow{\prod_{t = 1}^{q + 1}}V_to_t\right)\ket{\psi} \right\rVert & \leq (q + 1)\mathcal{O}\left(D^{-3/2}\right)\nonumber\\
    & = \mathcal{O}\left(D^{-3/2}\right),
\end{align}
resulting in the same upper bound on the expectation value error:
\begin{align}
    \left\lVert \bra{\psi}\left(\overleftarrow{\prod_{t = 1}^{q + 1}}V_te^{iO_t/\sqrt{D}}\right)\ket{\psi} - \bra{\psi}\left(\overleftarrow{\prod_{t = 1}^{q + 1}}V_to_t\right)\ket{\psi} \right\rVert & \leq \mathcal{O}\left(D^{-3/2}\right).
\end{align}
Note implicit constants only depend on $q$ and operator norms $\left\lVert O_t \right\rVert$ as required. We now expand the order $2$ approximation of near-identity unitaries in the approximating state:
\begin{align}
    \left(\overleftarrow{\prod_{t = 1}^{q + 1}}V_to_t\right)\ket{\psi} & = \left(\overleftarrow{\prod_{t = 1}^{q + 1}}V_t\left(I_{\mathcal{H}} + \frac{iO_t}{\sqrt{D}} - \frac{1}{2D}O_t^2\right)\right)\ket{\psi}\nonumber\\
    & = \sum_{\bm{k} = \left(k_t\right)_{t \in [q + 1]} \in \{0, 1, 2\}^{[q + 1]}}\left(\overleftarrow{\prod_{t = 1}^{q + 1}}V_t\left(I_{\mathcal{H}}\mathbf{1}\left[k_t = 0\right] + \frac{iO_t}{\sqrt{D}}\mathbf{1}\left[k_t = 1\right] - \frac{1}{2D}O_t^2\mathbf{1}\left[k_t = 2\right]\right)\right)\ket{\psi}\nonumber\\
    & = \sum_{\bm{k} \in \{0, 1, 2\}^{[q + 1]}}i^{q_1\left(\bm{k}\right)}\left(-1/2\right)^{q_2\left(\bm{k}\right)}D^{-q_1(\bm{k})/2 - q_2\left(\bm{k}\right)}\left(\overleftarrow{\prod_{t = 1}^{q + 1}}V_tO_t^{k_t}\right)\ket{\psi},\label{eq:second_order_correlations_expansion_step_1}
\end{align}
where in the final line, we introduced $q_1\left(\bm{k}\right)$ and $q_2\left(\bm{k}\right)$ counting the number of occurrences of $1$ and $2$ in $\bm{k}$ respectively:
\begin{align}
    q_k\left(\bm{k}\right) & := \left|\left\{t \in [q + 1]\,:\,k_t = k\right\}\right|, \qquad k = 0, 1, 2.
\end{align}
Intuitively, up to error $D^{-3/2}$, only the $\bm{k}$ for which 
\begin{align}
    \left(q_1\left(\bm{k}\right), q_2\left(\bm{k}\right)\right) \in \{(0, 0), (1, 0), (2, 0), (0, 1)\} =: Q_0.
\end{align}
contribute. Let us then compute the contribution from these $K$:
\begin{align}
    & \sum_{\substack{\bm{k} \in \{0, 1, 2\}^{[q + 1]}\\\left(q_1\left(\bm{k}\right),\,q_2\left(\bm{k}\right)\right) \in Q_0}}i^{q_1\left(\bm{k}\right)}\left(-1/2\right)^{q_2\left(\bm{k}\right)}D^{-q_1(\bm{k})/2 - q_2\left(\bm{k}\right)}\left(\overleftarrow{\prod_{t = 1}^{q + 1}}V_tO_t^{k_t}\right)\ket{\psi}\nonumber\\
    & = \sum_{\substack{\bm{k} \in \{0, 1, 2\}^{[q + 1]}\\\left(q_1\left(\bm{k}\right),\,q_2\left(\bm{k}\right)\right) = (0, 0)}}i^{q_1\left(\bm{k}\right)}\left(-1/2\right)^{q_2\left(\bm{k}\right)}D^{-q_1(\bm{k})/2 - q_2\left(\bm{k}\right)}\left(\overleftarrow{\prod_{t = 1}^{q + 1}}V_tO_t^{k_t}\right)\ket{\psi}\nonumber\\
    & \hspace*{20px} + \sum_{\substack{\bm{k} \in \{0, 1, 2\}^{[q + 1]}\\\left(q_1\left(\bm{k}\right),\,q_2\left(\bm{k}\right)\right) = (1, 0)}}i^{q_1\left(\bm{k}\right)}\left(-1/2\right)^{q_2\left(\bm{k}\right)}D^{-q_1(\bm{k})/2 - q_2\left(\bm{k}\right)}\left(\overleftarrow{\prod_{t = 1}^{q + 1}}V_tO_t^{k_t}\right)\ket{\psi}\nonumber\\
    & \hspace*{20px} + \sum_{\substack{\bm{k} \in \{0, 1, 2\}^{[q + 1]}\\\left(q_1\left(\bm{k}\right),\,q_2\left(\bm{k}\right)\right) = (2, 0)}}i^{q_1\left(\bm{k}\right)}\left(-1/2\right)^{q_2\left(\bm{k}\right)}D^{-q_1(\bm{k})/2 - q_2\left(\bm{k}\right)}\left(\overleftarrow{\prod_{t = 1}^{q + 1}}V_tO_t^{k_t}\right)\ket{\psi}\nonumber\\
    & \hspace*{20px} + \sum_{\substack{\bm{k} \in \{0, 1, 2\}^{[q + 1]}\\\left(q_1\left(\bm{k}\right),\,q_2\left(\bm{k}\right)\right) = (0, 1)}}i^{q_1\left(\bm{k}\right)}\left(-1/2\right)^{q_2\left(\bm{k}\right)}D^{-q_1(\bm{k})/2 - q_2\left(\bm{k}\right)}\left(\overleftarrow{\prod_{t = 1}^{q + 1}}V_tO_t^{k_t}\right)\ket{\psi}.\label{eq:second_order_expansion_time_correlations_leading_term}
\end{align}
The first of the above sum contains only a single term $\bm{k} = 0^{q + 1}$:
\begin{align}
    \sum_{\substack{\bm{k} \in \{0, 1, 2\}^{[q + 1]}\\\left(q_1\left(\bm{k}\right),\,q_2\left(\bm{k}\right)\right) = (0, 0)}}i^{q_1\left(\bm{k}\right)}\left(-1/2\right)^{q_2\left(\bm{k}\right)}D^{-q_1(\bm{k})/2 - q_2\left(\bm{k}\right)}\left(\overleftarrow{\prod_{t = 1}^{q + 1}}V_tO_t^{k_t}\right)\ket{\psi} & = \left(\prod_{t = 1}^{q + 1}V_t\right)\ket{\psi}.\label{eq:second_order_expansion_time_correlations_leading_term_contribution_1}
\end{align}
The second of the above sums is over $\bm{k}$ including a single $1$, with all other entries being $0$; hence, it is equivalent to summing over the choice of $l \in [q + 1]$ where $k_l = 1$:
\begin{align}
    \sum_{\substack{\bm{k} \in \{0, 1, 2\}^{[q + 1]}\\\left(q_1\left(\bm{k}\right),\,q_2\left(\bm{k}\right)\right) = (1, 0)}}i^{q_1\left(\bm{k}\right)}\left(-1/2\right)^{q_2\left(\bm{k}\right)}D^{-q_1(\bm{k})/2 - q_2\left(\bm{k}\right)}\left(\overleftarrow{\prod_{t = 1}^{q + 1}}V_tO_t^{k_t}\right)\ket{\psi} & = \frac{i}{\sqrt{D}}\sum_{l \in [q + 1]}\left(\overleftarrow{\prod_{t = l}^{q + 1}}V_t\right)O_l\left(\overleftarrow{\prod_{t = 1}^{l - 1}}V_t\right)\ket{\psi}.\label{eq:second_order_expansion_time_correlations_leading_term_contribution_2}
\end{align}
The third of the above sums is over $\bm{k}$ with exactly two $1$'s and the rest of entries being $0$; one may then sum over the distinct entries $0 \leq l < m \leq q + 1$ where $k_l = k_m = 1$:
\begin{align}
    & \sum_{\substack{\bm{k} \in \{0, 1, 2\}^{[q + 1]}\\\left(q_1\left(\bm{k}\right),\,q_2\left(\bm{k}\right)\right) = (2, 0)}}i^{q_1\left(\bm{k}\right)}\left(-1/2\right)^{q_2\left(\bm{k}\right)}D^{-q_1(\bm{k})/2 - q_2\left(\bm{k}\right)}\left(\overleftarrow{\prod_{t = 1}^{q + 1}}V_tO_t^{k_t}\right)\ket{\psi}\nonumber\\
    & = -\frac{1}{D}\sum_{1 \leq l < m \leq q + 1}\left(\overleftarrow{\prod_{t = m}^{q + 1}}V_t\right)O_m\left(\overleftarrow{\prod_{t = l}^{m - 1}}V_t\right)O_l\left(\overleftarrow{\prod_{t = 1}^{l - 1}}V_t\right)\ket{\psi}\nonumber\\
    & = -\frac{1}{2D}\sum_{\substack{1 \leq l, m \leq q + 1\\l \neq m}}\left(\overleftarrow{\prod_{t = m}^{q + 1}}V_t\right)O_m\left(\overleftarrow{\prod_{t = l}^{m - 1}}V_t\right)O_l\left(\overleftarrow{\prod_{t = 1}^{l - 1}}V_t\right)\ket{\psi}.\label{eq:second_order_expansion_time_correlations_leading_term_contribution_3}
\end{align}
Finally, the fourth contribution in equation \ref{eq:second_order_expansion_time_correlations_leading_term} is over the $\bm{k}$ with a single $2$ and all other entries $0$. Hence, it can be expressed as a summation over the single time step $l \in [q + 1]$ such that $k_l = 2$:
\begin{align}
    \sum_{\substack{\bm{k} \in \{0, 1, 2\}^{[q + 1]}\\\left(q_1\left(\bm{k}\right),\,q_2\left(\bm{k}\right)\right) = (0, 1)}}i^{q_1\left(\bm{k}\right)}\left(-1/2\right)^{q_2\left(\bm{k}\right)}D^{-q_1(\bm{k})/2 - q_2\left(\bm{k}\right)}\left(\overleftarrow{\prod_{t = 1}^{q + 1}}V_tO_t^{k_t}\right)\ket{\psi} & = -\frac{1}{2D}\sum_{1 \leq l \leq q + 1}\left(\overleftarrow{\prod_{t = l}^{q + 1}}V_t\right)O_l^2\left(\overleftarrow{\prod_{t = 1}^{l - 1}}V_t\right)\ket{\psi}.\label{eq:second_order_expansion_time_correlations_leading_term_contribution_4}
\end{align}
Adding equations \ref{eq:second_order_expansion_time_correlations_leading_term_contribution_1}, \ref{eq:second_order_expansion_time_correlations_leading_term_contribution_2}, \ref{eq:second_order_expansion_time_correlations_leading_term_contribution_3} and \ref{eq:second_order_expansion_time_correlations_leading_term_contribution_4} gives the leading term of the claimed expansion in equation \ref{eq:second_order_expansion_time_correlations}. We now prove that the remaining terms from equation \ref{eq:second_order_correlations_expansion_step_1}, characterized by $\left(q_1\left(\bm{k}\right),\,q_2\left(\bm{k}\right)\right) \notin Q_0$ cumulate an error at most $\mathcal{O}\left(D^{-3/2}\right)$, with implicit constant depending only on $q$ and the operator norms of the $O_t$.  We bound the norm of these remaining terms using the triangle inequality:
\begin{align}
    & \left\lVert \sum_{\substack{\bm{k} \in \{0, 1, 2\}^{q + 1}\\\left(q_1\left(\bm{k}\right),\,q_2\left(\bm{k}\right)\right) \notin Q_0}}i^{q_1\left(\bm{k}\right)}\left(-1/2\right)^{q_2\left(\bm{k}\right)}D^{-q_1\left(\bm{k}\right)/2 - q_2\left(\bm{k}\right)}\left(\overleftarrow{\prod_{t = 1}^{q + 1}}V_tO_t^{k_t}\right)\ket{\psi} \right\rVert_2\nonumber\\
    & \leq \sum_{\substack{\bm{k} \in \{0, 1, 2\}^{q + 1}\\\left(q_1\left(\bm{k}\right),\,q_2\left(\bm{k}\right)\right) \notin Q_0}}2^{-q_2\left(\bm{k}\right)}D^{-q_1\left(\bm{k}\right)/2 - q_2\left(\bm{k}\right)}\prod_{t = 1}^{q + 1}\left\lVert O_t \right\rVert^{k_t}\nonumber\\
    & \leq \sum_{\substack{\bm{k} \in \{0, 1, 2\}^{q + 1}\\\left(q_1\left(\bm{k}\right),\,q_2\left(\bm{k}\right)\right) \notin Q_0}}2^{-q_2\left(\bm{k}\right)}D^{-q_1\left(\bm{k}\right)/2 - q_2\left(\bm{k}\right)}\left(\max_{t \in [q + 1]}\left\lVert O_t \right\rVert\right)^{q_1\left(\bm{k}\right) + 2q_2\left(\bm{k}\right)}
\end{align}
The last summand only depends on the number of occurrences of $0, 1, 2$ in $(q + 1)$-tuple $\bm{k}$. One may then sum over these numbers of occurrences, denoted $q_0, q_1, q_2$, accounting for the
\begin{align}
    \binom{q + 1}{q_0, q_1, q_2} & = \frac{\left(q + 1\right)!}{q_0!q_1!q_2!}
\end{align}
tuples with the number of occurrences. Our previous bound becomes:
\begin{align}
    & \left\lVert \sum_{\substack{\bm{k} \in \{0, 1, 2\}^{q + 1}\\\left(q_1\left(\bm{k}\right),\,q_2\left(\bm{k}\right)\right) \notin Q_0}}i^{q_1\left(\bm{k}\right)}\left(-1/2\right)^{q_2\left(\bm{k}\right)}D^{-q_1\left(\bm{k}\right)/2 - q_2\left(\bm{k}\right)}\left(\overleftarrow{\prod_{t = 1}^{q + 1}}V_tO_t^{k_t}\right)\ket{\psi} \right\rVert_2\nonumber\\
    & \leq \sum_{\substack{q_0, q_1, q_2\\q_0 + q_1 + q_2 = q + 1\\\left(q_1, q_2\right) \notin Q_0}}\binom{q + 1}{q_0, q_1, q_2}2^{-q_2}D^{-q_1/2 - q_2}\left(\max_{t \in [q + 1]}\left\lVert O_t \right\rVert\right)^{q_1 + 2q_2}\nonumber\\
    & = \sum_{\substack{q_0, q_1, q_2\\q_0 + q_1 + q_2 = q + 1}}\binom{q + 1}{q_0, q_1, q_2}2^{-q_2}D^{-q_1/2 - q_2}\left(\max_{t \in [q + 1]}\left\lVert O_t \right\rVert\right)^{q_1 + 2q_2}\nonumber\\
    & \hspace*{30px} - \sum_{\substack{q_0, q_1, q_2\\q_0 + q_1 + q_2 = q + 1\\\left(q_1, q_2\right) \in Q_0}}\binom{q + 1}{q_0, q_1, q_2}2^{-q_2}D^{-q_1/2 - q_2}\left(\max_{t \in [q + 1]}\left\lVert O_t \right\rVert\right)^{q_1 + 2q_2}\nonumber\\
    & = \left(1 + \frac{1}{\sqrt{D}}\max_{t \in [q + 1]}\left\lVert O_t \right\rVert + \frac{1}{2D}\left(\max_{t \in [q + 1]}\left\lVert O_t \right\rVert\right)^2\right)^{q + 1} - 1 - \frac{q + 1}{\sqrt{D}}\max_{t \in [q + 1]}\left\lVert O_t \right\rVert\nonumber\\
    & \hspace*{20px} - \frac{q(q + 1)}{2D}\left(\max_{t \in [q + 1]}\left\lVert O_t\right\rVert\right)^2 - \frac{q + 1}{2D}\left(\max_{t \in [q + 1]}\left\lVert O_t\right\rVert\right)^2,
\end{align}
where the subtracted terms respectively correspond to $(q_1, q_2) = (0, 0), (1, 0), (2, 0), (0, 1)$. We observe the subtracted terms are precisely the contributions up to order $2$ in $1/\sqrt{D}$ in the expansion of the previous term
\begin{align}
    \left(1 + \frac{1}{\sqrt{D}}\max_{t \in [q + 1]}\left\lVert O_t \right\rVert + \frac{1}{2D}\left(\max_{t \in [q + 1]}\left\lVert O_t \right\rVert\right)^2\right)^{q + 1}.
\end{align}
This proves
\begin{align}
    \left\lVert \sum_{\substack{\bm{k} \in \{0, 1, 2\}^{q + 1}\\\left(q_1\left(\bm{k}\right),\,q_2\left(\bm{k}\right)\right) \notin Q_0}}i^{q_1\left(\bm{k}\right)}\left(-1/2\right)^{q_2\left(\bm{k}\right)}D^{-q_1\left(\bm{k}\right)/2 - q_2\left(\bm{k}\right)}\left(\overleftarrow{\prod_{t = 1}^{q + 1}}V_tO_t^{k_t}\right)\ket{\psi} \right\rVert_2 & \leq \mathcal{O}\left(D^{-3/2}\right),
\end{align}
with implicit constant only depending on $q$ and norms $\left\lVert O_t \right\rVert$ as desired.
\end{proof}
\end{lemma}
We now apply lemma \ref{lemma:second_order_expansion_time_correlations} to expectation
\begin{align}
    & \bra{+}\overrightarrow{\prod_{t = 1}^p}\exp\left(\frac{i\gamma_t}{\sqrt{D}}z_{\varnothing}^{[t]}Z_{\varnothing} + \frac{i\gamma_t}{\sqrt{D}}\sum_{\left(v,\,(v,\,k)\right) \in \mathcal{E}^{\left(\Delta,\,D\right)}}Z_vZ_{(v,\,k)}\right)\exp\left(i\beta_t\sum_{v \in \mathcal{V}^{\left(\Delta,\,D\right)}}X_v\right)\nonumber\\
    & \hspace*{18px} \overleftarrow{\prod_{t = 1}^p}\exp\left(-i\beta_t\sum_{v \in \mathcal{V}^{\left(\Delta,\,D\right)}}X_v\right)\exp\left(-\frac{i\gamma_t}{\sqrt{D}}z_{\varnothing}^{[t]}Z_{\varnothing} - \frac{i\gamma_t}{\sqrt{D}}\sum_{\left(v,\,(v,\,k)\right) \in \mathcal{E}^{\left(\Delta,\,D\right)}}Z_vZ_{(v,\,k)}\right)\ket{+}\nonumber\\
    & = \bra{+}\overrightarrow{\prod_{t = 1}^p}\exp\left(\frac{i\gamma_t}{\sqrt{D}}z_{\varnothing}^{[t]}Z_{\varnothing}\right)\exp\left(\frac{i\gamma_t}{\sqrt{D}}\sum_{\left(v,\,(v,\,k)\right) \in \mathcal{E}^{\left(\Delta,\,D\right)}}Z_vZ_{(v,\,k)}\right)\exp\left(i\beta_t\sum_{v \in \mathcal{V}^{\left(\Delta,\,D\right)}}X_v\right)\nonumber\\
    & \hspace*{30px} \overleftarrow{\prod_{t = 1}^p}\exp\left(-i\beta_t\sum_{v \in \mathcal{V}^{\left(\Delta,\,D\right)}}X_v\right)\exp\left(-\frac{i\gamma_t}{\sqrt{D}}\sum_{\left(v,\,(v,\,k)\right) \in \mathcal{E}^{\left(\Delta,\,D\right)}}Z_vZ_{(v,\,k)}\right)\exp\left(-\frac{i\gamma_t}{\sqrt{D}}z_{\varnothing}^{[t]}Z_{\varnothing}\right)\ket{+},\label{eq:subtree_expectation}
\end{align}
where the unitaries close to identity $e^{iO_t/\sqrt{D}}$ are the 
\begin{align}
    \exp\left(\frac{i\gamma_t}{\sqrt{D}}z^{[t]}_{\varnothing}\right), \quad \exp\left(-\frac{i\gamma_t}{\sqrt{D}}z^{[-t]}_{\varnothing}\right) \qquad t \in [p],
\end{align}
inserted between the QAOA unitaries playing the role of the $V_t$. More specifically, adapting the indexing of lemma \ref{lemma:second_order_expansion_time_correlations} for convenience, we rewrite expectation \ref{eq:subtree_expectation} as
\begin{align}
    & \bra{\psi}\left(\overrightarrow{\prod_{t = 1}^p}e^{iO_t/\sqrt{D}}V_t\right)\left(\overleftarrow{\prod_{t = 1}^p}V_{-t}e^{iO_{-t}/\sqrt{D}}\right)\ket{\psi}\nonumber\\
    & = \bra{\psi}e^{iO_1/\sqrt{D}}V_1e^{iO_2/\sqrt{D}}V_2e^{iO_3/\sqrt{D}}V_3\ldots V_{-3}e^{iO_{-3}/\sqrt{D}}V_{-2}e^{iO_{-2}/\sqrt{D}}V_{-1}e^{iO_{-1}/\sqrt{D}}\ket{\psi},
\end{align}
where
\begin{align}
    \mathcal{H} & := \left(\mathbb{C}^2\right)^{\otimes \mathcal{V}^{\left(\Delta,\,D\right)}},\\
    \ket{\psi} & := \ket{+}_{\mathcal{V}^{\left(\Delta,\,D\right)}},\\
    V_t & := \exp\left(i\gamma_t\sum_{(v,\,(v,\,j))\in \mathcal{E}^{\left(\Delta,\,D\right)}}Z_vZ_{(v, j)}\right)\exp\left(i\beta_t\sum_{v \in \mathcal{V}^{\left(\Delta,\,D\right)}}X_v\right), && t \in [p],\\
    O_t & = \gamma_tz^{[t]}_{\varnothing}Z_{\varnothing}, && t \in [p],\\
    V_{-t} & = \exp\left(-i\beta_t\sum_{v \in \mathcal{V}^{\left(\Delta,\,D\right)}}X_v\right)\exp\left(-i\gamma_t\sum_{(v,\,(v,\,j))\in \mathcal{E}^{\left(\Delta,\,D\right)}}Z_vZ_{(v, j)}\right), && t \in [p],\\
    O_{-t} & = -\gamma_tz^{[-t]}_{\varnothing}Z_{\varnothing}, && t \in [p].
\end{align}
Note that for all $t \in [p]$, $V_{-t} = V_t^{\dagger}$; however, $O_t$ and $O_{-t}$ may be opposite or equal to each other depending on the values of independent path integral bits $z^{[t]}_{\varnothing}, z^{[-t]}_{\varnothing}$. Let us compute moments $\mu^{(0)}, \bm{\mu}^{(1)}, \bm{\mu}^{(2)}$ with these adapted index notations. First,
\begin{align}
    \mu^{(0)} & = \bra{\psi}\left(\overrightarrow{\prod_{t = 1}^p}V_t\right)\left(\overleftarrow{\prod_{t = 1}^p}V_{-t}\right)\ket{\psi}\nonumber\nonumber\\
    & = \bra{\psi}\left(\overrightarrow{\prod_{t = 1}^p}V_t\right)\left(\overleftarrow{\prod_{t = 1}^p}V_t^{\dagger}\right)\ket{\psi}\nonumber\\
    & = \bra{\psi}\left(\overrightarrow{\prod_{t = 1}^p}V_t\right)\left(\overrightarrow{\prod_{t = 1}^p}V_t\right)^{\dagger}\ket{\psi}\nonumber\\
    & = \braket{\psi}{\psi}\nonumber\\
    & = 1.\label{eq:apply_second_order_correlations_expansion_mu0}
\end{align}
The first-order moment $\bm{\mu}^{(1)}$ vanishes by $\mathbf{Z}_2$-symmetry of the Hamiltonian. More explicitly, fixing $l \in [p]$ for definiteness (we also need to consider $l \in -[p]$),
\begin{align}
    \mu^{(1)}_l & = \bra{\psi}\left(\overrightarrow{\prod_{t = 1}^{l - 1}}V_t\right)O_l\left(\overrightarrow{\prod_{t = l}^p}V_t\right)\left(\overleftarrow{\prod_{t = 1}^p}V_{-t}\right)\ket{\psi}\nonumber\\
    & = \bra{+}_{\mathcal{V}^{\left(\Delta,\,D\right)}}\left(\overrightarrow{\prod_{t = 1}^{l - 1}}V_t\right)O_l\left(\overrightarrow{\prod_{t = l}^p}V_t\right)\left(\overleftarrow{\prod_{t = 1}^p}V_{-t}\right)\ket{+}_{\mathcal{V}^{\left(\Delta,\,D\right)}}\nonumber\\
    & = \bra{+}_{\mathcal{V}^{\left(\Delta,\,D\right)}}\left(\overrightarrow{\prod_{t = 1}^{l - 1}}V_t\right)O_l\left(\overrightarrow{\prod_{t = l}^p}V_t\right)\left(\overleftarrow{\prod_{t = 1}^p}V_{-t}\right)X^{\otimes \mathcal{V}^{\left(\Delta,\,D\right)}}\ket{+}_{\mathcal{V}^{\left(\Delta,\,D\right)}}\nonumber\\
    & = \bra{+}_{\mathcal{V}^{\left(\Delta,\,D\right)}}\left(\overrightarrow{\prod_{t = 1}^{l - 1}}V_t\right)O_lX^{\otimes \mathcal{V}^{\left(\Delta,\,D\right)}}\left(\overrightarrow{\prod_{t = l}^p}V_t\right)\left(\overleftarrow{\prod_{t = 1}^p}V_{-t}\right)\ket{+}_{\mathcal{V}^{\left(\Delta,\,D\right)}}\nonumber\\
    & = -\bra{+}_{\mathcal{V}^{\left(\Delta,\,D\right)}}\left(\overrightarrow{\prod_{t = 1}^{l - 1}}V_t\right)X^{\otimes \mathcal{V}^{\left(\Delta,\,D\right)}}O_l\left(\overrightarrow{\prod_{t = l}^p}V_t\right)\left(\overleftarrow{\prod_{t = 1}^p}V_{-t}\right)\ket{+}_{\mathcal{V}^{\left(\Delta,\,D\right)}}\nonumber\\
    & = -\bra{+}_{\mathcal{V}^{\left(\Delta,\,D\right)}}X^{\otimes \mathcal{V}^{\left(\Delta,\,D\right)}}\left(\overrightarrow{\prod_{t = 1}^{l - 1}}V_t\right)O_l\left(\overrightarrow{\prod_{t = l}^p}V_t\right)\left(\overleftarrow{\prod_{t = 1}^p}V_{-t}\right)\ket{+}_{\mathcal{V}^{\left(\Delta,\,D\right)}}\nonumber\\
    & = -\bra{+}_{\mathcal{V}^{\left(\Delta,\,D\right)}}\left(\overrightarrow{\prod_{t = 1}^{l - 1}}V_t\right)O_l\left(\overrightarrow{\prod_{t = l}^p}V_t\right)\left(\overleftarrow{\prod_{t = 1}^p}V_{-t}\right)\ket{+}_{\mathcal{V}^{\left(\Delta,\,D\right)}}\nonumber\\
    & = -\mu^{(1)}_l.\label{eq:apply_second_order_correlations_expansion_mu1_calculation}
\end{align}
Between the second and third line, we used that $\ket{+}^{\otimes \mathcal{V}^{\left(\Delta,\,D\right)}}$ is a $+1$ eigenstate of the $X^{\otimes \mathcal{V}^{\left(\Delta,\,D\right)}}$ operator. Between the third and fourth line, we used this operator commutes through the mixer and cost Hamiltonian ---for the latter, this follows from $\left[X \otimes X, Z \otimes Z\right] = 0$. Between the fourth and the fifth line, we used that the operator anticommutes with $O_l$, as $O_l \propto Z_{\varnothing}$ and $X$ anticommutes with $Z$. In the following lines, we used again commutation of the operator with the QAOA unitaries and stability of the $\ket{+}^{\otimes \mathcal{V}^{\left(\Delta,\,D\right)}}$ under this operator. All in all, this shows that
\begin{align}
    \mu^{\left(1\right)}_l & = 0.\label{eq:apply_second_order_correlations_expansion_mu1}
\end{align}
The reasoning for $l \in -[p]$ is similar. Let us now consider the second-order moments. Let $l, m$ be indices from $\pm [p]$. For definiteness, let us first consider the case where they are both positive and assume $l \leq m$. We then consider:
\begin{align}
    \mu^{\left(2\right)}_{l, m} & = \bra{\psi}\left(\overrightarrow{\prod_{t = 1}^{l - 1}}V_t\right)O_l\left(\overrightarrow{\prod_{t = l}^{m - 1}}V_t\right)O_m\left(\overrightarrow{\prod_{t = m}^p}V_t\right)\left(\overleftarrow{\prod_{t = 1}^p}V_{-t}\right)\ket{\psi}\nonumber\\
    & = \bra{+}_{\mathcal{V}^{\left(\Delta,\,D\right)}}\left(\overrightarrow{\prod_{t = 1}^{l - 1}}e^{i\gamma_mC^{\left(\Delta,\,D\right)}}e^{i\beta_tB^{\left(\Delta,\,D\right)}}\right)\left(\gamma_lz^{[l]}_{\varnothing}Z_{\varnothing}\right)\left(\overrightarrow{\prod_{t = l}^{m - 1}}e^{i\gamma_tC^{\left(\Delta,\,D\right)}}e^{i\beta_tB^{\left(\Delta,\,D\right)}}\right)\left(\gamma_mz^{[m]}_{\varnothing}Z_{\varnothing}\right)\nonumber\\
    & \hspace*{50px} \times \left(\overrightarrow{\prod_{t = m}^p}e^{i\gamma_tC^{\left(\Delta,\,D\right)}}e^{i\beta_tB^{\left(\Delta,\,D\right)}}\right)\left(\overleftarrow{\prod_{t = 1}^p}e^{-i\beta_tB^{\left(\Delta,\,D\right)}}e^{i\gamma_tC^{\left(\Delta,\,D\right)}}\right)\ket{+}_{\mathcal{V}^{\left(\Delta,\,D\right)}}\nonumber\\
    & = \gamma_l\gamma_mz^{[l]}_{\varnothing}z^{[m]}_{\varnothing}\bra{+}_{\mathcal{V}^{\left(\Delta,\,D\right)}}\left(\overrightarrow{\prod_{t = 1}^{l - 1}}e^{i\gamma_mC^{\left(\Delta,\,D\right)}}e^{i\beta_tB^{\left(\Delta,\,D\right)}}\right)Z_{\varnothing}\left(\overrightarrow{\prod_{t = l}^{m - 1}}e^{i\gamma_tC^{\left(\Delta,\,D\right)}}e^{i\beta_tB^{\left(\Delta,\,D\right)}}\right)Z_{\varnothing}\nonumber\\
    & \hspace*{110px} \times \left(\overrightarrow{\prod_{t = m}^p}e^{i\gamma_tC^{\left(\Delta,\,D\right)}}e^{i\beta_tB^{\left(\Delta,\,D\right)}}\right)\left(\overleftarrow{\prod_{t = 1}^p}e^{-i\beta_tB^{\left(\Delta,\,D\right)}}e^{i\gamma_tC^{\left(\Delta,\,D\right)}}\right)\ket{+}_{\mathcal{V}^{\left(\Delta,\,D\right)}}\nonumber\\
    & = \gamma_l\gamma_mz^{[l]}_{\varnothing}z^{[m]}_{\varnothing}\overline{\bra{+}_{\mathcal{V}^{\left(\Delta,\,D\right)}}\left(\overrightarrow{\prod_{t = 1}^p}e^{i\gamma_tC^{\left(\Delta,\,D\right)}}e^{i\beta_tB^{\left(\Delta,\,D\right)}}\right)\left(\overleftarrow{\prod_{t = m}^p}e^{-i\beta_tB^{\left(\Delta,\,D\right)}}e^{-i\gamma_tC^{\left(\Delta,\,D\right)}}\right)Z_{\varnothing}}\nonumber\\
    & \hspace*{115px} \overline{\left(\overleftarrow{\prod_{t = l}^{m - 1}}e^{-i\beta_tB^{\left(\Delta,\,D\right)}}e^{-i\gamma_tC^{\left(\Delta,\,D\right)}}\right)Z_{\varnothing}\left(\overleftarrow{\prod_{t = 1}^{l - 1}}e^{-i\beta_tB^{\left(\Delta,\,D\right)}}e^{-i\gamma_tC^{\left(\Delta,\,D\right)}}\right)\ket{+}_{\mathcal{V}^{\left(\Delta,\,D\right)}}}\nonumber\\
    & = \gamma_l\gamma_mz^{[l]}_{\varnothing}z^{[m]}_{\varnothing}\overline{G^{\left(\mathrm{sym},\,\Delta,\,D\right)}_{l, m}}.
\end{align}
The other possible signs of $l, m$ are reduced to this case by canceling unitaries $V_t, V_{-t}$ and using complex conjugation (similar to the example) to reflect the position of $Z_{\varnothing}$ operator insertions from left to right. For instance, still assuming $1 \leq l \leq m \leq p$ but now computing $\mu^{\left(2\right)}_{-l, m}$,
\begin{align}
    \mu^{(2)}_{-l, m} & = \bra{\psi}\left(\overrightarrow{\prod_{t = 1}^{m - 1}}V_t\right)O_m\left(\overrightarrow{\prod_{t = m}^p}V_t\right)\left(\overleftarrow{\prod_{t = l}^p}V_{-t}\right)O_{-l}\left(\overleftarrow{\prod_{t = 1}^{l - 1}}V_{-t}\right)\ket{\psi}\nonumber\\
    & = \bra{\psi}\left(\overrightarrow{\prod_{t = 1}^{m - 1}}V_t\right)O_m\left(\overrightarrow{\prod_{t = m}^p}V_t\right)\left(\overleftarrow{\prod_{t = m}^p}V_{-t}\right)\left(\overleftarrow{\prod_{t = l}^{m - 1}}V_{-t}\right)O_{-l}\left(\overleftarrow{\prod_{t = 1}^{l - 1}}V_{-t}\right)\ket{\psi}\nonumber\\
    & = \bra{\psi}\left(\overrightarrow{\prod_{t = 1}^{m - 1}}V_t\right)O_m\left(\overrightarrow{\prod_{t = m}^p}V_t\right)\left(\overrightarrow{\prod_{t = m}^p}V_t\right)^{\dagger}\left(\overleftarrow{\prod_{t = l}^{m - 1}}V_{-t}\right)O_{-l}\left(\overleftarrow{\prod_{t = 1}^{l - 1}}V_{-t}\right)\ket{\psi}\nonumber\\
    & =  \bra{\psi}\left(\overrightarrow{\prod_{t = 1}^{m - 1}}V_t\right)O_m\left(\overleftarrow{\prod_{t = l}^{m - 1}}V_{-t}\right)O_{-l}\left(\overleftarrow{\prod_{t = 1}^{l - 1}}V_{-t}\right)\ket{\psi}\nonumber\\
    & = \bra{\psi}\left(\overrightarrow{\prod_{t = 1}^{m - 1}}V_t\right)\left(\overrightarrow{\prod_{t = m}^{p}}V_t\right)\left(\overrightarrow{\prod_{t = m}^p}V_t\right)^{\dagger}O_m\left(\overleftarrow{\prod_{t = l}^{m - 1}}V_{-t}\right)O_{-l}\left(\overleftarrow{\prod_{t = 1}^{l - 1}}V_{-t}\right)\ket{\psi}\nonumber\\
    & = \bra{\psi}\left(\overrightarrow{\prod_{t = 1}^{m - 1}}V_t\right)\left(\overrightarrow{\prod_{t = m}^{p}}V_t\right)\left(\overleftarrow{\prod_{t = m}^p}V_{-t}\right)O_m\left(\overleftarrow{\prod_{t = l}^{m - 1}}V_{-t}\right)O_{-l}\left(\overleftarrow{\prod_{t = 1}^{l - 1}}V_{-t}\right)\ket{\psi}\nonumber\nonumber\\
    & = -\gamma_l\gamma_mz^{[-l]}_{\varnothing}z^{[m]}_{\varnothing}\bra{\psi}\left(\overrightarrow{\prod_{t = 1}^{m - 1}}V_t\right)\left(\overrightarrow{\prod_{t = m}^{p}}V_t\right)\left(\overleftarrow{\prod_{t = m}^p}V_{-t}\right)Z_{\varnothing}\left(\overleftarrow{\prod_{t = l}^{m - 1}}V_{-t}\right)Z_{\varnothing}\left(\overleftarrow{\prod_{t = 1}^{l - 1}}V_{-t}\right)\ket{\psi}\nonumber\\
    & = -\gamma_l\gamma_mz^{[-l]}_{\varnothing}z^{[m]}_{\varnothing}G^{\left(\mathrm{sym},\,\Delta,\,D\right)}_{l, m}
\end{align}
Similarly, still assuming $1 \leq l \leq m \leq p$,
\begin{align}
    \mu^{(2)}_{l, -m} & = -\gamma_l\gamma_mz^{[l]}_{\varnothing}z^{[-m]}_{\varnothing}\overline{G^{\left(\mathrm{sym},\,\Delta,\,\,D\right)}},\\
    \mu^{(2)}_{-l, -m} & = \gamma_l\gamma_mz^{[-l]}_{\varnothing}z^{[-m]}_{\varnothing}G^{\left(\mathrm{sym},\,\Delta,\,D\right)}_{l, m}.
\end{align}
The expressions for $\mu^{(2)}_{l, m}$ can be unified for all signs of $l, m$ as
\begin{align}
    \mu^{(2)}_{l, m} & = G^{\left(\Delta,\,D\right)}_{l, m}\Gamma_l\Gamma_mz^{[l]}_{\varnothing}z^{[m]}_{\varnothing} \qquad \forall l, m \in \pm [p],\label{eq:apply_second_order_correlations_expansion_mu2}
\end{align}
where we introduced symmetric matrix 
\begin{align}
    \bm{G}^{\left(\Delta,\,D\right)} & := \left(G^{\left(\Delta,\,D\right)}_{l, m}\right)_{l, m \in \pm [p]},
\end{align}
with entries defined as follows for $1 \leq l \leq m \leq p$:
\begin{align}
    G^{\left(\Delta,\,D\right)}_{l, m} & := \overline{G^{\left(\mathrm{sym},\,\Delta,\,D\right)}_{l, m}},\\
    G^{\left(\Delta,\,D\right)}_{-l, m} & := G^{\left(\mathrm{sym},\,\Delta,\,D\right)}_{l, m},\\
    G^{\left(\Delta,\,D\right)}_{l, -m} & := \overline{G^{\left(\mathrm{sym},\,\Delta,\,D\right)}_{l, m}},\\
    G^{\left(\Delta,\,D\right)}_{-l, -m}  &:= G^{\left(\mathrm{sym},\,\Delta,\,D\right)}_{l, m}.
\end{align}
The motivation for this notation is the following: by the induction hypothesis applied to $G^{\left(\mathrm{sym},\,\Delta,\,D\right)}$,
\begin{align}
    \bm{G}^{\left(\Delta,\,D\right)} & \xrightarrow[D \to \infty]{} \bm{G}^{\left(\Delta\right)},
\end{align}
where $\bm{G}^{\left(\Delta\right)}$ coincides with the step $\Delta$ iterate of the $\bm{G}$ matrix as defined in \cite{basso_et_al:LIPIcs.TQC.2022.7}. Now we computed moments $\mu^{\left(0\right)}, \bm{\mu}^{\left(1\right)}$ and $\bm{\mu}^{\left(2\right)}$, lemma \ref{lemma:second_order_expansion_time_correlations} allows to write an asymptotic expansion in the $D \to \infty$ limit of the expectation in equation \ref{eq:subtree_expectation}:
\begin{align}
    & \bra{+}\overrightarrow{\prod_{t = 1}^p}\exp\left(\frac{i\gamma_t}{\sqrt{D}}z_{\varnothing}^{[t]}Z_{\varnothing} + \frac{i\gamma_t}{\sqrt{D}}\sum_{\left(v,\,(v,\,k)\right) \in \mathcal{E}^{\left(\Delta,\,D\right)}}Z_vZ_{(v,\,k)}\right)\exp\left(i\beta_t\sum_{v \in \mathcal{V}^{\left(\Delta,\,D\right)}}X_v\right)\nonumber\\
    & \hspace*{18px} \overleftarrow{\prod_{t = 1}^p}\exp\left(-i\beta_t\sum_{v \in \mathcal{V}^{\left(\Delta,\,D\right)}}X_v\right)\exp\left(-\frac{i\gamma_t}{\sqrt{D}}z_{\varnothing}^{[t]}Z_{\varnothing} - \frac{i\gamma_t}{\sqrt{D}}\sum_{\left(v,\,(v,\,k)\right) \in \mathcal{E}^{\left(\Delta,\,D\right)}}Z_vZ_{(v,\,k)}\right)\ket{+}\nonumber\\
    & = \bra{\psi}\left(\overrightarrow{\prod_{t = 1}^p}e^{iO_t/\sqrt{D}}V_t\right)\left(\overleftarrow{\prod_{t = 1}^p}V_{-t}e^{iO_{-t}/\sqrt{D}}\right)\ket{\psi}\nonumber\\
    & = \mu^{(0)} + \frac{i}{\sqrt{D}}\sum_{t \in \pm [p]}\mu_t - \frac{1}{2D}\sum_{l, m \in \pm [p]}\mu^{[2]}_{l, m} + \mathcal{O}\left(D^{-3/2}\right)\nonumber\\
    & = 1 - \frac{1}{2D}\sum_{l, m \in \pm [p]}G^{\left(\Delta,\,D\right)}_{l, m}\Gamma_l\Gamma_mz^{[l]}_{\varnothing}z^{[m]}_{\varnothing} + \mathcal{O}\left(D^{-3/2}\right).
\end{align}
We recall the implicit constant in the error term only depends on $q = 2p$ and the norms of operators $O_t$, i.e. $\left\lVert O_t \right\rVert = |\gamma_t|$; in particular, this implicit constant has no dependence on $D$. Raising this asymptotic expansion to the power $D$, we reach the following estimate of the expectation over \textit{all} subtrees $\mathcal{T}^{\left(\Delta,\,D\right)}$ in equation \ref{eq:tree_root_correlation_cavity_expression_subtrees_expectation}:
\begin{align}
    & \left(\bra{+}\overrightarrow{\prod_{t = 1}^p}\exp\left(\frac{i\gamma_t}{\sqrt{D}}z_{\varnothing}^{[t]}Z_{\varnothing} + \frac{i\gamma_t}{\sqrt{D}}\sum_{\left(v,\,(v,\,k)\right) \in \mathcal{E}^{\left(\Delta,\,D\right)}}Z_vZ_{(v,\,k)}\right)\exp\left(i\beta_t\sum_{v \in \mathcal{V}^{\left(\Delta,\,D\right)}}X_v\right)\right.\nonumber\\
    & \hspace*{30px} \left.\overleftarrow{\prod_{t = 1}^p}\exp\left(-i\beta_t\sum_{v \in \mathcal{V}^{\left(\Delta,\,D\right)}}X_v\right)\exp\left(-\frac{i\gamma_t}{\sqrt{D}}z_{\varnothing}^{[t]}Z_{\varnothing} - \frac{i\gamma_t}{\sqrt{D}}\sum_{\left(v,\,(v,\,k)\right) \in \mathcal{E}^{\left(\Delta,\,D\right)}}Z_vZ_{(v,\,k)}\right)\ket{+}\right)^D\nonumber\\
    & = \left(1 - \frac{1}{2D}\sum_{l, m \in \pm [p]}G^{\left(\Delta,\,D\right)}_{l, m}\Gamma_l\Gamma_mz^{[l]}_{\varnothing}z^{[m]}_{\varnothing} + \mathcal{O}\left(D^{-3/2}\right)\right)^D\nonumber\\
    & = \exp\left(D\log\left\{1 - \frac{1}{2D}\sum_{l, m \in \pm [p]}G^{\left(\Delta,\,D\right)}_{l, m}\Gamma_l\Gamma_mz^{[l]}_{\varnothing}z^{[m]}_{\varnothing} + \mathcal{O}\left(D^{-3/2}\right)\right\}\right)\nonumber\\
    & = \exp\left(-\frac{1}{2}\sum_{l, m \in \pm [p]}G^{\left(\Delta,\,D\right)}_{l, m}\Gamma_l\Gamma_mz^{[l]}_{\varnothing}z^{[m]}_{\varnothing} + \mathcal{O}\left(D^{-1/2}\right)\right)\nonumber\\
    & = \exp\left(-\frac{1}{2}\sum_{l, m \in \pm [p]}G^{\left(\Delta\right)}_{l, m}\Gamma_l\Gamma_mz^{[l]}_{\varnothing}z^{[m]}_{\varnothing} + o_D\left(1\right)\right)\nonumber\\
    & = \exp\left(-\frac{1}{2}\sum_{l, m \in \pm [p]}G^{\left(\Delta\right)}_{l, m}\Gamma_l\Gamma_mz^{[l]}_{\varnothing}z^{[m]}_{\varnothing}\right) + o_D(1).
\end{align}
In the final two lines, $o_D\left(1\right)$ refers to a vanishing quantity in the limit $D \to \infty$ (we need not know how fast it vanishes). Plugging this estimate for the subtrees expectation (equation \ref{eq:tree_root_correlation_cavity_expression_subtrees_expectation}), as well as equation \ref{eq:tree_root_correlation_cavity_expression_root_expectation}, into equation \ref{eq:tree_root_correlation_cavity_expression} for $G^{\left(\mathrm{sym},\,\Delta + 1,\,D\right)}_{j, k}$, yields the following estimate for this quantity:
\begin{align}
    G^{\left(\mathrm{sym},\,\Delta + 1,\,D\right)}_{j, k} & = \sum_{\substack{\bm{z}_{\varnothing} \in \{1, -1\}^{2p + 1}}}z^{[-j]}_{\varnothing}z^{[-k]}_{\varnothing}f\left(\bm{z}_{\varnothing}\right)\left(\exp\left(-\frac{1}{2}\sum_{l, m \in \pm [p]}G^{\left(\Delta\right)}_{l, m}\Gamma_l\Gamma_mz^{[l]}_{\varnothing}z^{[m]}_{\varnothing}\right) + o_D\left(1\right)\right)\nonumber\\
    & = \sum_{\substack{\bm{z}_{\varnothing} \in \{1, -1\}^{2p + 1}}}z^{[-j]}_{\varnothing}z^{[-k]}_{\varnothing}f\left(\bm{z}_{\varnothing}\right)\exp\left(-\frac{1}{2}\sum_{l, m \in \pm [p]}G^{\left(\Delta\right)}_{l, m}\Gamma_l\Gamma_mz^{[l]}_{\varnothing}z^{[m]}_{\varnothing}\right) + o_D\left(1\right),
\end{align}
where the last simplification uses that the number of of terms in the sum and quantity $f\left(\bm{z}_{\varnothing}\right)$ are independent of $D$. The above estimate now shows
\begin{align}
    G^{\left(\mathrm{sym},\,\Delta + 1,\,D\right)}_{j, k} & \xrightarrow[D \to \infty]{}  \sum_{\substack{\bm{z}_{\varnothing} \in \{1, -1\}^{2p + 1}}}z^{[-j]}_{\varnothing}z^{[-k]}_{\varnothing}f\left(\bm{z}_{\varnothing}\right)\exp\left(-\frac{1}{2}\sum_{l, m \in \pm [p]}G^{\left(\Delta\right)}_{l, m}\Gamma_l\Gamma_mz^{[l]}_{\varnothing}z^{[m]}_{\varnothing}\right)\nonumber\\
    & = \sum_{\bm{a} \in \{1, -1\}^{2p + 1}}a_{-j}a_{-k}f\left(\bm{a}\right)\exp\left(-\frac{1}{2}\sum_{l, m \in \pm [p]}G^{\left(\Delta\right)}_{l, m}\Gamma_l\Gamma_la_la_m\right)\nonumber\\
    & = G^{\left(\Delta + 1\right)}_{-l, -m},
\end{align}
where the final step is by definition of the $\bm{G}$ matrix iteration \cite{basso_et_al:LIPIcs.TQC.2022.7}.
This finishes the proof of the induction hypothesis $\Delta \to \Delta + 1$ and the proof of proposition \ref{prop:relation_g_matrix}.

\subsection{Mapping of level-$p$ QAOA to $p$-modes spin-bosons system}
\label{sec:proof_mapping}

In this section, we derive a succession of approximation for the QAOA state on the tree (definition \ref{def:bethe_lattice}) of depth $\Delta$ and degree $D$
\begin{align}
    \ket{\Psi^{\left(\Delta,\,D\right)}\left(\bm{\gamma}, \bm{\beta}\right)} & = \left(\overleftarrow{\prod_{t = 1}^p}e^{-i\beta_tB^{\left(\Delta,\,D\right)}}e^{-i\gamma_tC^{\left(\Delta,\,D\right)}}\right)\ket{+}_{\mathcal{T}^{\left(\Delta,\,D\right)}},
\end{align}
as introduced in definition \ref{def:qaoa_state}. More specifically, the approximation will hold ``in the limit $D \to \infty$". Rigorously defining approximation of states in this limit is at least not immediate, as the Hilbert spaces states live are distinct, with dimensions depending on $D$. Fortunately, under some embedding, states can be made to ``live in the same space". We sketch the principle of this embedding in the rest of paragraph. First, similar to the proof of proposition \ref{prop:relation_g_matrix}, we note tree $\mathcal{T}^{\left(\Delta,\,D\right)}$ consists of the junction of $D$ copies of $\mathcal{T}^{\left(\Delta - 1,\,D\right)}$ and a root vertex. This translates to the following identification of quantum state spaces:
\begin{align}
    \left(\mathbb{C}^2\right)^{\otimes \mathcal{V}^{\left(\Delta,\,D\right)}} & \simeq \mathbb{C}^2 \otimes \left(\left(\mathbb{C}^2\right)^{\otimes \mathcal{V}^{\left(\Delta - 1,\,D\right)}}\right)^{\otimes D}.
\end{align}
We will then observe that $\ket{\Psi^{\left(\Delta,\,D\right)}\left(\bm{\gamma}, \bm{\beta}\right)}$, projected onto any state of the root vertex, lives in the symmetric subspace of the $\mathcal{T}^{\left(\Delta - 1,\,D\right)}$ subtrees
\begin{align}
    \left(\left(\mathbb{C}^2\right)^{\otimes \mathcal{V}^{\left(\Delta - 1,\,D\right)}}\right)^{\otimes D}.
\end{align}
The symmetric subspace can in turn be viewed as a subspace of the Fock space ---more specifically, of the Fock space truncated to a number of particles $D$. Under this embedding, QAOA states can be seen as living in a common space independent of $D$. We will finally show that the embedded state converges to a well-defined Fock state in the limit $D \to \infty$.

We now make these statements formal. For that purpose, we will need the following simple technical lemma:
\begin{lemma}[Approximation of tensor power]
\label{lemma:tensor_power_approximation}
Let $\mathcal{H}$ any Hilbert space. Then, for any pair of vectors $\bm{u}, \bm{v} \in \mathcal{H}$, the following approximation holds in the limit $D \to \infty$:
\begin{align}
    \left\lVert \left(\bm{u} + \bm{v}\right)^{\otimes D} - \bm{u}^{\otimes D} \right\rVert_2 & \leq D\left\lVert \bm{v} \right\rVert_2\left(\left\lVert \bm{u} \right\rVert_2 + \left\lVert \bm{v} \right\rVert_2\right)^{D - 1}.\label{eq:tensor_power_approximation_nonorthogonal}
\end{align}
Assuming orthogonality $\bm{u} \perp \bm{v}$, we obtain tighter $D$ dependence
\begin{align}
    \left\lVert \left(\bm{u} + \bm{v}\right)^{\otimes D } - \bm{u}^{\otimes D} \right\rVert_2 & \leq \sqrt{D}\left\lVert \bm{v} \right\rVert_2\left\lVert \bm{u} + \bm{v} \right\rVert_2^{D - 1}\\
    & = \sqrt{D}\left\lVert \bm{v} \right\rVert_2\left(\left\lVert \bm{u} \right\rVert_2^2 + \left\lVert \bm{v} \right\rVert_2^2\right)^{(D - 1)/2}\label{eq:tensor_power_approximation_orthogonal}
\end{align}
\begin{proof}
We start with the following telescopic sum representation of the tensor powers difference:
\begin{align}
    & \left(\bm{u} + \bm{v}\right)^{\otimes D} - \bm{u}^{\otimes D}\nonumber\\
    & = \sum_{1 \leq k \leq D}\left(\left(\bm{u} + \bm{v}\right)^{\otimes k} \otimes \bm{u}^{\otimes (D - k)} - \left(\bm{u} + \bm{v}\right)^{\otimes (k - 1)} \otimes \bm{u}^{\otimes \left(D + 1 - k\right)}\right)\nonumber\\
    & = \sum_{1 \leq k \leq D}\left(\left(\bm{u} + \bm{v}\right)^{\otimes (k - 1)} \otimes \left(\bm{u} + \bm{v}\right) \otimes \bm{u}^{\otimes (D - k)} - \left(\bm{u} + \bm{v}\right)^{\otimes (k - 1)} \otimes \bm{u} \otimes \bm{u}^{\otimes \left(D - k\right)}\right)\nonumber\\
    & = \sum_{1 \leq k \leq D}\left(\bm{u} + \bm{v}\right)^{\otimes (k - 1)} \otimes \bm{v} \otimes \bm{u}^{\otimes (D - k)}
\end{align}
In general, the triangle inequality applied to the previous sum gives
\begin{align}
    \left\lVert \left(\bm{u} + \bm{v}\right)^{\otimes D} - \bm{u}^{\otimes D} \right\rVert_2 & \leq \sum_{1 \leq k \leq D}\left\lVert \left(\bm{u} + \bm{v}\right)^{\otimes (k - 1)} \otimes \bm{v} \otimes \bm{u}^{\otimes \left(D - k\right)} \right\rVert_2\nonumber\\
    & \leq \sum_{1 \leq k \leq D}\left\lVert \bm{u} + \bm{v} \right\rVert_2^{k - 1}\left\lVert \bm{v} \right\rVert_2\left\lVert \bm{u} \right\rVert_2^{D - k}\nonumber\\
    & = \sum_{1 \leq k \leq D}\left(\left\lVert \bm{u} \right\rVert + \left\lVert \bm{v} \right\rVert_2\right)^{k - 1}\left\lVert \bm{v} \right\rVert_2\left\lVert \bm{u} \right\rVert_2^{D - k}\nonumber\\
    & \leq \sum_{1 \leq k \leq D}\left(\left\lVert \bm{u} \right\rVert + \left\lVert \bm{v} \right\rVert_2\right)^{D - 1}\left\lVert \bm{v} \right\rVert_2\nonumber\\
    & = D\left\lVert \bm{v} \right\rVert_2\left(\left\lVert \bm{u} \right\rVert_2 + \left\lVert \bm{v} \right\rVert_2\right)^{D - 1},
\end{align}
which is the desired bound (equation \ref{eq:tensor_power_approximation_nonorthogonal}). Now further assuming orthogonality $\bm{u} \perp \bm{v}$ makes the terms in telescopic sum pairwise orthogonal, meaning
\begin{align}
    \left\lVert \left(\bm{u} + \bm{v}\right)^{\otimes D} - \bm{u}^{\otimes D} \right\rVert_2^2 & = \sum_{1 \leq k \leq D}\left\lVert \left(\bm{u} + \bm{v}\right)^{\otimes (k - 1)} \otimes \bm{v} \otimes \bm{u}^{\otimes \left(D - k\right)} \right\rVert_2^2\nonumber\\
    & = \sum_{1 \leq k \leq D}\left\lVert \left(\bm{u} + \bm{v}\right)^{\otimes (k - 1)} \right\rVert_2^2\left\lVert \bm{v} \right\rVert_2^2 \left\lVert \bm{u}^{\otimes (D - k)} \right\rVert_2^2\nonumber\\
    & = \sum_{1 \leq k \leq D}\left(\left\lVert \bm{u} \right\rVert_2^2 + \left\lVert \bm{v} \right\rVert_2^2\right)^{k - 1}\left\lVert \bm{v} \right\rVert_2^2 \left(\left\lVert \bm{u} \right\rVert_2^2\right)^{D - k}\nonumber\\
    & \leq \sum_{1 \leq k \leq D}\left(\left\lVert \bm{u} \right\rVert_2^2 + \left\lVert \bm{v} \right\rVert_2^2\right)^{D - 1}\left\lVert \bm{v} \right\rVert_2^2\nonumber\\
    & = D\left\lVert \bm{v} \right\rVert_2^2\left(\left\lVert \bm{u} \right\rVert_2^2 + \left\lVert \bm{v} \right\rVert_2^2\right)^{D - 1},
\end{align}
which is the required bound in the case of orthogonality (equation \ref{eq:tensor_power_approximation_orthogonal}).
\end{proof}
\end{lemma}

We now derive a succession of approximations (lemmas \ref{lemma:qaoa_state_approximation_1} and \ref{lemma:qaoa_state_approximation_2}) for the QAOA state, where the error vanishes in the limit $D \to \infty$.

\begin{lemma}[The QAOA state in the large $D$ limit: a first approximation]
\label{lemma:qaoa_state_approximation_1}
Define for $l < m$ the QAOA stated modified by inserting $Z_{\varnothing}$ operators before layers $l$ and $m$:
\begin{align}
    \ket{\Psi_{l, m}^{(\Delta,\,D)}} & := \left(\overleftarrow{\prod_{t = m}^{p}}e^{-i\beta_tB}e^{-i\gamma_tC}\right)Z_{\varnothing}\left(\overleftarrow{\prod_{t = l}^{m - 1}}e^{-i\beta_tB}e^{-i\gamma_tC}\right)Z_{\varnothing}\left(\overleftarrow{\prod_{t = 1}^{l - 1}}e^{-i\beta_tB}e^{-i\gamma_tC}\right)\ket{+},
\end{align}
similar to definition \ref{def:qaoa_state_z_inserted} but with $Z_{\varnothing}$ inserted at two locations instead of one. Then, the following approximation holds for the QAOA state on the depth-$\Delta$ tree in terms of the states on the depth-$(\Delta - 1)$ tree:
\begin{align}
    \ket{\Psi^{(\Delta,\,D)}}  & = \sum_{\bm{a}_{1:p + 1} \in \{1, -1\}^{p + 1}}f_{1:p}\left(\bm{a}_{1:p + 1}\right)\ket{a_{p + 1}} \otimes \left(\left(1 - \sum_{1 \leq l \leq p}\frac{\gamma_l^2}{2D}\right)\ket{\Psi^{(\Delta - 1,\,D)}(\bm\gamma, \bm\beta)} + \sum_{1 \leq l \leq p}\frac{-i\gamma_la_l}{\sqrt{D}}\ket{\Psi_l^{(\Delta - 1,\,D)}(\bm\gamma, \bm\beta)}\right.\nonumber\\
    & \left. \hspace*{190px} + \sum_{1 \leq l < m \leq p}\frac{-\gamma_l\gamma_ma_la_m}{D}\ket{\Psi_{l, m}^{(\Delta - 1,\,D)}\left(\bm\gamma, \bm\beta\right)}\right)^{\otimes D}\nonumber\\
    & \hspace*{20px} + \mathcal{O}_{\left\lVert\cdot \right\rVert_2}\left(D^{-1/2}\right)\label{eq:qaoa_state_approximation_1}
\end{align}
In the above equation, the tensor powers forming the Hilbert space of qubits indexed by vertices $\mathcal{V}^{\left(\Delta,\,D\right)}$ of $\mathcal{T}^{\left(\Delta,\,D\right)}$ is assumed to be ordered as in the proof of proposition \ref{prop:relation_g_matrix}, i.e.:
\begin{align}
    \left(\mathbb{C}^2\right)^{\otimes \mathcal{V}^{\left(\Delta,\,D\right)}} & \simeq \underbrace{\mathbb{C}^2}_{\textrm{qubit $\varnothing$ Hilbert space}} \otimes \underbrace{\left(\left(\mathbb{C}^2\right)^{\otimes \mathcal{V}^{\left(\Delta - 1,\,D\right)}}\right)^{\otimes D}}_{\textrm{$D$ copies of $\mathcal{T}^{\left(\Delta - 1,\,D\right)}$}}.
\end{align}
\begin{proof}
The QAOA over the regular tree of depth $\Delta$ and degree $D$ as introduced in definition \ref{def:qaoa_state} is:
\begin{align}
    \ket{\Psi^{\left(\Delta,\,D\right)}} & = \left(\overleftarrow{\prod_{t = 1}^p}e^{-i\beta_tB^{\left(\Delta,\,D\right)}}e^{-i\gamma_tC^{\left(\Delta,\,D\right)}}\right)\ket{+}_{\mathcal{V}^{\left(\Delta,\,D\right)}}.
\end{align}
Similar to the proof of proposition \ref{prop:relation_g_matrix}, we insert $(p + 1)$ completeness relations over computational basis states of the root vertex $\varnothing$ qubits:
\begin{align}
    \sum_{z^{[l]}_{\varnothing} \in \{1, -1\}}\ket{z^{[l]}_{\varnothing}}\bra{z^{[l]}_{\varnothing}} & = \bm{I}_{\varnothing} \otimes \bm{I}_{\mathcal{V}^{\left(\Delta,\,D\right)} - \{\varnothing\}}, \qquad 1 \leq l \leq p + 1.
\end{align}
We collect the bits of these completeness relations into a bitstring
\begin{align}
    \bm{z}_{\varnothing} & := \left(z^{[p + 1]}_{\varnothing}, z^{[p]}_{\varnothing}, \ldots, z^{[2]}_{\varnothing}, z^{[1]}_{\varnothing}\right) \in \{1, -1\}^{p + 1}
\end{align}
This recasts the state to:
\begin{align}    & \ket{\Psi^{\left(\Delta,\,D\right)}}\nonumber\\
    & = \sum_{\bm{z}_{\varnothing} \in \{1, -1\}^{p + 1}}\ket{z^{[p + 1]}_{\varnothing}}_{\varnothing}\bra{z^{[p + 1]}_{\varnothing}}_{\varnothing}\left(\overleftarrow{\prod_{t = 1}^p}e^{-i\beta_tB^{\left(\Delta,\,D\right)}}e^{-i\gamma_tC^{\left(\Delta,\,D\right)}}\ket{z^{[t]}_{\varnothing}}_{\varnothing}\bra{z^{[t]}_{\varnothing}}_{\varnothing}\right)\ket{+}_{\mathcal{V}^{\left(\Delta,\,D\right)}}\nonumber\\
    & = \sum_{\bm{z}_{\varnothing} \in \{1, -1\}^{p + 1}}\braket{z^{[1]}_{\varnothing}}{+}\ket{z^{[p + 1]}_{\varnothing}}_{\varnothing}\left(\overleftarrow{\prod_{t = 1}^p}\bra{z^{[t + 1]}}_{\varnothing}e^{-i\beta_tB^{\left(\Delta,\,D\right)}}e^{-i\gamma_tC^{\left(\Delta,\,D\right)}}\ket{z^{[t]}_{\varnothing}}_{\varnothing}\right)\ket{+}_{\mathcal{V}^{\left(\Delta,\,D\right)} - \{\varnothing\}}\label{eq:qaoa_state_tree_decomposition_step_1}
\end{align}
Reasoning as in the proof of proposition \ref{prop:relation_g_matrix}, we can decompose the cost and mixer Hamiltonians by singling out the root vertex $\varnothing$:
\begin{align}
    C^{\left(\Delta,\,D\right)} & = \frac{1}{\sqrt{D}}\sum_{j \in [D]}Z_{\varnothing}Z_j + \sum_{j \in [D]}C^{\left(\Delta,\,D\right)}_{\succeq j}\\
    & = \frac{1}{\sqrt{D}}\sum_{j \in [D]}Z_{\varnothing}Z_j + \frac{1}{\sqrt{D}}\sum_{j \in [D]}\sum_{\left(v,\,\left(v, k\right)\right) \in \mathcal{E}^{\left(\Delta - 1,\,D\right)}_{\succeq j}}Z_{v}Z_{(v, k)},\\
    B^{\left(\Delta,\,D\right)} & = X_{\varnothing} + \sum_{j \in [D]}B^{\left(\Delta,\,D\right)}_{\succeq j}\\
    & = X_{\varnothing} + \sum_{j \in [D]}\sum_{v \in \mathcal{V}^{\left(\Delta,\,D\right)}_{\succeq j}}X_v.
\end{align}
In these two decompositions, the second term is supported outside of the root vertex $\varnothing$. This allows to evaluate the ``matrix elements" (more accurately, the partial traces) in the last equation as follows:
\begin{align}
    \bra{z^{[t + 1]}_{\varnothing}}_{\varnothing}e^{-i\beta_tB^{\left(\Delta,\,D\right)}}e^{-i\gamma_tC^{\left(\Delta,\,D\right)}}\ket{z^{[t]}_{\varnothing}}_{\varnothing} & = \bra{z^{[t + 1]}_{\varnothing}}e^{-i\beta_tX}\ket{z^{[t]}_{\varnothing}}\left(\prod_{j \in [D]}e^{-i\beta_t B^{\left(\Delta,\,D\right)}_{\succeq j}}e^{-i\gamma_tC^{\left(\Delta,\,D\right)}_{\succeq j}}e^{-i\gamma_t z^{[t]}_{\varnothing}Z_j/\sqrt{D}}\right).\label{eq:qaoa_state_tree_decomposition_matrix_element}
\end{align}
We then notice the operator in the final equation is a product over $j \in [D]$ of unitaries acting identically on subtrees $\mathcal{T}^{\left(\Delta,\,D\right)}_{\succeq j}$. Therefore, using the same identification as in the proof of proposition \ref{prop:relation_g_matrix}:
\begin{align}
    \mathcal{T}^{\left(\Delta,\,D\right)}_{\succeq j} \simeq \mathcal{T}^{\left(\Delta - 1,\,D\right)} & \implies \left(\mathbb{C}^2\right)^{\otimes \mathcal{V}^{\left(\Delta,\,D\right)}_{\succeq j}} \simeq \left(\mathbb{C}^2\right)^{\otimes \mathcal{V}^{\left(\Delta - 1,\,D\right)}}\nonumber\\
    & \implies \left(\mathbb{C}^2\right)^{\otimes \mathcal{V}^{\left(\Delta,\,D\right)}} \simeq \mathbb{C}^2 \otimes \left(\left(\mathbb{C}^2\right)^{\otimes \mathcal{V}^{\left(\Delta - 1,\,D\right)}}\right)^{\otimes D},
\end{align}
equation \ref{eq:qaoa_state_tree_decomposition_matrix_element} can be rewritten as:
\begin{align}
    & \bra{z^{[t + 1]}_{\varnothing}}_{\varnothing}e^{-i\beta_tB^{\left(\Delta,\,D\right)}}e^{-i\gamma_tC^{\left(\Delta,\,D\right)}}\ket{z^{[t]}_{\varnothing}}_{\varnothing} \simeq \bra{z^{[t + 1]}_{\varnothing}}e^{-i\beta_tX}\ket{z^{[t]}_{\varnothing}}\left(e^{-i\beta_t B^{\left(\Delta - 1,\,D\right)}}e^{-i\gamma_tC^{\left(\Delta - 1,\,D\right)}}e^{-i\gamma_t z^{[t]}_{\varnothing}Z_{\varnothing}/\sqrt{D}}\right)^{\otimes D}
\end{align}
Plugging this into equation \ref{eq:qaoa_state_tree_decomposition_step_1} gives the following expression for the QAOA state:
\begin{align}
    \ket{\Psi^{\left(\Delta,\,D\right)}} & \simeq \sum_{\bm{z}_{\varnothing} \in \{1, -1\}^{p + 1}}\braket{z^{[1]}_{\varnothing}}{+}\prod_{t = 1}^p\bra{z^{[t + 1]}_{\varnothing}}e^{-i\beta_t}\ket{z^{[t]}_{\varnothing}}\nonumber\\
    & \hspace*{80px} \times \ket{z^{[p + 1]}} \otimes \left(\overleftarrow{\prod_{t = 1}^p}e^{-i\beta_tB^{\left(\Delta - 1,\,D\right)}}e^{-i\gamma_tC^{\left(\Delta - 1,\,D\right)}}e^{-i\gamma_tz^{[t]}_{\varnothing}Z_{\varnothing}/\sqrt{D}}\ket{+}_{\mathcal{V}^{\left(\Delta - 1,\,D\right)}}\right)^{\otimes D}\nonumber\\
    & = \sum_{\bm{z}_{\varnothing} \in \{1, -1\}^{p + 1}}f_{1:p}\left(\bm{a}_{1:p + 1}\right)\ket{z^{[p + 1]}} \otimes \left(\overleftarrow{\prod_{t = 1}^p}e^{-i\beta_tB^{\left(\Delta - 1,\,D\right)}}e^{-i\gamma_tC^{\left(\Delta - 1,\,D\right)}}e^{-i\gamma_tz^{[t]}_{\varnothing}Z_{\varnothing}/\sqrt{D}}\ket{+}_{\mathcal{V}^{\left(\Delta - 1,\,D\right)}}\right)^{\otimes D}\label{eq:qaoa_state_tree_decomposition_step_2}
\end{align}
We now fix a bitstring $\bm{z}_{\varnothing}$ and consider an approximation of the state raised to the tensor power $D$. More specifically, we consider an approximation arising from the leading-order Taylor expansion of the near-identity unitary (in the limit $D \to \infty$):
\begin{align}
    \exp\left(-\frac{i\gamma_tz^{[t]}_{\varnothing}}{\sqrt{D}}Z_{\varnothing}\right) & = 1 - \frac{i\gamma_tz^{[t]}}{\sqrt{D}}Z_{\varnothing} - \frac{\gamma_t^2}{2D} + \mathcal{O}_{\left\lVert \cdot \right\rVert}\left(D^{-3/2}\right),
\end{align}
where $\mathcal{O}_{\left\lVert \cdot \right\rVert}\left(D^{-3/2}\right)$ refers to an operator with operator norm order bounded by a constant times $D^{-3/2}$. It is important (the contrary would make the statement vacuous) that this constant does not depend on $D$. We now use this asymptotic expansion of the operator in each of the $p$ layers
\begin{align}
    e^{-i\beta_tB^{\left(\Delta - 1,\,D\right)}}e^{-i\gamma_tC^{\left(\Delta - 1,\,D\right)}}e^{-i\gamma_tz^{[t]}_{\varnothing}Z_{\varnothing}/\sqrt{D}},
\end{align}
to obtain an following estimate for the state raised to the tensor power $D$ in equation \ref{eq:qaoa_state_tree_decomposition_step_2}. To lighten the notation, we temporarily introduce the following symbols for the QAOA ansatz unitaries over the tree of depth $\left(\Delta - 1\right)$ and degree $D$:
\begin{align}
    V_t & := e^{-i\beta_tB^{\left(\Delta - 1,\,D\right)}}e^{-i\gamma_tC^{\left(\Delta - 1,\,D\right)}}.
\end{align}
From these notations, the state raised to the tensor power $D$ in equation \ref{eq:qaoa_state_tree_decomposition_step_2} reads:
\begin{align}
    & \left(\overleftarrow{\prod_{t = 1}^p}e^{-i\beta_tB^{\left(\Delta - 1,\,D\right)}}e^{-i\gamma_tC^{\left(\Delta - 1,\,D\right)}}e^{-i\gamma_tz^{[t]}_{\varnothing}Z_{\varnothing}/\sqrt{D}}\right)\ket{+}_{\mathcal{V}^{\left(\Delta - 1,\,D\right)}}\nonumber\\
    & = \left(\overleftarrow{\prod_{t = 1}^p}V_te^{-i\gamma_tz^{[t]}_{\varnothing}Z_{\varnothing}/\sqrt{D}}\right)\ket{+}_{\mathcal{V}^{\left(\Delta - 1,\,D\right)}}\nonumber\\
    & = \left(\overleftarrow{\prod_{t = 1}^p}V_t\right)\ket{+}_{\mathcal{V}^{\left(\Delta - 1,\,D\right)}} - \frac{i}{\sqrt{D}}\sum_{1 \leq l \leq p}\gamma_lz^{[l]}_{\varnothing}\left(\overleftarrow{\prod_{t = l}^p}V_t\right)Z_{\varnothing}\left(\overleftarrow{\prod_{t = 1}^{l - 1}}V_t\right)\ket{+}_{\mathcal{V}^{\left(\Delta - 1,\,D\right)}}\nonumber\\
    & \hspace*{20px} - \frac{1}{2D}\sum_{1 \leq l \leq p}\gamma_l^2\left(\overleftarrow{\prod_{t = 1}^p}V_t\right)\ket{+}_{\mathcal{V}^{\left(\Delta - 1,\,D\right)}}\nonumber\\
    & \hspace*{20px} - \frac{1}{2D}\sum_{1 \leq l < m \leq p}\gamma_l\gamma_mz_{\varnothing}^{[l]}z_{\varnothing}^{[m]}\left(\overleftarrow{\prod_{t = m}^p}V_t\right)Z_{\varnothing}\left(\overleftarrow{\prod_{t = l}^{m - 1}}V_t\right)Z_{\varnothing}\left(\overleftarrow{\prod_{t = 1}^{l - 1}}V_t\right)\ket{+}_{\mathcal{V}^{\left(\Delta - 1,\,D\right)}}\nonumber\\
    & \hspace*{20px} + \mathcal{O}_{\left\lVert \cdot \right\rVert_2}\left(D^{-3/2}\right)\\
    & = \left(1 - \sum_{1 \leq l \leq p}\frac{\gamma_l^2}{2D}\right)\ket{\Psi^{\left(\Delta - 1,\,D\right)}} + \sum_{1 \leq l \leq p}\frac{-i\gamma_lz^{[l]}_{\varnothing}}{\sqrt{D}}\ket{\Psi_l^{\left(\Delta - 1,\,D\right)}} + \sum_{1 \leq l < m \leq p}\frac{-\gamma_l\gamma_mz^{[l]}_{\varnothing}z^{[m]}_{\varnothing}}{2D}\ket{\Psi^{\left(\Delta - 1,\,D\right)}_{l, m}} + \mathcal{O}_{\left\lVert \cdot \right\rVert_2}\left(D^{-3/2}\right)
\end{align}
where the $\mathcal{O}_{\left\lVert \cdot \right\rVert_2}\left(D^{-3/2}\right)$ denotes a vector bounded in the Hilbert space norm by a constant times $D^{-3/2}$, where the constant may depend on $\bm\beta, \bm\gamma, p$, but certainly not on $D$; the situation and reasoning is similar to the proof of lemma \ref{lemma:second_order_expansion_time_correlations}. We now use lemma \ref{lemma:tensor_power_approximation}, more specifically the general (nonorthogonal) case equation \ref{eq:tensor_power_approximation_nonorthogonal}, to raise the previous estimate to the tensor power $D$. More precisely, we rewrite this estimate in the form:
\begin{align}
    \left(\bm{u} + \bm{v}\right)^{\otimes D} & = \bm{u}^{\otimes D} + \mathcal{O}\left(D\left\lVert \bm{v} \right\rVert_2\left(\left\lVert \bm{u} \right\rVert_2 + \left\lVert \bm{v} \right\rVert_2 \right)^{D - 1}\right)\nonumber\\
    & = \bm{u}^{\otimes D} + \mathcal{O}\left(D\left\lVert \bm{v} \right\rVert_2\left(\left\lVert \bm{u} + \bm{v} \right\rVert_2 + 2\left\lVert \bm{v} \right\rVert_2 \right)^{D - 1}\right),\label{eq:tensor_power_approximation_nonorthogonal_applied}
\end{align}
and let 
\begin{align}
    \bm{u} + \bm{v} & := \left(\overleftarrow{\prod_{t = 1}^p}e^{-i\beta_tB^{\left(\Delta - 1,\,D\right)}}e^{-i\gamma_tC^{\left(\Delta - 1,\,D\right)}}e^{-i\gamma_tz^{[t]}_{\varnothing}Z_{\varnothing}/\sqrt{D}}\right)\ket{+}_{\mathcal{V}^{\left(\Delta - 1,\,D\right)}},\nonumber\\
    \bm{u} & := \left(1 - \sum_{1 \leq l \leq p}\frac{\gamma_l^2}{2D}\right)\ket{\Psi^{\left(\Delta,\,D\right)}} + \sum_{1 \leq l \leq p}\frac{-i\gamma_lz^{[l]}_{\varnothing}}{\sqrt{D}}\ket{\Psi_l^{\left(\Delta,\,D\right)}} + \sum_{1 \leq l < m \leq p}\frac{-\gamma_l\gamma_mz^{[l]}_{\varnothing}z^{[m]}_{\varnothing}}{2D}\ket{\Psi^{\left(\Delta,\,D\right)}_{l, m}},
\end{align}
so that
\begin{align}
    \bm{v} & = \mathcal{O}_{\left\lVert \cdot \right\rVert_2}\left(D^{-3/2}\right).
\end{align}
In this case, the error term in equation \ref{eq:tensor_power_approximation_nonorthogonal_applied} can be estimated:
\begin{align}
    \mathcal{O}_{\left\lVert \cdot \right\rVert_2}\left(D\left\lVert \bm{v} \right\rVert_2 \left(\left\lVert \bm{u} + \bm{v} \right\rVert_2 + \left\lVert \bm{v} \right\rVert_2\right)^{D - 1}\right) & = \mathcal{O}_{\left\lVert \cdot \right\rVert_2}\left(D\mathcal{O}\left(D^{-3/2}\right)\left(1 + \mathcal{O}\left(D^{-3/2}\right)\right)^D\right)\nonumber\\
    & = \mathcal{O}_{\left\lVert \cdot \right\rVert_2}\left(D^{-1/2}\left(1 + \mathcal{O}\left(D^{-1/2}\right)\right)\right)\nonumber\\
    & = \mathcal{O}_{\left\lVert \cdot \right\rVert_2}\left(D^{-1/2}\right),\label{eq:tensor_power_approximation_nonorthogonal_applied_error_term}
\end{align}
proving the lemma.
\end{proof}

\end{lemma}

\begin{lemma}[The QAOA state in the large $D$ limit: a second approximation]
\label{lemma:qaoa_state_approximation_2}
Let the setting and notation be as in lemma \ref{lemma:qaoa_state_approximation_1}. Then, the following other approximation holds on the QAOA state on the alternative Bethe lattice of depth $\Delta$ and degree $D$ in the large $D$ limit:
\begin{align}
    \ket{\Psi^{(\Delta,\,D)}} & = \sum_{\bm{a}_{1:p + 1} \in \{1, -1\}^{p + 1}}f_{1:p}\left(\bm{a}_{1:p + 1}\right)\exp\left(-\frac{1}{2}\sum_{1 \leq j, k \leq p}G^{(\mathrm{sym},\,\Delta - 1,\,D)}_{j, k}\gamma_j\gamma_ka_ja_k\right)\nonumber\\
    & \hspace*{90px} \times \ket{a_{p + 1}} \otimes \left(\ket{\Psi^{(\Delta - 1,\,D)}} + \sum_{1 \leq l \leq p}\frac{-i\gamma_la_l}{\sqrt{D}}\ket{\Psi_l^{(\Delta - 1,\,D)}}\right)^{\otimes D}\nonumber\\
    & \hspace*{20px} + \mathcal{O}_{\left\lVert \cdot \right\rVert_2}\left(D^{-1/2}\right)\label{eq:qaoa_state_approximation_2}
\end{align}
\begin{proof}
We start with the approximation of the state established in lemma \ref{lemma:qaoa_state_approximation_1}:
\begin{align}
    & \ket{\Psi^{(\Delta,\,D)}}\nonumber\\
    & = \sum_{\bm{a}_{1:p + 1} \in \{1, -1\}^{p + 1}}f_{1:p}\left(\bm{a}_{1:p + 1}\right)\ket{a_{p + 1}} \otimes \left(\left(1 - \sum_{1 \leq l \leq p}\frac{\gamma_l^2}{2D}\right)\ket{\Psi^{(\Delta - 1,\,D)}} + \sum_{1 \leq l \leq p}\frac{-i\gamma_la_l}{\sqrt{D}}\ket{\Psi_l^{(\Delta - 1,\,D)}}\right.\nonumber\\
    & \left. \hspace*{190px} + \sum_{1 \leq l < m \leq p}\frac{-\gamma_l\gamma_ma_la_m}{D}\ket{\Psi_{l, m}^{(\Delta - 1,\,D)}}\right)^{\otimes D}\nonumber\\
    & \hspace*{20px} + \mathcal{O}_{\left\lVert\cdot \right\rVert_2}\left(D^{-1/2}\right)\label{eq:qaoa_state_approximation_1_rewritten}
\end{align}
As suggested by the target formula equation \ref{eq:qaoa_state_approximation_2}, we wish to remove the states with two $Z$ insertions:
\begin{align}
    \ket{\Psi^{\left(\Delta - 1,\,D\right)}_{l, m}} & := \left(\overleftarrow{\prod_{t = m}^p}V_t\right)Z_{\varnothing}\left(\overleftarrow{\prod_{t = l}^{m - 1}}V_t\right)Z_{\varnothing}\left(\overleftarrow{\prod_{t = 1}^{l - 1}}V_t\right)\ket{+}_{\mathcal{V}^{\left(\Delta - 1,\,D\right)}}
\end{align}
inside the tensor power, where we used the same notation as in the proof of lemma \ref{lemma:qaoa_state_approximation_1} for the QAOA unitaries:
\begin{align}
    V_t & := \exp\left(-i\beta_tB^{\left(\Delta - 1,\,D\right)}\right)\exp\left(-i\gamma_tC^{\left(\Delta - 1,\,D\right)}\right).
\end{align}
To achieve that, we consider the consider the orthogonal projection of $\ket{\Psi^{\left(\Delta - 1,\,D\right)}_{l, m}}$ onto $\ket{\Psi^{\left(\Delta - 1,\,D\right)}}$
\begin{align}
    \ket{\Psi^{\left(\Delta - 1,\,D\right)}_{l, m}} & =: \ket{\Psi^{\left(\Delta - 1,\,D\right)}_{\parallel;\,l, m}} + \ket{\Psi^{\left(\Delta - 1,\,D\right)}_{\perp;\,l, m}},\label{eq:qaoa_state_double_z_insertion_orthogonal_decomposition}\\
    \ket{\Psi^{\left(\Delta - 1,\,D\right)}_{\parallel;\,l, m}} & := \braket{\Psi^{\left(\Delta - 1,\,D\right)}}{\Psi^{\left(\Delta - 1,\,D\right)}_{l, m}}\ket{\Psi^{\left(\Delta - 1,\,D\right)}},\\
    \ket{\Psi^{\left(\Delta - 1,\,D\right)}_{\perp;\,l, m}} & := \ket{\Psi^{\left(\Delta - 1,\,D\right)}_{l, m}} - \ket{\Psi^{\left(\Delta - 1,\,D\right)}_{\parallel;\,l, m}}\nonumber\\
    & \perp \ket{\Psi^{\left(\Delta - 1,\,D\right)}}.
\end{align}
We now observe the following orthogonality relations:
\begin{align}
    \ket{\Psi^{\left(\Delta - 1,\,D\right)}} & \perp \ket{\Psi^{\left(\Delta - 1,\,D\right)}_l} && \forall 1 \leq l \leq p,\\
    \ket{\Psi^{\left(\Delta - 1,\,D\right)}_l} & \perp \ket{\Psi^{\left(\Delta - 1,\,D\right)}_{m, n}} && \forall 1 \leq l \leq p,\,1 \leq m < n \leq p.
\end{align}
The first relation follows from
\begin{align}
    \braket{\Psi^{\left(\Delta - 1,\,D\right)}}{\Psi^{\left(\Delta - 1,\,D\right)}_l} & = \bra{+}_{\mathcal{V}^{\left(\Delta - 1,\,D\right)}}\left(\overrightarrow{\prod_{t = 1}^p}V_t^{\dagger}\right)\left(\overleftarrow{\prod_{t = l}^p}V_t\right)Z_{\varnothing}\left(\overleftarrow{\prod_{t = 1}^{l - 1}}V_t\right)\ket{+}_{\mathcal{V}^{\left(\Delta - 1,\,D\right)}}\nonumber\\
    & = 0
\end{align}
by $\mathbf{Z}_2$-symmetry of the QAOA ansatz ---see the explicit calculation in equation \ref{eq:apply_second_order_correlations_expansion_mu1_calculation}. The second claimed orthogonality relation follows similarly. From these two relations, combined with the orthogonal decomposition of $\ket{\Psi^{\left(\Delta - 1,\,D\right)}_{l, m}}$ in equation \ref{eq:qaoa_state_double_z_insertion_orthogonal_decomposition}, it follows the additional orthogonality relation:
\begin{align}
    \ket{\Psi^{\left(\Delta - 1,\,D\right)}_l} & \perp \ket{\Psi^{\left(\Delta - 1,\,D\right)}_{\perp;\,m, n}} && \forall 1 \leq l \leq p,\,1 \leq m < n \leq p.
\end{align}
Finally, we explicitly compute the orthogonal projection coefficient of $\ket{\Psi^{\left(\Delta - 1,\,D\right)}_{l, m}}$ onto $\ket{\Psi^{\left(\Delta - 1,\,D\right)}}$:
\begin{align}
    \braket{\Psi^{\left(\Delta,\,D\right)}}{\Psi^{\left(\Delta,\,D\right)}_{l, m}} & = \bra{+}_{\mathcal{V}^{\left(\Delta,\,D\right)}}\left(\overrightarrow{\prod_{t = 1}^p}V_t\right)\left(\overleftarrow{\prod_{t = m}^p}V_t\right)Z_{\varnothing}\left(\overleftarrow{\prod_{t = l}^{m - 1}}V_t\right)Z_{\varnothing}\left(\overleftarrow{\prod_{t = 1}^{l - 1}}V_t\right)\ket{+}_{\mathcal{V}^{\left(\Delta - 1,\,D\right)}}\nonumber\\
    & = G^{\left(\mathrm{sym},\,\Delta - 1,\,D\right)}_{l, m}.
\end{align}
All in all, we have reexpressed the vector raised to the tensor power $D$ in equation \ref{eq:qaoa_state_approximation_1_rewritten} as follows:
\begin{align}
    & \left(1 - \sum_{1 \leq l \leq p}\frac{\gamma_l^2}{2D}\right)\ket{\Psi^{\left(\Delta - 1,\,D\right)}} + \sum_{1 \leq l \leq p}\frac{-i\gamma_la_l}{\sqrt{D}}\ket{\Psi^{\left(\Delta - 1,\,D\right)}_l} + \sum_{1 \leq l < m \leq p}\frac{-\gamma_l\gamma_ma_la_m}{D}\ket{\Psi^{\left(\Delta - 1,\,D\right)}_{l, m}}\nonumber\\
    & = \left(1 - \sum_{1 \leq l \leq p}\frac{\gamma_l^2}{2D} + \sum_{1 \leq l < m \leq p}\frac{-\gamma_l\gamma_ma_la_m}{D}G^{\left(\mathrm{sym},\,\Delta - 1,\,D\right)}\right)\ket{\Psi^{\left(\Delta - 1,\,D\right)}} + \sum_{1 \leq l \leq p}\frac{-i\gamma_la_l}{\sqrt{D}}\ket{\Psi^{\left(\Delta - 1,\,D\right)}_l}\nonumber\\
    & \hspace*{20px} + \sum_{1 \leq l < m \leq p}\frac{-\gamma_l\gamma_ma_la_m}{D}\ket{\Psi^{\left(\Delta - 1,\,D\right)}_{\perp;\,l, m}}\nonumber\\
    & = \left(1 - \frac{1}{2D}\sum_{1 \leq l, m \leq p}G^{\left(\mathrm{sym},\,\Delta - 1,\,D\right)}\gamma_l\gamma_ma_la_m\right)\ket{\Psi^{\left(\Delta - 1,\,D\right)}} + \sum_{1 \leq l \leq p}\frac{-i\gamma_la_l}{\sqrt{D}}\ket{\Psi^{\left(\Delta - 1,\,D\right)}_l}\nonumber\\
    & \hspace*{20px} + \sum_{1 \leq l < m \leq p}\frac{-\gamma_l\gamma_ma_la_m}{D}\ket{\Psi^{\left(\Delta - 1,\,D\right)}_{\perp;\,l, m}}
\end{align}
We now observe the final term:
\begin{align}
    \sum_{1 \leq l < m \leq p}\frac{-\gamma_l\gamma_ma_la_m}{D}\ket{\Psi^{\left(\Delta - 1,\,D\right)}_{\perp;\,l, m}}
\end{align}
is orthogonal to the rest thanks to the orthogonality relations previously established. This suggests to drop it when taking the tensor power $D$, using the orthogonal version (equation \ref{eq:tensor_power_approximation_orthogonal}) of lemma \ref{lemma:tensor_power_approximation}. More specifically, we let:
\begin{align}
    \bm{u} & := \left(1 - \frac{1}{2D}\sum_{1 \leq l, m \leq p}G^{\left(\mathrm{sym},\,\Delta - 1,\,D\right)}\gamma_l\gamma_ma_la_m\right)\ket{\Psi^{\left(\Delta - 1,\,D\right)}} + \sum_{1 \leq l \leq p}\frac{-i\gamma_la_l}{\sqrt{D}}\ket{\Psi^{\left(\Delta - 1,\,D\right)}_l},\\
    \bm{v} & := \sum_{1 \leq l < m \leq p}\frac{-\gamma_l\gamma_ma_la_m}{D}\ket{\Psi^{\left(\Delta - 1,\,D\right)}_{\perp;\,l, m}}
\end{align}
and rewrite the estimate in equation \ref{eq:tensor_power_approximation_orthogonal} as:
\begin{align}
    \left(\bm{u} + \bm{v}\right)^{\otimes D} & = \bm{u}^{\otimes D} + \mathcal{O}_{\left\lVert \cdot \right\rVert_2}\left(D^{1/2}\left\lVert \bm{v} \right\rVert \left\lVert \bm{u} + \bm{v} \right\rVert^{D - 1}\right).
\end{align}
We estimate the error term in the present case:
\begin{align}
    \mathcal{O}\left(D^{1/2}\left\lVert \bm{v} \right\rVert\left\lVert \bm{u} + \bm{v} \right\rVert_2^{D - 1}\right) & = \mathcal{O}\left(D^{1/2}\left\lVert \bm{v} \right\rVert \left(1 + \mathcal{O}\left(D^{-1/2}\right)\right)\right)\nonumber\\
    & = \mathcal{O}\left(D^{1/2}\left\lVert \sum_{1 \leq l < m \leq p}\frac{-\gamma_l\gamma_la_la_m}{D}\ket{\Psi^{\left(\Delta - 1,\,D\right)}_{\perp;\,\l, m}} \right\rVert \left(1 + \mathcal{O}\left(D^{-1/2}\right)\right)\right)\nonumber\\
    & \leq \mathcal{O}\left(D^{1/2}\sum_{1 \leq l < m \leq p}\frac{|\gamma_l||\gamma_m|}{D}\left\lVert \ket{\Psi^{\left(\Delta - 1,\,D\right)}_{\perp;\,l, m}} \right\rVert \left(1 + \mathcal{O}\left(D^{-1/2}\right)\right)\right)\nonumber\\
    & \leq \mathcal{O}\left(D^{-1/2}\right),
\end{align}
where the implicit constant in the $\mathcal{O}$ depends on $p$ and the $\bm\gamma$ angles, but surely not on $D$. The estimate used from the first to second line for the norm of $\bm{u} + \bm{v}$:
\begin{align}
    \left\lVert \bm{u} + \bm{v} \right\rVert_2 & = 1 + \mathcal{O}\left(D^{-3/2}\right)
\end{align}
is from the proof of lemma \ref{lemma:qaoa_state_approximation_1} (calculation in equation \ref{eq:tensor_power_approximation_nonorthogonal_applied_error_term}). We have therefore proven
\begin{align}
    & \left(\left(1 - \sum_{1 \leq l \leq p}\frac{\gamma_l^2}{2D}\right)\ket{\Psi^{\left(\Delta - 1,\,D\right)}} + \sum_{1 \leq l \leq p}\frac{-i\gamma_la_l}{\sqrt{D}}\ket{\Psi^{\left(\Delta - 1,\,D\right)}_l} + \sum_{1 \leq l < m \leq p}\frac{-\gamma_l\gamma_ma_la_m}{D}\ket{\Psi^{\left(\Delta - 1,\,D\right)}_{l, m}}\right)^{\otimes D}\nonumber\\
    & = \left(\left(1 - \frac{1}{2D}\sum_{1 \leq l, m \leq p}G^{\left(\mathrm{sym},\,\Delta - 1,\,D\right)}\gamma_l\gamma_ma_la_m\right)\ket{\Psi^{\left(\Delta - 1,\,D\right)}} + \sum_{1 \leq l \leq p}\frac{-i\gamma_la_l}{\sqrt{D}}\ket{\Psi^{\left(\Delta - 1,\,D\right)}_l}\right)^{\otimes D}\nonumber\\
    & \hspace*{20px} + \mathcal{O}_{\left\lVert \cdot \right\rVert_2}\left(D^{-1/2}\right)\label{eq:qaoa_state_approximation_2_step_1}
\end{align}
Now, using approximation
\begin{align}
    \left(1 - \frac{1}{2D}\sum_{1 \leq l, m \leq p}G^{\left(\mathrm{sym},\,\Delta - 1,\,D\right)}\gamma_l\gamma_ma_la_m\right) & = \exp\left(-\frac{1}{2D}\sum_{1 \leq l, m \leq p}G^{\left(\mathrm{sym},\,\Delta - 1,\,D\right)}\gamma_l\gamma_ma_la_m\right) + \mathcal{O}\left(D^{-2}\right),
\end{align}
we write a further approximation of the vector raised to the tensor power $D$ in the previous formula. Namely,
\begin{align}
    & \left(1 - \frac{1}{2D}\sum_{1 \leq l, m \leq p}G^{\left(\mathrm{sym},\,\Delta - 1,\,D\right)}\gamma_l\gamma_ma_la_m\right)\ket{\Psi^{\left(\Delta - 1,\,D\right)}} + \sum_{1 \leq l \leq p}\frac{-i\gamma_la_l}{\sqrt{D}}\ket{\Psi^{\left(\Delta - 1,\,D\right)}_l}\nonumber\\
    & = \exp\left(-\frac{1}{2D}\sum_{1 \leq l, m \leq p}G^{\left(\mathrm{sym},\,\Delta - 1,\,D\right)}\gamma_l\gamma_ma_la_m\right)\left(\ket{\Psi^{\left(\Delta - 1,\,D\right)}} + \sum_{1 \leq l \leq p}\frac{-i\gamma_la_l}{\sqrt{D}}\ket{\Psi^{\left(\Delta - 1,\,D\right)}_l}\right) + \mathcal{O}_{\left\Vert \cdot \right\rVert_2}\left(D^{-3/2}\right),
\end{align}
where $\mathcal{O}_{\left\Vert \cdot \right\rVert_2}\left(D^{-3/2}\right)$ refers to a vector of 2-norm bounded by a constant times $D^{-3/2}$, with constant independent of $D$. Using again the non-orthogonal version of lemma \ref{lemma:tensor_power_approximation} (equation \ref{eq:tensor_power_approximation_nonorthogonal}), this estimate can be raised to the tensor power $D$ to give:
\begin{align}
    & \left(\left(1 - \frac{1}{2D}\sum_{1 \leq l, m \leq p}G^{\left(\mathrm{sym},\,\Delta - 1,\,D\right)}\gamma_l\gamma_ma_la_m\right)\ket{\Psi^{\left(\Delta - 1,\,D\right)}} + \sum_{1 \leq l \leq p}\frac{-i\gamma_la_l}{\sqrt{D}}\ket{\Psi^{\left(\Delta - 1,\,D\right)}_l}\right)^{\otimes D}\nonumber\\
    & = \exp\left(-\frac{1}{2}\sum_{1 \leq l, m \leq p}G^{\left(\mathrm{sym},\,\Delta - 1,\,D\right)}\gamma_l\gamma_ma_la_m\right)\left(\ket{\Psi^{\left(\Delta - 1,\,D\right)}} + \sum_{1 \leq l \leq p}\frac{-i\gamma_la_l}{\sqrt{D}}\ket{\Psi^{\left(\Delta - 1,\,D\right)}_l}\right)^{\otimes D} + \mathcal{O}_{\left\lVert \cdot \right\rVert_2}\left(D^{-1/2}\right)\label{eq:qaoa_state_approximation_2_step_2}
\end{align}
Plugging this estimate into equation \ref{eq:qaoa_state_approximation_2_step_1}, and the original estimate for the QAOA state equation \ref{eq:qaoa_state_approximation_1_rewritten} yields the result. 
\end{proof}
\end{lemma}

The main advantage of lemma \ref{lemma:qaoa_state_approximation_2} compared to lemma \ref{lemma:qaoa_state_approximation_1} is the removal of temporarily defined states $\ket{\Psi^{\left(\Delta - 1,\,D\right)}_{l, m}}$. Besides, the appearance of matrix $\bm{G}^{\left(\mathrm{sym},\,\Delta - 1,\,D\right)}$ already allows partial insight into the $D \to \infty$ limit. Indeed, by proposition \ref{prop:relation_g_matrix}, coefficient
\begin{align}
    \exp\left(-\frac{1}{2}\sum_{1 \leq j, k \leq p}G^{\left(\mathrm{sym},\,\Delta - 1,\,D\right)}\gamma_j\gamma_ka_ja_k\right) & \xrightarrow[D \to \infty]{} \exp\left(-\frac{1}{2}\sum_{1 \leq j, k \leq p}G^{\left(\Delta - 1\right)}_{-j,\,-k}\gamma_j\gamma_ka_ja_k\right)
\end{align}
in the state's expression converges to a well-defined limit as $D \to \infty$. In contrast, the limiting object for the state raised to the tensor power $D$:
\begin{align}
    \left(\ket{\Psi^{\left(\Delta - 1,\,D\right)}} + \sum_{1 \leq t \leq p}\frac{-i\gamma_ta_t}{\sqrt{D}}\ket{\Psi^{\left(\Delta - 1\right)}_j}\right)^{\otimes D}
\end{align}
remains unclear. We nonetheless observe this vector lives in the $D$-particles symmetric subspace, with single-particle space spanned by
\begin{align}
    \ket{\Psi^{\left(\Delta - 1\,D\right)}}, \left(\ket{\Psi^{\left(\Delta - 1,\,D\right)}_j}\right)_{1 \leq j \leq p}.
\end{align}
Intuitively, we regard the first of these states as the ``vacuum" and the others as excited states. It will then be more convenient to use an orthonormal basis $\left(\ket{\Psi^{\left(\Delta - 1,\,D\right)}_{\perp;\,l}}\right)_{1 \leq l \leq p}$ of excited states, defined by a basis change matrix $\bm{L}^{\left(\Delta - 1,\,D\right)} \in \mathbb{C}^{p \times p}$:
\begin{align}
     \ket{\Psi^{\left(\Delta - 1,\,D\right)}_j} & =: \sum_{1 \leq l \leq p}L^{\left(\Delta - 1,\,D\right)}_{l,\,j}\ket{\Psi^{\left(\Delta - 1,\,D\right)}_{\perp;\,l}}\label{eq:qaoa_maxcut_tree_excited_states_orthogonal_decomposition},\\
     \bm{L}^{\left(\Delta - 1,\,D\right)} & = \left(L^{\left(\Delta - 1,\,D\right)}_{l,\,t}\right)_{l, t \in [p]} \in \mathbb{C}^{p \times p},\\
     \braket{\Psi^{\left(\Delta - 1,\,D\right)}_{\perp;\,l}}{\Psi^{\left(\Delta - 1,\,D\right)}_{\perp;\,m}} & = \delta_{l, m}.
\end{align}
Computation of such a matrix only requires knowledge of the Gram matrix of original vectors $\left(\ket{\Psi^{\left(\Delta - 1,\,D\right)}_j}\right)_{1 \leq j \leq p}$. Recalling definition \ref{def:hermitian_g}, this Gram matrix is precisely the Hermitian matrix $\bm{G}^{\left(\mathrm{herm},\,\Delta - 1,\,D\right)}$, which according to proposition \ref{prop:relation_g_matrix} also coincides in the limit $D \to \infty$ with the ``Hermitian corner" of the $\bm{G}^{\left(\Delta - 1\right)}$ matrix computed in earlier work \cite{basso_et_al:LIPIcs.TQC.2022.7}. More precisely, by direct calculation any $\bm{L}$ satisfying equation \ref{eq:qaoa_maxcut_tree_excited_states_orthogonal_decomposition} must satisfy:
\begin{align}
    \bm{L}^{\left(\Delta - 1,\,D\right)\dagger}\bm{L}^{\left(\Delta - 1,\,D\right)} & = \bm{G}^{\left(\mathrm{herm},\,\Delta - 1,\,D\right)}.
\end{align}
Conversely, if $\bm{L}^{\left(\Delta - 1,\,D\right)}$ satisfies the latter equation and is nonsingular:
\begin{align}
    \ket{\Psi^{\left(\Delta - 1,\,D\right)}_{\perp;\,l}} & := \sum_{1 \leq j \leq p}\left[\left(\bm{L}^{\left(\Delta - 1,\,D\right)}\right)^{-1}\right]_{j,\,l}\ket{\Psi^{\left(\Delta - 1,\,D\right)}_j}
\end{align}
defines an orthonormal basis of the excited states. Hence, assuming $\bm{G}^{\left(\mathrm{herm},\,\Delta - 1,\,D\right)}$ non-singular, a concrete choice of $\bm{L}^{\left(\Delta - 1,\,D\right)}$ could be
\begin{align}
    \bm{L}^{\left(\Delta - 1,\,D\right)} & := \sqrt{\bm{G}^{\left(\mathrm{herm},\,\Delta - 1,\,D\right)}},
\end{align}
where the square-root is well-defined since the argument is a Gram matrix. Plugging this orthogonal decomposition into the state from equation \ref{eq:qaoa_state_approximation_2}:
\begin{align}
    \left(\ket{\Psi^{\left(\Delta - 1,\,D\right)}} + \sum_{1 \leq t \leq p}\frac{-i\gamma_ta_t}{\sqrt{D}}\ket{\Psi^{\left(\Delta - 1,\,D\right)}_t}\right)^{\otimes D} & = \left(\ket{\Psi^{\left(\Delta - 1,\,D\right)}} + \frac{1}{\sqrt{D}}\sum_{1 \leq l \leq p}\left(-i\sum_{1 \leq t \leq p}L_{l,\,t}\gamma_ta_t\right)\ket{\Psi^{\left(\Delta - 1,\,D\right)}_{\perp;\,l}}\right)^{\otimes D},\label{eq:qaoa_subtrees_state_orthonormal_basis}
\end{align}
this symmetric state can now be expressed in the $D$-particle subspace with
\begin{align}
    \ket{\Psi^{\left(\Delta - 1,\,D\right)}}, \left(\ket{\Psi_{\perp;\,l}^{\left(\Delta - 1,\,D\right)}}\right)_{1 \leq l \leq p}
\end{align}
as an orthonormal basis of the single particle space. For that purpose, we start by precisely defining a basis of this subspace:
\begin{definition}[Symmetric basis for subtrees state]
\label{def:finite_degree_symmetric_state}
Given integers
\begin{align}
    n_1, n_2, \ldots, n_{p - 1}, n_p \geq 0,
\end{align}
the symmetric states of occupation numbers $\left(n_l\right)_{0 \leq l \leq p}$ over subtrees $\mathcal{T}^{\left(\Delta - 1,\,D\right)}$ is defined as:
\begin{align}
    \ket{n_1, n_2, \ldots, n_{p - 1}, n_{p}}_{\Delta - 1,\,D} & := \binom{D}{\left(n_l\right)_{0 \leq l \leq p}}^{-1/2}\sum_{\substack{\left(l_j\right)_{j \in [D]} \in \{0,\,\ldots,\,p\}^D\\\forall 1 \leq l \leq p,\,\left|\{j \in [D]\,:\,l_j = l\}\right| = n_l}}\bigotimes_{j \in [D]}\ket{\Psi^{\left(\Delta - 1,\,D\right)}_{\perp;\,l_j}}.
\end{align}
In the above equation, we defined
\begin{align}
    n_0 & := D - \sum_{1 \leq l \leq p}n_l
\end{align}
and 
\begin{align}
    \ket{\Psi^{\left(\Delta - 1,\,D\right)}_{\perp;\,0}} := \ket{\Psi^{\left(\Delta - 1,\,D\right)}}.
\end{align}
Besides, we used the multinomial coefficient
\begin{align}
    \binom{D}{\left(n_l\right)_{0 \leq l \leq p}} & := \frac{D!}{\prod\limits_{0 \leq l \leq p}n_l!},
\end{align}
counting the number of ways of distributing $n_0$ states $\ket{\Psi^{\left(\Delta - 1,\,D\right)}_{\perp;\,0}} = \ket{\Psi^{\left(\Delta - 1,\,D\right)}}$, $n_1$ states $\ket{\Psi^{\left(\Delta - 1,\,D\right)}_{\perp;\,1}}$, $n_2$ states $\ket{\Psi^{\left(\Delta - 1,\,D\right)}_{\perp;\,2}}$, etc. among $D$ tensor powers. Since the sum in the formula is indeed over such tensor product states, and the single-particle states form an orthonormal basis, this combinatorial observation confirms $\ket{n_1, n_2, \ldots, n_{p - 1}, n_p}_{\Delta - 1,\,D}$ is properly normalized.
\end{definition}

We now express the coefficient of the state in equation \ref{eq:qaoa_subtrees_state_orthonormal_basis} in the basis introduced in definition \ref{def:finite_degree_symmetric_state}:

\begin{lemma}[Coefficient of subtrees state in symmetric basis]
\label{lemma:coefficient_subtree_states_symmetric_basis}
The subtrees state in equation \ref{eq:qaoa_subtrees_state_orthonormal_basis} can be expanded as follows is the symmetric states basis introduced in definition \ref{def:finite_degree_symmetric_state}:
\begin{align}
    & \left(\ket{\Psi^{\left(\Delta - 1,\,D\right)}} + \frac{1}{\sqrt{D}}\sum_{1 \leq l \leq p}\left(-i\sum_{1 \leq t \leq p}L_{l,\,t}\gamma_ta_t\right)\ket{\Psi^{\left(\Delta - 1,\,D\right)}_{\perp;\,l}}\right)^{\otimes D}\nonumber\\
    & = \sum_{\substack{\left(n_l\right)_{0 \leq l \leq p}\\\sum\limits_{0 \leq l \leq p}n_l = D}}\left(\prod_{1 \leq  l \leq p}\left(-i\sum_{1 \leq j \leq p}L_{l,\,j}\gamma_ja_j\right)^{n_l}\right)D^{-\frac{1}{2}\sum\limits_{1 \leq l \leq p}n_l}\binom{D}{\left(n_l\right)_{0 \leq l \leq p}}^{1/2}\ket{n_1, \ldots, n_p}_{\Delta - 1,\,D},\label{eq:qaoa_subtrees_state_finite_degree_fock_space_expansion}
\end{align}
Besides, for any fixed ($D$-independent) set of occupation numbers $\left(n_l\right)_{1 \leq l \leq p}$, the coefficient of the corresponding number state $\ket{n_1, \ldots, n_p}_{\Delta - 1,\,D}$ converges as $D \to \infty$:
\begin{align}
    & \bra{n_1, \ldots, n_p}_{\Delta - 1,\,D}\left(\ket{\Psi^{\left(\Delta - 1,\,D\right)}} + \frac{1}{\sqrt{D}}\sum_{1 \leq l \leq p}\left(-i\sum_{1 \leq t \leq p}L_{l,\,t}\gamma_ta_t\right)\ket{\Psi^{\left(\Delta - 1\right)}_{\perp;\,l}}\right)^{\otimes D}\nonumber\\
    & \xrightarrow[D \to \infty]{} \prod_{1 \leq l \leq p}\frac{1}{\sqrt{n_l!}}\left(-i\sum_{1 \leq j \leq p}L_{l,\,j}\gamma_ja_j\right)^{n_j}\label{eq:qaoa_subtrees_state_fock_coefficients_convergence}
\end{align}
\begin{proof}
We aim to express
\begin{align}
    \left(\ket{\Psi^{\left(\Delta - 1,\,D\right)}} + \frac{1}{\sqrt{D}}\sum_{1 \leq l \leq p}\left(-i\sum_{1 \leq t \leq p}L_{l,\,t}\gamma_ta_t\right)\ket{\Psi^{\left(\Delta - 1\right)}_{\perp;\,l}}\right)^{\otimes D}
\end{align}
in the symmetric basis. We then start by expanding the tensor product by distributivity. To slightly lighten the notation, write
\begin{align}
    \ket{\Psi^{\left(\Delta - 1,\,D\right)}} + \frac{1}{\sqrt{D}}\sum_{1 \leq l \leq p}\left(-i\sum_{1 \leq t \leq p}L_{l,\,t}\gamma_ta_t\right)\ket{\Psi^{\left(\Delta - 1\right)}_{\perp;\,l}} & := \sum_{1 \leq l \leq p}c_l\ket{\Psi^{\left(\Delta - 1,\,D\right)}_{\perp;\,l}},
\end{align}
where coefficients $c_l$ are defined by:
\begin{align}
    c_0 & := 1,\\
    c_l & := -\frac{i}{\sqrt{D}}\sum_{1 \leq t \leq p}L_{l,\,t}\gamma_ta_t, && 1 \leq l \leq p.
\end{align}
Note these depend on $D$. Then:
\begin{align}
    \left(\ket{\Psi^{\left(\Delta - 1,\,D\right)}} + \frac{1}{\sqrt{D}}\sum_{1 \leq l \leq p}\left(-i\sum_{1 \leq t \leq p}L_{l,\,t}\gamma_ta_t\right)\ket{\Psi^{\left(\Delta - 1\right)}_{\perp;\,l}}\right)^{\otimes D} & = \left(\sum_{0 \leq l \leq p}c_l\ket{\Psi^{\left(\Delta - 1,\,D\right)}_{\perp;\,l}}\right)^{\otimes D}\nonumber\\
    & = \bigotimes_{j \in [D]}\sum_{0 \leq l_j \leq p}c_{l_j}\ket{\Psi^{\left(\Delta - 1,\,D\right)}_{\perp;\,l_j}}\nonumber\\
    & = \sum_{\left(l_j\right)_{j \in [D]} \in \{0,\,\ldots,\,p\}^D}\left(\prod_{j \in [D]}c_{l_j}\right)\bigotimes_{j \in [D]}\ket{\Psi^{\left(\Delta - 1,\,D\right)}_{\perp;\,l_j}}.
\end{align}
We now group the terms of the sum based on the number of occurences of each integer $0, \ldots, p$ among sequence $\left(l_j\right)_{j \in [D]}$, denoting by $n_l$ ($0 \leq l \leq p$) the number of occurrences of $l$. Note that the $\left(n_l\right)_{0 \leq l \leq p}$ sum to $D$ by definition. This gives:
\begin{align}
    & \left(\ket{\Psi^{\left(\Delta - 1,\,D\right)}} + \frac{1}{\sqrt{D}}\sum_{1 \leq l \leq p}\left(-i\sum_{1 \leq t \leq p}L_{l,\,t}\gamma_ta_t\right)\ket{\Psi^{\left(\Delta - 1\right)}_{\perp;\,l}}\right)^{\otimes D}\nonumber\\
    & = \sum_{\substack{\left(n_l\right)_{0 \leq l \leq p}\\\sum\limits_{0 \leq l \leq p}n_l = D}}\hspace*{5px}\sum_{\substack{\left(l_j\right)_{j \in [D]}\\\forall 0 \leq l \leq p,\,\left|\{j \in [D]\,:\,l_j = l\}\right| = n_l}}\left(\prod_{j \in [D]}c_{l_j}\right)\bigotimes_{j \in [D]}\ket{\Psi^{\left(\Delta - 1,\,D\right)}_{\perp;\,l_j}}\nonumber\\
    & = \sum_{\substack{\left(n_l\right)_{0 \leq l \leq p}\\\sum\limits_{0 \leq l \leq p}n_l = D}}\left(\prod_{0 \leq  l \leq p}c_l^{n_l}\right)\sum_{\substack{\left(l_j\right)_{j \in [D]}\\\forall 1 \leq l \leq p,\,\left|\{j \in [D]\,:\,l_j = l\}\right| = n_l}}\bigotimes_{j \in [D]}\ket{\Psi^{\left(\Delta - 1,\,D\right)}_{\perp;\,l_j}}\nonumber\nonumber\\
    & = \sum_{\substack{\left(n_l\right)_{0 \leq l \leq p}\\\sum\limits_{0 \leq l \leq p}n_l = D}}\left(\prod_{0 \leq  l \leq p}c_l^{n_l}\right)\binom{D}{\left(n_l\right)_{0 \leq l \leq p}}^{1/2}\ket{n_1, \ldots, n_p}_{\Delta - 1,\,D}\nonumber\\
    & = \sum_{\substack{\left(n_l\right)_{0 \leq l \leq p}\\\sum\limits_{0 \leq l \leq p}n_l = D}}\left(\prod_{1 \leq  l \leq p}\left(-\frac{i}{\sqrt{D}}\sum_{1 \leq j \leq p}L_{l,\,j}\gamma_ja_j\right)^{n_l}\right)\binom{D}{\left(n_l\right)_{0 \leq l \leq p}}^{1/2}\ket{n_1, \ldots, n_p}_{\Delta - 1,\,D}\nonumber\\
    & = \sum_{\substack{\left(n_l\right)_{0 \leq l \leq p}\\\sum\limits_{0 \leq l \leq p}n_l = D}}\left(\prod_{1 \leq  l \leq p}\left(-i\sum_{1 \leq j \leq p}L_{l,\,j}\gamma_ja_j\right)^{n_l}\right)D^{-\frac{1}{2}\sum\limits_{1 \leq l \leq p}n_l}\binom{D}{\left(n_l\right)_{0 \leq l \leq p}}^{1/2}\ket{n_1, \ldots, n_p}_{\Delta - 1,\,D},
\end{align}
which is the claimed expansion (equation \ref{eq:qaoa_subtrees_state_finite_degree_fock_space_expansion}).
\end{proof}
\end{lemma}

Lemma \ref{lemma:coefficient_subtree_states_symmetric_basis} just established an expansion in the symmetric basis of the state defined in equation \ref{eq:qaoa_subtrees_state_orthonormal_basis}. We recall the state defined in this equation occurred in the approximation of the tree QAOA state derived in lemma \ref{lemma:qaoa_state_approximation_2}. Plugging the symmetric basis expansion into the last lemma then gives:
\begin{align}
    \ket{\Psi^{\left(\Delta,\,D\right)}} & = \ket{\Phi^{\left(\Delta,\,D\right)}} + \mathcal{O}_{\left\lVert \cdot \right\rVert_2}\left(D^{-1/2}\right),\label{eq:qaoa_state_approximation_symmetric_basis}
\end{align}
with
\begin{align}
    \ket{\Phi^{\left(\Delta,\,D\right)}} & := \sum_{a_{p + 1} \in \{1, -1\}}\sum_{\substack{\left(n_l\right)_{1 \leq l \leq p}\\\sum\limits_{1 \leq l \leq p}n_l \leq D}}c\left(a_{p + 1},\,\left(n_l\right)_{1 \leq l \leq p},\,D\right)\ket{a_{p + 1}} \otimes \ket{n_1,\ldots, n_l}_{\Delta - 1,\,D}\label{eq:qaoa_state_approximation_symmetric_basis_formula}
\end{align}
where 
\begin{align}
    c\left(a_{p + 1},\,\left(n_l\right)_{1 \leq l \leq p},\,D\right) & := \sum_{\bm{a}_{1:p} \in \{1, -1\}^p}f_{1:p + 1}\left(\bm{a}_{1:p}, a_{p + 1}\right)\exp\left(-\frac{1}{2}\sum_{l, m \in [p]}G^{\left(\mathrm{sym},\,\Delta - 1,\,D\right)}\gamma_l\gamma_ma_la_m\right)\nonumber\\
    & \hspace*{20px} \times D^{-\frac{1}{2}\sum\limits_{1 \leq l \leq p}n_l}\sqrt{\frac{D!}{\left(D - \sum\limits_{1 \leq l \leq p}n_l\right)!\prod\limits_{1 \leq l \leq p}n_l!}}\prod_{1 \leq l \leq p}\left(-i\sum_{1 \leq t \leq p}L_{l, t}\gamma_ta_t\right)^{n_l}.\label{eq:symmetric_basis_coefficient}
\end{align}
In the last two equations, we slightly revised the notation from lemma \ref{lemma:coefficient_subtree_states_symmetric_basis} for convenience, notably summing over $\left(n_l\right)_{1 \leq l \leq p}$ (starting from index $l = 1$) and adapting equality constraint $\sum_{0 \leq l \leq p}n_l = D$ in lemma \ref{lemma:coefficient_subtree_states_symmetric_basis} to equivalent constraint $\sum_{1 \leq l \leq p}n_l \leq D$ in the current context. We may now embed approximating state $\ket{\Phi^{\left(\Delta,\,D\right)}}$ into the $p$-particles Fock space tensor with a single qubit space: $\mathbb{C}^2 \otimes \bm{L}_2\left(\mathbf{R}\right)^{\otimes p}$ by linear extension of embedding:
\begin{align}
    \ket{x} \otimes \ket{n_1, \ldots, n_p}_{\Delta - 1,\,D} & \hookrightarrow \ket{x} \otimes \ket{n_1, \ldots, n_p} \qquad (x \in \{0, 1\}, n_1, \ldots, n_l \in \mathbf{N}),
\end{align}
where $\ket{n_1, \ldots, n_p}$ on the right-hand side refers to the Fock basis state of numbers $n_1, \ldots, n_p$ (without reference to the QAOA state), unlike state $\ket{n_1, \ldots, n_p}_{\Delta - 1,\,D}$ on the left-hand-side, which was defined with reference to QAOA on the alternative Bethe lattice of depth $(\Delta - 1)$ and degree $D$ (definition \ref{def:finite_degree_symmetric_state}). Under this embedding,
\begin{align}
    \ket{\Phi^{\left(\Delta,\,D\right)}} & \hookrightarrow \ket{\widetilde{\Phi}^{\left(\Delta,\,D\right)}},
\end{align}
where
\begin{align}
    \ket{\widetilde{\Phi}^{\left(\Delta,\,D\right)}} & := \sum_{a_{p + 1} \in \{1, -1\}}\sum_{\substack{\left(n_l\right)_{1 \leq l \leq p} \in \mathbf{N}^p\\\sum\limits_{1 \leq l \leq p}n_l \leq D}}c\left(a_{p + 1},\,\left(n_l\right)_{1 \leq l \leq p},\,D\right)\ket{a_{p + 1}} \otimes \ket{n_1,\ldots, n_l}\label{eq:qaoa_state_approximation_fock_basis}
\end{align}
We now show the last state converges to a well-defined state in the limit $D \to \infty$. In other words, after approximation by $\ket{\Phi^{\left(\Delta,\,D\right)}}$, and Fock space embedding as $\ket{\widetilde{\Phi}^{\left(\Delta,\,D\right)}}$, the tree QAOA state $\ket{\Psi^{\left(\Delta,\,D\right)}}$ converges in the limit $D \to \infty$:

\begin{theorem}[Convergence of embedded approximate tree state in $D \to \infty$]
\label{th:convergence_of_tree_state_to_spin_boson}
State $\ket{\widetilde{\Phi}^{\left(\Delta,\,D\right)}}$, defined by equations \ref{eq:qaoa_state_approximation_fock_basis}, \ref{eq:symmetric_basis_coefficient}, and embedding approximation $\ket{\Phi^{\left(\Delta,\,D\right)}}$ (equation \ref{eq:qaoa_state_approximation_symmetric_basis_formula}) to the QAOA state $\ket{\Psi^{\left(\Delta,\,D\right)}}$, strongly converges to a well-defined limit in $\mathbb{C}^2 \otimes \mathbf{L}_2\left(\mathbf{R}\right)^{\otimes p}$ as $D \to \infty$:
\begin{align}
    \ket{\widetilde{\Phi}^{\left(\Delta,\,D\right)}} & \xrightarrow[D \to \infty]{} \ket{\widetilde{\Phi}^{\left(\Delta\right)}},
\end{align}
where
\begin{align}
    \ket{\widetilde{\Phi}^{\left(\Delta\right)}} & := \sum_{\substack{a_{p + 1} \in \{1, -1\}\\\left(n_l\right)_{1 \leq l \leq p} \in \mathbf{N}^p}}c\left(a_{p + 1}, \left(n_l\right)_{1 \leq l \leq p}\right)\ket{a_{p + 1}} \otimes \ket{n_1, \ldots, n_p},\label{eq:qaoa_state_approximation_fock_basis_infinite_degree}
\end{align}
with
\begin{align}
    c\left(a_{p + 1}, \left(n_l\right)_{1 \leq l \leq p}\right) & := \sum_{\bm{a}_{1:p} \in \{1, -1\}^p}f\left(\bm{a}_{1:p}, a_{p + 1}\right)\exp\left(-\frac{1}{2}\sum_{l, m \in [p]}G^{\left(\mathrm{sym},\,\Delta - 1\right)}_{l, m}\gamma_l\gamma_ma_la_m\right) \prod_{1 \leq l \leq p}\frac{1}{\sqrt{n_l!}}\left(-i\sum_{1 \leq t \leq p}L_{l, t}\gamma_ta_t\right)^{n_l}.\label{eq:fock_basis_coefficient_infinite_degree}
\end{align}
\begin{proof}
We start by showing weak convergence of $\ket{\widetilde{\Phi}^{\left(\Delta,\,D\right)}}$ to $\ket{\widetilde{\Phi}^{\left(\Delta\right)}}$. This is equivalent to convergence of each coordinate in an orthogonal basis; in particular, taking the tensor product of the computational basis with the $p$-modes Fock basis, this amounts to
\begin{align}
    c\left(a_{p + 1},\,\left(n_l\right)_{1 \leq l \leq p},\,D\right) & \xrightarrow[D \to \infty]{} c\left(a_{p + 1},\,\left(n_l\right)_{1 \leq l \leq p}\right).
\end{align}
The above limit in turn follows immediately from the explicit formulae (equations \ref{eq:symmetric_basis_coefficient}, \ref{eq:fock_basis_coefficient_infinite_degree}) for both coefficients; the important point is that in these formulae, the sum over $\bm{a}_{1:p} \in \{1, 1\}^p$ has a number of terms independent of $D$. This proves weak convergence.

To deduce strong convergence from here, it suffices to show that 
\begin{align}
    \left\lVert \ket{\widetilde{\Phi}^{\left(\Delta\right)}} \right\rVert_2 & = 1,\\
    \left\lVert \ket{\widetilde{\Phi}^{\left(\Delta,\,D\right)}} \right\rVert_2 & \xrightarrow[D \to \infty]{} 1.
\end{align}
The first equality follows from writing $\ket{\widetilde{\Phi}^{\left(\Delta\right)}}$ as a spin-bosons stateproduced by a unitary quantum circuit acting on a qubit and bosonic modes; this is the substance of the spin-bosons construction detailed in appendix \ref{sec:sk_qaoa_iteration_spin_bosons}. As for the second limit, it follows from
\begin{align}
    \ket{\Phi^{\left(\Delta,\,D\right)}} & = \ket{\Psi^{\left(\Delta,\,D\right)}} + \mathcal{O}_{\left\lVert \cdot \right\rVert_2}\left(D^{-1/2}\right)
\end{align}
(lemma \ref{lemma:qaoa_state_approximation_2}),
\begin{align}
    \left\lVert \ket{\Psi^{\left(\Delta,\,D\right)}} \right\rVert_2 & = 1
\end{align}
(as a state produced by a quantum circuit) and the fact embedding
\begin{align}
    \ket{\Phi^{\left(\Delta,\,D\right)}} & \hookrightarrow \ket{\widetilde{\Phi}^{\left(\Delta,\,D\right)}}
\end{align}
is isometric.
\end{proof}
\end{theorem}

\section{Classical algorithms for SK model}

In this Section, we focus on the problem of producing a string of spins $\bm z$ that corresponds to near-optimal ($\epsilon$-approximate) value of energy $C(\bm z)$.
There are a few classical algorithms with provable approximation guarantees for this task, in the $n \rightarrow \infty$ limit\footnote{The classical algorithms also assume that the SK model does not have the overlap gap property (OGP). We remark that QAOA with constant $p$ is also obstructed by OGP. For additional discussion of this assumption, see Appendix \ref{app:amp_montanari}.}. 
These algorithms are all polynomial in $n$, yet, at least theoretically, appear to have poor $\epsilon$ dependence. 
We acknowledge that the goal of the classical works was to produce polynomial-time approximation schemes (PTAS), so there may not have been extensive effort put in to reducing the $\epsilon$ dependence. However, at least for the Approximate Message Passing (AMP) approaches, proving inverse-polynomial $\epsilon$ dependence, i.e. fully PTAS, seems to be a challenging task (see Appendix~\ref{app:amp_montanari}). Additionally, these classical algorithms require access to the Parisi measure and corresponding Parisi PDE solution, which are considered mild assumptions \cite{montanari2019, alaoui2020algorithmicthresholdsmeanfield}.

Montanari \cite{montanari2019} proposed an algorithm based on a variant of AMP that produces a sequence of spins with an energy that $\epsilon$-approximates
the SK infinite-size, free energy at inverse-temperature $\beta$. As Montanari showed, this can be used to produce an approximation scheme for the zero-temperature setting ($\beta \rightarrow \infty$), i.e. finding an $\epsilon$ approximate optimizer to SK as well. The dependence on $\epsilon$ was left unspecified in the original work, yet can be upper bounded with exponential dependence in $1/\epsilon^4$, as we show in Appendix \ref{app:amp_montanari}. Ivkov and Schramm \cite{ivkov2023semidefiniteprogramssimulateapproximate} also showed that AMP for SK can be simulated efficiently by semi-definite programming hierarchies.

Subsequently,  Alaoui, Montanari and Sellke \cite{alaoui2020optimizationmeanfieldspinglasses}, generalized the approach of Montanari to higher-order spin glasses, while also modifying the original AMP algorithm for SK. The approach of Ref.~\cite{alaoui2020optimizationmeanfieldspinglasses}, while also based on AMP, instead produces a sequence spins of with an energy that $\epsilon$ approximates
the zero-temperature Parisi solution directly, with no annealing of $\beta$. Unfortunately, it appears that the $\epsilon$ dependence of Ref.~\cite{alaoui2020optimizationmeanfieldspinglasses} is 
more challenging to bound due to an apparent singularity in the functional order parameter at zero-temperature \cite{auffinger2016parisiformulagroundstate}, which does not appear to be a barrier to analyzing Ref.~\cite{montanari2019}. We note that this algorithm was benchmarked in Ref.~\cite{alaoui2020algorithmicthresholdsmeanfield} but the $\epsilon$ dependence was not investigated in sufficient detail to enable a comparison with QAOA. 

While AMP has been applied to a variety of statistical inference problems \cite{feng2022unifying}, it seems unclear how to generalize the approach for SK to other optimization problems. We note that a belief propagation algorithm for \maxcut on high-girth, regular graphs has been shown to behave like the SK AMP iteration as the degree goes to infinity \cite{alaoui2023localalgorithmsmaximumcut}, which is explained by the universality of AMP~\cite{cheairi2024algorithmicuniversalitylowdegreepolynomials}. Furthermore, AMP has been used as a subroutine for sampling from the SK model at high-temperatures \cite{alaoui2024samplingsherringtonkirkpatrickgibbsmeasure}. Due to the high-temperature limitation this does not result in another approximate optimization algorithm.

Lastly, Jekel, Sandhu and Shi \cite{jekel2025potential} proposed the potential Hessian ascent approach, which instead performs an iterative ascent utilizing a spectral projection of the Hessian of a regularized version of the SK objective function. This approach tracks a system of  SDEs, associated with the finite-$\beta$ Parisi solution, that is dual to the one AMP tracks. Like \cite{montanari2019}, this approach does anneal $\beta$. Due to an exponentially small, in inverse $\epsilon$, step size, their approach has an iteration complexity that is exponential in $1/\epsilon^2$. Unlike AMP, the Hessian ascent approach seems more akin to the descent techniques used in high-dimensional, non-linear optimization \cite{nesterov2018lectures} and hence may be more widely applicable.

\section{On the Complexity of the Approximate Message-Passing Algorithm for SK}
\label{app:amp_montanari}

In this Section, we present a complete upper bound on the complexity of the approximate message-passing (AMP) algorithm for SK model.  We do so to enable a comparison between the empirical scaling of QAOA and the state-of-the-art algorithms for SK.  The classical algorithms have the guarantee that they output spins that achieve a $(1-\epsilon)$-approximation to the optimum with probability one as $n\rightarrow \infty$. For finite, $n$, there is an additional probability of failure. Even for finite $n$ in the average case,  there is an additional error of the order $\mathcal{O}(n^{-\alpha})$ that would need to be taken into account. In the case of AMP, this is because it attempts to track the Parisi solution, which is only valid in the infinite-size limit.

In prior work on AMP for SK, the $\epsilon$ dependence was left unspecified, being denoted as $C(\epsilon)$. This gave a depth of $\mathcal{O}(C(\epsilon)n)$ and total work of $\mathcal{O}(C(\epsilon)n^2)$. By depth, we mean the longest sequence of operations that cannot be parallelized.
By  reproducing the proofs of Refs.~\cite{montanari2019, alaoui2020optimizationmeanfieldspinglasses} we show that the depth of the AMP algorithm for SK (specifically the one in \cite{montanari2019}) can be upper-bounded by $\mathcal{O}\left(n\frac{e^{\mathcal{O}(1/\epsilon^4)}}{\epsilon^{16}}\right)$. The approach of \cite{alaoui2020optimizationmeanfieldspinglasses}  (Theorem 2) obtains an $\mathcal{O}(n/\epsilon^2)$ depth contingent on certain regularities of the zero-temperature Parisi solution. However, there is a current barrier to this due to an issue mentioned in prior literature \cite{auffinger2016parisiformulagroundstate} that we refer to as the Parisi singularity. Hence, it is unclear if an upper bound of $\mathcal{O}(\text{poly}(1/\epsilon))$ for AMP is attainable. Specifically, given the current analysis, this requires showing that the regularity (bounds, Lipschitzness and total variation) of the Parisi solution at zero-temperature, i.e. the functional order parameter $\gamma^{*}(t) : [0, 1) \rightarrow \mathbb{R}$ and the corresponding solution to the Parisi (partial differential equation (PDE) $\Phi^{(\infty)}_*$, is $\mathcal{O}(\text{poly}\log(1/\epsilon))$ for $t \in [0, 1 -\epsilon]$.

We note that the aforementioned depth for AMP is worse than the Potential Hessian Ascent algorithm of \cite{jekel2025potential}, which has a depth  of $\widetilde{\mathcal{O}}\left(n^{1+\delta}e^{\mathcal{O}(1/\epsilon^2)}\right)$ for $\delta \leq \frac{1}{22}$. Furthermore, we highlight the folklore notion that the complexity of the AMP algorithm is at least $\Theta(1/\epsilon^2)$. This is due to the AMP
tracking a system of stochastic differential equations (SDEs) utilizing that Euler-Maruyama scheme, which \emph{in the worst case}, has a convergence rate of $\mathcal{O}(\sqrt{\delta})$ with the number of iterations.

In the process of determining the $\epsilon$ dependence, we obtain a (somewhat) concise explanation of the how and why the incremental AMP algorithm for SK works.

\subsection{The algorithm of \cite{montanari2019, alaoui2020optimizationmeanfieldspinglasses}}

Recall the Sherrington-Kirkpatrick(SK) Hamiltonian $H_n$ for $n$ spins:
\begin{align*}
    H_n(\sigma) = \sum_{1 \leq i,j \leq n} \frac{J_{ij}}{2}\sigma_i \sigma_j = \frac{1}{2}\langle W , \sigma^{\otimes 2}\rangle,
\end{align*}
where $J_{ij} \sim \mathcal{N}(0, \frac{1}{\sqrt{n}})$, and $W$ is a vectorization of $J$, resulting in an order-two, symmetric tensor. For inverse-temperature $\beta$, the infinite-volume log, Gibbs partition function $F(\beta)$,  
\begin{align*}
    F(\beta) := \lim_{n\rightarrow \infty}\frac{1}{n}\mathbb{E}_{J}\log \sum_{\sigma \in \{-1, 1\}^n} e^{-\beta H_n(\sigma)},
\end{align*}
is characterized by the famous Parisi variational formula \cite{talagrand2006parisi, panchenko2013sherrington}
\begin{align}
\label{eqn:parisi_formula}
    F(\beta) = \inf_{\alpha_{\beta} \in \mathcal{M}}P_{\beta}(\alpha_{\beta}),
\end{align}
where
\begin{align}
\label{eqn:paris_functional}
    \mathcal{P}_{\beta}(\alpha_{\beta}) = \Phi(0, 0) - \frac{\beta^2}{2}\int_0^1\alpha_{\beta}(t)t dt,
\end{align}
and $\mathcal{M}$ is the set of cumulative distribution functions (CDFs) over $[0, 1]$.
Note $\Phi$ solves the following two-dimensional PDE:%
\begin{align}
\label{eqn:parisi_pde}
&\partial_t\Phi(t, x) + \frac{\beta^2}{2}\partial_x^2\Phi(t,x ) + \frac{\beta^2}{2}\alpha_{\beta}(t)(\partial_x\Phi(t, x))^2 = 0 \\
&\Phi(1, x) = \log(2\cosh x), \nonumber\\
& t \in [0, 1], \nonumber
\end{align}
which is referred to as the Parisi PDE.
When $\alpha_{\beta}$ is a step function, the above can be solved via recursive Cole-Hopf transformations \cite{Auffinger_2014}, and continuously extended to all of $\mathcal{M}$. Note that Equation \eqref{eqn:parisi_formula} has a unique minimizer, which we will denote $\alpha_{\beta}^{*}$.

It is well known \cite{panchenko2013sherrington, montanari2019, Auffinger_2014} that the following limits hold
\begin{align*}
&\lim_{\beta \rightarrow \infty}\inf_{\alpha_{\beta} \in \mathcal{M}}\frac{\mathcal{P}_{\beta}(\alpha_{\beta})}{\beta} = \lim_{\beta \rightarrow \infty}\frac{F(\beta)}{\beta} = \lim_{n\rightarrow \infty}\mathbb{E}_{J}\max_{\sigma \in \{-1, 1\}^n} \frac{H_n(\sigma)}{N}\\
&\max_{\sigma \in \{-1, 1\}^n}\frac{H_n(\sigma)}{n} \underset{ n\rightarrow \infty, \text{a.s}}{\rightarrow} \lim_{\beta \rightarrow \infty}\inf_{\alpha_\beta \in \mathcal{M}}\mathcal{P}_{\beta}(\alpha_{\beta}).
\end{align*}
It was also shown by \cite{auffinger2016parisiformulagroundstate} that an analogous variational formula holds at $\beta \rightarrow \infty$. Let $\mathcal{U}$ be the set of all non-decreasing, positive, integrable functions over $[0, 1)$. Then
\begin{align*}
    \lim_{\beta \rightarrow \infty}\frac{F(\beta)}{\beta} = \inf_{\gamma \in \mathcal{U}}\mathcal{P}_{\infty}(\gamma),
\end{align*}
where
\begin{align*}
    \mathcal{P}_{\infty}(\gamma) = \Phi^{(\infty)}_{\gamma}(0, 0) - \frac{1}{2}\int_0^1 \gamma(t)t dt,
\end{align*}
and $\Phi^{(\infty)}$ solves
\begin{align}
\label{eqn:zero-temp-parisi-pde}
&\partial_t\Phi^{(\infty)}(t, x) + \frac{1}{2}\partial_x^2\Phi^{(\infty)}(t,x ) + \frac{1}{2}\gamma(t)(\partial_x\Phi^{(\infty)}(t, x))^2 = 0 \\
&\Phi^{(\infty)}(1, x) = \lvert x \rvert, \nonumber\\
& t \in [0, 1]. \nonumber
\end{align}
The infinite-beta (zero-temperature) variational problem was also shown to have a unique minimizer, which we will refer to as $\gamma^*(t)$ \cite{Chen_2017}.

The approach in \cite{auffinger2016parisiformulagroundstate} was to consider the weak $\beta \rightarrow \infty$ limit of the distribution functions $\beta\alpha_{\beta}^{*}(t)$, which can be shown to exist over $[0, 1)$ but not $[0, 1]$ due to an apparent singularity at $t=1$, i.e. $\lim_{t \rightarrow 1^{-}}\gamma(t)$ can be infinite. While we have nice regularity properties for $\Phi$ when $\beta < \infty$ and $t$ is far from $1$ \cite{Chen_2017}, it is hard to say anything about the whole $[0, 1]$ range. Hence, we will refer to this issue as the Parisi singularity.

For the AMP algorithm to produce a  polynomial-time approximation scheme (PTAS) for the SK model, we require the assumption that SK is Full Replica Symmetry Breaking (FRSB) at low temperatures or equivalently does not have the overlap gap property (OGP) \cite{gamarnik2021overlap}. FRSB implies that for $\beta$ above some critical value, $\alpha^*_{\beta}$ is strictly increasing over some interval $[0, q^*]$, $q^* \leq 1$ \cite{auffinger2013propertiesparisimeasures}. It has already been shown that the number of steps in $\alpha_\beta^*$ diverges as $\beta \rightarrow \infty$ \cite{auffinger2021skmodelinfinitestep}, hinting at the possibility of FRSB.

There are currently two versions of the AMP algorithm for SK with important differences: Montanari's \cite{montanari2019} and  Alaoui et al. \cite{alaoui2020optimizationmeanfieldspinglasses}. The goal of the latter paper was also to extend the algorithm of \cite{montanari2019} to reach the algorithmic threshold of higher-order spin glasses. The goal of both AMP algorithms is to  produce spins $\sigma$ such that as $n\rightarrow \infty$, $n^{-1}H_n(\sigma)$ tracks the quantity
\begin{align}
\label{eqn:ideal-gen}
\int_0^{t^*} \mathbb{E}[u(s, X_s)]ds
\end{align}
associated with the following SDE system
\begin{align}
&dZ_t = s\cdot dB_t\\
\label{eqn:ac_sde}
&dX_t = v(t, x)dt + \beta dB_t\\
&dM_t =u(t, x)dB_t,
\end{align}
where there is one Brownian Motion $B_t$. The latter with specific forms of $v$ and $u$ (shown below) is known as Auffinger and Chen SDE \cite{auffinger2013propertiesparisimeasures}.  Both show that under the assumption of FRSB in the SK model, Equation \eqref{eqn:ideal-gen} can be made arbitrarily close to the minimum of the Parisi functional. Hence, the main goal of the AMP algorithm is to track \eqref{eqn:ideal-gen} such that for fixed $\epsilon$ the total work is polynomial in $n$, i.e. that is a PTAS for sufficiently large $n$.

The algorithms differ in their choice of $v, u$ and $s$. Specifically, \cite{alaoui2020optimizationmeanfieldspinglasses} selects
\begin{align}
\label{eqn:zero-temp-controls}
& s = 1 \nonumber\\
&v(t, x) = \gamma^*(t)\partial_x\Phi_{*}^{(\infty)}(t,x) \nonumber\\
&u(t, x) = \partial_x^2\Phi_{*}^{(\infty)}(t, x)
\end{align}
following the Parisi formula at zero-temperature, and $t^* = 1-\epsilon$. The algorithm has a depth of $\mathcal{O}(n\mathcal{A}(\epsilon)/\epsilon^2)$, where $\mathcal{A}(\epsilon)$ depends on the the Lipschitz constants and total variation of $u$ and $v$. As mentioned earlier, the zero-temperature Parisi variational problem encounters the Parisi singularity. Hence, it is possible that as $t^* \rightarrow 1$ from below, $\gamma^*(t)$ diverges. Unlike the finite $\beta$ case, this prevents us from obtaining bounds on regularity of $u$ and $v$. Specifically, \cite{Chen_2017} shows that when $t^*$ is far from $1$ we can bound the regularity, where it is apparent that the proof requires $\gamma(t^*)$ to be finite. This works for fixed $t^*$ but  $t^* = 1 - \epsilon$ for varying $\epsilon$. The above turns out not to be an issue for the approach of \cite{montanari2019}, but at the cost of a significantly worse-than $1/\epsilon^2$ upper bound. Hence, it seems that the Parisi singularity presents a barrier to obtaining the folklore $1/\epsilon^2$ iteration complexity for AMP.

Montanari \cite{montanari2019} considers the following choices for $u, v$ and $s$
\begin{align*}
&s = \beta\\
&v(t, x) = \beta^2\alpha_{\beta}^{*}(t)\partial_x\Phi_{*}(t,x)\\
&u(t, x) = \beta\partial_x^2\Phi_{*}(t, x),
\end{align*}
which correspond to the Parisi variational problem for finite $\beta$, i.e. $\mathcal{P}_{\beta}(\alpha_{\beta}^*)$. Hence for a fixed $\beta$, the algorithm only tracks the minimizer of the Parisi functional for finite $\beta$. Then Lemma 3.6, 3.7 of  \cite{montanari2019} show that for $\beta = \mathcal{O}(1/\epsilon)$ we can make \eqref{eqn:ideal-gen}, for $t^* = q^*$, $\epsilon$-close to the minimum of the zero-temperature Parisi functional. It appears that we can in fact determine an upper bound on the complexity of this approach, although it was not explicitly shown in \cite{montanari2019}. The goal of this appendix to clarify this.

\subsection{The Approximate Message Passing Algorithm for SK}

In this subsection, we review the AMP iteration. We will need to assume that the algorithm already has access to $\alpha_{\beta}^*$ minimizing \eqref{eqn:parisi_formula} and the corresponding $\Phi_*(t, x)$, along with its first and second spatial derivatives. These assumptions are standard \cite{montanari2019, alaoui2020algorithmicthresholdsmeanfield, jekel2025potential} and effectively amount to a pre-processing step that could be done once.

The iterates $\mathbf{z}^{(j)} \in \mathbb{R}^n$ for AMP for SK are:
\begin{align*}
&\mathbf{z}^{(\ell+1)} = \frac{1}{2}W\{f_{\ell}(\mathbf{z}^{(0)}, \dots, \mathbf{z}^{(\ell)})\} - \sum_{j=0}^{\ell} d_{\ell, j}f_{j-1}(\mathbf{z}^{(0)}, \dots, \mathbf{z}^{(j-1)})\\
&d_{\ell, j} =  \frac{1}{n}\sum_{i=1}^{n}\frac{\partial f_{\ell}}{\partial z^{j}_i}({z}^{(0)}_i, \dots, {z}^{(\ell)}_i),
\end{align*}
where $d_{\ell, j}$ form the Onsager correction terms. Also
\begin{align}
\label{eqn:mat-vec}
     (W\{\mathbf{u}\})_i:= \sum_{1 \leq j \leq n}W_{i, j}u_j.
\end{align}

The output of the algorithm is the sequence 
\begin{align*}
 \mathbf{m}^{(j)} := f_{j}(\mathbf{z}^{(0)}, \dots, \mathbf{z}^{(j)}) \in \mathbb{R}^{n}, 0 \leq j \leq \ell
\end{align*}
where $f_{j}$ is a nonlinearity. Thus based on an earlier discussion, the goal is to ensure that for $\ell = \lfloor q^*/\delta \rfloor$, we suppress the following error
\begin{align}
\label{eqn:tracking_error}
    \lvert  \underset{N\rightarrow \infty}{\text{p-}\lim} \frac{H_N(\mathbf{m}^{\ell})}{N} - \int_0^{\ell\delta} \mathbb{E}[u(t,X_t)]dt \rvert.
\end{align}
The functions $f_{\ell}$ act component-wise, i.e. 
\begin{align*}
    f_{j}(\mathbf{z}^{(0)}, \dots, \mathbf{z}^{(j)}) := (f_{j}(z_i^{(0)}, \dots, z_i^{(j)}))_{1\leq i \leq n}.
\end{align*}
The nonlinearities are chosen such that $f_{k}(\mathbf{z}^0, \dots, \mathbf{z}^{k})$ performs
\begin{align*}
&(\mathbf{z}^0, \dots, \mathbf{z}^{k}) \rightarrow (\mathbf{m}^0, \dots, \mathbf{m}^{k}) \\
&x^{(j+1)}_k = x^{(j)}_k + v(j\delta, x^{(j)}_k)\delta + (z^{(j+1)}_k - z^{(j)}_k), \quad 0 \leq j \leq \ell-1, 1 \geq k \geq N.\\
& m^{(j)}_k = m^{(0)}_k + \sum_{r=0}^{j-1} u^{(\delta)}(x^{(r)}_k)(z^{(r+1)}_k - z^{(r)}_k), \quad k \geq 1, m^{(0)}_k = \sqrt{\delta},
\end{align*}
where %
\begin{align*}
&u_0^{(\delta)} = 1\\
&u_{\ell}^{(\delta)}(x_k) = \frac{u(\ell\delta, x^{(\ell)}_k)}{a^{(\delta)}_\ell}\\
&(a_{\ell}^{(\delta)})^2 = \frac{1}{N}\sum_{i=1}^{N}(u(\ell \delta, x^{(\ell)}_i))^2\\
\end{align*}
Note that ignoring the $\alpha_{\ell}^{(\delta)}$, the $f_{\ell}$ are Lipschitz in $\mathbf{z}$. One will also note that each step of AMP makes one query to the gradient of $H_n$, and the total amount of work for a given iteration is $\mathcal{O}(n^2)$. However, one can easily observe from the iteration that the depth of the algorithm is actually $\mathcal{O}(n)$, as the most time-consuming step \eqref{eqn:mat-vec} can be parallelized. There is also an additional rounding procedure to cause the output $\mathbf{m}^{(\ell)}$ to be mapped to spins in $\{-1, 1\}^{N}$.

The key result for AMP is a limit theorem characterizing its iterates over $\ell+1$ steps $\mathbf{z}^{(0)}, \dots \mathbf{z}^{(\ell)}$, known as the state evolution. First we require a definition:
\begin{definition}[Pseudo-Lipschitz \cite{montanari2019}]
We say that $\psi : \mathbb{R}^{D} \rightarrow \mathbb{R}$ is pseudo-Lipschitz with constant $L > 0$ if 
\begin{align*}
    \lvert \psi(x) - \psi(y)\rvert \leq L \lVert x -y \rVert (1 + \lVert x \rVert + \lVert y \rVert), \quad \forall x, y \in \mathbb{R}^{D}.
\end{align*}
\end{definition}

\begin{theorem}[State Evolution : Proposition 2.1 \cite{montanari2019} \& Proposition 3.1 \cite{alaoui2020optimizationmeanfieldspinglasses}]
\label{thm:state_evolution}
Suppose we start with an initial prior $p_0$ (with finite second moment) on $Z_0$ and draw our initial iterate $\mathbf{z}^{(0)}$ from the distribution $p_0$ with i.i.d. coordinates. Then for any pseudo-Lipschitz function $\psi : \mathbb{R}^{\ell+1} \rightarrow \mathbb{R}$ we have
\begin{align*}
     \frac{1}{N}\sum_{i=1}^{N}\psi (\mathbf{z}^{0}_i, \dots \mathbf{z}^{\ell}_i) \underset{N\rightarrow \infty}{\rightarrow} \mathbb{E}[\psi(Z_0, \dots, Z_{\ell})],
\end{align*}
where the convergence is in probability, and  $Z_1, \dots, Z_{\ell}$ are centered Gaussians independent of $Z_0$  with recursive covariance
\begin{align*}
   \mathbb{E}[Z_{j+1}Z_{k+1}] = \mathbb{E}[f_j(Z_0, \dots , Z_j)f_k(Z_0, \dots, Z_k)],
\end{align*}
for Lipschitz $f_{\ell}$.
\end{theorem}

To emphasize the usefulness of the above theorem, note that it implies, at least for constant $\alpha^{(\delta)}_{\ell}$, that the AMP iteration converges in the $n\rightarrow \infty$ limit (in probability) to the iteration:
\begin{align}
\label{eqn:discretization}
&X^{(\delta)}_{j+1} = X^{(\delta)}_{j} + v(j\delta, X^{(\delta)}_{j})\delta + (Z^{(\delta)}_{j+1} - Z^{(\delta)}_{j}) \nonumber\\
&M^{(\delta)}_{j} = M_0 + \sum_{j=0}^{\ell-1}\frac{u(j,\delta, X^{(\delta)}_j)}{\Sigma_j^{(\delta)}}(Z^{(\delta)}_{j+1} - Z^{(\delta)}_{j}),
\end{align}
where
\begin{align}
\label{eqn:Sigma}
(\Sigma_{\ell}^{(\delta)})^2 = \mathbb{E}[(u(\ell \delta, X^{(\delta)}_\ell))^2].
\end{align}
The above represents a time-discretization of the earlier mentioned SDE system.

We note that in \cite{montanari2019, alaoui2020optimizationmeanfieldspinglasses}, the empirical means in the computation of $a_{\ell}^{(\delta)}$ and $q_{\ell}^{(\delta)}$ are replaced with their limiting expectations according to the state evolution, hence making $f_{\ell}$ Lipschitz. However, as the inverse function is not pseudo-Lipschitz of any order, it is not clear that the state evolution result applies to $f_{\ell}$ composed with the scaling by $a_{\ell}^{(\delta)}$. Additionally, the limiting expectations of  $a_{\ell}^{(\delta)}$ are clearly not something we have access to ahead of time, unlike the empirical means. However, this issue is something that is not addressed in the theoretical literature and may contribute to the overall complexity. Note that in the numerical implementation of \cite{alaoui2020algorithmicthresholdsmeanfield}, the authors also utilized empirical means for $\alpha_\ell^{(\delta)}$.

\subsection{Smoothness of drift and diffusion coefficients}

\label{sec:smoothness}

The following observations have already been referenced in \cite{montanari2019}, and are key to determining the overall runtime.

According to \cite[Proposition 2.ii]{Auffinger_2014}, we have that.
\begin{align}
\label{eqn:parisi_deriv_bounds}
\lvert \partial^{j}_x\Phi_{*}(t, x) \rvert \leq 4, j \in \{1, 2, 3\}.
\end{align}

We recall the definition of bounded strong total variation.
\begin{definition}[Definition 2.1 \cite{alaoui2020optimizationmeanfieldspinglasses}]
We say that $f : [a, b] \times \mathbb{R} \rightarrow \mathbb{R}$ has bounded strong total variation  if there exists a constant $C < \infty$ such that
\begin{align*}
    \sup_{n} \sup_{a \leq t_0 \leq \cdots \leq t_n \leq b}\sup_{x_1, \dots, x_n \in \mathbb{R}} \sum_{i=1}^n \lvert f(t_i, x_i) - f(t_{i-1}, x_i)\rvert \leq C,
\end{align*}
for all partitions of $[a, b]$ and sequences $x_i$ in $\mathbb{R}$.
\end{definition}

Recall that 
\begin{align*}
&v(t, x) = \beta^2\alpha_{\beta}^{*}(t)\partial_x\Phi_{*}(t,x)\\
&u(t, x) = \beta\partial_x^2\Phi_{*}(t, x).
\end{align*}

Since for finite $\beta$, $\alpha_{\beta}^*$ is a distribution function, we of course have $\lVert \alpha_{\beta}^*\rVert_{\infty} = 1$.

Hence  by \eqref{eqn:parisi_deriv_bounds} we have that
\begin{align*}
&\lvert \partial_x\Phi_{*}(t, z) - \partial_x\Phi_{*}(t, y) \rvert \leq 4\lvert z -y \rvert, \forall t \in [0, 1], z,y \in \mathbb{R}\\
&\lvert \partial^2_x\Phi_{*}(t, z) - \partial^2_x\Phi_{*}(t, y) \rvert \leq 4\lvert z -y \rvert, \forall t \in [0, 1], z,y \in \mathbb{R}.
\end{align*}

\begin{align}
&\lvert v(t, z) - v(t, y) \rvert \leq 4\beta^2\lvert z -y \rvert, \forall t \in [0, 1], z,y \in \mathbb{R}\label{eq:amp_v_space_lipschitzness}\\
&\lvert u(t, z) - u(t, y) \rvert \leq 4\beta\lvert z -y \rvert, \forall t \in [0, 1], z,y \in \mathbb{R}.\label{eq:amp_u_space_lipschitzness}
\end{align}
Similar inequalities hold for Lipschitzness in time, due to $\partial^j\Phi_x$ being continuously differentiable in time over $t \in [0 ,1]$.

By Lemma 16 \cite{Jagannath_2015}
\begin{align}
&\lVert v(t,x)\rVert_{L_{\infty}([0,1] \times \mathbb{R})} \leq \beta^2\label{eq:amp_v_boundedness}\\
& \lVert u(t, x) \rVert_{L_{\infty}([0,1] \times \mathbb{R})} \leq \beta.\label{eq:amp_u_boundedness}
\end{align}

By Lemma 6.14 \cite{alaoui2020optimizationmeanfieldspinglasses} if $g$ is $L$ Lipschitz in time, uniformly in space and $h : [a, b] \rightarrow \mathbb{R}$ has bounded total variation, then $f = gh$ has bounded strong total variation, i.e.
\begin{align*}
\sum_{i=1}^{n}\lvert f(t_i, x_i) - f(t_{i-1}, x_i)\rvert \leq L(b-a)\lVert h \rVert_{\infty} + \lVert g \rVert_{\infty}\lVert h \rVert_{\text{TV}}.
\end{align*}

Note that $\alpha_{\beta}^*(t)$ has bounded total variation for a universal constant $C$, which follows from $\alpha_{\beta}^*$ being monotonic. Hence
$h = \alpha_{\beta}^*(t)$, $g = \partial_x\Phi_{*}(t,x)$. \cite[Proposition 2.iii]{Auffinger_2014}, the partial time derivatives are continuous, hence
$L = \mathcal{O}(1)$. One can also see this by inspecting the PDE for $\partial_t\partial_x\Phi_*$ in \cite[Theorem 4]{Jagannath_2015}. Hence the strong total variation of $v$ is 
\begin{align}
\label{eqn:total_variation_v}
\sum_{i=1}^{n}\lvert v(t_i, x_i) - v(t_{i-1}, x_i)\rvert \leq  \beta^2 C_1,
\end{align}
where $C_1$ is a universal constant.

\cite[Proposition 2.iii]{Auffinger_2014} also implies that the partial time derivative of $\partial_x^2\Phi$ is continuous over $[0, 1]$, hence the time Lipschitz constant is $O(1)$. This gives 
\begin{align}
\label{eqn:total_variation_u}
\sum_{i=1}^{n}\lvert u(t_i, x_i) - u(t_{i-1}, x_i)\rvert \leq  \beta C_2,
\end{align}
for some universal constant $C_2$.

\subsection{Upper Bounding The $\epsilon$ Dependence}

The following is basically a reproduction and mixture of the relevant proofs of \cite{alaoui2020optimizationmeanfieldspinglasses, montanari2019} with some slight simplifications in the presentation and catered towards SK only. The goal is to show that the previous work actually implies a complete complexity upper bound in terms of $\epsilon$.

To start, we recall the following guarantees that we have the on sequence $(Z^{(\delta)}_j, M^{(\delta)}_j)_{j\geq 0}$ based on the chosen scaling $a^{(\delta)}_\ell$ for the diffusion term $u$.

\begin{lemma}[Lemma 5.1 \& Lemma 5.1 \cite{alaoui2020optimizationmeanfieldspinglasses}]
\label{lem:gauss_proccess_lemma}
Suppose $\Sigma_{\ell}^{(\delta)}$ are chosen as in Equation \eqref{eqn:Sigma}.  Then the sequence $(Z_{\ell})^{(\delta)})_{\ell \geq 0}$ is a Gaussian process starting at $Z_0^{(\delta)} =0$. It's increments $\Delta_{\ell}^{(\delta)} := Z_{\ell}^{(\delta)} - Z_{\ell-1}^{(\delta)}$ are independent, have zero mean and  a variance of
\begin{align*}
&\mathbb{E}[(\Delta_{\ell}^{(\delta)})^2] = \beta^2\delta.
\end{align*}
Also for the Martingale $M_{\ell}^{(\delta)}$
\begin{align*}
&\mathbb{E}[(M_{\ell}^{(\delta)})^2] = (\ell+1)\delta
\end{align*}
\end{lemma}
Due to the Gaussian nature, the above completely characterizes the limiting distribution  output by incremental AMP, i.e. what we use as our driving Brownian motion in the discretization of the SDE system.

The following Lemma bounds the SDE tracking error (strong convergence error). Instead of the usual $L_2$ error, we will obtain a bound on the Wasserstein-$2$ error by selecting a particular coupling.

\begin{lemma}[Proposition 5.3 \cite{alaoui2020optimizationmeanfieldspinglasses} with Explicit $\beta$ Dependence]
\label{lem:convergence_bound_error}
There exists a coupling between $\{ (Z_{\ell}^{\delta}, X_{\ell}^{(\delta)}\}_{\ell \geq 0}$ and $(Z_{t}, X_{t}\}_{t \geq 0}$ such that the following holds. For $0 < \delta < 1$ and $\ell \leq \delta^{-1}$:

\begin{align*}
&\max_{1\leq j \leq \ell}\mathbb{E}[\lvert X_j^{(\delta)} - X_{\delta j}\rvert^2] = \mathcal{O}\left(\beta^8 e^{\mathcal{O}(\beta^4)}\delta\right)
\end{align*}
\end{lemma}
\begin{proof}

Starting with the standard Brownian motion $(B_t)_{t\in [0, 1]}$, we couple the increments $Z_{\ell}^{(\delta)} - Z_{\ell-1}^{(\delta)}$ to $B_t$ as follows:
\begin{align*}
    Z_{\ell}^{\delta} - Z_{\ell-1}^{\delta} := \int_{\delta(\ell-1)}^{\delta\ell} \beta dB_s, \ell \geq 1.
\end{align*}
From It\^o isometry, the Gaussian nature, and the previous lemma, we have the increments are distributed as desired when marginalizing. This way of defining also ensures that $Z^{(\delta)}_{\ell} = Z_{\delta \ell}$. Let $\Delta^{X}_j = X_j^{(\delta)} - X_{\delta j}$ correspond to the tracking error in $X$. It can be shown that with $\Delta_0^{X} =0$ and \eqref{eqn:discretization} that
\begin{align*}
\Delta_{\ell}^{X} &= \sum_{j=1}^{\ell}\Delta_j^{X} - \Delta_{j-1}^{X} \\&= \sum_{j=1}^{\ell}\int_{(j-1)\delta}^{j\delta} (v((j-1)\delta, X_{j-1}^{(\delta)}) - v((j-1)\delta, X_t) )dt\\
&+\sum_{j=1}^{\ell}\int_{(j-1)\delta}^{j\delta} (v((j-1)\delta, X_t) - v(t; X_t))dt.
\end{align*}

For the first term, recalling space Lipschitzness of $v$ from Eqn.~\ref{eq:amp_v_space_lipschitzness},
\begin{align*}
\int_{(j-1)\delta}^{j\delta} (v((j-1)\delta, X_{j-1}^{(\delta)}) - v((j-1)\delta, X_t) )dt \leq 4\beta^2\int_{(j-1)\delta}^{j\delta}\lvert X_{j-1}^{(\delta)} - X_t\rvert dt.
\end{align*}

The second term is bounded via the total variation bound (Eqn.~\ref{eqn:total_variation_v}). This leads to that
\begin{align*}
    \sum_{j=1}^{\ell} \int_{(j-1)\delta}^{j\delta} \lvert v((j-1)\delta, X_t) - v(t, X_t)\rvert \leq C_1 \beta^2\delta.
\end{align*}
This gives that
\begin{align*}
    \lvert \Delta_{\ell}^{X} \rvert \leq  4 \beta^2\sum_{j = 1}^{\ell}\int_{(j-1)\delta}^{j\delta}\lvert X_{j-1}^{(\delta)} - X_t\rvert dt + C_1 \beta^2\delta.
\end{align*}
Then we have 
\begin{align*}
\mathbb{E}[(\Delta_{\ell}^{X})^2] \leq 32\beta^4\ell\delta\sum_{j = 1}^{\ell}\int_{(j-1)\delta}^{j\delta}\mathbb{E}[\lvert X_{j-1}^{(\delta)} - X_t\rvert^2] dt + 2C_1^2\beta^4\delta^2.
\end{align*}
We also have
\begin{align*}
\mathbb{E}[\lvert X_j^{(\delta)} - X_t\rvert^2] &\leq 2\mathbb{E}[\lvert X_j^{(\delta)} - X_{j\delta }\rvert^2]  +  2\mathbb{E}[\lvert X_{j\delta} - X_{t}\rvert^2]\\
&= 2\mathbb{E}[(\Delta^{X}_j)^2]  +  2\mathbb{E}[\lvert X_{j\delta} - X_{t}\rvert^2].
\end{align*}
The second term can be bounded by
\begin{align}
\label{eqn:process-time-lipschitz}
\mathbb{E}[(X_t - X_s)^2] &= \mathbb{E}[\left(\int_{s}^{t}v(r,X_r) dr + \int_s^t \beta dB_r\right)^2] \nonumber\\
&\leq 2\mathbb{E}[\left(\int_{s}^{t}v(r,X_r) dr\right)^2] + 2\mathbb{E}[\left(\int_s^t \beta dB_r\right)^2] \nonumber\\
&=2\mathbb{E}[\left(\int_{s}^{t}v(r,X_r) dr\right)^2] + 2\int_s^t \beta^2 dr\nonumber\\
&\leq  2\beta^4\left(\int_{s}^{t} dr\right)^2 + 2\beta^2\int_s^t dr\nonumber\\
& \leq 2\beta^4\left|t - s\right|^2 + 2\beta^2\left|t - s\right|\nonumber\\
&\leq 4\beta^4\lvert t -s \rvert.
\end{align}

Plugging in the above two bounds, gives
\begin{align*}
\mathbb{E}[(\Delta_{\ell}^{X})^2] \leq 64\beta^4\ell\delta^2 \sum_{j=1}^{\ell}\mathbb{E}[(\Delta_j^{X})^2] + 256\beta^8\ell \delta \sum_{j=1}^{\ell}\int_{(j-1)\delta}^{j\delta}(t- (\ell-1)\delta)dt + 2 C_1^2\beta^4\delta^2,
\end{align*}
using that $\ell\delta \leq 1$ we have
\begin{align*}
\mathbb{E}[(\Delta_{\ell}^{X})^2] \leq (128 + 2C_1^2)\beta^8\delta + 64\beta^4\delta \sum_{j=1}^{\ell}\mathbb{E}[(\Delta_j^{X})^2] .
\end{align*}

We can now apply a discrete version of Gronwall's inequality to get that
\begin{align*}
\mathbb{E}[(\Delta_{\ell}^{X})^2]  \leq (128 + 2C_1^2)\exp\left(64\beta^4\right)\beta^8\delta.
\end{align*}

Hence 
\begin{align*}
\mathbb{E}[\lvert X_{\ell}^{(\delta)} - X_{\ell\delta}\rvert^2] \leq (128 + 2C_1^2)\exp\left(64\beta^4\right)\beta^8\delta.
\end{align*}
\end{proof}

\begin{lemma}[Extracted From Proof of Proposition 5.3 \cite{alaoui2020optimizationmeanfieldspinglasses}]
\label{lemm:inv_sigma_bound_lemma}
Suppose that the true $M_t$ satisfies $\mathbb{E}[M_t^2] = t, \forall t \in [0, q^*]$. For the SK model
\begin{align*}
    \lvert \frac{1}{\Sigma_j^{(\delta)}} - 1\rvert = \mathcal{O}\left(\beta^6 e^{\mathcal{O}(\beta^4)}\sqrt{\delta}\right)
\end{align*}
\end{lemma}
\begin{proof}

From Equation \eqref{eqn:Sigma}
\begin{align*}
(\Sigma_j^{(\delta)})^2 = \mathbb{E}[(u(\delta j, X_j^{(\delta)}))^2].
\end{align*}
For
\begin{align*}
\lvert (u(t, z))^2 - (u(t, y))^2 \rvert &\leq \lvert u(t, z) + u(t, y) \rvert  \lvert u(t, z) - u(t, y) \rvert \\
&\leq 8\beta^2\lvert z -y \rvert, \forall t \in [0, 1], z,y \in \mathbb{R}.
\end{align*}

Hence 
\begin{align*}
\lvert (\Sigma_j^{(\delta)})^2 - \mathbb{E}[(u(\delta j, X_{j{\delta}}))^2] \rvert =\lvert \mathbb{E}[(u(\delta j , X_j^\delta))^2] - \mathbb{E}[(u(\delta j, X_{j{\delta}}))^2] \rvert.
\end{align*}

Suppose we replace  each one-dimensional expectation by an expectation with respect to the coupling between the two processes  defined in the hypothesis of the proposition. Then we get that with Lemma \ref{lem:convergence_bound_error}:
\begin{align*}
\lvert (\Sigma_j^{(\delta)})^2 - \mathbb{E}[(u(\delta j, X_{j{\delta}}))^2] \rvert &\leq  8\beta^2\mathbb{E}[\lvert X_j^{(\delta)}- X_{j{\delta}}\rvert] \\
& =\mathcal{O}\left(\beta^6 e^{\mathcal{O}(\beta^4)}\sqrt{\delta}\right).
\end{align*}

Recall that we assumed that $\mathbb{E}[M_t^2] = t, \forall t \in [0, q^*]$, and 
\begin{align*}
    t = \mathbb{E}[M_t^2] = \int_{0}^{t} \mathbb{E}[(u(s, X_s))^2] ds,
\end{align*}
by It\^o isometry. By Lebesgue's differentiation theorem and continuity of $u$, we have
\begin{align*}
   \mathbb{E}[(u(t, X_t))^2] = 1,
\end{align*}
for all $t \in [0, q^*]$. Hence
\begin{align*}
\lvert (\Sigma_j^{(\delta)})^2 - 1 \rvert  =   \mathcal{O}\left(\beta^6 e^{\mathcal{O}(\beta^4)}\sqrt{\delta}\right).
\end{align*}
Note that for SK
\begin{align*}
(\Sigma_j^{(\delta)})^2 \geq  \mathbb{E}[(u(\delta j, X_j^{(\delta)})^2] \geq (C')^{-2},
\end{align*}
for some universal constant $C'$ by Lemma 16 \cite{Jagannath_2015}. Hence, we can say
\begin{align*}
    (\Sigma_j^{(\delta)})^{-1} \leq C'.
\end{align*}

Hence,
\begin{align}
\label{eqn:inv_sigma_bound}
\lvert \frac{1}{\Sigma_j^{(\delta)}} - 1\rvert = [\Sigma_j^{(\delta)}(1+\Sigma_j^{\delta})]^{-1}\lvert (\Sigma_j^{\delta})^2 - 1\rvert  =\mathcal{O}\left(\beta^6 e^{\mathcal{O}(\beta^4)}\sqrt{\delta}\right).
\end{align}

\end{proof}

Then let us define
\begin{align*}
& \langle W, \mathbf{u} \otimes \mathbf{v} \rangle_n := \frac{1}{n}\sum_{1 \leq j \leq i \leq n}W_{i, j}v_iu_j\\
&\langle \mathbf{u}, \mathbf{v}\rangle_n = \frac{1}{n}\sum_{1 \leq j \leq n}u_j v_j\\
&B^{(k)} := \langle W, ( (\mathbf{m}^{(k)}) + \mathbf{m}^{(k-1)}) \otimes (\mathbf{m}^{(k)} - \mathbf{m}^{(k-1)})\rangle_{n},
\end{align*}
where 
\begin{align*}
N^{-1}(H_N(\mathbf{m}^{(k)}) - H_N(\mathbf{m}^{(k-1)})) =  \langle W, ( \mathbf{m}^{(k)} + \mathbf{m}^{(k-1)}) \otimes (\mathbf{m}^{(k)} - \mathbf{m}^{(k-1)})\rangle_{N}.
\end{align*}
We then have that
\begin{align*}
    \underset{N\rightarrow \infty}{\text{p-}\lim}\sum_{k=1}^{\ell} B^{(k)}  = \underset{N\rightarrow \infty}{\text{p-}\lim}\frac{H_N(\mathbf{m}^{(\ell)})}{N}.
\end{align*}

\begin{lemma}[Lemma 5.5 in \cite{alaoui2020optimizationmeanfieldspinglasses}]
\label{lem:performance_lemma}
For $\ell = \lfloor q^*/\delta\rfloor$, we have for the SK model 
\begin{align}
\label{eqn:first}
&\lvert \underset{N\rightarrow \infty}{\text{p-}\lim} \sum_{k=1}^{\ell} B^{(k)} - \int_0^{\ell\delta} \mathbb{E}[u(t, X_t)] dt \rvert = \mathcal{O}\left(\beta^7 e^{\mathcal{O}(\beta^4)}\sqrt{\delta}\right)
\end{align}
\end{lemma}

Before proceeding to the proof, note that in the original proof, which catered towards arbitrary order spin models, there was additional error term that needed to be bounded. For SK this error term is zero.

\begin{proof}

Since $q^* \leq 1$, we will leave it out in the proof below for simplicity.
\begin{align*}
B^{(k)} &= \frac{1}{2}\langle W\{\mathbf{m}^{(k)}\}, \mathbf{m}^{(k)} - \mathbf{m}^{(k-1)}\rangle_N +  \frac{1}{2}\langle W\{\mathbf{m}^{(k-1)}\}, \mathbf{m}^{(k)} - \mathbf{m}^{(k-1)}\rangle_N \\
&:= \frac{1}{2}(S_{1,N} + S_{2, N}).
\end{align*}

Using the AMP iteration and taking inner products, we can write
\begin{align*}
&S_{1, N} = \langle \mathbf{z}^{(k+1)}, \mathbf{m}^{(k)} - \mathbf{m}^{(k-1)}\rangle_N + \sum_{j=0}^{k} d_{k, j}\langle \mathbf{m}^{(j-1)}, \mathbf{m}^{(k)} - \mathbf{m}^{(k-1)}\rangle_N\\
&S_{2, N} = \langle \mathbf{z}^{(k)}, \mathbf{m}^{(k)} - \mathbf{m}^{(k-1)}\rangle_N + \sum_{j=0}^{k - 1} d_{k-1, j}\langle \mathbf{m}^{(j-1)}, \mathbf{m}^{(k)} - m^{(k-1)}\rangle_N
\end{align*}

Invoking the state evolution and pseudo-lipschitzness
\begin{align*}
&\underset{N\rightarrow \infty}{\text{p-}\lim} S_{1,N}  = \mathbb{E}[Z_{k+1}^{(\delta)}(M_k^{(\delta)} - M_{k-1}^{(\delta)})] + \sum_{j=0}^{k} d_{k, j}\mathbb{E}[M_{j-1}^{(\delta)}(M_k^{(\delta)} - M_{k-1}^{(\delta)})]\\
&\underset{N\rightarrow \infty}{\text{p-}\lim} S_{2,N}  = \mathbb{E}[Z_{k}^{(\delta)}(M_k^{(\delta)} - M_{k-1}^{(\delta)})] + \sum_{j=0}^{k} d_{k-1, j}\mathbb{E}[M_{j-1}^{(\delta)}(M_k^{(\delta)} - M_{k-1}^{(\delta)})].
\end{align*}
Since $M_{t}^{(\delta)}$ is a martingale the rightmost terms above vanish.
Note that
\begin{align*}
 \mathbb{E}[Z_{k+1}^{(\delta)}(M_k^{(\delta)} - M_{k-1}^{(\delta)}] &=  \mathbb{E}[Z_{k}^{(\delta)}(M_k^{(\delta)} - M_{k-1}^{(\delta)})] +  \mathbb{E}[(Z_{k+1}^{(\delta)} - Z_{k}^{(\delta)}) (M_k^{(\delta)} - M_{k-1}^{(\delta)}]\\
 &=\mathbb{E}[Z_{k}^{(\delta)}(M_k^{(\delta)} - M_{k-1}^{(\delta)})]
\end{align*}
by independence. Hence, we get
\begin{align*}
\underset{N\rightarrow \infty}{\text{p-}\lim}\frac{1}{2}\left(S_{1,N} + S_{2,N}\right) & = \mathbb{E}\left[Z_k^{\left(\delta\right)}\left(M^{(\delta)}_k - M^{(\delta)}_{k - 1}\right)\right]\nonumber\\
& = \mathbb{E}\left[\left(Z^{(\delta)}_k - Z^{(\delta)}_{k - 1}\right)\left(M^{(\delta)}_k - M^{(\delta)}_{k - 1}\right)\right] + \mathbb{E}\left[Z^{(\delta)}_{k - 1}\left(M^{(\delta)}_k - M^{(\delta)}_{k - 1}\right)\right]\nonumber\\
& = \mathbb{E}\left[\left(Z^{(\delta)}_k - Z^{(\delta)}_{k - 1}\right)\left(M^{(\delta)}_k - M^{(\delta)}_{k - 1}\right)\right]\nonumber\\
& = \mathbb{E}[u^{(\delta)}_{k-1}(X_{k-1}^{(\delta)})(Z_k^{(\delta)} - Z_{k-1}^{(\delta)})^2]
\end{align*}
Then summing over $k$,
\begin{align*}
\underset{N\rightarrow \infty}{\text{p-}\lim} \sum_{k=1}^{\ell} B^{(k)} &= \sum_{k=1}^{\ell}\mathbb{E}[u^{(\delta)}_{k-1}(X_{k-1}^{(\delta)})(Z_k^{(\delta)} - Z_{k-1}^{(\delta)})^2]\\
&= \sum_{k=1}^{\ell}\mathbb{E}[u^{(\delta)}_{k-1}(X_{k-1}^{(\delta)})]\delta\\
&=\sum_{k=1}^{\ell}\frac{\mathbb{E}[u(\delta(k-1), X_{k-1}^{(\delta)})]}{\Sigma^{(\delta)}_{k-1}}\delta
\end{align*}

From Lemma \ref{lemm:inv_sigma_bound_lemma}
\begin{align*}
\lvert \frac{1}{\Sigma_j^{\delta}} - 1\rvert  \leq \widetilde{C}\beta^6 e^{\mathcal{O}(\beta^4)}\sqrt{\delta},
\end{align*}
for universal constant $\widetilde{C}$.

Hence for SK
\begin{align*}
\sum_{k=1}^{\ell}\left\lvert \frac{\mathbb{E}[u(\delta(k-1), X_{k-1}^{(\delta)})]}{\Sigma^{(\delta)}_{k-1}}\delta - \mathbb{E}[u(\delta(k-1), X_{k-1}^{(\delta)})]\delta\right\rvert &\leq \beta \lvert \frac{1}{\Sigma_j^{\delta}} - 1\rvert  \\
&\leq \widetilde{C}\beta^7 e^{\mathcal{O}(\beta^4)}\sqrt{\delta}.
\end{align*}

Also
\begin{align*}
\sum_{j=0}^{\ell-1}\int_{j\delta}^{(j+1)\delta}\mathbb{E}[\lvert (u(\delta j, X_j^{\delta}) - u(t, X_t))\rvert]dt 
&\leq 4\beta\sum_{j=0}^{\ell-1}\int_{j \delta}^{(j+1)\delta}\mathbb{E}[\lvert X_j^{\delta} - X_{j\delta}\rvert] dt\\
&+4\beta\sum_{j=0}^{\ell-1}\int_{j \delta}^{(j+1)\delta}\mathbb{E}[\lvert  X_{j\delta} -  X_{t}\rvert] dt \\
&+\sum_{j=0}^{\ell-1}\int_{j \delta}^{(j+1)\delta}\mathbb{E}[\lvert u(\delta j, X_t) - u(t, X_t) \rvert ] dt 
\end{align*}

The first term is bounded by $C\beta^5 e^{\mathcal{O}(\beta^4)}\sqrt{\delta}$ by Lemma \ref{lem:convergence_bound_error}. For the second term we have from \eqref{eqn:process-time-lipschitz} that
\begin{align*}
(\mathbb{E}[\lvert X_{s} - X_t\rvert])^2 &\leq \mathbb{E}[(X_t - X_s)^2] \\
&\leq 4 \beta^4\lvert t -s \rvert.
\end{align*}
Hence, the second term is $C_2\beta^3\sqrt{\delta}$ for some universal constant $C_2$. For the third term, we can use the total variation bound (Equation \eqref{eqn:total_variation_u})
\begin{align*}
& \sum_{j=0}^{\ell-1}\int_{j \delta}^{(j+1)\delta}\mathbb{E}[\lvert u(\delta j; X_t) - u(t; X_t)\rvert ] dt\nonumber\\
&\leq  \mathbb{E}\left[\sum_{j=0}^{\ell-1}\int_{j \delta}^{(j+1)\delta}\lvert u(\delta j, X_t) - u(t, X_t)\rvert\right]\\
& \leq \mathbb{E}\left[\sum_{j = 0}^{\ell - 1}\int_{j\delta}^{(j + 1)\delta}\left(\left|u\left(\delta j, X_t\right) - u\left(t, X_t\right)\right| + \left|u\left(\delta(j + 1), X_t\right) - u\left(t, X_t\right)\right|\right)\right] \\
& \leq \mathbb{E}\left[\sum_{j = 0}^{\ell - 1}\delta\sup_{t \in \left[j\delta, (j + 1)\delta\right]}\quad\sup_{x \in \mathbf{R}}\left(\left|u\left(j\delta, x\right) - u\left(t, x\right)\right| + \left|u(t, x) - u\left((j + 1)\delta, x\right)\right|\right)\right]\\
& \leq \delta\sup_{0 \leq t_0 \leq \ldots \leq t_{2\ell} \leq 1}\quad \sup_{x_0, \ldots, x_{2\ell - 1} \in \mathbf{R}}\sum_{j = 0}^{2\ell - 1}\left|u\left(t_{j + 1}, x_j\right) - u\left(t_j, x_j\right) \right|\\
&\leq C_2 \beta\delta 
\end{align*}

Hence putting all of the results together,

\begin{align*}
\lvert \underset{N\rightarrow \infty}{\text{p-}\lim} \sum_{k=1}^{\ell} B^{(k)} - \int_0^{\ell\frac{q^*}{\delta}}\mathbb{E}[u(s, X_s)]ds \rvert &\leq C\left(\beta^7 e^{\mathcal{O}(\beta^4)}\sqrt{\delta}+ \beta^5 e^{\mathcal{O}(\beta^4)}\sqrt{\delta} +\beta^3\sqrt{\delta} + \beta\delta\right)\\
&\leq C'\beta^7 e^{\mathcal{O}(\beta^4)}\sqrt{\delta}
\end{align*}
\end{proof}

Now we move to the thresholding and  rounding procedure to ensure feasibility. The entry-wise thresholding \cite{alaoui2020optimizationmeanfieldspinglasses} is the following
\begin{align*}
\widehat{m}_i := \begin{cases}
    m_i^{(\ell)}  & \text{if}~\lvert m_i^{\ell} \rvert \leq 1\\
    \text{sign}(m_i^{\ell}) &  \text{o.w.}
\end{cases}    
\end{align*}

\begin{lemma}[Lemma 5.7 \cite{alaoui2020optimizationmeanfieldspinglasses}]
\label{lem:thresholding}
For SK
\begin{align*}
     \underset{n\rightarrow \infty}{\text{p-}\lim} \lvert \frac{1}{n}H_n(\mathbf{m}^{(\ell)}) - \frac{1}{n}H_n(\widehat{\mathbf{m}})\rvert =\mathcal{O}\left(\beta^7 e^{\mathcal{O}(\beta^4)}\sqrt{\delta}\right)
\end{align*}
\end{lemma}
\begin{proof}
Define the function $\psi : \mathbb{R} \rightarrow \mathbb{R}$ as 
\begin{align*}
    \psi(x) = \begin{cases}
        (\lvert x \rvert - 1)^2 & \text{if}~\lvert x \rvert > 1\\
        0 & \text{o.w.}
    \end{cases}
\end{align*}

State evolution implies that
\begin{align*}
    \underset{N\rightarrow \infty}{\text{p-}\lim}\frac{1}{N}\sum_{i=1}^N \psi(\mathbf{m}_i^{(\ell)}) = \mathbb{E}[\psi(M_{\ell}^{(\delta)})].
\end{align*}

We also have that
\begin{align*}
\mathbb{E}[\psi(M_{\ell}^{(\delta)})] &\leq \mathbb{E}[ (\lvert M_{\ell}^{(\delta)}\rvert - 1) ^2]\\ 
&=\mathbb{E}[(\lvert M_{\ell \delta}\rvert - 1)^2] + \mathbb{E}[(\lvert M_{\ell}^{(\delta)}\rvert - 1)^2 - (\lvert M_{\ell \delta}\rvert - 1)^2]\\
&\leq 2\mathbb{E}[ (M_{\ell \delta})^2  - (M_{\ell}^{(\delta)})^2]\\
&\leq 2\mathbb{E}[\lvert M_{\ell \delta}  - M_{\ell}^{(\delta)} \rvert]\cdot (1+ \beta)\\
&\leq C\beta^7 e^{\mathcal{O}(\beta^4)}\sqrt{\delta}
\end{align*}
from  Lemma \ref{lem:performance_lemma}, and using that $M_{\delta \ell}$ is in $[-1, 1]$ almost surely.

The rest follows the proof of \cite{alaoui2020optimizationmeanfieldspinglasses} identically, where all constants are universal. Hence the result holds for a universal constant.
\end{proof}
The rounding procedure is modified similarly as it only depends on the AMP iteration through the bound in Lemma \ref{lem:thresholding}, which dominates the error. Specifically, this is for line (d) in the proof of Lemma 5.8 of \cite{alaoui2020optimizationmeanfieldspinglasses}.

Hence we obtain the following result.
\begin{theorem}[\cite{montanari2019} \& \cite{alaoui2020optimizationmeanfieldspinglasses}]
If the SK model is FRSB for all $\beta$ beyond some critical threshold, then there is an algorithm with depth $\mathcal{O}\left(n\cdot\frac{e^{\mathcal{O}(1/\epsilon^4)}}{\epsilon^{16}}\right)$ and $\mathcal{O}\left(n^2\cdot\frac{e^{\mathcal{O}(1/\epsilon^4)}}{\epsilon^{16}}\right)$ work  that outputs spins $\sigma_{\textup{alg}}\in \{-1, 1\}^{n}$ such that
\begin{align*}
    \underset{n\rightarrow \infty}{\text{p-}\lim}\frac{H(\sigma_{\textup{alg}})}{N} > \inf_{\gamma \in \mathcal{U}}\mathcal{P}^{(\infty)}(\gamma) - \epsilon.
\end{align*}
\end{theorem}

Here we see that the current upper bound is exponentially worse than the desired $1/\epsilon^2$. As mentioned earlier, the approach of \cite{alaoui2020optimizationmeanfieldspinglasses} avoids the additional $\beta$ factors that cause this blow-up. However, it is unclear how to deal with the Parisi singularity and completely bound the $\epsilon$ dependence for the approach of \cite{alaoui2020optimizationmeanfieldspinglasses}. It is possible that the singularity also causes the exponential blow-up in $1/\epsilon$. However, if the bounds on the Lipschitz constants and total variation in Section \ref{sec:smoothness} for \eqref{eqn:zero-temp-controls} can be shown to be $\mathcal{O}(\text{poly}\log(1/\epsilon))$ when $t^* = 1- \epsilon$, then there is an AMP algorithm with $\mathcal{O}(n\cdot \text{poly}(1/\epsilon))$ depth.

\stoptocentries
\putbib[bibliography]
\end{bibunit}

\end{document}